%% file: DissertationDhiraj.tex
\documentclass[11pt,papersize=8.5in,11in]{report}

\usepackage{comment}

\usepackage[dvips,letterpaper,top=1.0in,bottom=1.25in,left=1.5in,right=1.0in]{geometry}
\usepackage{graphicx}
\usepackage{titlesec}
\usepackage{epsfig}
\usepackage{dcolumn}

\usepackage{amsmath}
\usepackage{amsfonts}
\usepackage{amssymb}
\usepackage{amsthm}
\usepackage[adobe-utopia]{mathdesign}
\usepackage{bm}
\usepackage{float}
\usepackage[usenames,dvipsnames]{xcolor}

\usepackage{subcaption}

\usepackage{slashed}

\usepackage[font = singlespacing]{caption} 
\usepackage{setspace} 
\usepackage{enumitem} 
\usepackage{hyperref} 
\usepackage{appendix} 
\usepackage{multirow} 

\usepackage{tocloft}  
\usepackage{hyperref}
\usepackage{color}
\usepackage{grffile}
\usepackage{fancybox}
\usepackage{multicol}
\renewcommand{\cftchapafterpnum}{\vspace{\cftbeforechapskip}}

\usepackage{cite}

\hypersetup{colorlinks=true, citecolor=blue, linkcolor=black, urlcolor=blue}
\setlist{nolistsep}
\titleformat{\chapter}[display]{\normalfont\large\bfseries}{\MakeUppercase{\chaptertitlename}\ \thechapter}{0.5em}{}
\titlespacing{\chapter}{0pt}{-10pt}{0pt}
\titleformat{\section}{\normalfont\large\bfseries}{\thesection}{0.5em}{}

\usepackage{slashed}

\usepackage[all]{nowidow}

\usepackage[nottoc, notlof, notlot]{tocbibind}
\usepackage[nonumberlist, acronym]{glossaries}
\usepackage{setspace}
\begin{document}


\doublespacing
\pagenumbering{roman}

\input{./0/0.tex}


\input{./1/1.tex}

\input{./2/2.tex}

\input{./3/3.tex}

\input{./4/4.tex}

\input{./5/5.tex}

\input{./Appendix/pdpd.tex}
\input{./Appendix/NonRelativistic2LCwfs.tex}
\input{./Appendix/electronPolarization.tex}
\input{./Appendix/phiIntegrationUnpolarized.tex}
\input{./Appendix/zetaProducts.tex}
\bibliographystyle{unsrt}

\input{./0/vita.tex}

\doublespacing

\end{document}

%% file: 0/0.tex
\input{./0/title.tex}

\input{./0/signature.tex}


\input{./0/dedication.tex}

\input{./0/acknowledgments.tex}

\input{./0/abstract.tex}

\newpage
\input{./0/tableContent.tex} 

\renewcommand\cftchapfont{\mdseries}
\renewcommand\cftchappagefont{\mdseries}

\renewcommand{\cftchapafterpnum}{\vspace{\cftbeforechapskip}}


\renewcommand{\cftchapleader}{\cftdotfill{\cftdotsep}} 


\renewcommand\contentsname{\normalfont\MakeUppercase{Table of contents}\hfill}
\setcounter{tocdepth}{1}
\singlespacing
\renewcommand{\cftchapafterpnum}{\vspace{\cftbeforechapskip}}
                            
\addtocontents{toc}{\normalfont\MakeUppercase{Chapter}\hfill\normalfont\MakeUppercase{Page}\par}
\doublespacing


\newpage
\singlespacing
\listoftables 
\addtocontents{lot}{\normalfont\MakeUppercase{Table}\hfill\normalfont\MakeUppercase{Page}\par}
\doublespacing

\newpage
\newpage
\singlespacing
\listoffigures 
\addtocontents{lof}{\normalfont\MakeUppercase{Figure}\hfill\normalfont\MakeUppercase{Page}\par}
\doublespacing

%% file: 0/title.tex
\begin{titlepage}
  \begin{center}
  FLORIDA INTERNATIONAL UNIVERSITY \\
  Miami, Florida
	~\\
    ~\\
    ~\\
    ~\\
    ~\\
	QCD PROCESSES IN FEW NUCLEON SYSTEMS \\
    ~\\
    ~\\
    ~\\
    ~\\
    ~\\
       A dissertation submitted in partial fulfillment for the \\
    requirements of the degree of \\
    DOCTOR OF PHILOSOPHY \\
    in \\
    PHYSICS\\
    by \\
    Dhiraj Maheswari \\
    ~\\
    2018
  \end{center}
\end{titlepage}

\setcounter{page}{2}

%% file: 0/signature.tex
\newpage
\begin{singlespace}
\setlength{\parindent}{0.0cm}
To: Dean Michael R. Heithaus\\
\hangindent= 0.6cm{College of Arts, Sciences and Education}\\

This dissertation, written by
Dhiraj Maheswari,
and entitled
QCD Processes in Few Nucleon Systems,
having been approved in respect to style and intellectual content, is referred to you for judgment.\\

We have read this dissertation and recommend that it be approved.

\flushright{
\vspace{0.3cm}
\rule{7.8cm}{0.4pt}\\
Werner Boeglin\\
\vspace{0.8cm}
\rule{7.8cm}{0.4pt}\\
Lei Guo\\
\vspace{0.8cm}
\rule{7.8cm}{0.4pt}\\
Mirroslav Yotov\\
\vspace{0.8cm}
\rule{7.8cm}{0.4pt}\\
Misak M. Sargsian, Major Professor\\
}
\flushleft{
Date of Defense: June 20, 2018\\
\vspace{0.3cm}
The dissertation of Dhiraj Maheswari is approved.\\
}
\flushright{
  \rule{7.8cm}{0.4pt}\\
  Dean Michael R. Heithaus\\
  College of Arts, Sciences and Education\\
  \vspace{1cm}
  \rule{7.8cm}{0.4pt}\\
  Andr\'{e}s G.\ Gil\\
  Vice President for Research and Economic Development\\
  and Dean of the University Graduate School\\}
\vspace{2cm}
\begin{center}
  Florida International University, 2018
\end{center}
\end{singlespace}

%% file: 0/dedication.tex
\newpage
\vspace*{8cm}
\begin{center}
  DEDICATION
\end{center}

This dissertation is dedicated to my family, who supported me with every situations and guided me during my entire life.

%% file: 0/acknowledgments.tex
\newpage
\begin{center}
	ACKNOWLEDGMENTS
\end{center}

I would like to acknowledge all the  teachers and professors who taught me during all years of my life as a student, both in Nepal and in the USA.  

I am indebted to the members of my dissertation committee: 
Werner Boeglin, Lei Guo, and Mirroslav Yotov, who helped me during all my research. My special acknowledgment to my advisor, Misak Sargsian, who guided and helped me with lot of physics and was always patient with my mistakes and ignorance. 

I wish to acknowledge all the graduate program directors during my time at Department of Physics, FIU, Brian Raue, Rajamani Narayanan, and Jorge Rodriguez, 
as well the administrative staff of the  the FIU physics department,
specially Elizabeth Bergano-Smith, Omar Tolbert, Maria Martinez, John Omara, Ofelia Adan-Fernandez, and Robert Brown
who support me  with  all their  administrative and technical expertise.

I also thank Dr. Oswaldo Artiles, Frank Vera, and Christopher Leon, my fellow PhD students of the theoretical nuclear physics group, for the very helpful 
physics and mathematics discussions that I have had with them. I am also thankful to Dr. Wim Cosyn for his help and prompt responses whenever I contacted him.

This research would not have been possible without financial support.
I would like to acknowledge the Department of Energy for providing the grant that supported my research assistantship.

%% file: 0/abstract.tex
\newpage
\begin{center}
  ABSTRACT OF THE DISSERTATION \\
  QCD PROCESSES IN FEW NUCLEON SYSTEMS\\
  by\\
  Dhiraj Maheswari\\
  Florida International University, 2018\\
  Miami, Florida\\
  Professor Misak M. Sargsian, Major Professor
\end{center}
\vspace{-4mm}

One of the important issues of Quantum Chromodynamics (QCD) - the fundamental theory of 
strong interaction,  is the understanding of the role of the quark-gluon interactions  in the processes involving nuclear targets.  One direction in  such studies is  to explore the  onset of the quark gluon degrees of freedom in nuclear dynamics.
The other direction is using the nuclear targets as a ``micro-labs'' in studies of the QCD processes involving protons and neutrons bound in the nucleus. In the proposed research, we work in both directions considering high energy photo- and electro-production reactions involving deuteron and $^3$He nuclei.   

In the first half of the research, we study the high energy break-up of the   $^3$He nucleus, caused by a incoming photon, into a  proton-deuteron pair  at the large  center of mass scattering angle. The main motivation of the research is the theoretical interpretation of recent experimental data which revealed the unprecedentedly large exponent $s^{-17}$,
for the energy dependence of the differential cross section. In the present research, we extend the theoretical formalism of  the hard QCD rescattering model to calculate energy and angular dependences of the absolute cross section of the $\gamma ^3\text{He}\rightarrow pd$ reaction in high momentum transfer  limit.

The second half of the research explores the deep-inelastic scattering of a polarized electron off the polarized deuteron and $^3$He nuclei,  to explore the quark-gluon structure of polarized neutron. The main reason of using deuteron is that it is the most simple and best understood nucleus. While the reason of using polarized $^3$He as an effective polarized neutron target is that  because of the Pauli-principle, the two protons in the target are in the opposite spin states and thus the neutron has all the polarization of the $^3$He nucleus. However this approximation is exact only for the $S$-state and becomes less accurate with the increase of the internal momentum of the bound nucleons in the nucleus. There are several planned experiments  which  will be performed during next few years at the kinematics in which the internal momenta of the probed neutron cannot be neglected. Therefore, for the reliable interpretation of the data, all the nuclear effects,
especially the effects related to the relativistic treatment of high momentum component of the nuclear wave function, should be taken into account.
In this work, we developed a comprehensive theoretical framework for calculation of the all relevant nuclear effects that will allow the accurate extraction  of the neutron data from deep-inelastic scattering involving deuteron and $^3$He targets.

%% file: 0/tableContent.tex

\singlespacing

\hspace*{140pt} TABLE OF CONTENTS\\\\
CHAPTER \hspace*{355pt} PAGE\\\\
1 \hspace*{6pt}  Introduction \dotfill 1\\
\hspace*{12pt} 1.1 \hspace*{6pt} Hard Rescattering Mechanism- Background  \dotfill  6\\
\hspace*{12pt} 1.2 \hspace*{6pt} DIS on a nuclear target: general discussion  \dotfill  7\\\\
2 \hspace*{6pt} Hard photodisintegration of $^3$He into proton and deuteron \dotfill   11\\
\hspace*{12pt}  2.1 \hspace*{6pt} Kinematics and reference frame  \dotfill   11\\
\hspace*{12pt}  2.2 \hspace*{6pt} Development of model \dotfill  16\\
\hspace*{12pt}  2.3 \hspace*{6pt} Hard scattering kernel  \dotfill  26\\
\hspace*{12pt}  2.4 \hspace*{6pt} Including $pd \rightarrow pd$ amplitude  \dotfill 30\\
\hspace*{12pt}  2.5 \hspace*{6pt} The differential cross section \dotfill 32\\\\
3 \hspace*{6pt} Numerical approximations and results \dotfill   34\\
\hspace*{12pt} 3.1 \hspace*{6pt} Calculation of light-front transition spectral function \dotfill    34\\
\hspace*{12pt} 3.2 \hspace*{6pt} Hard elastic $pd \rightarrow pd$ scattering cross section \dotfill    35\\
\hspace*{12pt} 3.3 \hspace*{6pt} Estimation of effective charge $Q_{eff}$ \dotfill    38\\
\hspace*{12pt} 3.4 \hspace*{6pt} Results \dotfill    39\\\\
4 \hspace*{6pt} Spin structure of neutron and the problem of nuclear medium effects \dotfill   45\\
\hspace*{12pt} 4.1 \hspace*{6pt} Deep inelastic scattering from a polarized nucleus \dotfill   46\\
\hspace*{12pt} 4.2 \hspace*{6pt} Jacobian of the differentials \dotfill    52\\
\hspace*{12pt} 4.3 \hspace*{6pt} Cross section in plane wave impulse approximation \dotfill    54\\
\hspace*{12pt} 4.4 \hspace*{6pt} Numerical estimates in PWIA- unpolarized part \dotfill   65\\
\hspace*{12pt} 4.5 \hspace*{6pt} Numerical estimates in PWIA- including the polarized part \dotfill  72\\
\hspace*{12pt} 4.6 \hspace*{6pt} Polarized cross section asymmetry \dotfill    73\\
\hspace*{12pt} 4.7 \hspace*{6pt} Extraction of $g_1(x_N,Q^2)$ of neutron \dotfill    77\\
\hspace*{12pt} 4.8 \hspace*{6pt} Results \dotfill   81\\\\
5 \hspace*{6pt} Summary and Conclusion \dotfill    88\\\\
Appendices \dotfill  90\\\\
Bibliography \dotfill   111\\\\
VITA \dotfill   114\\

%% file: 1/1.tex
\chapter{Introduction}

\pagenumbering{arabic}

\input{./1/introQCDgeneral.tex}

 \input{./1/HRM.tex}
 

 \input{./1/DISNuclearGeneral.tex}

%

%% file: 1/introQCDgeneral.tex


\pagenumbering{arabic}


The theory of Quantum Chromodynamics (QCD) is the fundamental theory of strong nuclear interactions. Quantum Chromodynamics is a non-Abelian quantum field theory with a SU(3, $\mathbb{C}$) group structure.
According to the theory of QCD, nuclear matter is made of two fundamental kinds of fields, which are the \textit{quark}  and  \textit{gluon} fields. The quark fields transform under the fundamental representation of SU(3, $\mathbb{C}$). Quarks are known to come in six flavors: up (u), down (d), strange (s), charm (c), bottom (b), and top (t).
Each quark flavor has a corresponding anti-quark. Each quark flavor possesses a three-state internal degree of freedom, which is called \textit{color}, in analogy to RGB color space, and anti-quarks are considered to
have an \textit{anti-color}. For example, the anti-quark of a red ``u'' quark would be an anti-red ``$\bar{u}$'' anti-quark.

The gluons transform under the adjoint representation of SU(3, $\mathbb{C}$).
Because of this, the gluon has a $3^2 - 1 = 8$-state internal degree of freedom, which is also called the color of the gluon.

The colorless mixtures of quarks, anti-quarks and gluons give rise to the physical hadrons. Every known hadron transforms under a singlet representation of SU(3,$\mathbb{C}$). The basic known hadrons are categorized into two groups- baryons and mesons. A baryon is known to consist of three valence quarks and have a color state of:
\begin{equation}
\label{colorstatebaryon}
\frac{1}{\sqrt{6}}(|rgb \rangle + |gbr \rangle + brg \rangle - |bgr \rangle - grb \rangle - rbg \rangle).
\end{equation}
A meson is known to consist of a valence quark and a valence anti-quark and have a color state of:
\begin{equation}
\label{colorstatemeson}
\frac{1}{\sqrt{3}}(|r \bar{r} \rangle + |g \bar{g} \rangle + |b \bar{b} \rangle).
\end{equation} 

Anti-baryons, consisting of three valence anti-quarks, also exist as the anti-particles of
baryons.
The only known stable baryon is the \textit{proton}, which has valence quark flavor content $uud$. Its anti-particle, the anti-proton (valence quark content $\bar{u}\bar{u}\bar{d}$) is the only
stable anti-baryon. The \textit{neutron} (valence quark content $udd$) is a long-lived baryon, with a mean lifetime of 15 minutes. All other known baryons have lifetimes in the
order of microseconds or less.

Protons and neutrons (collectively called as \textit{nucleons}) are known to form stable bound states called \textit{nuclei}. More specifically, $Z$ protons and $(A - Z)$ neutrons can bind together to form a nucleus with total charge $Z$ and mass number $A$. Typically, theoretical and experimental studies of nuclei focus on nuclear structure in terms of its nucleonic degrees of freedom. One postulates that nucleons interact via a phenomenological potential, or the exchange of certain mesons, and proceeds to calculate nuclear properties in terms of various models. Under a \textit{mean field approximation}, in which each nucleon is treated as moving independently under the average
influence of the other $(A - 1)$ nucleons, the distances between nucleons are larger than both the size of the nucleon, and the distance scales at which QCD descriptions have successfully been applied. Accordingly, descriptions of the nucleus in terms of QCD often amount to treating the nucleus as a collection of quasi-free nucleons, each of which is individually described using QCD.

Nucleons on the fundamental level consist of valence and sea quarks as well as gluons. It is the valence quark composition which defines the quantum numbers of the nucleons. However, if we look inside a nucleus, we only see the nucleons but not the quarks. It is very rare when quark degrees of freedom are revealed in nuclear processes. On the other hand, such  observations are important for understanding the emergence of nuclear forces from QCD. There are limited number of experiments in which quark-gluon degrees of freedom are observed in nuclear processes.
One such process is the \textit{hard nuclear scattering}, in which the energy-momentum transferred to the nucleus is much larger than the nucleon masses. In hard scattering kinematic regime, we expect that only the minimal Fock components dominate in the wave function of the particles involved in the scattering. This expectation results in the prediction of the constituent (or quark) counting rule, according to which the energy dependence of two-body hard reaction is defined by the number of fundamental constituents participating in the reaction \cite{BF73,MMT72}. For example, if we consider a general two-body hard reaction of the type $ a + b \rightarrow c+ d$, then, according to the constituent counting rule, the cross section of the hard scattering process would have an energy dependence of the following form:
\begin{equation} \label{scaling}
\frac{d\sigma^{(ab\rightarrow cd)}}{dt} \sim	\frac{1}{s^{n_a + n_b + n_c + n_d -2}} ,
\end{equation}
where $n_i, i = a, b,c,d$ represent the number of the fundamental fields associated with respective particles involved in the scattering process, $s$ is the invariant energy square involved in the process and $t$ is related to the momentum transferred in the process. 
More specifically, if $a$ is a proton, $n_a$ will equal three and if it is a photon, $n_a$ would be one. Although not everything about the nucleons is known, the above counting rule can be used as a tool in verifying the role of the quark-gluon degrees of freedom in nuclear reactions.

In 1976, it was suggested\cite{BC76} to use the concept of quark-counting rule to explore the QCD degrees of freedom in the deuteron. One of the best candidate reactions  was the hard photodisintegration of the deuteron ($d$) into a proton ($p$) and neutron ($n$), $\gamma + d \rightarrow p +n$, for which, according to Eq.~(\ref{scaling}), the cross section should have an energy dependence of $s^{-11}$.  
The first such experiments  being carried out at the Stanford Linear Accelerator Center (SLAC)\cite{Napolitano:1988uu,Freedman:1993nt,Belz:1995ge} and 
Jefferson Lab\cite{Bochna:1998ca,Schulte01se,Schulte:2002tx,Mirazita:2004rb,Rossi:2004qm} revealed  $s^{-11}$ energy dependence in the cross section for photon energies 
 at $E_\gamma \ge 1$ GeV and the center of mass~(cm) scattering angle $\theta_{\text{cm}}=90^{\circ}$.
 
With the success in describing the energy dependence in the cross section for the photodisintegration of deuteron into a proton and neutron, it became tempting to use the constituent counting rule to nuclei heavier than deuteron too.  
For this, the two-body breakup reactions were extended to  $^3$He target, 
in which case two fast outgoing protons and slow neutron were detected in 
 $ \gamma + ^3\text{He}\rightarrow (pp) + n$ reactiont\cite{gheppn2004}.
The results of such an experiment\cite{Pomerantz:2009sb} were consistent  with the $s^{-11}$ scaling in the two-proton hard beak-up channel, but at much larger  photon energies ($E_{\gamma} > 2$ GeV) than  in the case of $pn$ break-up.  
Recently, the hard two-body break up reaction  has been measured for the more complex,  $\gamma + ^3\text{He}\rightarrow p+d$, channel\cite{Pomerantz13}. 
According to Eq.~(\ref{scaling}), such a reaction in the hard scattering regime should scale as $s^{-17}$, and surprisingly the experiment observed a scaling consistent with the exponent of 17 - an unprecedentedly large number  in the two-body hard processes.  A part of the dissertation aims to explain in detail the theoretical framework of generation of $s^{-17}$ scaling in the hard break up of $^3$He into a proton and deuteron pair.

Another significant aspect of the theory of QCD is its application to understand the \textit{spin} of nucleons, a fundamental characteristic of elementary particles. It is well known that the nucleons have spin of $\frac{1}{2}$. As the nucleons are also known to be a composite of quarks and gluons, a fundamental QCD question is how a nucleon gets its spin from quarks and gluons.
 According to the original quark model by Gell-Mann \cite{GellMann:1964nj} and Zweig \cite{Zweig:1964jf}, $100 \%$ of the hadronic spin was accounted for by the quarks. However, the results from the experiments performed by the European Muon Collaboration (EMC) \cite{Ashman:1989ig}, suggested that contribution of quarks to the spin of the proton is only 20 $\%$. The discrepancy was really shocking to the physicists at that time and the problem of where the missing proton spin comes from gave rise to the so-called \textit{proton spin crisis}. With time, experiments studying the deep inelastic scattering (process of probing inside of hadrons by using electrons, muons and neutrinos-collectively called as leptons) gave more information on the spin structure of the nucleons. In such experiments, the polarized beams and targets are used to measure the spin structure of the nucleon via scattering of charged leptons from the nucleons through the exchange of virtual photon.

The experiments\cite{Alguard, Hughes, Ashman} with polarized electrons and protons, performed at SLAC in the late 70s and early 80s in a limited Bjorken $x$ (momentum fraction of the nucleon carried by the quark) range revealed large spin effects in deep inelastic electron-proton scattering. These large effects were predicted by Bjorken\cite{Bjorken} and by simple SU(6) quark models. These experiments showed that the quark contribution to the total spin of the proton is very small $\approx$ 20$\%$, which was in contrast to the simple relativistic valence quark model prediction in which the spin of valence quarks contributed about 75$\%$ to the proton spin and remaining 25$\%$ from their orbital angular momentum. However, the current understanding\cite{Filippone} suggests that the total contribution to the nucleon spin comes from the valence quarks, quark anti-quark sea, their angular momenta, and gluons. This contribution is called the nucleon spin sum rule, written as:
\begin{equation}\label{spinsumrule}
S_z^N = S_z^q + L_z^q + J_z^g = \frac{1}{2},
\end{equation}
where $S_z^N$ is the nucleon spin, $S_z^q$ and $L_z^q$ represent the quark's spin and angular momentum respectively and $J_z^q$ is the total  momentum of the gluons.

The second part of the dissertation aims at studying the deep inelastic scattering of a polarized electron from a polarized deuteron and $^3$He nuclei to extract the spin structure of the neutron. The detailed theoretical framework for accounting of the nuclear effects in the extraction of polarized neutron structure function is given in the second part of the dissertation. As deuteron and $^3$He nuclei are vital part of our theoretical consideration, a brief information about these nuclei are given below.

The deuteron is nucleus of the one of the two stable isotopes (Deuterium) of Hydrogen, the other one being Hydrogen-1. Deuterium accounts for approximately 0.0156 $\%$ of all the naturally occurring hydrogen in the oceans.
The deuteron is  known to be the composite of a proton and neutron, with a total spin of 1. It is the simplest source to explore the interaction between a proton and neutron inside the nucleus. However, it becomes very complicated to extract the neutron spin information from deuteron, since it is not trivial to extract spin structure of spin $\frac{1}{2}$ particle from spin 1 particle. On the other hand, the $^3$He nucleus is a light and stable composition of two protons and one neutron. Its hypothetical existence was first proposed by an Australian physicist Mark Olipant\cite{MarkOlip} in 1934. The nucleus of $^3$He was first hypothesized to be a radioactive isotope of Helium until it was found in the composition of natural Helium gas. The availability of $^3$He is about 10,000 times rarer than $^4$He in the composition of natural Helium gas. Incidentally, only the nucleus of $^1$H and $^3$He are the known stable nuclei with more number of protons than neutrons. In the ground state, the two protons in $^3$He are spin paired while the neutron is unpaired, thus making the spin of $^3$He same as the spin of the neutron. Since free neutron is naturally unavailable, the  $^3$He nucleus becomes even more significant as it enables us to study the neutron's fundamental characteristics.

This dissertation is divided into following parts: Chapter 2 discusses the development of theoretical framework for calculation of the hard photodisintegration of $^3$He into a proton and deuteron pair. In Chapter 3, we discuss in details the numerical approximations used to calculate the differential cross section of hard photodisintegration of $^3$He. We then proceed by presenting the comparison of our calculations with the existing experimental data. In Chapter 4, we discuss in detail the deep inelastic scattering of a polarized electron from  polarized light targets to extract the spin strucutre of neutron. In Chapter 4, results obtained from this study are also presented. Finally, in Chapter 5, we summarize results of our study and present the conclusions.

In the remainder of this chapter, a background on Hard Rescattering Mechanism (HRM) is presented. Also a general discussion on Deep Inelastic Scattering (DIS) is presented.

%% file: 1/HRM.tex
\section{Hard Rescattering Mechanism- Background}
\label{sec_HRM}
Hard Rescattering Mechanism (HRM) is a theoretical framework which can be used to study the nuclear hard break-up processes. It was first developed to study the photodisintegration of the deuteron into a proton and a neutron\cite{gdpn}, represented as
\begin{equation}
\label{deutBrekup}
\gamma + d \rightarrow p + n.
\end{equation}
In the HRM model, the hard photodisintegration takes place in two stages. First, the incoming photon knocks out  a quark from one of the nucleons. 
Then in the second step, the outgoing fast quark undergoes a high momentum transfer hard scattering with a quark of the other nucleon sharing its large momentum among the constituents in the final state of the reaction. These two stages of HRM can be schematically represented in the following diagram:

\begin{figure}[ht]
\centering
\includegraphics[scale = 1.0]{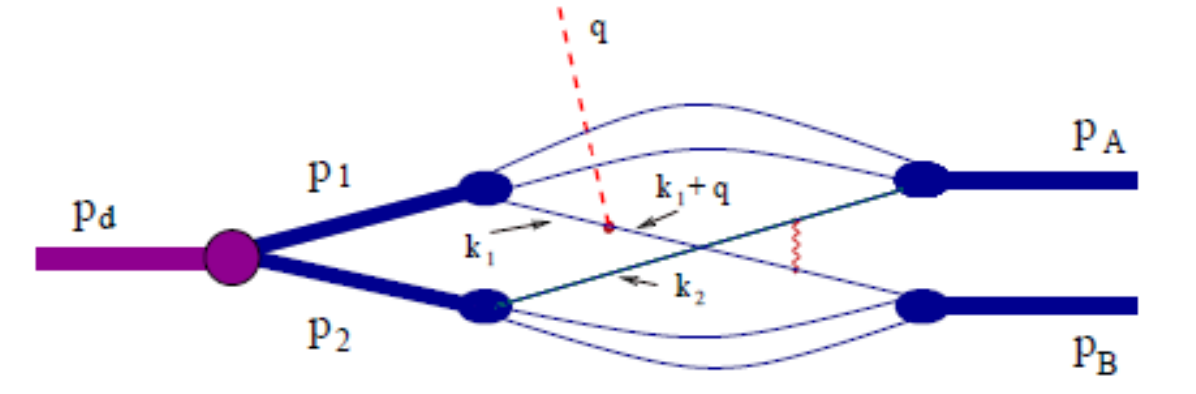}
\caption{Photodisintegration of deuteron into proton and neutron.}
\label{deuteronBreakup}
\end{figure}

\begin{figure}[ht]
\centering
\includegraphics[scale = 0.52]{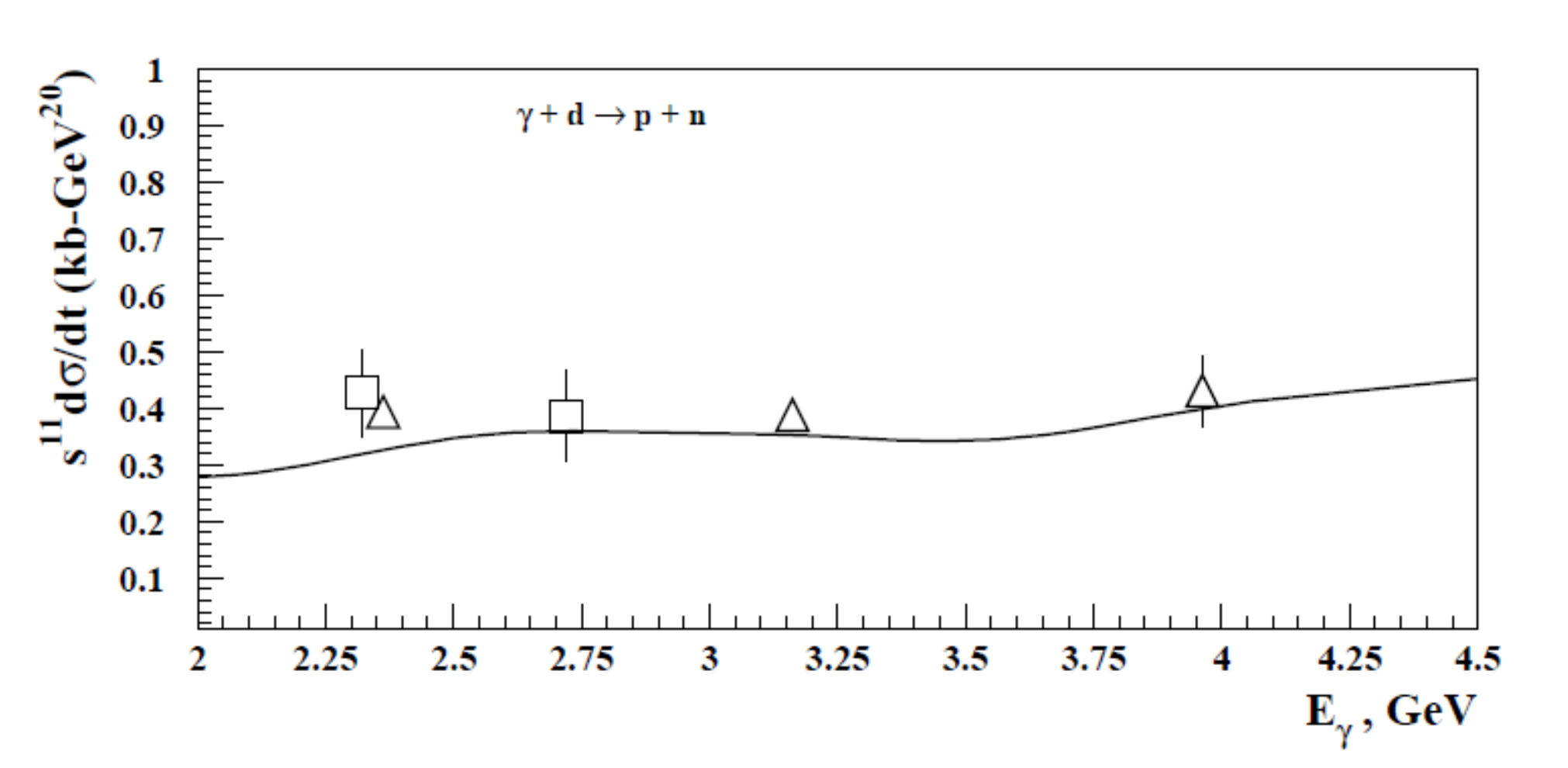}
\caption{Energy dependence of $s^{-11}$ scaled differential cross section of $\gamma + d \rightarrow p + n$ reaction at $\theta_{\text{cm}} = 90^\circ$. Data are from the experiments\cite{Bochna:1998ca,Schulte01se,Schulte:2002tx} and calculation from\cite{Sargsian:2002tj}.}
\label{deuteronScaling}
\end{figure}

In Fig.~(\ref{deuteronBreakup}), an energetic photon with momentum $q$ knocks out a quark with initial momentum $k_1$, which then undergoes a high momentum transfer (hard) scattering with a quark with momentum $k_2$ from the other nucleon, and thus producing a final two body state of proton with momentum $p_A$ and neutron with momentum $p_B$. For the nuclear process described by Eq.~(\ref{deutBrekup}), the constituent counting rule of, Eq.~(\ref{scaling}) predicts the cross section to scale like $s^{-11}$. The application of HRM to the photodisintegration of deuteron  not only predicted $s^{-11}$ energy dependence, it also reproduced the absolute cross section\cite{Sargsian:2002tj}, as shown in Fig.~(\ref{deuteronScaling}).
 
The agreement of the HRM with the experimental data as seen in Fig.~(\ref{deuteronScaling}) was the main motivation to extend the model to other nuclei, such as $^3$He. In Chapter 2, we discuss in detail the extension of HRM to study the hard photodisintegration of $^3$He into a proton and a deuteron pair.

%% file: 1/DISNuclearGeneral.tex
\section{DIS on a nuclear target: general discussion}
\label{sec_DISgenNuc}
We consider the situation in which a electron is scattered off a nuclear target, described by the following reaction:
\begin{equation}
e + A \rightarrow e' + X,
\end{equation} 
where $e$ and $A$ represent the incoming electron and nuclear target respectively, $e'$ is the scattered electron, and $X$ is the product of the scattering. These reactions are commonly referred as inclusive processes, in which only the scattered electron is detected.
The process is shown in Fig.~(\ref{scattering_nuclear_target_0}) below.
\begin{figure}[ht]
\centering
\includegraphics[scale=0.6]{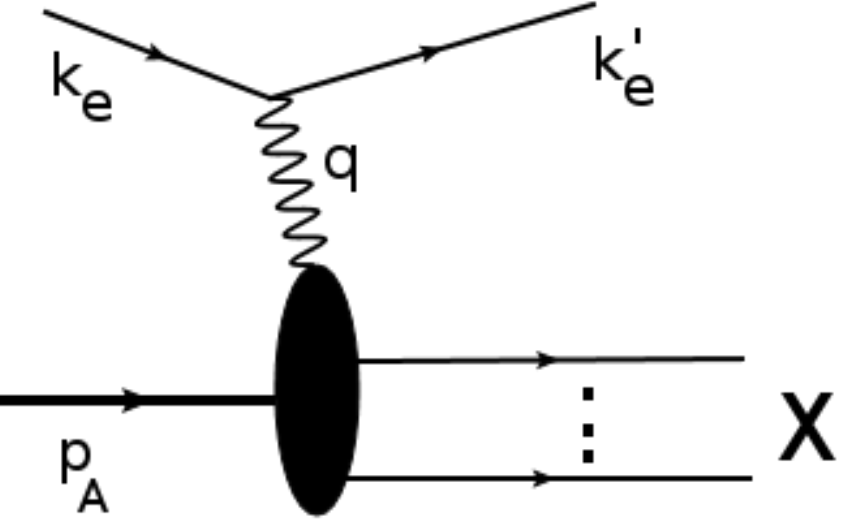}
\caption{DIS on a nuclear target.}
\label{scattering_nuclear_target_0}
\end{figure}

In Fig.~(\ref{scattering_nuclear_target_0}), $k_e$ and $k_e'$ represent the four momentum of the incoming and scattered electron respectively, $p_A$ represents the four momentum of the nuclear target, and $q_\mu \equiv k_e - k_e'$ is the four momentum transferred to the target in the scattering. 
If we define the four momenta of the incoming and scattered electron as $k_\mu = (E, \mathbf{k})$ and  $k_\mu'= (E^\prime, \mathbf{k'})$ respectively, the differential cross section for the scattering can be written as
\begin{equation} \label{cs}
\frac{d\sigma}{d\Omega dE^\prime} = \frac{\alpha_{em}^2}{Q^4} \frac{E^\prime}{E} L^{\mu \nu}W_{\mu \nu}^A,
\end{equation}
where  $\alpha_{em}$ is the fine structure constant, $ E, E^\prime $ are the initial and final energies of the electron, $ Q^2 \equiv -q^2 =  (k - k')^2 $ is the negative four momentum square transferred by the incoming electron and $ L^{\mu \nu}, W_{\mu \nu}^{A} $ are the leptonic and hadronic tensors respectively which are defined as follows:
\begin{equation} \label{leptonictensor}
L_{\mu \nu} = 4\Bigg[ \Big(k_\mu - \frac{q_\mu}{2}\Big) \Big(k_\nu - \frac{q_\nu}{2}\Big) \Bigg] + Q^2 \Bigg[-g_{\mu \nu} - \frac{q_{ \mu} q_{\nu}}{Q^2} \Bigg],
\end{equation}
and
\begin{eqnarray} \label{hadronictensor}
W^{\mu \nu}_A &&= \Bigg[ -g^{\mu \nu} - \frac{q^{\mu} q^{\nu}}{Q^2} \Big] W_{1A}(x_A, Q^2) + \Big[p^{\mu}_A + \frac{p_A \cdot q}{Q^2}{ q^{\mu}} \Big] \Big[p^{\nu}_A + \frac{p_A \cdot q}{Q^2}{ q^{\nu}} \Bigg] \frac{W_{2A}(x_A, Q^2)}{M_A^2}.
\end{eqnarray}
In Eq.~(\ref{hadronictensor}), $p_A\equiv (M_A, \mathbf{0})$ is the four-momentum of the nuclear target at rest with $M_A$ being its mass, $x_A = \dfrac{A Q^2}{2 p_A \cdot q}$ is the Bjorken variable. Here, $W_{1A}(x_A, Q^2)$ and $W_{2A}(x_A, Q^2)$ are the unpolarized structure functions, which characterize the nuclear target.

To calculate the product of $L_{\mu \nu}$ and $W_A^{\mu \nu}$, we make use of the gauge invariance condition that $q_{\mu} L^{\mu \nu} = 0 = q_{\nu} L^{\mu \nu}$ and obtain:
\begin{eqnarray}\label{contraction}
L_{\mu \nu}W^{\mu \nu}_{A} &&  = -g^{\mu \nu} \Bigg[ 4 \Big(k_\mu - \frac{q_\mu}{2}\Big) \Big(k_\nu - \frac{q_\nu}{2}\Big) + Q^2 \Big(-g_{\mu \nu} - \frac{q_{ \mu} q_{\nu}}{Q^2} \Big)\Bigg] W_{1A}   \nonumber \\
&& +  p_A^{\mu} p_A^{\nu} \Bigg[ 4\Big( k_\mu - \frac{q_\mu}{2}\Big) \Big(k_\nu - \frac{q_\nu}{2}\Big)  + Q^2 \Big(-g_{\mu \nu} -
\frac{q_{ \mu} q_{\nu}}{Q^2} \Big) \Bigg]\frac{W_{2A}}{M_A^2}.
\end{eqnarray}
Also, noting that $g_{\mu \nu} g^{\mu \nu} = 4$ and $g_{\mu \nu}k^{\mu} = k_{\nu}$, Eq.~(\ref{contraction}) can be written as
\begin{eqnarray} \label{contraction_simplified}
L_{\mu \nu} W^{\mu \nu}_A && = \Bigg[ -4 \Big( k - \frac{q}{2} \Big)^2 + Q^2 \Big( 4 + \frac{q^2}{Q^2} \Big) \Bigg] W_{1A}  \nonumber \\
&& + \Bigg[ 4 \Big(p_A \cdot k - \frac{p_A \cdot q}{2} \Big)^2 + Q^2 \Big(-p_{A}{^2} - \frac{(p_A \cdot q)^2}{Q^2} \Big) \Bigg]\frac{W_{2A}} {M_A^2}.
\end{eqnarray}
Defining $y_A = \frac{p_A \cdot q}{p_A \cdot k}$, Eq.~(\ref{contraction_simplified}) can be written as
\begin{eqnarray}
L_{\mu \nu}W^{\mu \nu}_A &&= 2 Q^2 W_{1A} +  \frac{(p_A \cdot k)^2}{M_A^2}  \Big[(2 - y_A)^2 - y_A^2\Big(1 + \frac{4 x_A^2 M_A^2}{Q^2}\Big) \Big]\frac{W_{2A}}{y_A^2}. 
\label{contraction_withyA}
\end{eqnarray}
Furthermore, using Eqs.~(\ref{cs}) and (\ref{contraction_withyA}), the inclusive nuclear cross section can be written in most general form as
\begin{eqnarray}\label{DIScsgeneral}
\frac{d\sigma}{d\Omega dE^\prime} && = \frac{\alpha_{em}^2}{Q^4} \frac{E^\prime}{E} \Bigg[2 Q^2 W_{1A} +  \frac{(p_A \cdot k)^2}{M_A^2}  \Big[(2 - y_A)^2 - y_A^2\Big(1 + \frac{4 x_A^2 M_A^2}{Q^2}\Big) \Big]\frac{W_{2A}}{y_A^2} \Bigg].
\end{eqnarray}
The inclusive processes are referred as deep-inelastic when $Q^2$ is large enough that scattered electrons resolve individual quarks in the target. Phenomenologically, it is observed that DIS regime is established at $Q^2 \geq 4$ GeV$^2$. In DIS regime, it is convenient to consider DIS structure functions $F_{1A}(x_A, Q^2), F_{2A}(x_A, Q^2)$, which are related to $W_{1A}$ and  $W_{2A}$ through the relation
\begin{eqnarray}
\label{F1AF2A_intermsof_W1AW2A}
F_{1A}(x_A, Q^2) = M_A W_{1A}(x_A, Q^2) \text{ and } F_{2A}(x_A, Q^2) = q_0 W_{2A}(x_A, Q^2).
\end{eqnarray}
Using the above relations, Eq.~(\ref{DIScsgeneral}) can be written as
\begin{eqnarray}\label{DIScsgeneral_intermsF1F2}
\frac{d\sigma}{d\Omega dE^\prime} && = \frac{\alpha_{em}^2}{Q^4} \frac{E^\prime}{E} \Bigg[2 Q^2 \frac{F_{1A}}{M_A} +  \frac{(p_A \cdot k)^2}{M_A^2 y_A^2}  \Big[(2 - y_A)^2 - y_A^2\Big(1 + \frac{4 x_A^2 M_A^2}{Q^2}\Big) \Big]\frac{F_{2A}}{q_0} \Bigg].
\end{eqnarray}
The above expression we obtained is relevant for electron scattering from unpolarized target. A similar expression can be obtained for polarized electron scattering from polarized target, which will be discussed in Chapter 4. The cross section of such process is described by four structure functions in the following form:

\begin{eqnarray} \label{dsigdomdeprime_form}
\frac{d^2\sigma}{d\Omega dE_k'} &&= \frac{\alpha_{em}^2}{Q^4}\frac{E_k'}{E_k} ~ \Bigg\{2 Q^2 \frac{F_{1A}}{M_A} +  \frac{p_A \cdot q}{M_A} \frac{F_{2A}}{y_A^2} \Bigg[(2 - y_A)^2 - y_A^2 \Big(1 + \frac{4 x_A^2 M_A^2 }{ Q^2 } \Big) \Bigg] \nonumber \\
&&  - 2 h_e Q^2 \frac{g_{1A}}{p_A \cdot q} \Big[\zeta_A \cdot q - 2 k \cdot \zeta_A \Big] - 4 h_e Q^2 \frac{g_{2A}}{p_A \cdot q} \Big[ \frac{q \cdot \zeta_A}{y_A} - \zeta_A \cdot k \Big] \Bigg\},
\end{eqnarray}
where $\zeta_A$ is the nuclear polarization vector. The additional structure functions $g_{1A}(x_A, Q^2)$ and $g_{2A}(x_A, Q^2)$ characterize the polarization properties of quarks in the nucleus in the DIS regime.

%% file: 2/2.tex
\chapter{Hard photodisintegration of $^3$He into proton and deuteron}


\input{./2/Chp2intro.tex}

\input{./2/KinematicsrefFrame.tex}
\input{./2/ReferenceFrame.tex}
\input{./2/modelDevelopment.tex}

\input{./2/Nuclearwfs.tex}
\input{./2/Nucleonicwfs.tex}
\input{./2/HardKernel.tex}
\input{./2/including_pdpd_amp.tex}
\input{./2/finalCSDerivation.tex}

%% file: 2/Chp2intro.tex
\label{sec:Ch2Intro}
Herein, we present the further development of the hard rescattering model to describe hard photodisintegration of the $^3$He into a proton-deuteron pair.
In Section \ref{sec_KinRef}, the kinematics and the reference frame of the two-body break-
up reaction will be discussed.
In Sections \ref{sec_ModelDev}-\ref{sec_four}, we develop the hard rescattering model for the $\gamma + ^3\text{He} \rightarrow p + d$
reaction discussing in detail the nuclear amplitude, which according to HRM, provides the
main contribution to the hard break-up cross section. 
In Section \ref{sec_CompleteDerivationCS}, we complete the derivation
by calculating the cross section and considering the methods of estimation of nuclear and
$pd \rightarrow pd$ rescattering parts entering in the cross section.

%% file: 2/KinematicsrefFrame.tex
\section{Kinematics and reference frame} 
\label{sec_KinRef}
\subsection{Kinematics}\label{sec_Kinematics}
\begin{figure}[ht]
\centering
\includegraphics[scale=0.5]{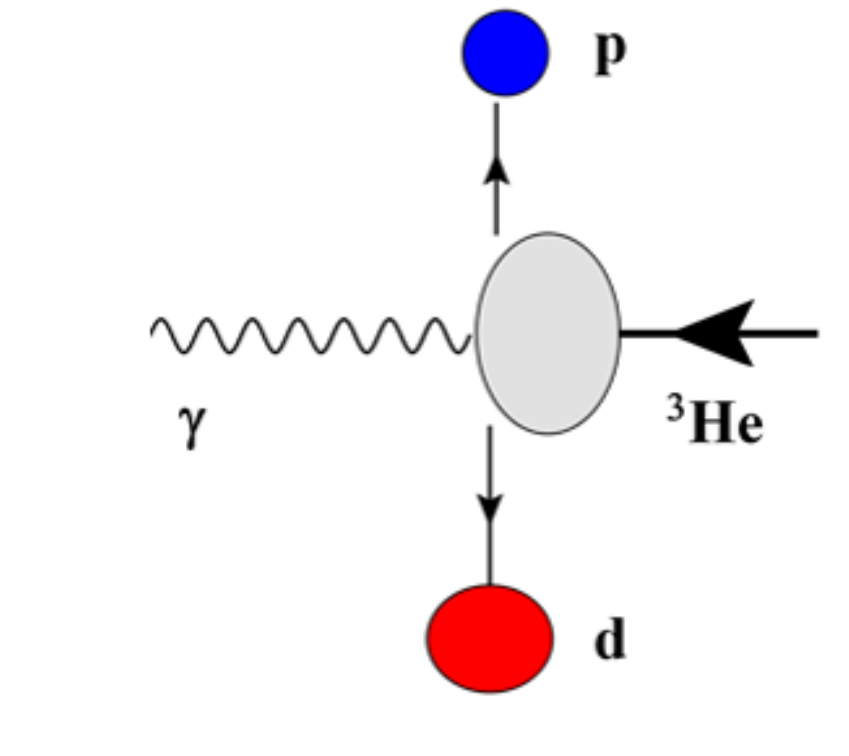}
\caption{Schematic diagram of photodisintegration of $^3$He into proton and deuteron.}
\label{scattering_cartoon}
\end{figure}

We are considering the following two-body photodisintegration reaction,
\begin{equation} \label{reaction}
\gamma + ^3\text{He} \rightarrow p + d,
\end{equation}
where the proton and deuteron are produced at large angles  measured in the center of mass  reference frame of the  reaction. One such process, at $90^\circ$ center of mass angle is shown in Fig.~(\ref{scattering_cartoon}).

\begin{figure}[ht]
\begin{subfigure}{0.5\textwidth}
\centering
    \includegraphics[width=1.0\linewidth]{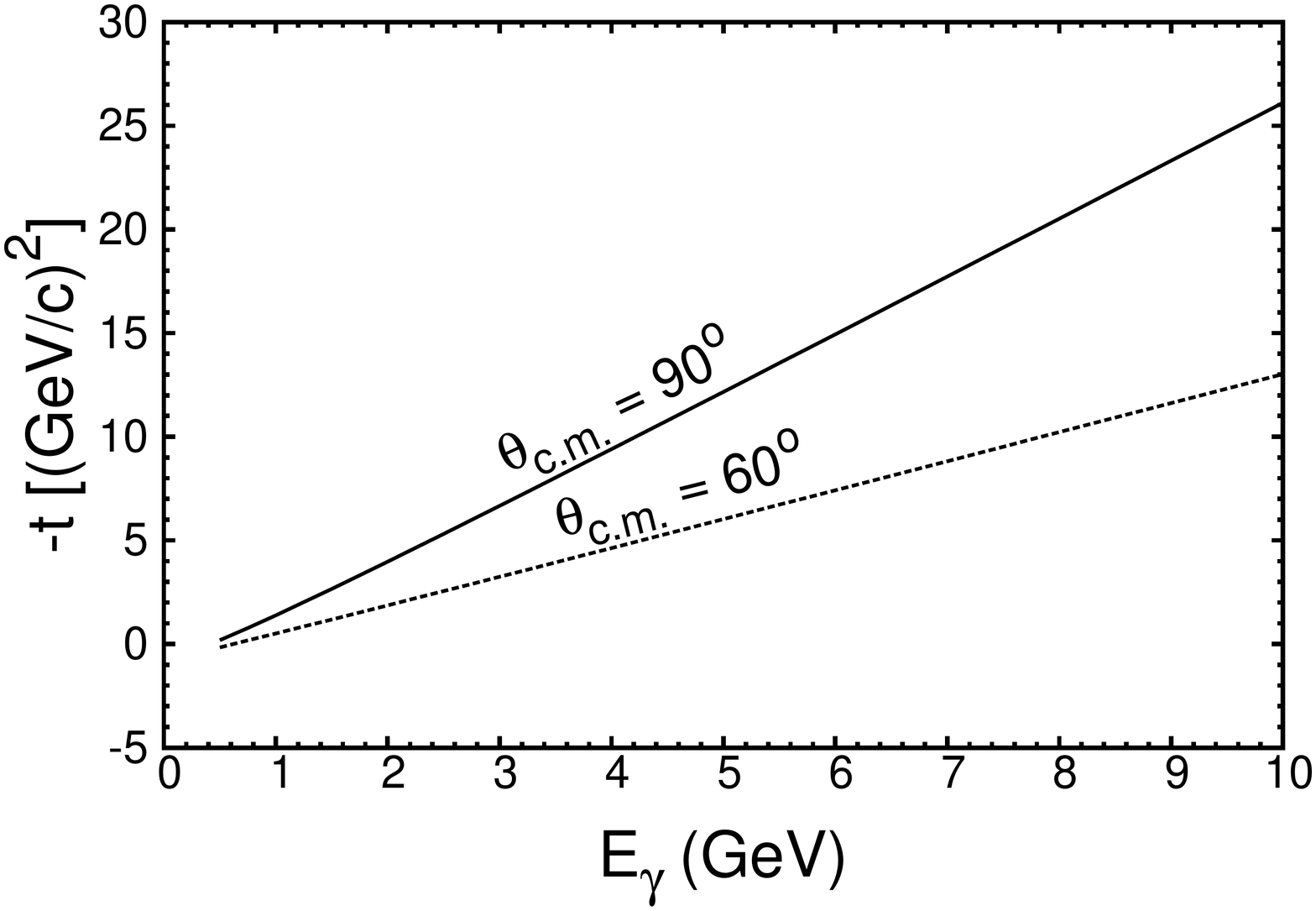}
    \caption{Photon energy dependence of invariant momentum transfer  $-t$.}
    \label{fig:dependenceOf_t_in_egamma}
\end{subfigure}%
\begin{subfigure}{0.5\textwidth}
\centering
    \includegraphics[width=1.0\linewidth]{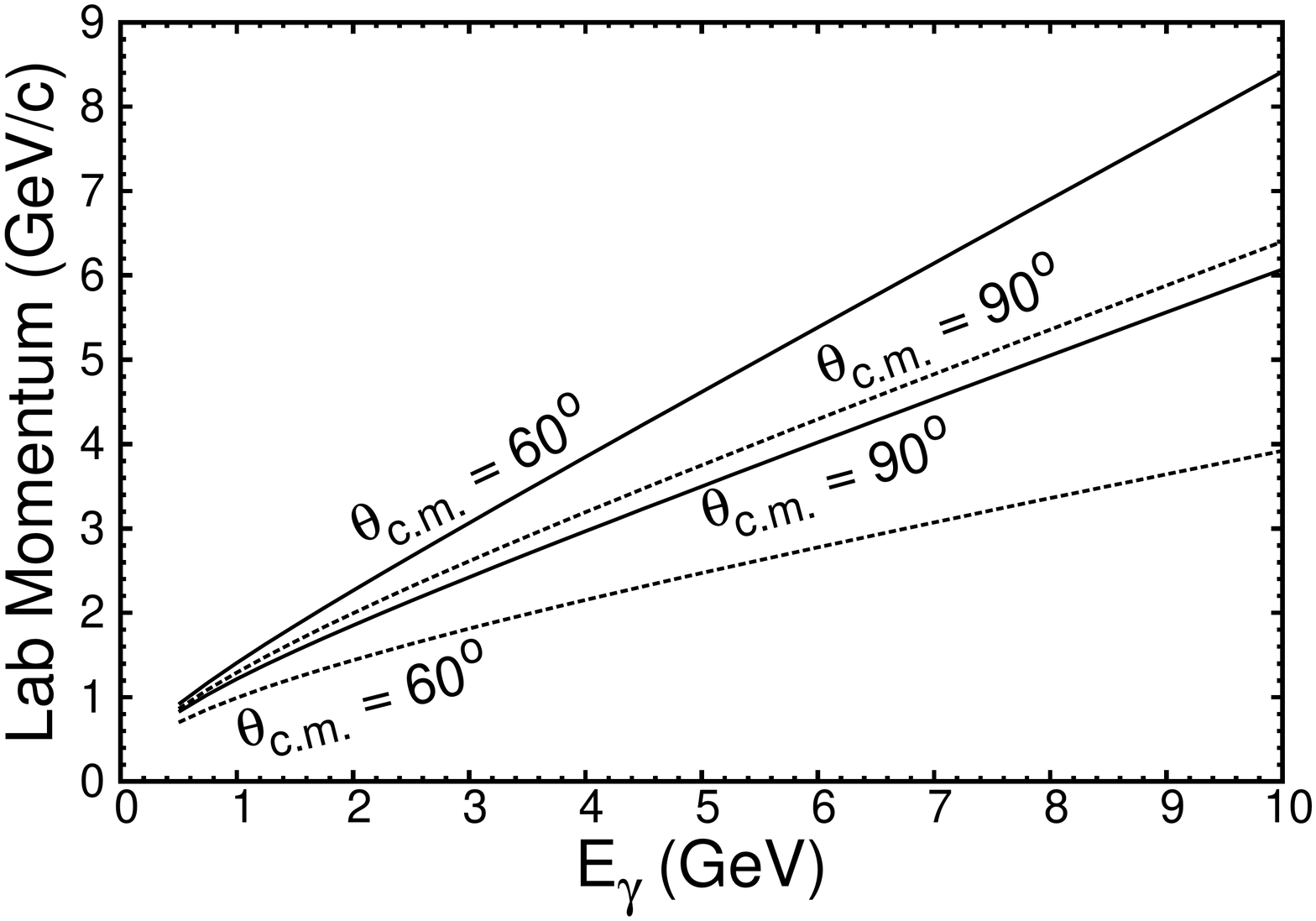}
    \caption{Lab momenta of outgoing proton and deuteron. Solid - proton, dashed - deuteron.}
    \label{fig:momentumDependence}
\end{subfigure}
\caption{Momentum transfer and lab momentum dependence over photon energy. Calculations are done for $\theta_{\text{cm}}=90^\circ$ and $60^\circ$.}
\label{fig:momentumDepenceOnEgamma}
\end{figure}

If we define the four momenta of incoming photon, $^3$He target, proton and deuteron as $q, p_{_{^3\text{He}}}, p_p \text{ and } p_d$ respectively, then the invariant energy $s$ and the invariant momentum transfer $t$ can be defined as
\begin{eqnarray}
s & = &  (q + p_{_{^3\text{He}}})^2  = m_{_{^3\text{He}}}^2 + 2 q \cdot p_{_{^3\text{He}}} = m_{^3\text{He}}^2 + 2 E_\gamma m_{_{^3\text{He}}} = (p_p + p_d)^2 \text{ and } \nonumber \\
t & =  & (q - p_p)^2 = m_p^2 - 2 q\cdot p_p = m_p^2 - 2E_\gamma^{\text{cm}}(E_p^{\text{cm}} - p_p^{\text{cm}}\cos \theta_{\text{cm}}) = (p_d - p_{_{^3\text{He}}})^2,
 \label{mandelstams}
\end{eqnarray} 
where $m_p$ and $m_{^3\text{He}}$ are masses of the proton and $^3$He target respectively, $E_\gamma$ is the incoming photon energy in the lab system, $E_\gamma^{\text{cm}}, E_p^{\text{cm}}$ are the photon and proton energies respectively in the center of mass system and $\theta_{\text{cm}}$ is the center of mass scattering angle. In the Lab system, we consider the $^3$He target to be at rest, while in the center of mass system, the incoming three-momentum of photon and $^3$He nucleus add up to zero, i.e., $\mathbf{q^{\text{cm}}} + \mathbf{p_{_{^3\text{He}}}^{\text{cm}}} = 0$. We at once observe that the invariant energy can be also written as
\begin{equation}
\label{sIntermsofCMenergies}
s = (E^{\text{cm}}_\gamma + E^{\text{cm}}_{_{^3\text{He}}})^2.
\end{equation}
The center of mass energies of the photon and the proton can be calculated using the conservation of four momentum, which requires
\begin{equation}
\label{fourMomentumConservation}
q + p_{_{^3\text{He}}} = p_p + p_d.
\end{equation}
Rearranging Eq.~(\ref{fourMomentumConservation}) and noting that $p_d^2 = m_d^2$, where $m_d$ is the mass of the deuteron, we obtain
\begin{eqnarray}
&&(q + p_{_{^3\text{He}}} - p_p)^2 = p_d^2 = m_d^2 \nonumber \\
\text{or,} \quad &&(q + p_{_{^3\text{He}}})^2 + m_p^2 - 2p_p \cdot (q + p_{_{^3\text{He}}}) = m_d^2 \nonumber \\
\text{or,} \quad && s + m_p^2 - 2[E_p^{\text{cm}}(E_\gamma^{\text{cm}} + p_{_{^3\text{He}}}^{\text{cm}}) - \mathbf{p_p^{\text{cm}}}\cdot \mathbf{(q^{\text{cm}} + p_{_{^3\text{He}}}^{\text{cm}})}] = m_d^2, 
\end{eqnarray}
where $m_p$ is the mass of the proton. 
Identifying from Eq.~(\ref{sIntermsofCMenergies}), $E_\gamma^{\text{cm}} + E_{^3\text{He}}^{\text{cm}} = \sqrt{s} $, the above equation reduces to 
\begin{equation}
\label{protonCMenergy}
 E_p^{\text{cm}}=\frac{1}{2\sqrt{s}}\Big(s + m_p^2 - m_d^2 \Big). 
\end{equation}
A similar procedure can be used to obtain the center of mass energy of the deuteron as
 \begin{equation}
 \label{deuteronCMenergy}
  E_d^{\text{cm}}=\frac{1}{2\sqrt{s}}\Big(s+m_d^2-m_p^2\Big).
 \end{equation}
 Also, rewriting the four momentum conservation as $(p_p + p_d - q)^2 = p_{_{^3\text{He}}}^2$ and noticing $q^2 = 0$, a similar derivation will give us the center of mass energy of the photon as
 \begin{equation}
 \label{photonCMenergy}
 E_\gamma^{\text{cm}} = \frac{1}{2\sqrt{s}}\Big(s - m_{_{^3\text{He}}}^2\Big).
\end{equation}
It can be similarly shown that 
\begin{equation}
\label{he3CMenergy}
E_{_{^3\text{He}}}^{\text{cm}} = \frac{1}{2\sqrt{s}}\Big(s + m_{_{^3\text{He}}}^2\Big).
\end{equation}  
One of the interesting features of Eq.~(\ref{mandelstams}) observed in Ref.\cite{Holt}, is the possibility to generate large center of mass energy $s$ even with moderate  energy of photon beams. The generation of  large center of mass energy can be explicitly seen from the expression of $s$, where photon 
energy is multiplied by the mass of the target. To be more specific, for reaction (\ref{reaction}), for example, for the photon energy, $E_\gamma  = 1$~GeV, one generates $s$ as large as it is generated by $6$~GeV/c proton beam in $pp$ scattering. This property was one of the reasons why the quark-counting scaling was observed in $\gamma d\rightarrow pn$ reaction for photon energies as low as $1.2$ GeV at $90^\circ$ center of mass break-up kinematics\cite{Mirazita:2004rb,Rossi:2004qm}.
Using Eqs.~(\ref{photonCMenergy}) and (\ref{protonCMenergy}), we can express the invariant momentum transfer $t$ in Eq.~(\ref{mandelstams}) as
\begin{equation}
\label{tInFull}
t=m_p^2-\frac{1}{2s}\Big(s-m_{_{^3\text{He}}}^2 \Big) \Big[(s+m_p^2-m_d^2)-\sqrt{\left\{ s-(m_p+m_d)^2\right\} \left\{ s-(m_p - m_d)^2 \right\}} \cos \theta_{\text{cm}} \Big].
\end{equation}
It follows from Eq.~(\ref{tInFull}) that in the high energy limit, $t\sim -\frac{s}{2}(1-cos\theta_{\text{cm}})$, which indicates that at large and fixed values of 
$\theta_{\text{cm}}$ one can achieve hard scattering regime,  $-t(-u)\gg m_N^2$ (where $m_N$ is the mass of the nucleon), providing large values of $s$.  For the latter, as it follows from the 
expression of $s$ in Eq.~(\ref{mandelstams}),  the  photon energy $E_\gamma$ is 
multiplied by the rest mass of the $^3$He nucleus because of which  even for moderate 
value of $E_\gamma$, the hard scattering condition,  ($-t(-u)\gg m_N^2$), is  easily achieved. 
This is seen in Fig.~(\ref{fig:dependenceOf_t_in_egamma}) where the dependence of $-t$ is presented as a function of incoming photon energy $E_{\gamma}$ at large and 
fixed values of $\theta_{\text{cm}}$.  As Fig.~(\ref{fig:dependenceOf_t_in_egamma}) shows, even at $E_\gamma \sim 1$~GeV the 
$-t\sim  1$ (GeV/c)$^2$, which is sufficiently large in order the reaction to be considered hard.

That the reaction (\ref{reaction})  at $E_\gamma\gtrsim 1$~GeV and $\theta_{\text{cm}}\sim 90^\circ$ can not be considered 
as conventional nuclear process with knocked-out nucleon and recoiled residual nuclear system follows
from Fig.~(\ref{fig:momentumDependence}), where the  lab momenta of outgoing proton and deuteron are shown for large 
$\theta_{\text{cm}}$. In this case, one  observes that starting at $E_\gamma \sim 1$~GeV/c, the momenta of outgoing proton and  deuteron are $> 1$GeV/c. 
Such a large momentum of the deuteron significantly exceeds the 
characteristic Fermi momentum  in the $^3$He nucleus, thus the deuteron can not be considered as residual (pre-existing in $^3$He).  These  momenta of the deuteron are also out of the kinematic range of eikonal, small angle rescattering\cite{gea,ms01,noredepn}, further diminishing the possibility of describing reaction~(\ref{reaction})  within the framework of conventional nuclear scattering.

Finally, another important property of the large center of mass break-up kinematics is the early onset of QCD degrees of freedom that result from the large inelasticities or large masses produced in the intermediate state of the reaction.  As it was shown 
in Ref.\cite{GilmanGross} for the photodisinegration of the deuteron, for photon energies at 1~GeV,   
one needs  around 15 channels of resonances in the intermediate state to describe the process within the  hadronic approach. This situation is similar for the case of the  $^3$He target in 
which one can estimate the mass of the intermediate state produced as $m_R \approx \sqrt{s}- m_d$. From 
the latter relation, one observes that for $E_\gamma=1$~GeV, $m_R \approx 1.8$~GeV, which is close to the deep inelastic threshold of $2$ GeV, which assures that the QCD degrees of freedom become more relevant for the description of the reaction.
  
Overall, the above kinematical discussion gives us a justification for the theoretical description using 
QCD degrees of freedom to be increasingly valid starting at photon energies of $\sim 1$~GeV.

%% file: 2/ReferenceFrame.tex
\subsection{Reference frame}\label{sec_RefFrame}
The theoretical framework used to study the reaction (\ref{reaction}) is developed in the light-front coordinate system. Below in Table~(\ref{LCDefns}), a general comparison between the light-front coordinates and the Minkowski spacetime is shown.
\begin{table}[ht]
\caption{Quantities in light-front coordinates and Minkowski spacetime.}
\centering
\begin{tabular}{c c c}
\hline\hline
Properties	& Minkowski spacetime	& Light-front \\
\hline
Momentum & $p^\mu = (E,p_x, p_y, p_z)$ & $p^\mu=(p_{-}, p_{+}, p_x, p_y), \text{ with } p_{\pm}=E \pm p_z $ \\
Self Product & $p^2 = E^2 - p^2$ & $p^2 =  p_{+}p_{-} - p_\perp^2$ \\
Product of two vectors & $p_1 \cdot p_2 = E_1E_2 - \mathbf{p_1} \cdot \mathbf{p_2}$ & $p_{1} \cdot p_{2} = \frac{1}{2} p_{{1+}}p_{{2-}} + \frac{1}{2} p_{{1-}}p_{{2+}} -p_{1\perp}\cdot p_{2\perp}$ \\
Differentials & $d^4p = dE d^3p$ & $d^4p = \frac{1}{2}dp_{+}dp_{-}d^2p_\perp$ \\
\hline
\label{LCDefns}
\end{tabular}
\end{table}
With the light-front coordinates defined, we now define the reaction reference frame, which is chosen such that the ``+" 
 and the transverse components of incoming photon are zero, i.e., $q_+ = q_\perp =  0$. The four momenta of incoming photon and the target nucleus are defined as:
\begin{eqnarray}
 q^\mu & = &  (q_+, q_-, q_\perp) =\Big(0,\sqrt{s'_{_{^3\text{He}}}},0\Big) \nonumber \\
 p^\mu_{_{^3\text{He}}} & = & (p_{_{^3\text{He}+}}, p_{_{^3\text{He}-}},p_{_{^3\text{He}\perp}}) = \Bigg(\sqrt{s'_{_{^3\text{He}}}}, \frac{m_{_{^3\text{He}}}^2}{\sqrt{s'_{_{^3\text{He}}}}},0 \Bigg)
 \label{InitialMomentaofPhotonHelium}
\end{eqnarray}
where $s'_{_{^3\text{He}}} = s - m_{_{^3\text{He}}}^2$.

%% file: 2/modelDevelopment.tex
\section{Development of model}
\label{sec_ModelDev}
The  HRM model originally developed to study the $\gamma d \rightarrow pn$ reaction\cite{gdpn}, which was successful in verifying the $s^{-11}$ dependence and also in reproducing  the absolute magnitude of the cross sections without free parameters for incoming photon energies  $\gtrsim 1$~GeV and large center of mass angles\cite{gdpn,gdpn2,gdpnang}. 
The HRM model allowed for the calculation of the polarization observables for the $\gamma d\rightarrow pn$ reaction\cite{gpdnpol}  and its prediction for the large magnitude of transferred polarization was confirmed by  the experiment of Ref.\cite{Jiang:2007ge}.
Subsequently, the HRM model was applied to the $\gamma + ^3$He $\rightarrow pp + n$ reactions\cite{gheppn}, in which two protons were produced in the hard break-up process while the neutron was soft.  The model described the scaling properties and the cross section reasonably well and were able to explain the observed smaller cross section as compared to the deuteron break-up reaction.  In Ref.\cite{gdBB}, it was also shown that HRM model 
 can be extended to the hard break-up of the nucleus to  any two baryonic state which can be produced from the NN scattering through the quark-interchange 
 interaction. 
 We assume in HRM that  the quark interchange  is the dominant mechanism for the hard rescattering of two outgoing energetic nucleons. The latter assumption is essential for factorization of the hard kernel of the scattering from the soft incalculable part of the scattering amplitude.
 
\begin{figure}[ht]
\centering
\includegraphics[width=1.0\textwidth]{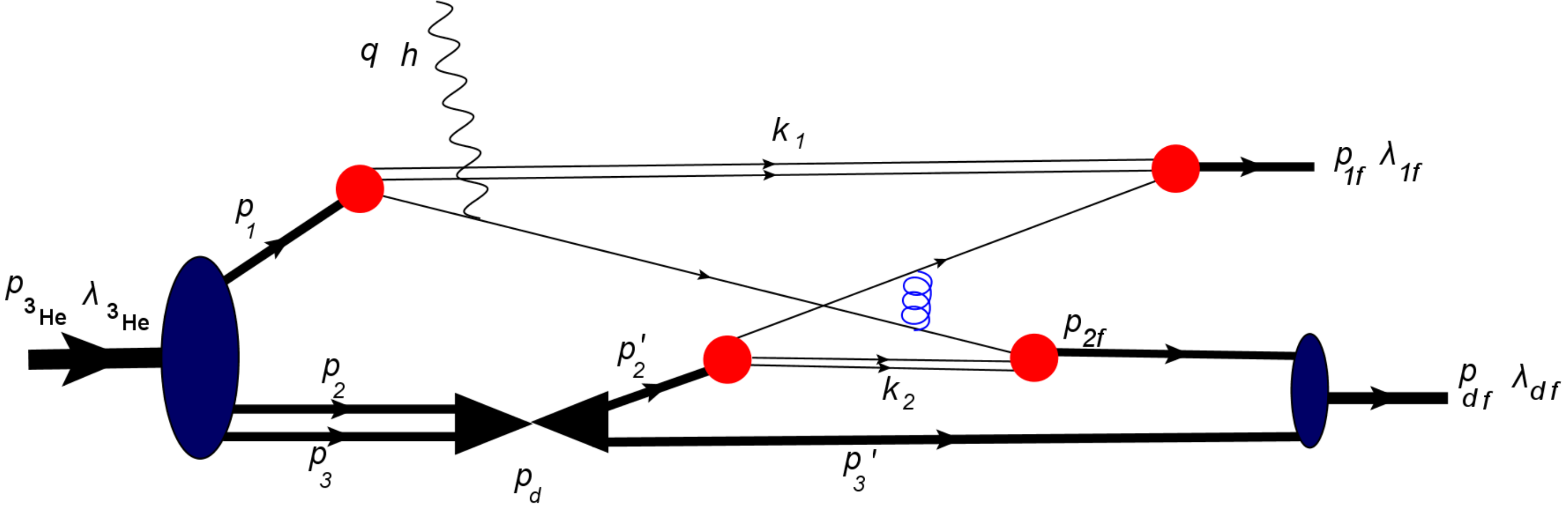}
\caption{Typical diagram of hard rescattering 
mechanism of $\gamma ^3\text{He}\rightarrow pd$ reaction.}
\label{diagram}
\end{figure}

In the HRM model, the hard photodisintegration takes place in two stages. First, the incoming photon knocks out  a quark from 
one of the nucleons. 
Then in the second step the outgoing fast quark undergoes a high momentum transfer hard scattering with the quark of 
the other nucleon sharing its large 
momentum among the constituents in the final state of the reaction. Since HRM utilizes the small momentum part of the target wave function which has large component of the initial $pd$ state,  
it is assumed that the energetic photon is  absorbed by any of the quarks belonging to  the protons in the nucleus  with the subsequent hard rescattering of struck quark off the quarks in the ``initial" $d$ system   producing hard final $pd$  state.
Within such scenario, the total scattering amplitude can be represented  as a sum of the multitude of the diagrams of type of Fig.~(\ref{diagram}) with all possibilities of  struck and rescattered quarks combining into a  fast outgoing $pd$ system. However instead of summing  all  the possible diagrams,  
the idea of HRM  is to factorize the hard $\gamma q$ scattering and sum the remaining parts into the amplitude of hard elastic $pd\rightarrow pd$ scattering.  
In this way  all the complexities related to the large number of diagrams and  non-perturbative quark wave function of the nucleons are absorbed into the $pd\rightarrow pd$ amplitude, which  can be taken from experiment.
To demonstrate the above described concept of HRM, we consider a typical 
scattering diagram of Fig.~(\ref{diagram}). Here, the incoming photon  knocks out a quark from one of the protons in the nucleus. 
The struck quark that now carries almost the whole momentum of the photon will share its momentum with a quark from the other nucleon through 
the quark-interchange. The resulting two energetic quarks will recombine with the residual quark-gluon systems to produce proton and deuteron with 
large relative momentum. 
Note that the  assumption,  that the nuclear spectator system is represented by intermediate deuteron state.  
is justified based on our previous studies of HRM\cite{gdpn,gheppn} in which it was found that the scattering amplitude is dominated by small initial momenta of interacting  nucleons. For the case of reaction (\ref{reaction}), because of the presence of the deuteron in the final state, the small momentum of initial proton   in the $^3\text{He}$ nucleus will originate predominantly from a two-particle $pd$ state. 

In Fig.~(\ref{diagram}), $h$, $\lambda_{_{^3\text{He}}}, \lambda_{1f} \text{ and } \lambda_{df}$ are the helicities of the incoming photon, $^3\text{He}$ nucleus, outgoing proton and deuteron respectively. Similarly, $q$, $p_{_{^3\text{He}}}, p_1,p_{1f},p_{d} \text{ and } p_{df}$ are the four momenta of the photon, $^3\text{He}$ nucleus, initial and outgoing protons, intermediate deuteron and the final deuteron respectively. The $k$'s define the four momenta of the \textit{spectator} quark systems. The four-momenta defined in Fig.~(\ref{diagram}) satisfy the following relations
\begin{displaymath}
p_{_{^3\text{He}}} = p_1 + p_2 + p_3;\quad p_2 + p_3 = p_{d} = p_2^{\prime} + p_3^{\prime};\quad 
 p_{2f} + p_{3}^{\prime} = p_{df} ;\quad 
p_{_{^3\text{He}}} + q = p_{1f} + p_{df},
\end{displaymath}
where $p_2$, $p_3$, $p_2^\prime$ and $p_3^\prime$ are four-momenta of the nucleons in the intermediate state deuteron.

We now write the Feynman amplitude 
corresponding to the diagram of Fig.~(\ref{diagram}), identifying terms corresponding to 
nuclear and nucleonic parts as follows:
\begin{eqnarray} \label{StartDerivationAmplitude}
&&\mathcal{M}^{\lambda_{df},\lambda_{1f};\lambda_{^3\text{He}},h} =  \sum_{\lambda_d'} \int \chi_{d}^{*\lambda_{d}'} (-i\Gamma_{_{DNN}}^{\dagger}) \frac{i(\slashed{p}_{2f} + m)}{p_{2f}^{2} - m_N^2 + i\epsilon} \frac{i(\slashed{p}_{3}'+ m)}{p_{3}'^{2}-m_N^2+i\epsilon} \frac{i(\slashed{p}_{2}' + m)}{p_{2}'^{2}-m_N^2 + i\epsilon} \nonumber \\
&&A:\qquad  i \frac{\Gamma_{_{DNN}} \chi_{d}^{\lambda_{d}} \chi_{d}^{*\lambda_{d}}}{p_{d}^2 - m_{_d}^2 + i\epsilon} (-i)\Gamma_{_{DNN}}^{\dagger} \frac{i(\slashed{p}_{3}+m)}{p_{3}^{2}-m_N^2+i\epsilon} \frac{i(\slashed{p}_{2}+m)}{p_{2}^{2}-m_N^2+i\epsilon} \frac{i(\slashed{p}_{1}+m)}{p_{1}^{2}-m_N^2+i\epsilon}\nonumber \\
&& \qquad i\Gamma_{_{^3\text{He}}} \chi_{_{^3\text{He}}}^{\lambda_{^3\text{He}}} \dfrac{d^{4}p_{2}'}{(2\pi)^{4}} \dfrac{d^{4}p_3}{(2\pi)^{4}} \dfrac{d^{4}p_{3}'}{(2\pi)^{4}} \nonumber \\
&& N1:  \int \chi_{p_{1f}} (-i)\Gamma_{N1}^{\dagger} \frac{i(\slashed{p}_{1f} - \slashed{k}_1 + m)}{(p_{1f} - k_1)^{2} - m_q^{2} + i\epsilon} \Big[-igT_{c}^{\beta} \gamma_{\mu}\Big] \frac{iS(k_{1})}{k_{1}^{2} - m_s^{2} + i\epsilon} \nonumber \\
&& \qquad \frac{i(\slashed{p}_{1} - \slashed{k}_1 + m_q)}{(p_{1} - k_1)^{2} - m_q^{2} + i\epsilon} i\Gamma_{n1}\dfrac{d^{4}k_{1}}{(2\pi)^{4}} \nonumber  \\
&& N2: \int (-i)\Gamma_{N}^{\dagger} \frac{i(\slashed{p}_{2f} - \slashed{k}_2 + m_q)}{(p_{2f} - k_2)^{2} - m_q^{2} + i\epsilon} \frac{iS(k_{2})}{k_{2}^{2} - m_s^{2} + i\epsilon} \frac{i(\slashed{p}_{2}' - \slashed{k}_2 + m_q)}{(p_{2}' - k_2)^{2} - m_q^{2} + i\epsilon}  i\Gamma_{n2'}\dfrac{d^{4}k_{2}}{(2\pi)^{4}} \nonumber \\
&& \gamma: -igT_c^\alpha \gamma_\nu \frac{i(\slashed{p}_1 + \slashed{q} - \slashed{k_1}+m_q)}{(p_1 - k_1 + q)^2 -m_q^2 + i\epsilon} \Big[-ie\gamma^\mu \epsilon^\mu_h\Big] \nonumber \\
&& g: \frac{i d_{\mu \nu}\delta_{\alpha \beta}}{q_{q}^{2}}.
\end{eqnarray}

In Eq.~(\ref{StartDerivationAmplitude}), the label $A$: identifies the nuclear part of the scattering amplitude characterized 
by the transition vertices $\Gamma_{^3\text{He}}$ (for  the  $^3\text{He}\rightarrow N_1,N_2,N_3$ transition) and 
$\Gamma_{DNN}$ (for $D\rightarrow N_2N_3$ transitions). 
The parts $N1$: and $N2$: identify the transition of nucleons $N_1$ and $N_2$ to quark-spectator system (characterized by the vertex $\Gamma_{N}$) with recombination to the final $N_{1f}$ and $N_{2f}$ nucleons.
Also, $S(k_1)$ and $S(k_2)$ denote the propagators of the spectator quark-gluons system.  
The $\gamma:$ identifies  the part in which  the photon with polarization  $\epsilon^\mu_h$   
interacts  with  the quark with ($p_1-k_1$) four-momentum, followed by the struck quark propagation.   
The label $g:$  represents the gluon propagator. Everywhere, $\chi$'s denote the spin wave functions of the nuclei and nucleons with $\lambda$'s defining 
the helicities. The summation over the $\lambda_{d}'$ represents the sum over the helicities of the intermediate deuteron.
The factor $g$ is the QCD coupling  constant with $T_c$ being color matrices.

The hard rescattering model  which allows to calculate the sum of the all diagrams similar to Fig.~(\ref{diagram}) builds on the following three assumptions:
\begin{enumerate}
\label{HRMassumptions}
\item The dominant contribution comes from the soft $^3\text{He} \rightarrow pd$ transition defined by small initial momentum of the proton. As a result, this transition can be calculated using non-relativistic wave functions of the $^3\text{He}$ and deuteron.
\item The high energy $\gamma q$ scattering can be factorized from the final state quark interchange rescattering.
\item  All quark-interchange rescatterings  can be summed into the elastic $pd\rightarrow pd$ amplitude. 
\end{enumerate}
A detailed derivation to show how the amplitude can be expressed in terms of the nuclear and nucleonic light-front wavefunctions is presented next.

%% file: 2/Nuclearwfs.tex
\subsection{Nuclear wavefunctions}
We begin by considering a part $A_1$ in label ${A:}$  in Eq.~(\ref{StartDerivationAmplitude}) related to the $^3\text{He}\rightarrow d$ nuclear transition as:
\begin{eqnarray} \label{A1}
 &&A_1  = \int (-i) \frac{\chi_{d}^{*^{\lambda_{d}}} \Gamma_{_{DNN}}^{\dagger}}{p_{d}^2 - m_{_d}^2 + i\epsilon} \frac{i(\slashed{p}_{3}+m)}{p_{3}^{2}-m_N^2+i\epsilon} \frac{i(\slashed{p}_{2}+m)}{p_{2}^{2}-m_N^2+i\epsilon} \frac{i(\slashed{p}_{1}+m)}{p_{1}^{2}-m_N^2+i\epsilon} \times \nonumber \\
&& \qquad  i\Gamma_{_{^3\text{He}}} \chi_{_{^3\text{He}}}^{\lambda_{^3\text{He}}} \frac{1}{2}\frac{dp_{_{2+}}dp_{_{2-}}d^2p_{_{2\perp}}}{(2\pi)^4} \frac{1}{2}\frac{dp_{_{3+}}dp_{_{3-}}d^2p_{_{3\perp}}}{(2\pi)^4},
\end{eqnarray}
where the differentials are expressed in the light front coordinates.
Using the fact that $p^2 = p_+p_- - p_\perp^2 = m^2$ in the light-front coordinates, the denominator of the propagator can be expressed in general as
\begin{equation}
\label{propDenomExmpansionGeneral}
p^2 - m^2 + i\epsilon = p_{_+} \Bigg(p_{_-} - \frac { m^2 + p_{_\perp}^2 }{p_+} + i\epsilon'\Bigg).
\end{equation}
Using Eq.~(\ref{propDenomExmpansionGeneral}) and also the sum rule satisfied by the spinors, $(\slashed{p} + m) = \sum_{\lambda} u(p, \lambda)\bar{u}(p, \lambda)$, we can rewrite Eq.~(\ref{A1}) as

\begin{eqnarray}
\label{A1detailed}
A_1 &&= \sum_{\lambda_d, \lambda_3, \lambda_2, \lambda_1} \int (-i) \frac{\chi_{d}^{*^{\lambda_{d}}} \Gamma_{_{DNN}}^{\dagger}}{p_{d}^2 - m_{_d}^2 + i\epsilon} \frac{u(p_3, \lambda_3)\bar{u}(p_3,\lambda_3)}{p_{3+} \Bigg(p_{3-} - \frac { m_N^2 + p_{3\perp}^2 }{p_{3+}} + i\epsilon'\Bigg)} \frac{u(p_2, \lambda_2)\bar{u}(p_2,\lambda_2)}{p_{2+} \Bigg(p_{2-} - \frac { m_N^2 + p_{2\perp}^2 }{p_{2+}} + i\epsilon'\Bigg)} \times \nonumber \\
&& \frac{u(p_1, \lambda_1)\bar{u}(p_1,\lambda_1)}{p_{1+} \Bigg(p_{1-} - \frac { m_N^2 + p_{1\perp}^2 }{p_{1+}} + i\epsilon'\Bigg)} 
 \quad  i\Gamma_{_{^3\text{He}}} \chi_{_{^3\text{He}}}^{\lambda_{^3\text{He}}} \frac{1}{2}\frac{dp_{_{2+}}dp_{_{2-}}d^2p_{_{2\perp}}}{(2\pi)^4} \frac{1}{2}\frac{dp_{_{3+}}dp_{_{3-}}d^2p_{_{3\perp}}}{(2\pi)^4}.
\end{eqnarray}
Noting that $p_{_{^3\text{He}}} = p_1 + p_d$ and using the conservation for the minus component, we find that
\begin{eqnarray}
\label{p1denominator}
p_{1-} - \frac{m_N^2 + p_{1\perp}^2}{p_{1+}} &=& p_{_{^3\text{He}-}} - p_{d-} - \frac{m_N^2 + p_{1\perp}^2}{p_{1+}} \nonumber\\
&=& \frac{m_{_{^3\text{He}}}^2}{p_{_{^3\text{He}+}}} - \frac{m_d^2 + p_{d\perp}^2}{p_{d+}} - \frac{m_N^2 + p_{1\perp}^2}{p_{1+}} \nonumber \\
&=& \frac{1}{p_{_{^3\text{He}+}}}\Bigg(m_{_{^3\text{He}}}^2 - \frac{m_d^2 + p_{d\perp}^2}{\beta_d} - \frac{m_N^2 + p_{1\perp}^2}{\beta_1}  \Bigg),
\end{eqnarray}
where $\beta_d = \frac{p_{d+}}{p_{_{^3\text{He}+}}}$ and $\beta_1 = \frac{p_{1+}}{p_{_{^3\text{He}+}}}$. The quantities $\beta_1$ and $\beta_d$ represent the fraction of the light front momentum of $^3$He carried by the initial proton and deuteron respectively. It is interesting to note that $\beta_1 + \beta_d = 1$. Combining Eqs.~(\ref{p1denominator}) and (\ref{A1detailed}), we obtain:
\begin{eqnarray}
\label{A1readyforHe3wfIntro}
A_1  &&= \sum_{\lambda_d, \lambda_3, \lambda_2, \lambda_1} \int (-i) \frac{\chi_{d}^{*^{\lambda_{d}}} \Gamma_{_{DNN}}^{\dagger}}{p_{d}^2 - m_{_d}^2 + i\epsilon} \frac{u(p_3, \lambda_3)\bar{u}(p_3,\lambda_3)}{p_{3+} \Bigg(p_{3-} - \frac { m_N^2 + p_{3\perp}^2 }{p_{3+}} + i\epsilon'\Bigg)} \frac{u(p_2, \lambda_2)\bar{u}(p_2,\lambda_2)}{p_{2+} \Bigg(p_{2-} - \frac { m_N^2 + p_{2\perp}^2 }{p_{2+}} + i\epsilon'\Bigg)} \times \nonumber \\
&& \frac{u(p_1, \lambda_1)\bar{u}(p_1,\lambda_1)}{\beta_1 \Bigg(m_{_{^3\text{He}}}^2 - \frac{m_d^2 + p_{d\perp}^2}{\beta_d} - \frac{m_N^2 + p_{1\perp}^2}{\beta_1}  \Bigg)} 
 \quad  i\Gamma_{_{^3\text{He}}} \chi_{_{^3\text{He}}}^{\lambda_{^3\text{He}}} \frac{1}{2}\frac{dp_{_{2+}}dp_{_{2-}}d^2p_{_{2\perp}}}{(2\pi)^4} \frac{1}{2}\frac{dp_{_{3+}}dp_{_{3-}}d^2p_{_{3\perp}}}{(2\pi)^4}.
\end{eqnarray}
We are now able to define the light-front wavefunction of $\bf{^3\text{He}}$ as follows(see e.g.\cite{FS81,pAjj,multisrc}):
\begin{equation} \label{He3wf}
\Psi_{_{^3\text{He}}}^{^{\lambda_{^3\text{He}}}}(\beta_1,\lambda_1,p_{{1\perp}},\beta_2,p_{{2\perp}}\lambda_2,\lambda_3)=\frac{\bar{u}(p_3,\lambda_3)\bar{u}(p_2,\lambda_2)\bar{u}(p_1,\lambda_1)}{\Big(m_{_{^3\text{He}}}^2 - \frac{m_{d}^2 + p_{d\perp}^2}{\beta_{d}}- \frac{m_N ^2 + p_{{1\perp}}^2 }{\beta_1}\Big)} \Gamma_{_{^3\text{He}}} \chi_{_{^3\text{He}}}^{\lambda_{^3\text{He}}}.
\end{equation}  
The light front $^3$He wave function gives the probability amplitude of finding 
the $\lambda_{^3\text{He}}$ helicity $^3\text{He}$ nucleus consisting of nucleons  with momenta $p_i$  and helicities $\lambda_i$, $i=1,2,3$.
The integration over the ``minus'' components of the momenta are then carried out by taking the pole value of the integral, according to the following scheme:
\begin{equation}
\label{polevalueint_scheme}
\int \frac{dp_-}{p_- - \frac{m_N^2 + p_\perp^2}{p_+} + i\epsilon} = -2\pi i |_{p_- = \frac{m_N^2 + p_\perp^2}{p_+}}.
\end{equation}
Introducing the light-front $^3$He wavefunction and performing the $p_{3-}$ integration according to Eq.~(\ref{polevalueint_scheme}), we find that Eq.~(\ref{A1readyforHe3wfIntro}) becomes:
\begin{eqnarray} \label{A1withHe3wf}
&& A_1 =  -i  \sum_{\lambda_{d},\lambda_{3},\lambda_{2},\lambda_{1}} \int \frac{\chi_{d}^{*^{\lambda_{d}}} \Gamma_{_{DNN}}^{\dagger}}{p_{d}^2 - m_{d}^2 + i\epsilon} u(p_3,\lambda_3) u(p_2,\lambda_2) u(p_1,\lambda_1)  \frac{1}{p_{{2+}}\Big(p_{{2-}} - \frac{m_N^2 + p_{{2\perp}}^2}{p_{{2+}}}\Big)} \nonumber \\
  && \frac{\Psi_{_{^3\text{He}}}^{^{\lambda_{^3\text{He}}}}(\beta_1,\lambda_1,p_{{1\perp}},\beta_2,p_{{2\perp}}\lambda_2,\lambda_3)}{\beta_1} \frac{1}{2} \frac{dp_{{2-}}dp_{{2+}}d^{2}p_{{2\perp}}}{(2\pi)^4} \frac{1}{2} \frac{d\beta_3}{\beta_3}\frac{d^2p_{{3\perp}}}{(2\pi)^3},
\end{eqnarray} where we introduce $\beta_3 = \frac{p_{3+}}{p_{^3\text{He}+}}$, which gives the fraction of light-front momentum of $^3$He carried by the nucleon ``3''.
Noting that the momentum of intermediate deuteron, $p_d = p_2 + p_3$, we next consider the term $p_{{2+}}\Big(p_{{2-}} - \frac{m_N^2 + p_{{2\perp}}^2}{p_{{2+}}}\Big)$ in Eq.~(\ref{A1withHe3wf}):
\begin{eqnarray}
\label{p2denominator}
p_{{2+}}\Big(p_{{2-}} - \frac{p_{{2\perp}}^2 + m_N^2}{p_{{2+}}}\Big) &&=  p_{{2+}} \Big( p_{{d-}} - p_{{3-}} - \frac{p_{{2\perp}}^2 + m_N^2}{p_{{2+}}}\Big) \nonumber \\
&& = p_{{2+}} \Big(\frac{m_{d}^{2} + p_{{d\perp}}^{2}}{p_{{d+}}} - \frac{m_N^2 + p_{{3\perp}}^{2}}{p_{{3+}}} -\frac{m_N^2 + p_{{2\perp}}^{2}}{p_{{2+}}}\Big)\nonumber \\
&& = \frac{p_{{2+}}}{p_{{d+}}}\Big( m_{d}^2 + p_{{d\perp}}^{2}- \frac{m_N^2 + p_{{3\perp}}^{2}}{\alpha_{3}} - \frac{m_N^2 + p_{{2\perp}}^{2}}{\alpha_{2}}     \Big) \nonumber \\
&& = (1 - \alpha_{{3}})\Big(m_{{d}}^{2} + p_{{d\perp}}^{2} - \frac{m_N^2 + p_{{3\perp}}^{2}}{\alpha_{3}} - \frac{ m_N^2 + p_{{2\perp}}^{2}}{1- \alpha_{3}}\Big),
\end{eqnarray}
where we define $ \alpha_{3} = \frac{p_{{3+}}}{p_{{d+}}} = \frac{\beta_3}{\beta_{d}} $ and $\alpha_{2} = \frac{p_{{2+}}}{p_{{d+}}} = 1 - \alpha_{3} = \frac{\beta_2}{\beta_d}$ as  the fractions of the momentum of the \textit{intermediate} deuteron carried by the nucleons 3 and 2 respectively. It can be seen that $ \alpha_{3} + \alpha_{2}=1$. 
Similar to Eq.~(\ref{He3wf}), we introduce  the light-front   wave function of the deuteron\cite{FS81, pAjj, multisrc}:
\begin{equation}
 \label{DeuteronWF}
\Psi_{_{d}}^{\lambda_{d}}(\alpha_{3},p_{{3\perp}},p_{{d\perp}}) = \frac{\bar{u}(p_2,\lambda_2) \bar{u}(p_3,\lambda_3)}{\Big(m_{{d}}^{2} + p_{{d\perp}}^{2} - \frac{m_N^2 + p_{{3\perp}}^{2}}{\alpha_{3}} - \frac{ m_N^2 + p_{{2\perp}}^{2}}{1- \alpha_{3}}\Big)} \Gamma_{_{DNN}}\chi_{_d}^{^{\lambda_{d}}},
\end{equation}
which describes the probability amplitude of finding in the $\lambda_{d}$   helicity  deuteron two nucleons with  momenta $p_i$ and  helicities $\lambda_i$, $i=1,2$.
Also, since $p_2 + p_3 = p_d$, the differentials with respect to $p_2$ can be written as:
\begin{equation} \label{p2minustopdminus}
\frac{dp_{{2-}}dp_{{2+}}dp_{{2\perp}}}{(2\pi)^4} = \frac{dp_{{d-}}dp_{{d+}}dp_{{d\perp}}}{(2\pi)^4}.
\end{equation}
The Eq.~(\ref{p2minustopdminus}) allows us, making use of Eq.~(\ref{polevalueint_scheme}), to integrate the $p_{d-}$ component of the intermediate deuteron momentum. Thus, using Eqs.~(\ref{p2denominator}), (\ref{p2minustopdminus}) and (\ref{DeuteronWF}), Eq.~(\ref{A1withHe3wf}) becomes:
\begin{eqnarray}
\label{A1withDeuteronHe3wfs}
& A_1 = & -i  \sum_{\lambda_{d},\lambda_{3},\lambda_{2},\lambda_{1}} \int \frac{\Psi_{_{d}}^{\dagger\lambda_{d}:\lambda_2, \lambda_3}(\alpha_{3},p_{{3\perp}},p_{{d\perp}})}{1-\alpha_{3}} u(p_1,\lambda_1) \frac{\Psi_{_{^3\text{He}}}^{^{\lambda_{^3\text{He}}}}(\beta_1,\lambda_1,p_{{1\perp}},\beta_2,p_{{2\perp}}\lambda_2,\lambda_3)}{\beta_1} \times \nonumber \\
&& \frac{1}{2} \frac{d\beta_{d}d^{2}p_{{d\perp}}}{\beta_{d}(2\pi)^3} \frac{1}{2} \frac{d\beta_3}{\beta_3}\frac{d^2p_{{3\perp}}}{(2\pi)^3}.
\end{eqnarray}
Next, we consider the  second part of the expression ``$A$" in Eq.~(\ref{StartDerivationAmplitude}) related to the transition of the deuteron from intermediate to the final state:
\begin{eqnarray} \label{B1part}
& A_2 &=  -\int \chi_{d}^{*\lambda_{d}}(\Gamma_{_{DNN}}^{\dagger}) \frac{i(\slashed{p}_{2f} + m)}{p_{2f}^{2} - m_N^2 + i\epsilon} \frac{i(\slashed{p}_{3}'+m)}
{p_{3}'^{2}-m_N^2+i\epsilon} \frac{i(\slashed{p}_{2}'+m)}{p_{2}'^{2}-m_N^2+i\epsilon} i\Gamma_{_{DNN}} \chi_{d}^{\lambda_{d}} \frac{d^4p_{3}'}{(2\pi)^4} \nonumber \\ 
&&= -\sum_{\substack{\lambda_{2f} \\ \lambda_{2}', \lambda_{3}'}} \int \chi_{d}^{*\lambda_{d}} \Gamma_{_{DNN}}^{\dagger} \frac{u(p_{2f},\lambda_{2f})
\bar{u}(p_{2f},\lambda_{2f})}{p_{2f}^{2} - m_N^2 + i\epsilon}  \frac{u(p_{2}',\lambda_{2}')\bar{u}(p_{2}',\lambda_{2}')}{p_{{2+}}'
\Big(p_{{2-}}' - \frac{m_N^2 + p_{2\perp}'^{2}}{p_{{2+}}'}+i\epsilon\Big)} \times \nonumber  \\ 
&& \qquad u(p_{3}',\lambda_{3}')\bar{u}(p_{3}',\lambda_{3}') i\Gamma_{_{DNN}} \chi_{d}^{\lambda_{d}} \frac{1}{2} 
\frac{dp_{{3+}}'}{p_{{3+}}'}\frac{d^2p_{{3\perp}}'}{(2\pi)^3},
\end{eqnarray}
where the $dp_{3-}'$ is integrated according to Eq.~(\ref{polevalueint_scheme}).
To estimate the denominator, $p_{2f}^2 - m_N^2$, making use of $p_{df} = p_{2f} + p_{3}\prime$, we find:
\begin{eqnarray}
p_{2f}^2 - m_N^2  &&= p_{{2f+}}\Bigg(p_{{2f-}} - \frac{p_{{2f\perp}}^{2} + m_N^2}{p_{{2f+}}}\Bigg) \nonumber \\
&& = {p_{{2f+}}\Bigg(p_{{df-}} - p_{{3-}}^\prime - \frac{p_{{2f\perp}}^{2} + m_N^2}{p_{{2f+}}}\Bigg)} \nonumber \\
&& = p_{{2f+}}\Bigg(\frac{m_{d}^{2} + p_{df\perp}^2}{p_{df+}} - \frac{m_N^2 + p\prime_{3\perp}^2}{p\prime_{3+}} - \frac{m_N^{2} + p_{2f\perp}^2}{p_{2f+}}\Bigg) \nonumber \\
&& = \frac{p_{{2f+}}}{p_{{df+}}} \Bigg( m_d^2 + p_{{df\perp}}^2 - \frac{m_N^2 + p\prime_{{3\perp}}^2}{p\prime_{{3+}}/p_{{df+}}} - \frac{m_N^2 + p_{{2f\perp}}^2}{p_{{2f+}}/p_{{df+}}}\Bigg).
\end{eqnarray}
Defining $ \frac{p_{_{3+}}'}{p_{_{df+}}}  = \frac{p_{_{3+}}'/p_{_{d+}}}{p_{_{df+}}/p_{_{d+}}} = \frac{\alpha_{3}'}{\gamma_{d}} $ and $\frac{p_{_{2f+}}}{p_{_{df+}}}  = 1 - \frac{p_{_{3+}}'}{p_{_{df+}}} = 1 - \frac{\alpha_{3}'}{\gamma_{d}}$,  the above equation reduces to:
\begin{equation} \label{p2fplusdenomfinal}
 p_{2f}^2 - m_N^2 = \Bigg(1 - \frac{\alpha_{3}'}{\gamma_{d}}\Bigg)\Bigg( m_d^2 + p_{{df\perp}}^2 - \frac{m_N^2 + p_{{3\perp}}'^{2}}
 {\alpha_{3}'/\gamma_{d}} - \frac{m_N^2 + p_{{2f\perp}}^2}{1 - \alpha_{3}'/\gamma_{d}}\Bigg),
\end{equation}
where the quantity $\gamma_{d} = \frac{p_{{df+}}}{p_{{d+}}} $ is the fraction of the momentum of the intermediate deuteron carried by the final deuteron. Using Eqs.~(\ref{p2fplusdenomfinal}) and (\ref{DeuteronWF}) in Eq.~(\ref{B1part}), we have
\begin{eqnarray}
& A_2 &= \sum_{\substack{\lambda_{df},\lambda_{2f} \\ \lambda_{2}', \lambda_{3}'}} \int \frac{{\Psi_{d}^{\dagger \lambda_{df}:\lambda_{3}', \lambda_{2f}}}(\alpha_{2f}/\gamma_{d},p_{2\perp},\alpha_{3}'/\gamma_{d},p_{3\perp}')}{1-\alpha_{3}'/\gamma_{d}} \bar{u}(p_{2f},\lambda_{2f}) u(p_{2}',\lambda_{2}') \times \nonumber \\
&&  \qquad \frac{\Psi_d^{\lambda_{d}:\lambda_{2}', \lambda_{3}'}(\alpha_{3}',p_{d\perp},p_{3\perp}')}{1-\alpha_{3}'} 
\frac{1}{2} \frac{d\alpha_{3}'}{\alpha_{3}'}	\frac{d^2p_{3\perp}'}{(2\pi)^3}.
\label{A2part}
\end{eqnarray}


%% file: 2/Nucleonicwfs.tex
\subsection{Nucleonic wavefunctions}
Now, we consider the $``N1"$ part of the amplitude in Eq.~(\ref{StartDerivationAmplitude}), which describes the transition of nucleon with momentum $p_1$ to the final nucleon with the momentum $p_{1f}$.  Using on-shell sum-rule relations for the numerators of the quark propagators for the $N1$ part, one has:
\begin{eqnarray}
\label{N1start}
&& N1 =\sum_{\substack{\lambda_1 \\ \eta_{1f},\eta_1, s }}\int \bar{u}(p_{1f}, \lambda_{1f}) (-i)\Gamma_{n1f}^{\dagger} 
\frac{u_q(p_{1f}-k_1,\eta_{1f})\bar{u}_q(p_{1f}-k_1,\eta_{1f})}{(p_{1f} - k_1)^{2} - m_q^{2} + i\epsilon} \Big[-igT_{c}^{\beta} \gamma_{\mu}\Big] \times \nonumber \\ 
&&\frac{\Psi_s(k_1)\bar{\Psi}_s(k_1)}{k_{1}^{2} - m_s^{2} + i\epsilon} 
  \frac{u_q(p_1-k_1,\eta_1)\bar{u}_q(p_1-k_1,\eta_1)}{(p_{1} - k_1)^{2} - m_q^{2} + i\epsilon} i\Gamma_{n1} 
  u(p_1,\lambda_1)\frac{1}{2}\dfrac{dk_{1+} dk_{1-} d^2k_{1\perp}}{(2\pi)^4}, 
\end{eqnarray}
where we sum over the initial helicity ($\eta_1$) of the quark of mass $m_q$ before being struck by the incoming photon and the final helicity ($\eta_{1f}$) of the quark that recombines to form the final state proton and express the numerator of the spectator propagator as $S(k_1) = \sum_{s}\Psi_s(k_1)\bar{\Psi}_s(k_1)$.
We expand the denominator $(p_{1f} - k_1)^2 - m_q^2$ in Eq.~(\ref{N1start}) as follows:
\begin{eqnarray}
(p_{1f} - k_{1})^2 - m_q^2 &&= (p_{1f} - k_{1})_{+} (p_{1f} - k_{1})_{-} - (p_{1f} - k_1)_{\perp}^{2} - m^2 \nonumber \\
&& = (p_{1f+} - k_{1+}) (p_{1f-} - k_{1-}) - (p_{1f} - k_{1})_{\perp}^{2} - m^2,
\end{eqnarray}
where, using $k_{1-} = \frac{m_{s}^{2} + k_{1\perp}^2}{k_{1+}}$, $m_s$ as mass of the spectator system and $ p_{1f-} = \frac{m_N^2 + p_{1f\perp}^2}{p_{1f+}}$, we obtain:
\begin{eqnarray}
 (p_{1f} - k_{1})^2 - m_q^2 &&= (p_{1f+} - k_{1+}) \Bigg( \frac{m_N^2 + p_{1f\perp}^2}{p_{1+}} - \frac{m_{s}^{2} +k_{1\perp}^2}{k_{1+}} \Bigg) - (p_1 - k_1)_{\perp}^2 - m_q^2 \nonumber \\
&& = (p_{1f+} - p_{1f+}x_{s1}) \Bigg(\frac{m_N^2 + p_{1\perp}^{2}}{p_{1f+}} - \frac{m_{s}^2 + k_{1\perp}^2}{x_{s1}p_{1f+}}\Bigg) - (p_{1f}-k_1)_{\perp}^2 - m_q^2 \nonumber \\
&& = (1 - x_{s1}) \Bigg(m_N^2 + p_{1f\perp}^2 - \frac{m_{s}^2 + k_{1\perp}^2}{x_{s1}} \Bigg) - (p_{1f} - k_1)_{\perp}^2 - m_q^2 \nonumber \\
&& = (1 - x_{s1}) \Bigg(m_N^2 + p_{1f\perp}^2 - \frac{m_{s}^2 + k_{1\perp}^2}{x_{s1}} - \frac{m_q^2 + (p_{1f} - k_1)_{\perp}^2}{1 - x_{s1}} \Bigg),
\label{N1denominatorFirst} 
\end{eqnarray}
where $x_{s1}=\frac{k_{1+}}{p_{1f+}}$, is interpreted as the momentum fraction of the final nucleon ``1'' carried by
the spectator quark system. A similar derivation allows us to express
\begin{equation}
(p_1 - k_1)^2 - m_q^2 = (1 - x_1)\Bigg( m_N^2 + p_{1\perp}^2 - \frac{m_{s}^2 + k_{1\perp}^2}{x_{1}} - \frac{m_q^2 + (p_1 - k_1)_{\perp}^2}{1-x_{1}} \Bigg),
\label{N1denominatorSecond}
\end{equation}
where $x_1 = \frac{k_{1+}}{p_{1+}}$, is interpreted as the momentum fraction of the initial nucleon ``1'' carried by the spectator quark system. Performing the  $dk_{1-}$ integration at the $k_{1-}$ pole value of the spectator system  allows us to introduce a single quark wave function of the nucleon in the following form:
\begin{equation} \label{QuarkWF}
\Psi_n^{\lambda;\eta}(X, k_\perp, p_\perp) = \frac{\bar{u}_{q}(p-k,\eta) \bar{\Psi}_s(k)}{m_N^2 + p_\perp^2 - \frac{m_s^2 + k_{\perp}^2}{X} - \frac{m_q^2 +  (p - k)_{\perp}^2 }{1-X}}\Gamma_n u(p,\lambda),
\end{equation}
which describes the probability amplitude of finding  a quark with helicity $\eta$ and momentum fraction $1-x$ in the $\lambda$ helicity nucleon with momentum $p$.
With this definition of quark wave function of the nucleon, for the $N1$ part, we obtain:
\begin{eqnarray} \label{N1withQuarkWF}
&N1& = i\sum_{\substack{ \lambda_1 \\ \eta_{1f}, \eta_1}} \int	\frac{\Psi_{n1f}^{\dagger \lambda_{1f};\eta_{1f}} (x_{s1},k_{1\perp}, p_{1f\perp})}{1 - x_{s1}} 
\bar{u}_q(p_{1f} - k_1,\eta_{1f}) u_q(p_1 - k_1,\eta_1) \times \nonumber \\ 
&& \Big[ -igT_{c}^\beta \gamma_{\mu}\Big]\frac{\Psi_{n1}^{\lambda_1;\eta_1}(x_1, k_{1\perp},p_{1\perp}) }{1 - x_1} \frac{1}{2}\dfrac{dx_1}{x_1} \dfrac{d^2k_{1\perp}}{(2\pi)^3}.
\end{eqnarray}
Performing  similar calculations for the $N2$ part of  Eq.~(\ref{StartDerivationAmplitude}), we obtain:
\begin{eqnarray} \label{N2}
&N2&=-\sum_{\substack{\lambda_{2f},\lambda_{2}' \\ \eta_{2f}, \eta_{2}'}}\int \frac{\Psi_{n2f}^{\dagger \lambda_{2f};\eta_{2f}}(x_{s2},k_{2\perp}, p_{2f\perp})}{1-x_{s2}} \bar{u}_q(p_{2f}-k_2,\eta_{2f})u_q(p_{2}'-k_2,\eta_{2}') \times \nonumber \\
&&\frac{\Psi_{n2'}^{\lambda_{2}';\eta_{2}'}(x_{2}',k_{2\perp}, p_{2\perp}')}{1-x_{2}'} 
 \frac{1}{2} \dfrac{dx_{2}'}{x_{2}'} \dfrac{d^2k_{2\perp}}{(2\pi)^3},
\end{eqnarray}
where $x_{s2} = \frac{k_{2+}}{p_{2f+}}$ and $x_{2}' = \frac{k_{2+}}{p_{2+}'}$ and are defined same way as $x_{s1}$ and $x_1$.

Substituting  Eqs.~(\ref{A1withDeuteronHe3wfs}), (\ref{A2part}), (\ref{N1withQuarkWF}) and (\ref{N2}) into Eq.~(\ref{StartDerivationAmplitude}),we have the following expression for the scattering amplitude: \vspace{0em}
\begin{eqnarray} \label{amp_beforePhotonQuarkInteraction}
&&\mathcal{M}^{\lambda_{df},\lambda_{1f};\lambda_{^3\text{He}},h}=  \sum_{\substack{(\lambda_{2f})(\lambda_{2}',\lambda_{3}') (\lambda_{d}) \\ (\lambda_1,\lambda_2,\lambda_3) \\ (\eta_{1f}, \eta_{2f}) (\eta_1, \eta_{2}' ) }}  \int  \frac{{\Psi_{d}^{\dagger \lambda_{df}:\lambda_{3}', \lambda_{2f}}}
(\alpha_{2f}/\gamma_{d},p_{2\perp},\alpha_{3}'/\gamma_{d},p_{3\perp}')}{1-\alpha_{3}'/\gamma_{d}} \Bigg
\{ \frac{\Psi_{n2f}^{\dagger \lambda_{2f};\eta_{2f}}(x_{s2}, k_{2\perp}, p_{2f\perp})}{1-x_{s2}} \times \nonumber \\
&& \bar{u}_q(p_{2f}-k_2, \eta_{2f}) [-igT_c^\alpha \gamma_\nu]  \Big[\frac{i(\slashed{p}_1 + \slashed{q} - \slashed{k_1}+m_q)}
{(p_1 - k_1 + q)^2 -m_q^2 + i\epsilon} \Big] [-ie\epsilon^\mu \gamma\mu] u_q(p_1 - k_1, \eta_1) \times \nonumber \\
&& \frac{\Psi_{n1}^{\lambda_1;\eta_1}(x_1, k_{1\perp},p_{1\perp}) }{1 - x_1} \Bigg\}_{1} 
 \Bigg\{ \frac{\Psi_{n1f}^{\dagger \lambda_{1f};\eta_{1f}} (x_{s1},k_{1\perp}, p_{1f\perp})}{1 - x_{s1}} \bar{u}_q(p_{1f}-k_1, \eta_{1f}) 
 [-igT_c^\beta\gamma_\mu] u_q(p_{2}'-k_2,\eta_{2}')  \times \nonumber \\
 && 
 \frac{\Psi_{n2'}^{\lambda_{2}';\eta_{2}'}(x_{2}' ,k_{2\perp}, p_{2\perp}')}{1-x_{2}'} \Bigg\}_{2} G^{\mu\nu}(r)  
 \frac{\Psi_d^{\lambda_{d}:\lambda_{2}', \lambda_{3}'}(\alpha_{3}',p_{d\perp},p_{3\perp}')}{1-\alpha_{3}'} \frac{\Psi_{_{d}}^{\dagger \lambda_{d}:\lambda_{2}, \lambda_{3}}(\alpha_{3},p_{_{3\perp}},p_{_{d\perp}})}{(1-\alpha_{3})} \times \nonumber \\
&& \frac{\Psi_{_{^3\text{He}}}^{^{\lambda_{^3\text{He}}}}(\beta_1,\lambda_1,p_{_{1\perp}},\beta_2,p_{_{2\perp}}\lambda_2,\lambda_3)}{\beta_1}  \frac{d \beta_{d}}{\beta_d} \frac{d^2p_{d\perp}}{2(2\pi)^3} \frac{d \beta_{3}}{\beta_3} 
  \frac{d^2p_{3\perp}}{2(2\pi)^3}
\frac{d \alpha_{3}'}{\alpha_{3}'} \frac{d^2p_{3\perp}'}{2(2\pi)^3}  \frac{dx_1}{x_1}\frac{d^2k_{1\perp}}{2(2\pi)^3} \frac{dx_{2}'}{x_{2}'}\frac{d^2k_{2\perp}}{2(2\pi)^3}.
\end{eqnarray}
The scattering amplitude in Eq.~(\ref{amp_beforePhotonQuarkInteraction}) can be described by the following blocks:
\begin{itemize}
\item In the initial state, the $^3\text{He}$ wave function describes the transition of the 
$^3\text{He}$ nucleus with helicity $\lambda_{^3\text{He}}$ to the three nucleon 
intermediate state  with helicities $\lambda_1, \lambda_2 \text{ and } \lambda_3$. The nucleons ``2" and ``3" combine to form an intermediate deuteron, 
which is described by the deuteron wave function.
\item The terms in $\{ ... \}_1$ describe the knocking out of a quark with helicity $\eta_1$ from the proton ``1" by the  photon, with helicity $h$. 
The struck quark then interchanges with a quark from one of the nucleons in the intermediate deuteron state  recombining into the nucleon with helicity 
$\lambda_{2f}$. This nucleon then combines with the nucleon with helicity $\lambda_3$ and produces the final $\lambda_{df}$ helicity deuteron.
\item The terms in $\{ ...\}_2$ describe the emerging of a quark with helicity $\eta_{2}'$  from the $\lambda_{2}'$ helicity nucleon, 
which then interacts with the knocked out quark by exchanging gluon and  producing a quark with helicity $\eta_{1f}$. The $\eta_{1f}$-helicity quark then combines 
with the spectator quarks and produces a final nucleon with helicity $\lambda_{1f}$.
\end{itemize}

%% file: 2/HardKernel.tex
\section{Hard scattering kernel}
In Eq.~(\ref{amp_beforePhotonQuarkInteraction}), the expression in $\{ \}_1 \{ \}_2 G^{\mu \nu}(r)$ describes the hard photon-quark interaction followed by a quark interchange through the gluon exchange. We first study the propagator of the struck quark, $\frac{i(\slashed{p}_1 + \slashed{q} - \slashed{k_1}+m_q)}
{(p_1 - k_1 + q)^2 -m_q^2 + i\epsilon}$.
The denominator in the struck quark propagator can be expressed as \vspace{0em}
\begin{eqnarray}
\label{denom_struckQuarkProp}
(p_1 - k_1 + q)^2 - m_q^2 &&= (p_{1+} - p_{1+} x_1)(p_{1-} - k_{1-} + q_-) - (p_1 - k_1)_{\perp}^{2} - m_q^2 \nonumber \\
&& =p_{1+}(1 - x_1)\Bigg( \frac{m_N^2 + p_{1\perp}^2}{p_{1+}} - \frac{m_{s}^{2} + k_{1\perp}^{2}}{k_{1+}} + \sqrt{s^{\prime}_{_{^3\text{He}}}} \Bigg) - (p_{1\perp} - k_{1\perp})^2 - m_{q}^{2}  \nonumber \\
&& =\beta_1 \sqrt{s^{\prime}_{_{^3\text{He}}}} ( 1 - x_1) \Bigg( \frac{m_N^2 + p_{1\perp}^2}{\beta_1 \sqrt{s^{\prime}_{_{^3\text{He}}}}} - \frac{m_{s}^{2} + k_{1\perp}^{2}}{x_1 \beta_1 \sqrt{s^{\prime}_{_{^3\text{He}}}}} + \sqrt{s^{\prime}_{_{^3\text{He}}}} \Big) - (p_{1\perp} - k_{1\perp})^2 - m_{q}^{2} \nonumber \\
&& = \sqrt{s^{\prime}_{_{^3\text{He}}}}(1 - x_1) \Bigg( \frac{m_N^2 + p_{1\perp}^2}{\sqrt{s^{\prime}_{_{^3\text{He}}}}} - \frac{m_{s}^{2} + k_{1\perp}^{2}}{x_1\sqrt{s^{\prime}_{_{^3\text{He}}}}} + \beta_1\sqrt{s^{\prime}_{_{^3\text{He}}}} \Bigg) - (p_{1\perp} - k_{1\perp})^2 - m_{q}^{2}  \nonumber \\
&& = s^{\prime}_{_{^3\text{He}}} (1 - x_1)\Bigg( \frac{m_N^2 + p_{1\perp}^2}{s^{\prime}_{_{^3\text{He}}}} - \frac{m_{s}^{2} + k_{1\perp}^2}{x_1 s^{\prime}_{_{^3\text{He}}}} + \beta_1 - \frac{m_{q}^{2} +(p_{1\perp} - k_{1\perp})^2}{s^{\prime}_{_{^3\text{He}}}(1-x_1)} \Bigg) \nonumber \\
&& = s^{\prime}_{_{^3\text{He}}} (1 - x_1)(\beta_1 - \beta_s),
\end{eqnarray}
where 
\begin{equation}
\label{PoleValueofPropagator}
\beta_s = -\frac{1}{s'_{^3\text{He}}}\Bigg(m_N^2 + p_{1\perp}^2 - \frac{m_s^2 + k_{1\perp}^2}{x_1} - \frac{m_q^2 + (p_{1\perp} - k_{1\perp})^2 }{1-x_1}\Bigg).
\end{equation}
Using the sum rule relation ($\slashed{p}+m = \sum\limits_\lambda u(p,\lambda)\bar u(p,\lambda)$) for the numerator of the struck quark propagator together with 
Eq.~(\ref{denom_struckQuarkProp}), we can rewrite Eq.~(\ref{amp_beforePhotonQuarkInteraction}) as follows:

\begin{eqnarray} \label{amplitudeBeforePhotonPolarization}
&&\mathcal{M}^{\lambda_{df},\lambda_{1f};\lambda_{^3\text{He}},h}= \sum_{\substack{(\lambda_{2f})(\lambda_{2}',\lambda_{3}') (\lambda_{d}) \\ (\lambda_1,\lambda_2,\lambda_3) \\ (\eta_{1f}, \eta_{2f}) (\eta_1, \eta_{2}' ) (\eta_{q1})}} 
 \int  \frac{{\Psi_{d}^{\dagger \lambda_{df}:\lambda_{3}', \lambda_{2f}}}(\alpha_{2f}/\gamma_{d},p_{2\perp},\alpha_{3}'/\gamma_{d},p_{3\perp}')}{1-\alpha_{3}'/\gamma_{d}} \times \nonumber \\ 
&&\Bigg\{ \frac{\Psi_{n2f}^{\dagger \lambda_{2f};\eta_{2f}}(x_{s2} ,k_{2\perp},p_{2f\perp})}{1-x_{s2}} \bar{u}_q(p_{2f}-k_2, \eta_{2f})
 [-igT_c^\alpha \gamma_\nu] 
 \Big[\frac{u_q(p_1 + q- k_1,\eta_{q1}) \bar{u}_q(p_1 + q - k_1,\eta_{q1}) }{s'_{^3\text{He}} (1 - x_1)(\beta_1 - \beta_s)} \Big]  \times \nonumber \\
 && [-ie\epsilon^\mu \gamma_\mu] u_q(p_1 - k_1, \eta_1)  
\frac{\Psi_{n1}^{\lambda_1;\eta_1}(x_1, k_{1\perp},p_{1\perp}) }{1 - x_1} \Bigg\}_{1} \Bigg\{ \frac{\Psi_{n1f}^{\dagger \lambda_{1f};\eta_{1f}} (x_{s1},k_{1\perp}, p_{1f\perp})}{1 - x_{s1}} \bar{u}_q(p_{1f}-k_1, \eta_{1f}) \times \nonumber \\
&&  [-igT_c^\beta\gamma_\mu] 
u_q(p_{2}'-k_2,\eta_{2}')
\frac{\Psi_{n2'}^{\lambda_{2}';\eta_{2}'}(x_{2}' ,k_{2\perp},p_{2\perp}')}{1-x_{2}'} \Bigg\}_{2} G^{\mu\nu}(r) \frac{\Psi_d^{\lambda_{d}:\lambda_{2}', \lambda_{3}'}(\alpha_{3}',p_{d\perp},p_{3\perp}')}{1-\alpha_{3}'}  \times \nonumber \\
&& \frac{\Psi_{_{d}}^{\dagger \lambda_{d}:\lambda_{2}, \lambda_{3}}(\alpha_{3},p_{_{3\perp}},p_{_{d\perp}})}{(1-\alpha_{3})} \frac{\Psi_{_{^3\text{He}}}^{^{\lambda_{^3\text{He}}}}(\beta_1,\lambda_1,p_{_{1\perp}},\beta_2,p_{_{2\perp}}\lambda_2,\lambda_3)}{\beta_1}  \frac{d \beta_{d}}{\beta_d} \frac{d^2p_{d\perp}}{2(2\pi)^3} \frac{d \beta_{3}}{\beta_3} \frac{d^2p_{3\perp}}{2(2\pi)^3}\frac{d \alpha_{3}'}{\alpha_{3}'} \frac{d^2p_{3\perp}'}{2(2\pi)^3} \times \nonumber \\
&& \frac{dx_1}{x_1}\frac{d^2k_{1\perp}}{2(2\pi)^3} \frac{dx_{2}'}{x_{2}'}\frac{d^2k_{2\perp}}{2(2\pi)^3}.
\end{eqnarray}
The sum rule for the numerator of the struck quark propagator is valid for on-shell spinors only. Our use of the sum rule 
is justified because of the fact of using the peaking approximation (to be discussed later) in evaluating  Eq.~(\ref{amplitudeBeforePhotonPolarization})  in which the denominator of the struck quark is estimated at its pole value.

\subsection{ Photon quark interaction}
We now consider the term:
\begin{equation}
\bar{u}_q(p_1-k_1+q,\eta_{q1})[-ie\epsilon_h^\mu \gamma^\mu]u_q(p_1-k_1,\eta_1) ,
\label{ugu}
\end{equation}
where the incoming photon with helicity $h$ is described by polarization vectors:  $\epsilon_{R/L} = \mp\sqrt{\frac{1}{2}} (\epsilon_1 \pm i\epsilon_2)$ for 
$h=1/(-1)$ respectively. Here $\epsilon_1 \equiv (1,0,0)$  and   $\epsilon_2 \equiv (0,1,0)$.
Using these definitions we express:
\begin{equation}
 -\epsilon_h^\mu \gamma^\mu  =  \epsilon^\perp \gamma^\perp = -\epsilon_R \gamma_{L} + \epsilon_L \gamma_{R},
\label{epsgamma}
\end{equation}
where $\gamma_{R/L} = \frac{\gamma_x \pm i\gamma_y}{\sqrt{2}}$. 
We also resolve  the spinor of the quark with spin $\alpha$ to the  $\pm$ helicity states as follows:
\begin{equation}
u(p,\alpha) = u^{+}(p,\alpha) + u^{-}(p,\alpha)= \frac{1}{2}(1 + \gamma^5) u(p,\alpha) + \frac{1}{2}(1 - \gamma^5) u(p,\alpha).
\end{equation}
Finally, in the reference frame of  Eq.~(\ref{InitialMomentaofPhotonHelium}) the light-cone four-momenta ($(p_+,p_-,p_\perp)$) of the initial and final quarks 
in Eq.~(\ref{ugu}),  in the massless limit, are:
\begin{eqnarray} \label{momenta_of_Quarks}
 &&\text{Initial Momentum: }p_1 - k_1 =\Big(\beta_1(1 - x_1)\sqrt{s'_{^3\text{He}}}, 0,0 \Big), \nonumber \\
 &&\text{ Final Momentum: }p_1 - k_1 + q = \Big( \beta_1 (1 - x_1)\sqrt{s'_{^3\text{He}}}, \sqrt{s'_{^3\text{He}}},0\Big),
\end{eqnarray}
where we use the the relations $q_+ = 0$, $p_{1+} = \beta_1 p_{^3\text{He}+}$ and $k_{1+} = x_1 p_{1+}$. Because of the finite $\beta_1\sim\frac{1}{3}$ and small 
$x_1\ll 1$ entering in the amplitude (see Sec.\ref{PeakApp}) we can also neglect the {``-''} component of 
the initial quark: $(p_1-k_1)_- \approx \frac{ (p_1-k_1)_{\perp}^2 + m_q^2}{\beta_1(1-x_1)\sqrt{s^\prime_{^3\text{He}}}}\sim 0$.

Using Eq.~(\ref{momenta_of_Quarks}) and above definitions of photon polarization, $\gamma$-matrices and quark helicity states, we find that in the massless quark limit, the only non-vanishing matrix elements of $\bar u \gamma_\pm u$ are:
\begin{eqnarray} \label{matrixelements}
&&\bar{u}^{-}_q(p_1-k_1+q,-\frac{1}{2}) \gamma_{+} u^{-}_q(p_1-k_1,-\frac{1}{2}) = -2\sqrt{2E_1E_2} \nonumber \\
&&\bar{u}^{+}_q(p_1-k_1+q, \frac{1}{2}) \gamma_{-} u^{+}_q(p_1-k_1, \frac{1}{2}) = 2\sqrt{2E_1E_2}, 
\end{eqnarray}
where $E_1 = \beta_1(1 - x_1)\frac{\sqrt{s'_{^3\text{He}}}}{2} \text{ and } E_2 = (1 -  \beta_1(1 - x_1))\frac{\sqrt{s'_{^3\text{He}}}} {2}$ are the initial and final energies of the struck quark respectively.

Using the above relations for Eq.~(\ref{ugu}) we obtain:  
\begin{equation} \label{matrix_elems}
\bar{u}_q (p_1 - k_1 + q,\eta_{q1}) [ie\epsilon_h^\perp \gamma^\perp]u_q(p_1 - k_1, \eta_{1}) = ie Q_i2 \sqrt{2E_1E_2}(-h) \delta^{\eta_{q1} h} \delta^{\eta_1 h},
\end{equation}
where $Q_i$ is the charge of the struck quark in units of $e$. The above result indicates that incoming  $h$- helicity  photon selects the quark with the same 
helicity ($h=\eta_1$) conserving it during the interaction ($h = \eta_{q_1}$).  

\subsection{ Peaking Approximation}
\label{PeakApp}

We now consider the $d\beta_d$ integration in Eq.~(\ref{amplitudeBeforePhotonPolarization})  noticing that,  $d\beta_d = d\beta_1$ and 
separating the pole and principal value parts in the propagator of the struck quark as follows: 
\begin{equation} \label{IntegralWithPVpart_again}
 \frac{1}{\beta_1 -{\beta_s} + i\epsilon} = -i\pi \delta(\beta_1-{\beta_s}) + \text{P.V.} \int \frac{d\beta_1}{\beta_1 - {\beta_s}}.
\end{equation}
Furthermore, we neglect by P.V. part of the propagator since its contribution comes from the high momentum part of the nuclear wave function 
$p_1\sim \sqrt{s^\prime_{^3\text{He}}}$ which is strongly suppressed\cite{gdpn}. The integration with the pole part of the propagator will 
fix the value of $\beta_1 = \beta_s$ and the latter in the massless quark limit and negligible transverse component of $\vec p_1$ can be expressed as follows:
\begin{equation} \label{beta_s_lim}
\beta_s = \frac{1}{s'_{^3\text{He}}}\Big[\frac{m_s^2(1-x_1) + k_{1\perp}^2}{x_1(1-x_1)} - m_N^2 \Big].
\end{equation}

Now, using the fact that $^3\text{He}$ wave function strongly peaks at $ \beta_1 = \frac{1}{3}$ (thus making $\beta_d = \frac{2}{3}$), one can estimate the ``peaking'' value of the amplitude in Eq.~(\ref{amplitudeBeforePhotonPolarization})  taking $\beta_s= \frac{1}{3}$.
The latter condition results in  $x_1 \rightarrow \frac{3(m_s^2 + k_{1\perp}^2)}{s'_{^3\text{He}}} \sim  0$ since $s'_{^3\text{He}}$ is very large in comparison with the transverse momentum $k_{1\perp}$ of the spectator system, which allows us to approximate  $(1-x_1) \approx 1$. With these approximations, we find that the initial and final energies of the struck quark become:
\begin{eqnarray}\label{stuck_quark_energy_simplified}
&&E_1 = \beta_1 (1-x_1)\frac{\sqrt{s'_{^3\text{He}}}}{2} = \frac{1}{3}\frac{\sqrt{s'_{^3\text{He}}}}{2}  \nonumber \\ 
&&E_2=\Big(1-\beta_1(1-x_1)\Big)\frac{\sqrt{s'_{^3\text{He}}}}{2}= \frac{2}{3}\frac{\sqrt{s'_{^3\text{He}}}}{2}.
\end{eqnarray}
\vspace{0em}
Using Eq.~(\ref{stuck_quark_energy_simplified}) in Eq.~(\ref{matrix_elems}) and setting $\beta_1 = 1/3$ everywhere for 
Eq.~(\ref{amplitudeBeforePhotonPolarization}) we obtain the following for the amplitude: \vspace{0em}
\begin{eqnarray} \label{AmpToIntroducePDamp}
&& \mathcal{M}^{\lambda_{df},\lambda_{1f};\lambda_{^3\text{He}},h}=  \frac{3}{4} (-h)\frac{1}{\sqrt{s'_{^3\text{He}}}} \sum_{i}eQ_i
 \sum_{\substack{(\lambda_{2f})(\lambda_{2}',\lambda_{3}') (\lambda_{d}) \\ (\lambda_1,\lambda_2,\lambda_3) \\ (\eta_{1f}, \eta_{2f}) (\eta_{2}' ) }} \int  \frac{{\Psi_{d}^{\dagger \lambda_{df}:\lambda_{3}',\lambda_{2f}}}(\alpha_{2f}/\gamma_{d},p_{2\perp},\alpha_{3}'/\gamma_{d},p_{3\perp}')}{1-\alpha_{3}'/\gamma_{d}}  \nonumber \\
 && \qquad \Bigg\{ \frac{\Psi_{n2f}^{\dagger \lambda_{2f};\eta_{2f}}(x_{s2},p_{2f\perp},k_{2\perp})}{1-x_{s2}}  
\bar{u}_q(p_{2f}-k_2, \eta_{2f})
[-igT_c^\alpha \gamma_\nu] 
 \Big[u_q(p_1+q-k_1,h)   \Big] \times \nonumber \\
&& \frac{\Psi_{n1}^{\lambda_1;h}(x_1, k_{1\perp},p_{1\perp}) }{1 - x_1} \Bigg\}_{1}
 \Bigg\{ \frac{\Psi_{n1f}^{\dagger \lambda_{1f};\eta_{1f}} (x_{s1},k_{1\perp}, p_{1f\perp})}{1 - x_{s1}}
 \bar{u}_q(p_{1f}-k_1, \eta_{1f}) [-igT_c^\beta\gamma_\mu] u_q(p_{2}'-k_2,\eta_{2}') \times \nonumber \\
&& 
 \frac{\Psi_{n2'}^{\lambda_{2}';\eta_{2}'}(x_{2}',p_{2\perp}',k_{2\perp})}{1-x_{2}'} \Bigg\}_{2} G^{\mu\nu}(r) 
 \frac{\Psi_d^{\lambda_{d}:\lambda_{2}',\lambda_{3}'}(\alpha_{3}',p_{d\perp},p_{3\perp}')}{1-\alpha_{3}'} 
 \frac{\Psi_{_{d}}^{\dagger \lambda_{d}:\lambda_{2},\lambda_{3}}(\alpha_{3},p_{_{3\perp}},p_{_{d\perp}})}{1-\alpha_{3}} \times \nonumber \\
 &&\Psi_{_{^3\text{He}}}^{^{\lambda_{^3\text{He}}}}(\beta_1=1/3,\lambda_1,p_{_{1\perp}},\beta_2,p_{_{2\perp}}\lambda_2,\lambda_3) 
  \frac{d^2p_{d\perp}}{(2\pi)^2} 
\frac{d \beta_{3}}{\beta_3} \frac{d^2p_{3\perp}}{2(2\pi)^3}\frac{d \alpha_{3}'}{\alpha_{3}'} \frac{d^2p_{3\perp}'}{2(2\pi)^3} \times \nonumber \\
&&  \frac{dx_1}{x_1}\frac{d^2k_{1\perp}}{2(2\pi)^3} \frac{dx_{2}'}{x_{2}'}\frac{d^2k_{2\perp}}{2(2\pi)^3}.
\end{eqnarray}

The above expression corresponds to the amplitude of Fig.~(\ref{diagram}).  To be able to calculate the total scattering amplitude of $\gamma ^3\text{He} \rightarrow pd$ scattering, we need to sum the multitude of similar diagrams representing all possible combinations of photon coupling to quarks in one of the protons followed by quark interchanges or possible multi-gluon exchanges between outgoing nucleons, thus producing final $pd$ system with large relative momentum. The latter rescattering is inherently nonperturbative. The same is true for the quark wave function of the nucleon which is largely unknown.  
The main idea behind HRM is that, instead of calculating all the amplitudes explicitly,  we note that the hard kernel in 
Eq.~(\ref{AmpToIntroducePDamp}), 
$ \{\cdots \}_1 \{\cdots\}_2$,  together with the gluon propagator is similar to that of the  hard $pd \rightarrow pd$ scattering, as shown in Appendix A. 

%% file: 2/including_pdpd_amp.tex
\section{Including $pd \rightarrow pd$ amplitude}
\label{sec_four}
The amplitude of $pd \rightarrow pd$ scattering, given by Eq.~(\ref{pd_amplitude}) is derived in the $pd$ center of mass reference frame, in which the final momenta $p_{1f}$ and $p_{df}$ are chosen to be  the same as in reaction (\ref{reaction}).
As a result, the  $pd\rightarrow pd$ amplitude  is defined at the same invariant energy $s = (p_{1f} + p_{df})^2$ as  in Eq.~(\ref{mandelstams}), but at a different invariant momentum transfer defined as $t_{pd} = (p_{df} - p_{d})^2$. 
For the  further derivation, it is important to observe that within the peaking approximation, the momentum transfer entering in the rescattering part of the amplitude in Eq.~(\ref{AmpToIntroducePDamp}) is approximately equal to $t_{pd}$:
\begin{equation} 
t_N  =  t_{pd}  \approx   (p_{df} - p_d)^2,
\label{t_N}
\end{equation}
where $p_d$ is the four-momentum of the deuteron in the intermediate state of the reaction (Fig.\ref{diagram}).

Furthermore, because $q_+=0$, the spinor $u_q(p_1-k_1 + q,h)$ in Eq.~(\ref{AmpToIntroducePDamp}) is defined at the same 
momentum fraction $1-x_1$  and transverse momentum as the spinor $u_q(p_1 - k_1)$ in Eq.~(\ref{pd_amplitude}). 
The final step that allows us to replace the quark-interchange part of the Eq.~(\ref{AmpToIntroducePDamp}) by  the 
$pd\rightarrow pd$ amplitude is the observation that due to  the condition of  $\beta_1 = \beta_s \approx \frac{1}{3}$,  it follows from Eq.~(\ref{PoleValueofPropagator}) that the momentum fraction of the struck quark $1-x_1 \sim 1 - \frac{m_s^2}{ s^\prime_{^3\text{He}}} \sim 1$, which justifies the additional assumption according to which 
the helicity of  the struck quark is the same as the nucleon's from which it originates, i.e. $\eta_1 = \lambda_1$.  
With this assumption one can sum 
over $\eta_1$ in Eq.(\ref{pd_amplitude}), which allows us to substitute it into Eq.(\ref{AmpToIntroducePDamp}), thus yielding
\begin{eqnarray} \label{simplifiedAmplitude}
&\mathcal{M}^{\lambda_{df},\lambda_{1f};\lambda_{^3\text{He}},h} =&  \frac{3}{4}\frac{1}{\sqrt{s'_{^3\text{He}}}} \sum_{i}eQ_i (h)
 \sum_{\substack{\lambda_{d} \\  \lambda_2,\lambda_3 }} \int  \mathcal{M}_{pd}^{\lambda_{df},\lambda_{1f} ;\lambda_{d},h} (s, t_N) \frac{\Psi_{_{d}}^{\dagger \lambda_{d}:\lambda_2,\lambda_3}(\alpha_{3},p_{_{3\perp}},\beta_d,p_{_{d\perp}})}{1-\alpha_{3}} \times \nonumber \\
&& \qquad \qquad \Psi_{_{^3\text{He}}}^{{\lambda_{^3\text{He}}}: h, \lambda_2, \lambda_3}(\beta_1=1/3,  p_{_{1\perp}},\beta_2,p_{_{2\perp}})
\frac{d^2p_{d\perp}}{(2\pi)^2}\frac{d \beta_3}{\beta_{3}} \frac{d^2p_{3\perp}}{2(2\pi)^3}.
\end{eqnarray}

We can further simplify Eq.~(\ref{simplifiedAmplitude}) using the fact that the  momentum transfer in the $pd\rightarrow pd$ scattering amplitude significantly exceeds the momenta of bound nucleons in the center of mass of the nucleus.
As a result, we can factorize the  $pd\rightarrow pd$ amplitude from the integral in Eq.~(\ref{simplifiedAmplitude}) at $t_{pd}$ approximated as
\begin{equation}
t_{pd} \approx  (p_{df} - m_d)^2 = (p_{1f} - (m_N+q))^2,
\label{tpd_fact}
\end{equation}
resulting in:
\begin{eqnarray} 
\mathcal{M}^{\lambda_{df},\lambda_{1f};\lambda_{^3\text{He}},h} &=  &  \frac{3}{4} \frac{1}{\sqrt{s'_{^3\text{He}}}} \sum_{i} \sum_{\substack{\lambda_{d} \\  }}eQ_i (h) \mathcal{M}_{pd}^{\lambda_{df},\lambda_{1f} ;\lambda_{d},h} (s, t_{pd})  \nonumber \\
& &   \times  \int   \Psi_{^3\text{He}/d}^{ \lambda_{^3\text{He}}:\lambda_1,\lambda_d}(\beta_1 = 1/3, p_{1\perp})  \frac{d^2p_{1\perp}}{(2\pi)^2},
\label{pdAmplitudeFactored}
\end{eqnarray}
where we introduced  the light-front nuclear transition wave function as:
\begin{eqnarray}
\Psi^{\lambda_{^3\text{He}}:\lambda_1, \lambda_{d}}_{^3\text{He}/d}(\beta_1, p_{1\perp}) &=  & \sum_{\lambda_2, \lambda_3} \int \frac{\Psi_{_{d}}^{ \dagger\lambda_{d}: \lambda_2, \lambda_3}  (\alpha_3, p_{3\perp}, \beta_d, p_{d\perp} )} {2 (1 - \alpha_3)} \Psi_{_{^3\text{He}}}^{{\lambda_{^3\text{He}}}: \lambda_1, \lambda_2, \lambda_3}(\beta_1, p_{1\perp}, \beta_2, p_{2\perp}) \nonumber \\
& & \ \ \ \ \ \ \ \  \times  \frac{d\beta_3}{\beta_3}\frac{d^2p_{3\perp}}{2(2\pi)^3}.	
 \label{LCNTF}
\end{eqnarray}
The above defined light-front nuclear transition wave function gives the probability amplitude of the $^3\text{He}$  nucleus 
transitioning to a proton and  deuteron with respective momenta  $p_1$ and $p_d$ and  helicities $\lambda_1$ and $\lambda_d$.

In Eq.~(\ref{pdAmplitudeFactored}), we sum over all the valence quarks in the bound proton that interact with incoming photon. 
To calculate such a sum, an underlying model for hard nucleon interaction based on the explicit quark degrees of freedom is needed. 
Such a model will allow us to simplify further the amplitude of  Eq.~(\ref{pdAmplitudeFactored}) representing it 
through the product of an effective charge $Q_{eff}$ that is probed by the incoming photon in the reaction and 
the hard $pd \rightarrow pd$ amplitude in the form
\begin{equation} 
\mathcal{M}^{\lambda_{df},\lambda_{1f};\lambda_{^3\text{He}},h} =   \frac{3}{4} \frac{eQ_{eff}(h)}{\sqrt{s'_{^3\text{He}}}}  \sum_{\substack{\lambda_{d} \\  }} \mathcal{M}_{pd}^{\lambda_{df},\lambda_{1f} ;\lambda_{d},h} (s, t_{pd}) 
  \int   \Psi_{^3\text{He}/d}^{ \lambda_{^3\text{He}}:\lambda_1,\lambda_d}
(\beta_1 = 1/3, p_{1\perp}) 
 \frac{d^2p_{1\perp}}{(2\pi)^2}.
\label{Amplitude}
\end{equation}

%% file: 2/finalCSDerivation.tex
\section{The differential cross section}
\label{sec_CompleteDerivationCS}
The  differential cross section of  reaction (\ref{reaction}) can be presented in the standard form
\begin{equation} \label{CS}
\frac{d\sigma}{dt} = \frac{1}{16\pi} \frac{1}{s'^2_{^3\text{He}}} |\overline{\mathcal{M}}|^2,
\end{equation}
where for the case of unpolarized scattering,
\begin{equation} \label{squard_averaged_amp}
|\overline{\mathcal{M}}|^2 = \frac{1}{2} \frac{1}{2} \sum_{\lambda_{^3\text{He}}, h} \sum_{\lambda_{df}, \lambda_{1f}} \left|\mathcal{M}^{\lambda_{df},\lambda_{1f};\lambda_{^3\text{He}},h} \right|^2.
\end{equation}
The squared amplitude is summed by the final helicities and averaged by the helicities of $^3\text{He}$ and incoming photon.
The factorization approximation of Eq.~(\ref{Amplitude}) allows to express Eq.~(\ref{squard_averaged_amp}) through the
convolution of the averaged square of $\mathcal{M}_{pd}$ amplitude in the form
\begin{equation} \label{squaredAmpWithSquaredpdAmp}
|\overline{\mathcal{M}}|^2 = \frac{9}{16} \frac{e^2Q_{eff}^2}{s'_{^3\text{He}}} \frac{1}{2} |\overline{\mathcal{M}_{pd}}|^2 S_{^3\text{He}/d}(\beta_1 = 1/3),
\end{equation}
where
\begin{equation} \label{M_pd2def}
|\overline{\mathcal{M}_{pd}}|^2 = \frac{1}{3} \frac{1}{2} \sum_{\lambda_{df}, \lambda_{1f}; \lambda_d, \lambda_1} \left|\mathcal{M}_{pd}^{\lambda_{df}, \lambda_{1f}; \lambda_d, \lambda_1}(s, t_{pd})\right|^2,
\end{equation}
and the nuclear light-front transition spectral function is defined as:
\begin{equation} \label{LCTSF}
S_{^3\text{He}/d} (\beta_1) = \frac{1}{2} \sum_{\lambda_{^3\text{He}}; \lambda_1, \lambda_d} \left|\int \Psi_{^3\text{He}/d}^{\lambda_{^3\text{He}}:\lambda_1, \lambda_d} (\beta_1, p_{1\perp}) \frac{d^2p_{1\perp}}{(2\pi)^2} \right|^2.
\end{equation}
Substituting Eq.~(\ref{squaredAmpWithSquaredpdAmp}) into (\ref{CS}), we can express the differential cross section through the
differential cross section of elastic $pd \rightarrow pd$ scattering in the form:
\begin{equation} \label{differentialCS}
\frac{d\sigma}{dt} = \frac{9}{32} \frac{e^2Q_{eff}^2}{s'_{^3\text{He}}} \Big(\frac{s'_N}{s'_{^3\text{He}}}\Big) \frac{d\sigma_{pd}}{dt}(s, t_{pd}) S_{^3\text{He}/d}(\beta_1 = 1/3),
\end{equation}
where $s'_N = s - m_N^2$.

In the next chapter, a discussion on the numerical approximations to compute the differential cross section is presented.

%% file: 3/3.tex
\chapter{Numerical approximations and results}

In Chapter 3, a discussion on the calculation of light-front transition spectral function is presented. We then discuss the parametrization of the hard elastic $pd \rightarrow pd$ differential cross section which enters into the expression of the cross section of Eq.~(\ref{differentialCS}). The discussion on the calculation of the effective charge $Q_{eff}$ is presented, followed by the final form of the cross section which is used to calculate the cross section. The chapter is concluded by presenting the results of our calculation, where we compare our calculations with the existing experimental data.

\input{./3/LFspecFunction.tex}
\input{./3/pdpdCrossSection.tex}
\input{./3/Qeff.tex}
\input{./3/finalCS.tex}

%% file: 3/LFspecFunction.tex
\section{Calculation of light-front transition spectral function}
The calculation of the light-front  transition spectral function of Eq.~(\ref{LCTSF}) uses the peaking approximation, which maximizes the nuclear wave function's contribution to the 
scattering amplitude, $\beta_1 \approx \frac{1}{ 3}$ and $\beta_d \approx \frac{2}{3}$. These values of light-cone momentum 
fractions correspond to a small internal momenta of the nucleons in the nucleus. Additionally, since the deuteron wave 
function strongly peaks at small relative momenta between two spectator (``2" and ``3" in Fig.~(\ref{diagram}))  nucleons, 
the  integral in Eq.~(\ref{LCNTF}) is dominated at $\beta_3 \approx \frac{1}{3}$ and $\alpha_3\approx \frac{1}{2}$. 
This justifies the application of non-relativistic approximation in the calculation of the transition spectral function of Eq.~(\ref{LCTSF}).

In the non-relativistic limit, using the  boost invariance of the momentum fractions, $\beta_i$ ($i=1,2,3$),  we can relate 
them to the three-momenta  of the constituent nucleons in the 
lab frame of the nucleus as follows
\begin{equation}
\beta_i  = \frac{p_{i+}}{p_{^3\text{He}+}} \approx \frac{1}{3} + \frac{p^{lab}_{i,z}}{ 3 m_N}.
\label{beta_pz}
\end{equation}

Using above relations, we can approximate  $\frac{d\beta_3 }{\beta_3}\approx \frac{dp^{lab}_{3z}}{ m_N}$ and 
$\alpha_3\approx \frac{1}{2} + \frac{p_{2,z}}{2 m_N}$ in Eq.~(\ref{LCNTF}).
Introducing also  the relative three-momentum in the $2,3$ nucleon system as:
 \begin{equation}
\mathbf{p_{rel}} = \frac{1}{2}(\mathbf{ p^{\ lab}_3} - \mathbf{ p^{\ lab}_2)},
\end{equation}
\vspace{0em}
and using the relation between light-front and non-relativisitc nuclear wave functions in the small momentum limit (see Appendix B)\vspace{0em},:
\begin{equation}
\Psi^{LC}_A(\beta,p_\perp) = \frac{1}{\sqrt{A}}  \big(m_N 2(2\pi)^3 \big)^\frac{A-1}{ 2}\Psi_{A}^{NR}(\mathbf{p}),
\end{equation}
we can express the light-front nuclear transition wave function of Eq.~(\ref{LCNTF}) through the 
non-relativistic  $^3\text{He}$ to $d$ transition wave function as follows:
\begin{equation}
\Psi^{\lambda_{3He}:\lambda_1, \lambda_d}_{^3\text{He}/d}(\beta_1, p_{1\perp})  =
\sqrt{\frac{1}{6}} \sqrt{m_N 2(2\pi)^3} \cdot
\Psi^{\lambda_{3He}:\lambda_1, \lambda_{d}}_{^3\text{He}/d,NR}(\mathbf{p_1}),
\label{LCtoNRwft}
\end{equation}
where non-relativistic transition wave function is defined as:
\begin{equation}
\Psi^{\lambda_{^3\text{He}}:\lambda_1, \lambda_d}_{^3\text{He}/d,NR}(\mathbf{p_1}) 
= \sum_{\lambda_2, \lambda_3} \int 
\Psi_{_{d},NR}^{ \dagger\lambda_{d}: \lambda_2, \lambda_3}  (p_{rel}) 
\Psi_{_{^3\text{He}},NR}^{{\lambda_{^3\text{He}}}: \lambda_1, \lambda_2, \lambda_3}(p_{1}, p_{rel}) d^3p_{rel}.	
\label{NRtranswf}
\end{equation}
Using Eq.~(\ref{LCtoNRwft}), we  express the light-front spectral function through the non-relativistic counterpart in the form:
\begin{equation}\label{LCspectralNR}
S_{^3\text{He}/d} (\beta_1) = \frac{ m_N 2(2\pi)^3}{6} N_{pd} \ S_{^3\text{He}/d}^{NR} (p^{lab}_{1z}),
\end{equation}
where $\beta_1$ and $p_{1z}^{lab}$ are related according to Eq.~(\ref{beta_pz}) 
and $N_{pd}=2$ is the number of the effective $pd$ pairs.  The non-relativistic spectral function is 
defined as:
\begin{equation}
S_{^3\text{He}/d}^{NR} (p_{1z}) =  
\frac{1}{2} \sum_{\lambda_{^3\text{He}}; \lambda_1, \lambda_d} \left| 
\int \Psi_{^3\text{He}/d,NR}^{\lambda_{^3\text{He}}:\lambda_1, \lambda_d} (p_{1z}, p_{1\perp}) 
\frac{d^2p_{1\perp}}{(2\pi)^2}\right| ^2,
\end{equation}
where both the $^3\text{He}$ and $d$ wave functions are renormalized to unity.
In the above expressions, all the momenta  entering in the 
non-relativisitic wave functions are considered in the laboratory frame of the $^3\text{He}$ nucleus.

%% file: 3/pdpdCrossSection.tex
\section{Hard elastic $pd\rightarrow pd$ scattering cross section}
The hard  $ pd \rightarrow pd $ elastic scattering cross section entering in Eq.~(\ref{differentialCS}) is 
defined at the same invariant energy $s$ as the reaction (\ref{reaction}) but at different (from Eq.~(\ref{mandelstams}))
 invariant momentum transfer, $t_{pd}$ defined in Eq.~(\ref{tpd_fact}). Comparing Eqs.~(\ref{tpd_fact}) and (\ref{mandelstams})  $t_{pd}$ can be expressed through $t$ in the following form:
\begin{equation} \label{t_pd}
t_{pd} = \frac{1}{3} m_d^2 - \frac{2}{9} m_{^3\text{He}}^2 + \frac{2}{3} t.
\end{equation} and
\begin{equation} \label{cmangles}
\frac{1}{3} (1 - \cos\theta_{\text{cm}}) \approx  \frac{1}{2} (1 - \cos\theta_{\text{cm}}^*).
\end{equation}
\begin{figure}[ht]
\centering
\includegraphics[width = 12cm, height =8cm]{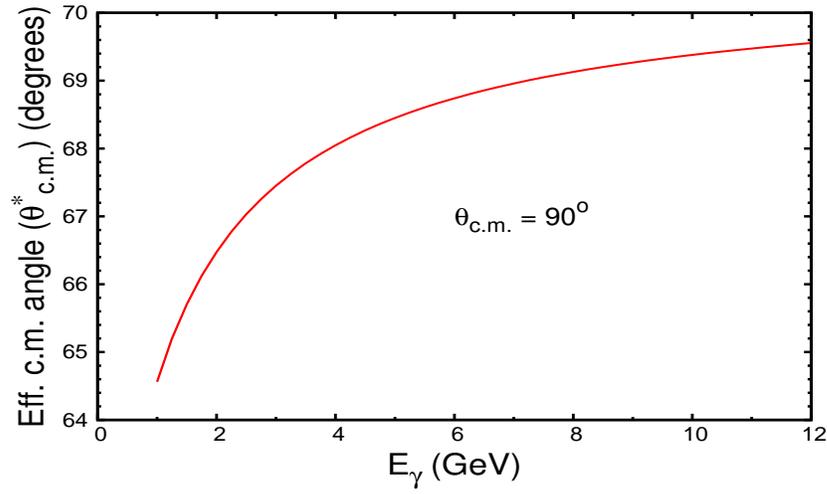}
\caption{Effective center of mass angle vs the incident photon energy.}
\label{eff_cm_angle}
\end{figure} 
As it follows from  the above equation for large momentum transfer, $|t_{pd}| < |t|$, therefore for the same 
$s$, the $pd\rightarrow pd$ 
scattering will take place at smaller angles in the $pd$ center of mass reference frame.  To evaluate this difference we 
introduce the $\theta_{\text{cm}}^*$ which represents the CM scattering angle for $pd\rightarrow pd$ reaction in the form:
\begin{equation}\label{t_pd_explicit}
t_{pd} = (p_{df} - p_{di})^2 = 2 (m_d^2 - E_{d,\text{cm}}^2)(1 - \cos\theta_{\text{cm}}^*), 
\end{equation} 
where $E_{d,\text{cm}}  = \frac{s+m_{d}^2 - m_N^2}{ 2\sqrt{s}}$. 
Then, comparing this equation with Eq.~(\ref{tInFull}) in  the asymptotic limit of 
high energies, we find that for the center of mass scattering angle $\theta_{\text{cm}} = 90^o$ for reaction (\ref{reaction}),  the asymptotic limit of
the effective center of mass scattering angle of  $pd\rightarrow pd $ scattering is  $\theta_{\text{cm}}^* \approx 70^o$. 
The dependence  of $\theta_{\text{cm}}^*$  at finite energies of incoming photon is shown in  Fig.~(\ref{eff_cm_angle}).  The figure indicates that for 
realistic comparison of HRM prediction with the data, one needs the $pd\rightarrow pd$ cross section for the range of center of mass scattering angles.

\begin{figure} [ht]
\begin{subfigure}{0.5\textwidth}
\centering
    \includegraphics[width=1.0\linewidth]{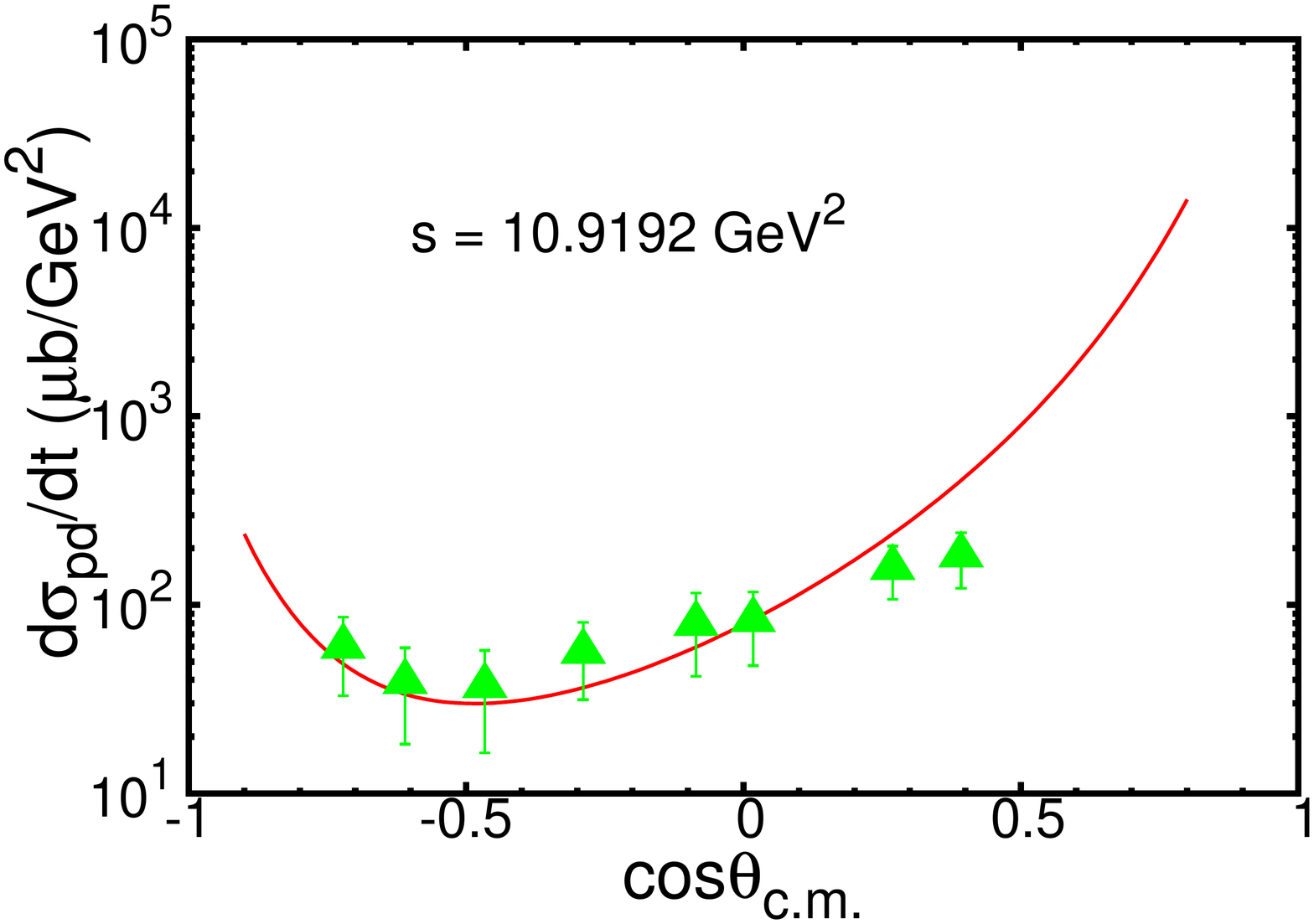}
    \caption{}
    \label{fig:pdfit1}
\end{subfigure}%
\begin{subfigure}{0.5\textwidth}
\centering
    \includegraphics[width=1.0\linewidth]{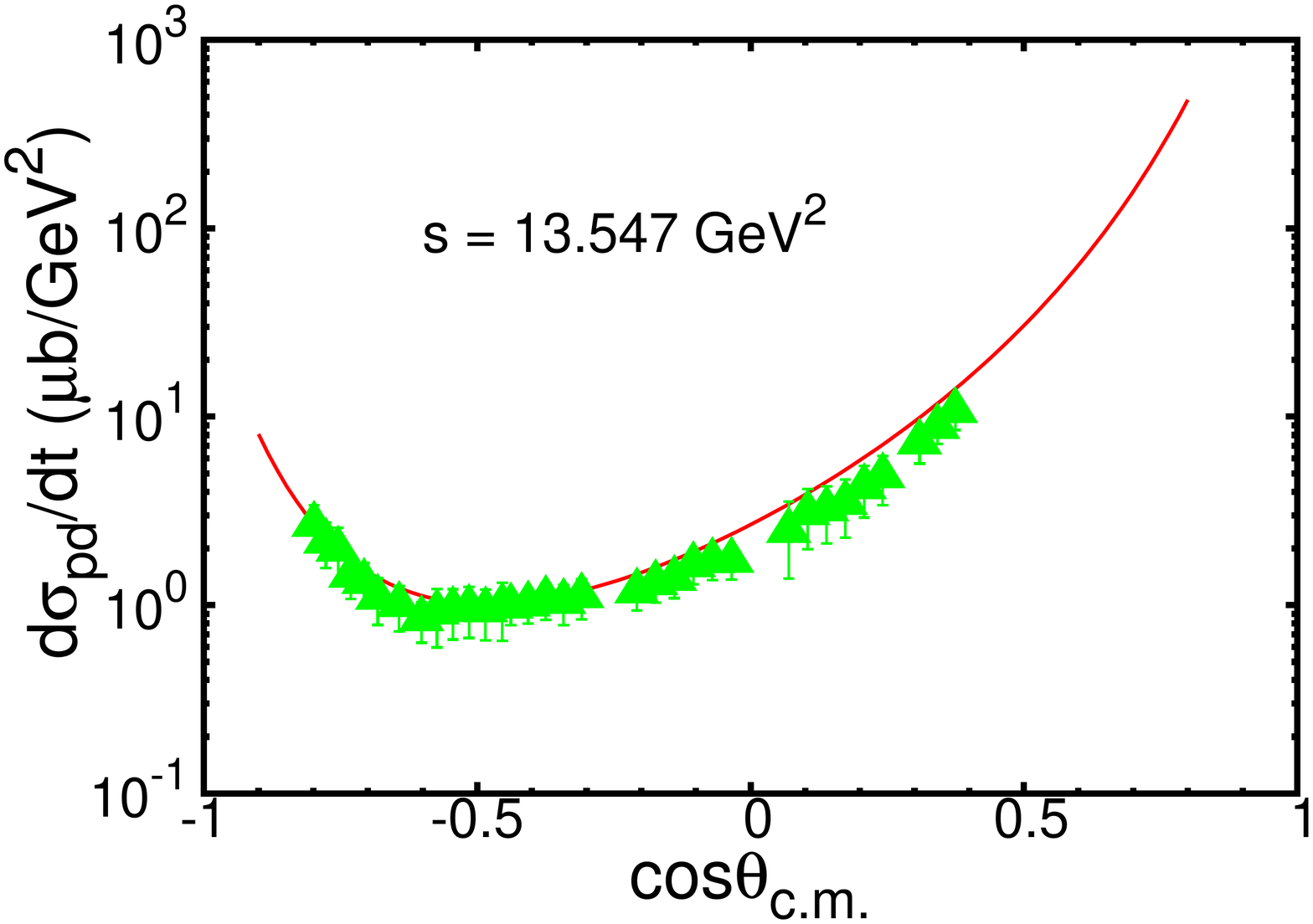}
    \caption{}
    \label{fig:pdfit2}
\end{subfigure}
\begin{subfigure}{0.5\textwidth}
\centering
    \includegraphics[width=1.0\linewidth]{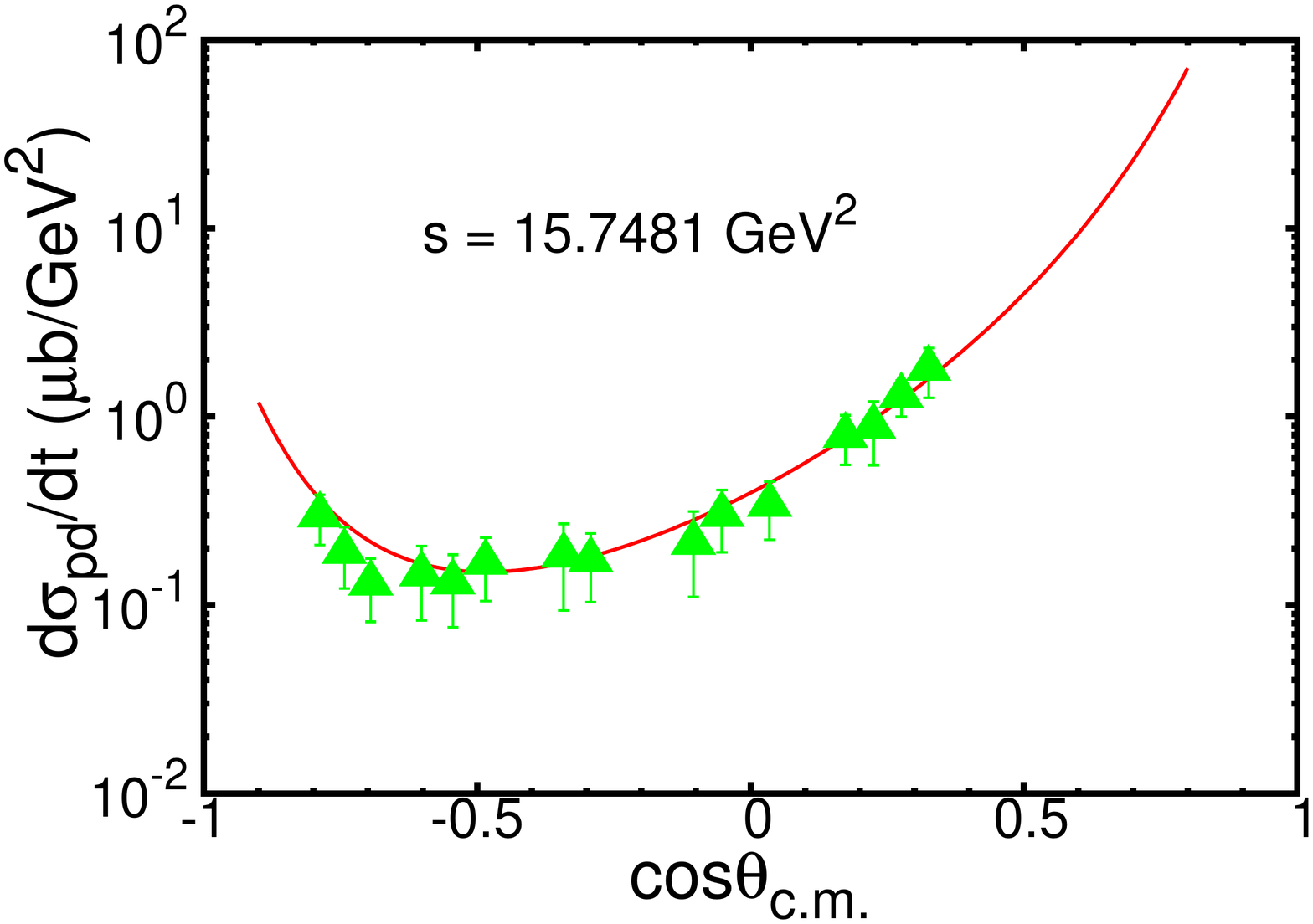}
    \caption{}
    \label{fig:pdfit3}
\end{subfigure}%
\begin{subfigure}{0.5\textwidth}
\centering
    \includegraphics[width=1.0\linewidth]{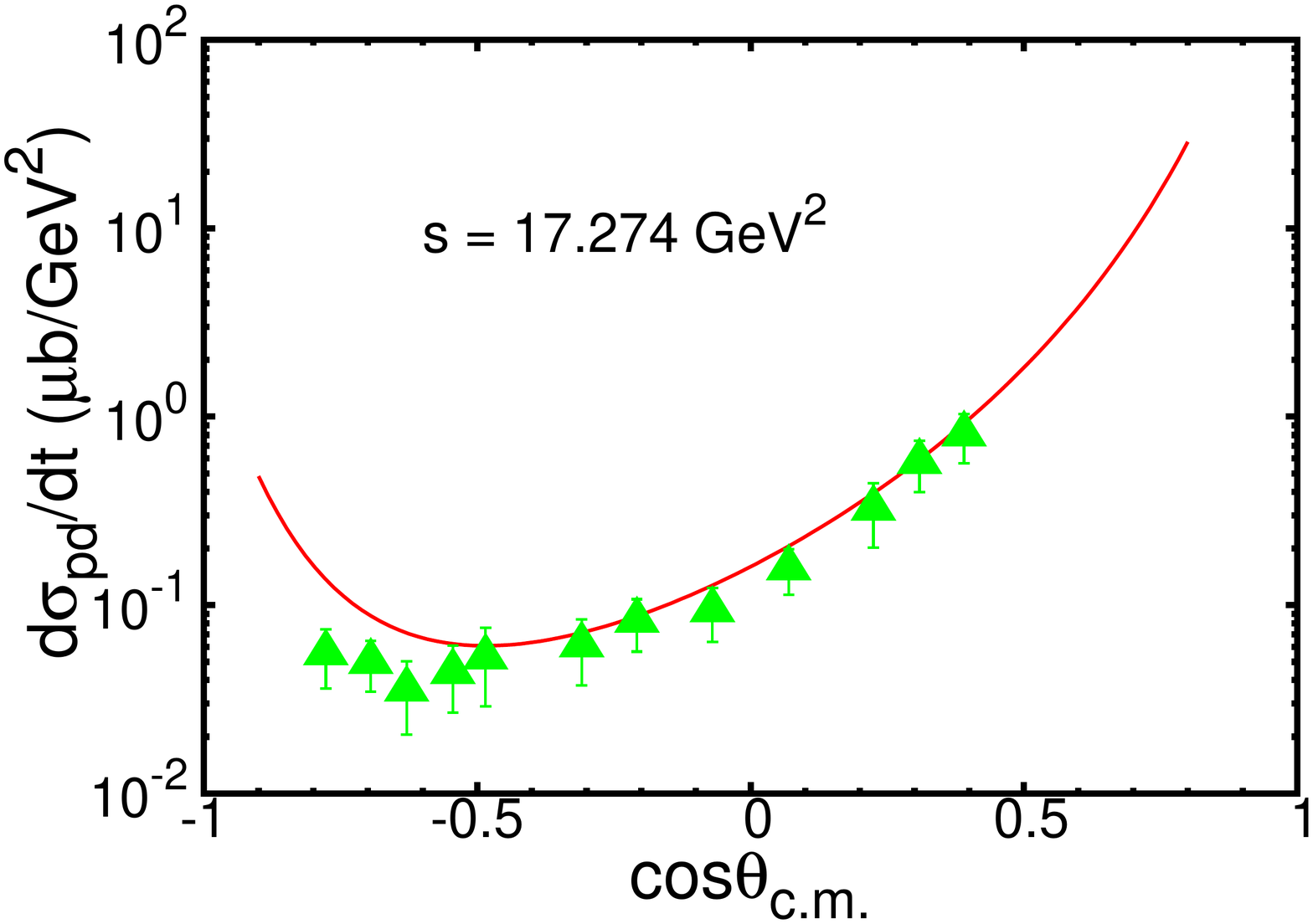}
    \caption{}
    \label{fig:pdfit4}
\end{subfigure}%
\caption{Fits of elastic $pd \rightarrow pd$ scattering cross section. The data in (a) are from \cite{800MeV} and 
in (b),(c) and (d) are from \cite{4.50GeV}. The curves are the fits to the  data obtained using Eq.~(\ref{pd_paramterization}) with fit parameters from  Table (\ref{table_fitparams}).}
\label{Figpdfits}
\end{figure}

To achieve this,  we  parametrized the existing experimental data on elastic $pd\rightarrow pd$ 
scattering \cite{1.0GeV,582MeV,641.3MeV,800MeV,4.50GeV} which covers the invariant energy range of 
($s \sim$ 9.5 GeV$^2$ - 17.3 GeV$^2$). 
The analytic form  used for  the parametrization of the $pd \rightarrow pd$ cross section is
\begin{equation} 
\label{pd_paramterization}
\frac{d\sigma_{pd}}{dt}(s,\cos\theta_{\text{cm}}^*) = \frac{1}{(s/10)^{16}}  \frac{ A(s) e^{B(\cos\theta_{\text{cm}}^*)} } {(1 - \cos^2\theta_{\text{cm}}^*)^3},
 \end{equation}  	
where $A(s) = Ce^{(a_1s + a_2s^2)}$, and  $B(x) =   b x + c x^2$, with the fit parameters given in Table~\ref{table_fitparams}.
The samples of fits obtained for the elastic $pd \rightarrow pd$ hard scattering are presented in Fig.~(\ref{Figpdfits}). \\ 

\begin{table}[h!]
\centering
\begin{tabular}{| c | c | c | c | c | } 
\hline
  $C$~($\mu$b GeV$^{30}$) & $a_1$~(GeV$^{-2}$) & $a_2$~(GeV$^{-4}$) & b &  c  \\ 
 \hline
  (9.72 $\pm$ 1.33)E+04  & -0.98 $\pm$ 0.05  & 0.04 $\pm$ 0.001  &  3.45 $\pm$ 0.02 &  -0.83 $\pm$ 0.05 \\ 
 \hline 
\end{tabular}
\caption{Fit Parameters for $p d \rightarrow p d$ cross sections.}
\label{table_fitparams}
\end{table}

The errors quoted in the table for the fitting parameters result in a  overall error in the $pd\to pd$ cross section 
on the level of 22-37\%.  Note that the  form of the ansatz used in Eq.~(\ref{pd_paramterization}) is in agreement with the 
energy and angular dependence following from the quark interchange mechanism of the $pd$ elastic scattering.  As a result
the ansatz is strictly valid for  large center of mass angles $|cos(\theta_{\text{cm}}^*)| \le 0.6$.  However, the 
fitting  procedure was extended beyond this angular range by introducing an additional function $e^{B(\cos\theta_{\text{cm}}^*)}$.

%% file: 3/Qeff.tex
\section{Estimation of the effective charge $Q_{eff}$}
To calculate the effective quark charge associated with the hard rescattering amplitude, we notice that from Eq.~(\ref{simplifiedAmplitude}) it follows that the $Q_{eff}$ 
should satisfy the following relation
\begin{equation}
 \sum\limits_{i \in p} Q_i\langle d'p' \mid M_{pd,i}\mid dp\rangle     =  Q_{eff}\langle d'p' |M_{pd}\mid dp\rangle,
\end{equation}
 where by $i$ we sum by  the quarks in the proton that was struck by incoming photon.  To use  the above equation, we need a specific model for $pd$ 
 elastic scattering which explicitly accounts for the underlying  quark degrees of freedom of  $pd$ scattering.  
 For such a model we use the quark-interchange 
 mechanism~(QIM) in hard $pd$ scattering.  The consideration of a quark-interchange mechanism is justified if one works in the regime in which the $pd$ elastic scattering exhibits  scaling in agreement with quark counting rule i.e. $s^{-16}$.  
 
 Similar to Refs.\cite{gdpn,gheppn,gdBB} within QIM, the $Q_{eff}$ can be estimated using the relation,
\begin{equation}
Q_{eff}= \frac{N_{uu}(Q_u) + N_{dd}(Q_d) + N_{ud}(Q_u+Q_d)}{ N_{uu} + N_{dd} + N_{ud}},
\label{QF}
\end{equation}
where $Q_i$ is the charge of the $u$ and $d$ valence quarks  in the proton $P$ and $N_{ii}$ represents 
the number of quark interchanges for $u$ and $d$ flavors   necessary to 
produce a given helicity $pd$ amplitude. We note that for the particular case of elastic $pd$ scattering, $N_{ud} = 0$ and 
one obtains $Q_{eff} = \frac{1}{3}$.

%% file: 3/finalCS.tex
\section{Results}
Substituting Eq.~(\ref{LCspectralNR}) into Eq.~(\ref{differentialCS})  and taking into accounts the above estimation of 
$Q_{eff}$, we arrive at the final expression for the differential cross section which will be used for 
the numerical estimates:
\begin{equation}
\frac{d\sigma}{dt}(s,t) = \frac{2 \pi^4 \alpha_{em}}{3 s'_{^3\text{He}}} \Big(\frac{s'_N}{s'_{^3\text{He}}}\Big) \frac{d\sigma_{pd}}{dt}(s, t_{pd}) \cdot m_N S^{NR}_{^3\text{He}/d}(p_{1z}=0),
\label{diffeq_b}
\end{equation}
where $\alpha_{em} = \frac{e^2}{4\pi}$ is the fine structure constant.
For the evaluation of the transition spectral function $S^{NR}_{^3\text{He}/d}$,  we use  the realistic $^3\text{He}$~\cite{Nogga:2002} and deuteron~\cite{Wiringa:1994wb}
wave functions  that use the  V18 potential\cite{Wiringa:1994wb} of NN interaction, which yields\cite{eheppnII}  $S^{NR}_{^3\text{He}/d}(p_{1z}=0) = 4.1\times 10^{-4}$~GeV. 
For the differential cross section of 
the large center of mass  elastic $pd\rightarrow pd$ scattering, $\frac{d\sigma_{pd}}{dt}(s, t_{pd})$, we use the parametrization  of Eq.~(\ref{pd_paramterization}) 
which covers the invariant energy range of up to $s=17.3$~GeV$^2$, corresponding to $E_\gamma = 1.67$~GeV for the reaction~(\ref{reaction}).  
\begin{figure}[h!]
\centering
\includegraphics[width = 12 cm, height = 8 cm]{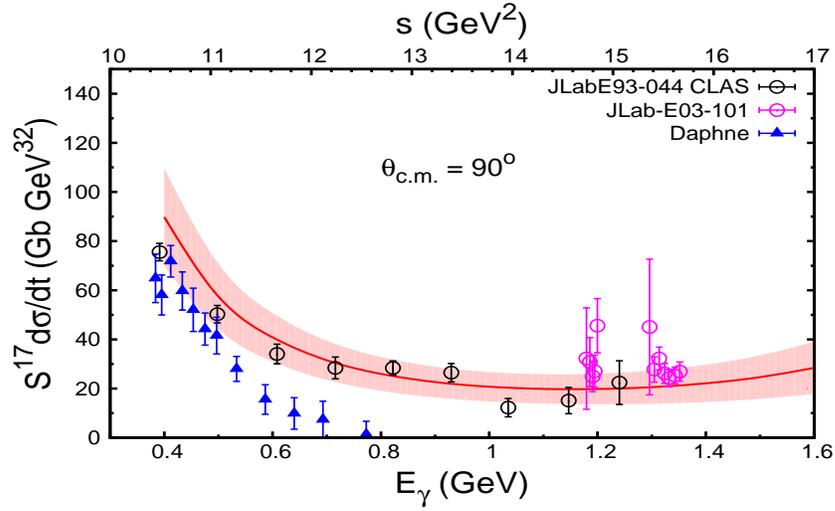}
\caption{Energy dependence of the differential cross section at $\theta_{\text{cm}} = 90^\circ$ scaled by the factor of $s^{17}$. The solid 
curve is the calculation according to Eq.~(\ref{diffeq_b}).  The experimental data are from Refs.\cite{DAPHNE,Pomerantz13,Berman}. See also discussion in Ref.\cite{Pomerantz13} on disagreement between DAPHNE and JLAB/CLAS data. }
\label{comp}
\end{figure}
In Fig.~(\ref{comp}), we present the comparison of  our calculation of the energy dependence of  the $s^{17}$ scaled differential cross section  
at $\theta_{\text{cm}} = 90^0$ with the data of Ref.\cite{Pomerantz13}. The shaded area represents the error due to the accuracy and above discussed fitting of the elastic $pd\rightarrow pd$ cross sections.

As the comparison shows,  Eq.~(\ref{diffeq_b}) describes surprisingly well the Jefferson Lab data considering the 
fact that the cross section between $E_{\gamma} = 0.4$~GeV and $E_{\gamma} = 1.3$~GeV drops by a factor of $\sim 4000$.
It is interesting that  the HRM model describes data reasonably well even for the range of 
$E_{\gamma} < 1$~GeV, for which the general conditions for the onset of QCD degrees of freedom is not satisfied.  This situation is specific to the HRM model in which there is another scale $t_{pd}$, the invariant momentum transfer in the hard rescattering amplitude. The $-t_{pd} > 1$ GeV$^2$  condition is necessary for factorization of the  hard scattering kernels from the soft nuclear parts.  As it follows from Eq.~(\ref{t_pd}), such a threshold for $t_{pd}$ is already reached for  incoming photon  energies of 0.7~GeV. What concerns to the photon energies below 0.7 GeV, then the qualitative agreement of the HRM model with the data is an indication  of the smooth transition from the hard to the soft regime of the interaction. 
 \vspace{0em} 
\begin{figure}[h!]
\centering
\includegraphics[width = 12 cm, height = 8 cm]{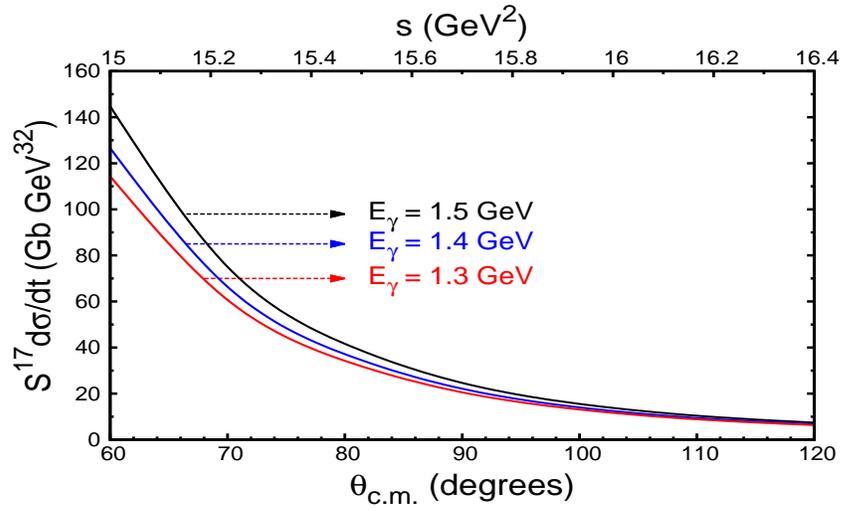}
\captionsetup{format = plain, justification = raggedright}
\caption{The angular dependence  of $s^{17}$ scaled differential cross section for incoming photon energies of 1.3~GeV, 1.4~GeV and 1.5~GeV.}
\label{ang_deps}
\end{figure} 
\vspace{0em}
The HRM model allows also to calculate the angular distribution of the differential cross section for fixed values of $s$. In Fig.~(\ref{ang_deps}), we present the prediction for the angular distribution of the energy scaled differential cross section at largest photon energies for which there are available data\cite{YI}. 
The interesting feature of the HRM prediction is that, because of the fact that the magnitude of  invariant momentum transfer of the 
reaction~(\ref{reaction}), $t$ is larger than that of   the $pd\rightarrow pd$ scattering, $t_{pd}$ (Eq.~(\ref{t_pd})) the 
effective center of mass angle in the latter case, $\theta^*_{\text{cm}}< \theta_{\text{cm}}$ (see Eq.~(\ref{cmangles}) and Fig.~(\ref{eff_cm_angle})) and as a result HRM predicts angular distributions  monotonically decreasing with an increase of $\theta_{\text{cm}}$ for up to $\theta_{\text{cm}}\approx 120^\circ$.
\vspace{0em}
\begin{figure}[h!]
\centering
\includegraphics[width = 12 cm, height = 8 cm]{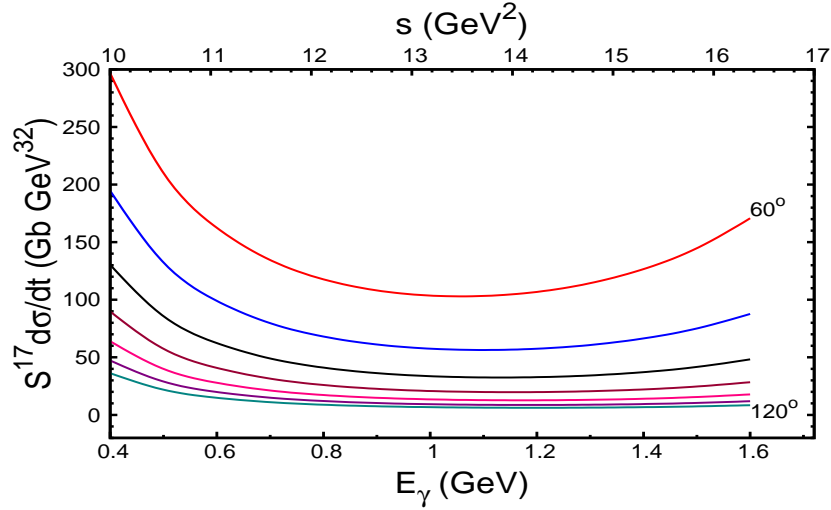}
\captionsetup{format = plain, justification = raggedright}
\caption{Energy dependence of  $s^{17}$ scaled differential cross section  for different values of $\theta_{\text{cm}}$.  
The upper curve corresponds to $\theta_{\text{cm}} = 60^\circ$, with the following curves corresponding to the increment of the center of mass angle by $10^\circ$. }
\label{diff_angles}
\end{figure}
\vspace{0em} 
Finally in Fig.~(\ref{diff_angles}), we present the calculation of $s^{17}$ scaled differential cross section as a function of incoming photon energy for 
different fixed  and large center of mass angles, $\theta_{\text{cm}}$.  
Note that in both Fig.~(\ref{ang_deps}) and Fig.~(\ref{diff_angles}) the  accuracy of the theoretical predictions is similar to that of the energy dependence at $\theta_{\text{cm}}=90^0$ presented in Fig.~(\ref{comp}).

The possibility of comparing these calculations with the experimental data will allow 
to ascertain the range of  validity of the HRM mechanism.  These comparisons will allow us to identify the minimal momentum transfer in these nuclear reactions for which one observes  the onset of the QCD degrees of freedom. At this time, even though we do not have excessive experimental data to provide more profound validation to our model, below, we present the comparison of our calculations to a set of ``preliminary" experimental data. 
\begin{figure} [ht]
\begin{subfigure}{0.5\textwidth}
\centering
    \includegraphics[width=1.0\linewidth]{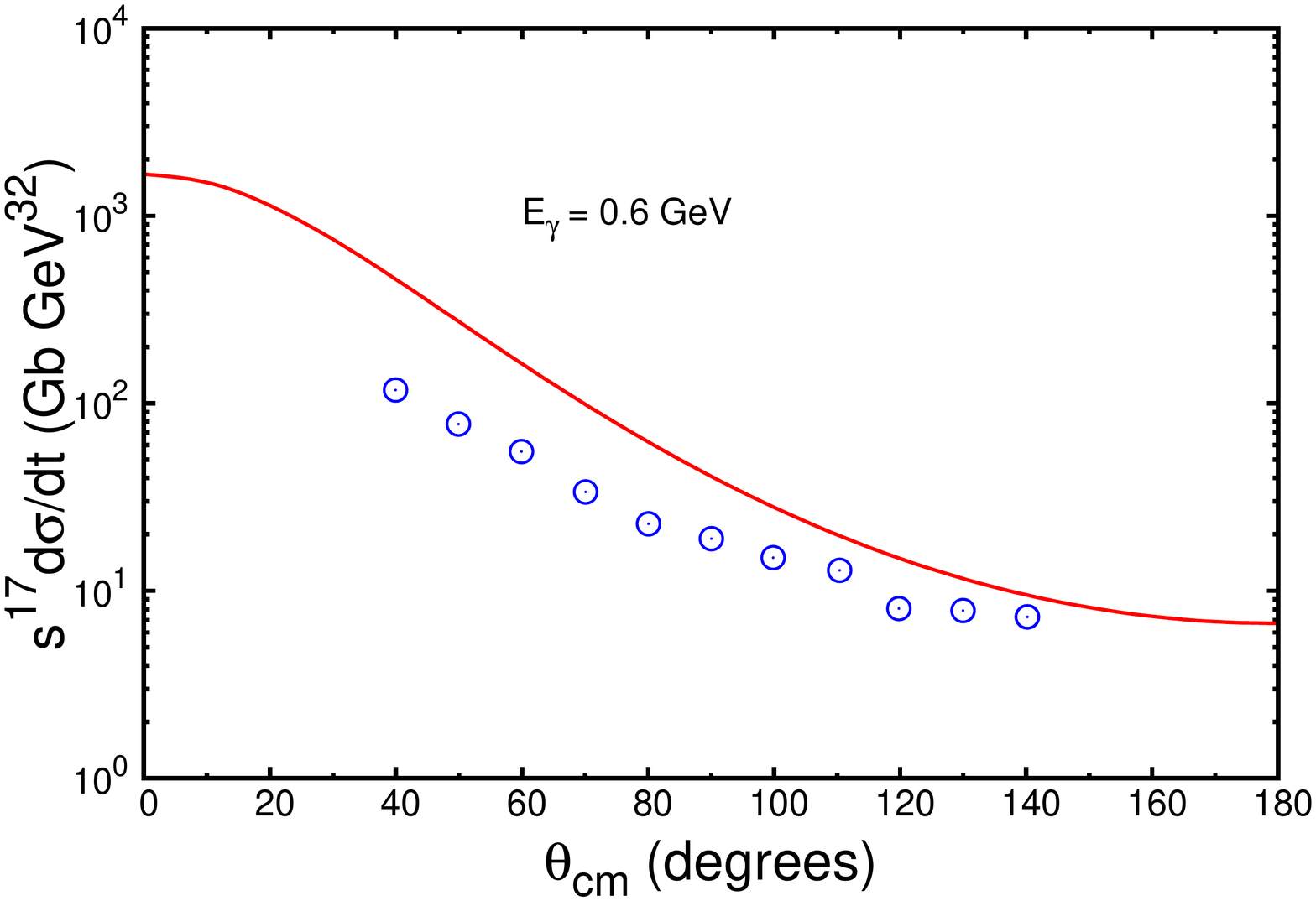}
    \caption{}
    \label{fig:angdep1}
\end{subfigure}%
\begin{subfigure}{0.5\textwidth}
\centering
    \includegraphics[width=1.0\linewidth]{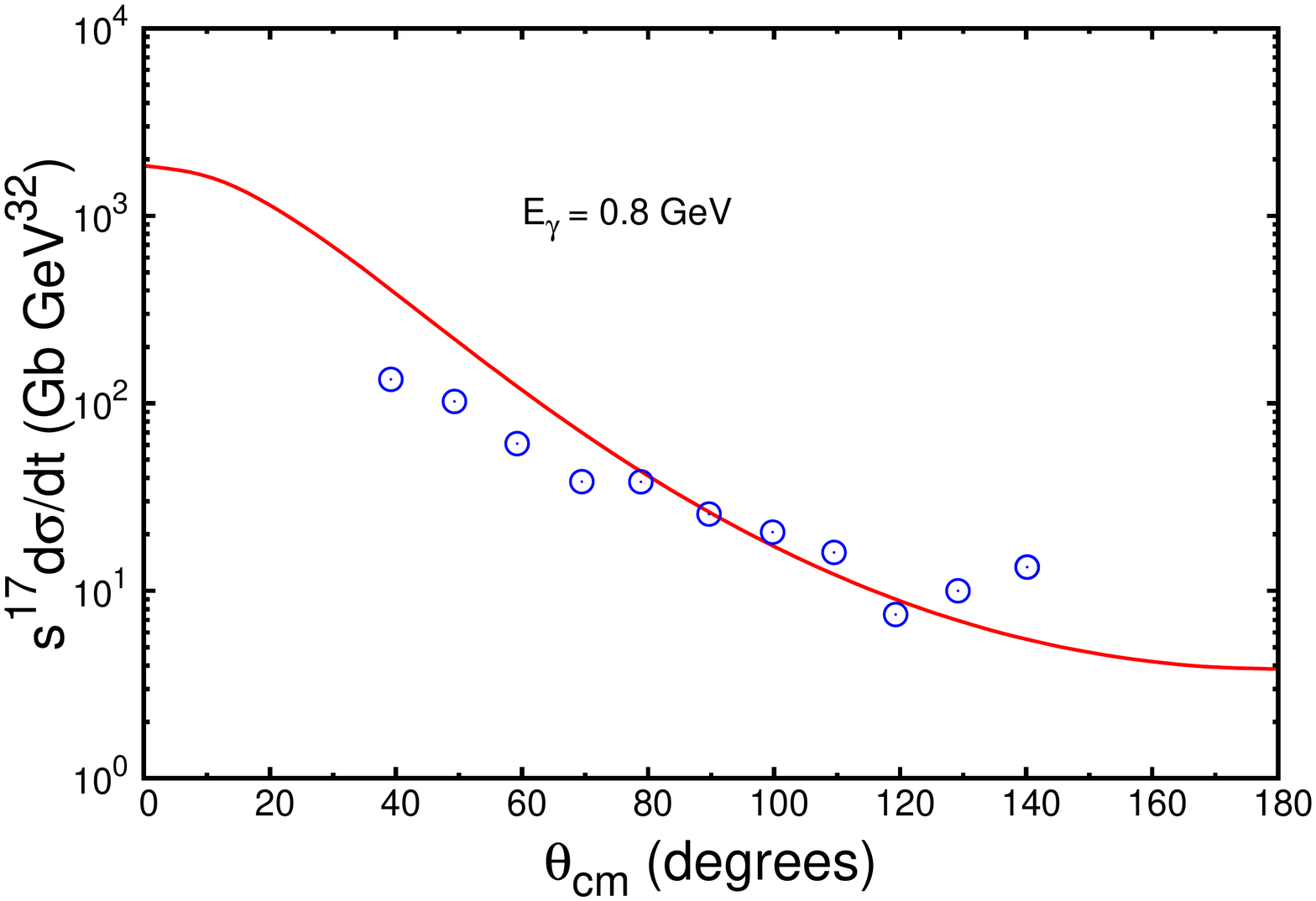}
    \caption{}
    \label{fig:angdep2}
\end{subfigure}
\begin{subfigure}{0.5\textwidth}
\centering
    \includegraphics[width=1.0\linewidth]{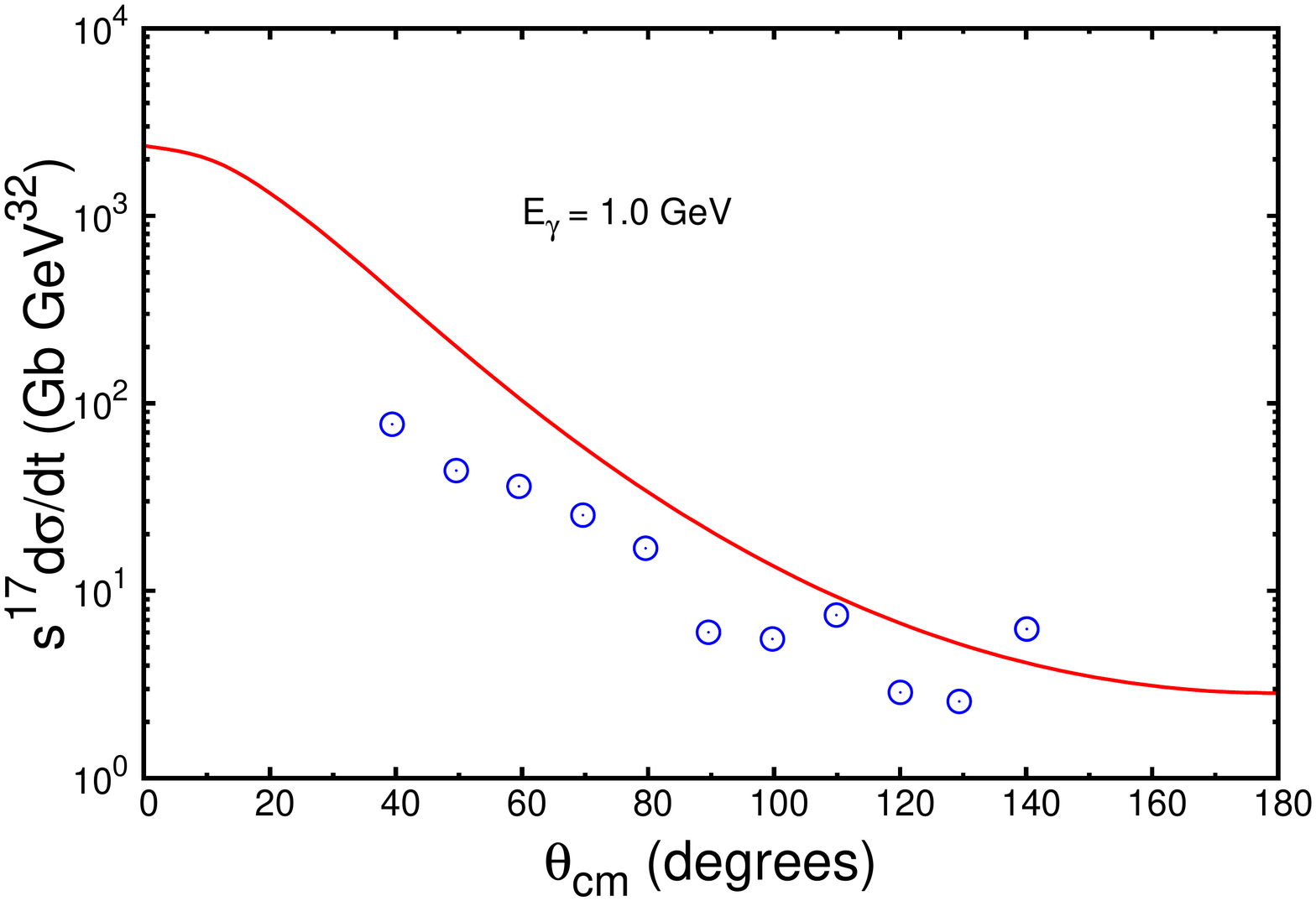}
    \caption{}
    \label{fig:angdep3}
\end{subfigure}%
\begin{subfigure}{0.5\textwidth}
\centering
    \includegraphics[width=1.0\linewidth]{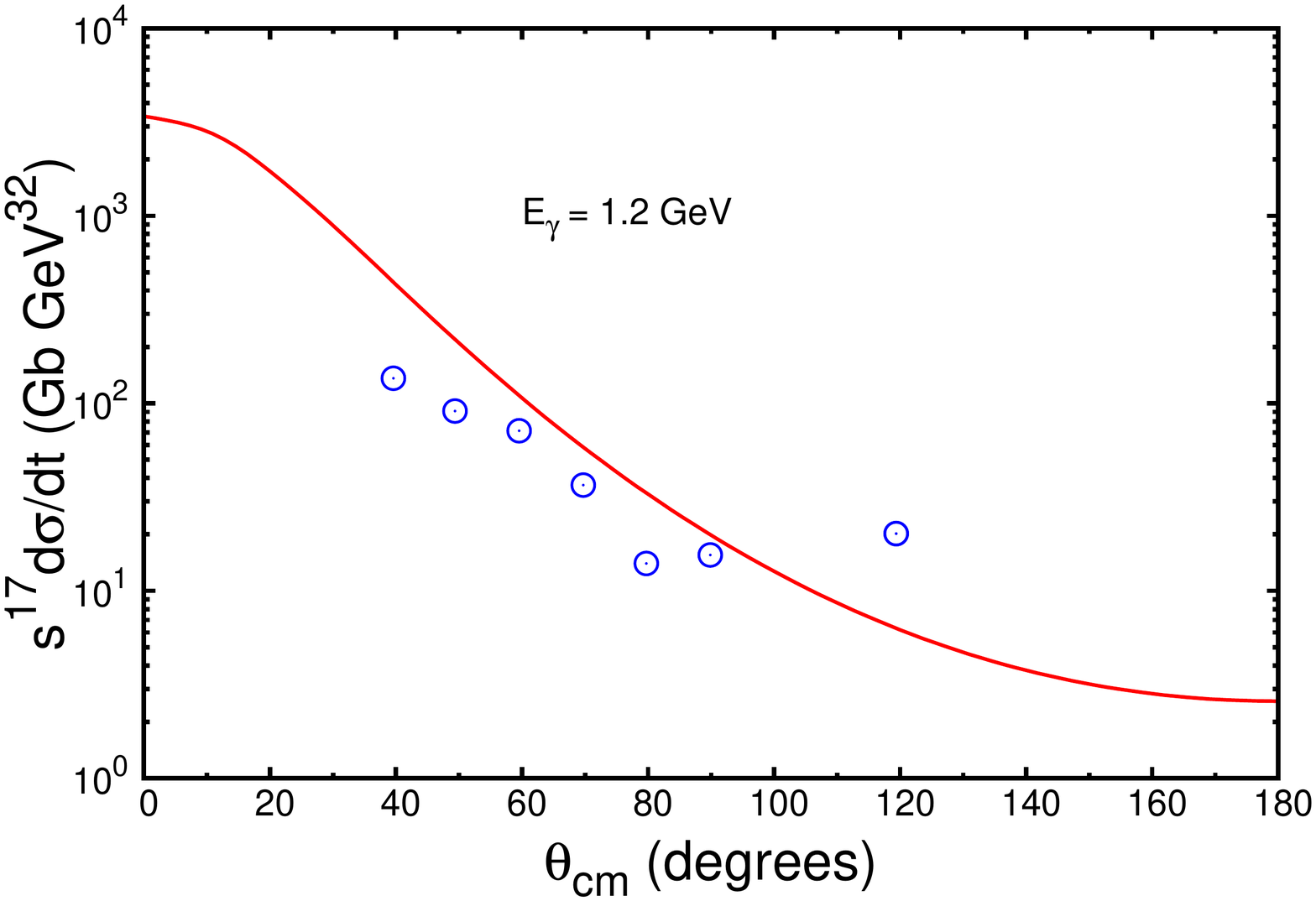}
    \caption{}
    \label{fig:angdep4}
\end{subfigure}%
\caption{Angular dependence of $s^{17}$ scaled differential cross section for various incident photon energies. Data shown in these figures are preliminary, and obtained from \cite{YI}.}
\label{YordankaAngularDependences}
\end{figure}

As can be seen from these comparisons, the usage of HRM becomes more meaningful at larger center of mass angles, which again justifies our study of the hard photodisintegration of $^3$He at $90^\circ$ center of mass angle. The accuracy of the theoretical  calculations shown in Fig.~(\ref{YordankaAngularDependences}) is again similar to that as in the case of $\theta_{\text{cm}} = 90^\circ$, shown in Fig.~(\ref{comp}).

We also present the comparison of energy dependence of the cross section at various center of mass scattering angles with ``preliminary'' experimental data. 
\vspace{0em}
\begin{figure} [h!]
\begin{subfigure}{0.5\textwidth}
\centering
    \includegraphics[width=1.0\linewidth]{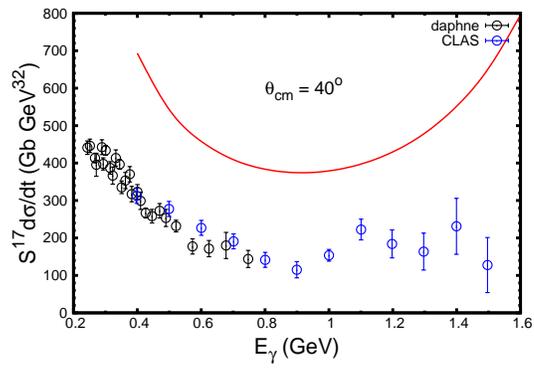}
    \caption{}
    \label{fig:engdep1}
\end{subfigure}
\begin{subfigure}{0.5\textwidth}
\centering
    \includegraphics[width=1.0\linewidth]{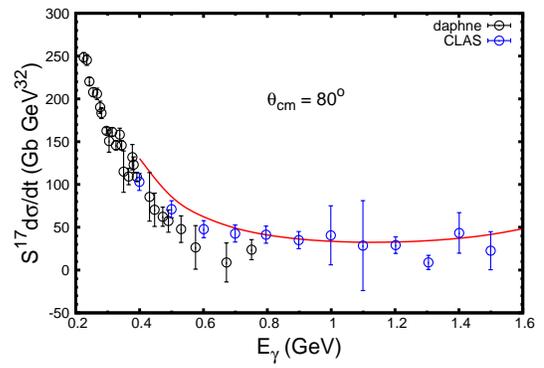}
    \caption{}
    \label{fig:engdep2}
\end{subfigure}
\begin{subfigure}{0.5\textwidth}
\centering
    \includegraphics[width=1.0\linewidth]{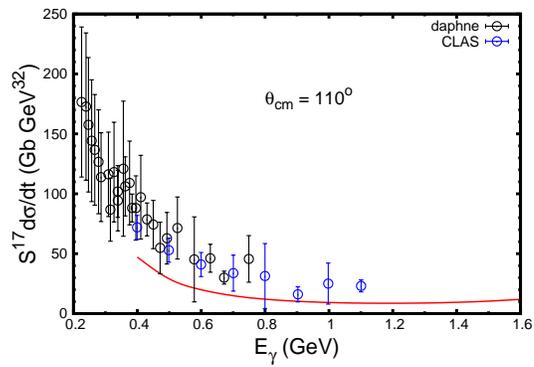}
    \caption{}
    \label{fig:engdep3}
\end{subfigure}
\begin{subfigure}{0.5\textwidth}
\centering
    \includegraphics[width=1.0\linewidth]{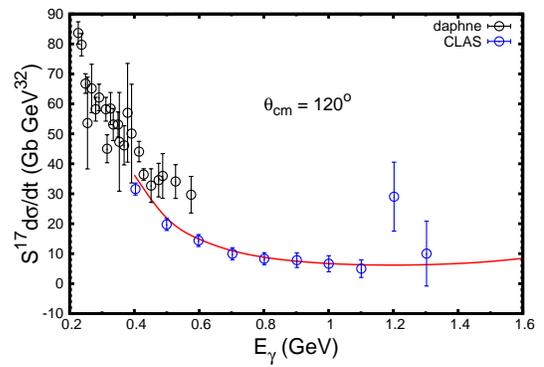}
    \caption{}
    \label{fig:engdep4}
\end{subfigure}
\begin{subfigure}{0.5\textwidth}
\centering
    \includegraphics[width=1.0\linewidth]{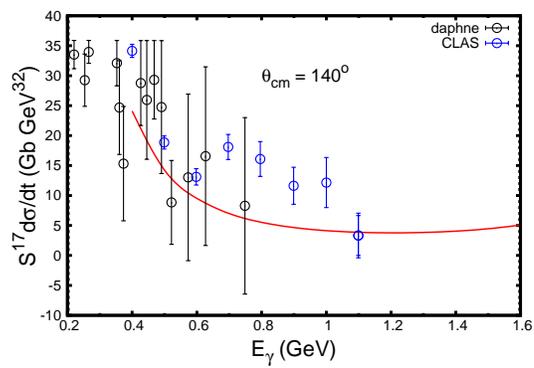}
    \caption{}
    \label{fig:engdep5}
\end{subfigure}%
\caption{Energy dependence of $s^{17}$ scaled differential cross section for various center of mass scattering angles. Data shown in these figures are preliminary, and obtained from \cite{YI}.}
\label{YordankaEnergyDependences}
\end{figure}

As it is seen in Figs.~(\ref{YordankaEnergyDependences}), the theoretical calculations of the cross section using the HRM model is in very good agreement with the ``preliminary'' experimental data at larger center of mass scattering angles. The availability of more experimental data will only allow us to be more certain regarding the accuracy with which our model is able to calculate the cross section of hard photodisintegration reaction at different kinematical conditions.

%% file: 4/4.tex
\chapter{Spin structure of neutron and the problem of nuclear medium effects}

\input{./4/intro4.tex}

\input{./4/PolDisNuclear.tex}

\input{./4/coefficientsDeriv.tex}

\input{./4/Jacobian.tex}

\input{./4/PWIA.tex}

\input{./4/PWIAcsDeriv.tex}

\input{./4/NumericalEstimatesPWIA.tex}

\input{./4/NumericalEstimatesPolarization.tex}

\input{./4/Asymmetries.tex}

\input{./4/g1nresults.tex}

%% file: 4/intro4.tex
\label{sec:Ch4Intro}
 
As briefly discussed in the Introduction of the dissertation, one of the most fundamental properties of elementary particles is their spin because it helps to understand their symmetry behavior under space-time transformations. The spin degrees of freedom for composite particles can be used to study their internal dynamics. In this respect, the simultaneous study of spin structure of proton and neutron allows to probe the interaction of quarks and gluons in unique way that can clarify the longstanding issue of the spin composition of the nucleon in QCD.
~However, study of the spin of the nuetron is significantly difficult compared to that of the proton. The most important reason for this is the existence of free proton but not that of the neutron.
To study neutron, one usually uses the lightest nuclei such as deuteron and $^3$He and considers deep-inelastic scattering with or without polarization of the target. Then, during the analysis of the data one needs to subtract by proton contribution, as well as correct for binding and Fermi motion effects.

\begin{figure}[h!]
\begin{subfigure}{0.5\textwidth}
    \includegraphics[width=0.9\linewidth]{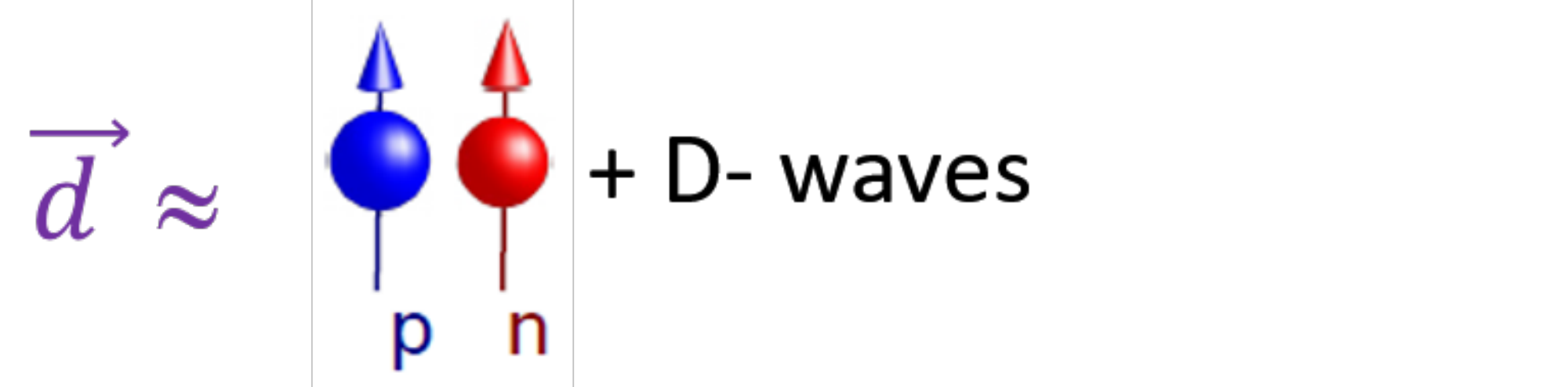}
    \caption{Spin of nucleons in deuteron.}
    \label{fig:spinNucleonsDeuteron}
\end{subfigure}%
\begin{subfigure}{0.5\textwidth}
\centering
    \includegraphics[width=0.9\linewidth]{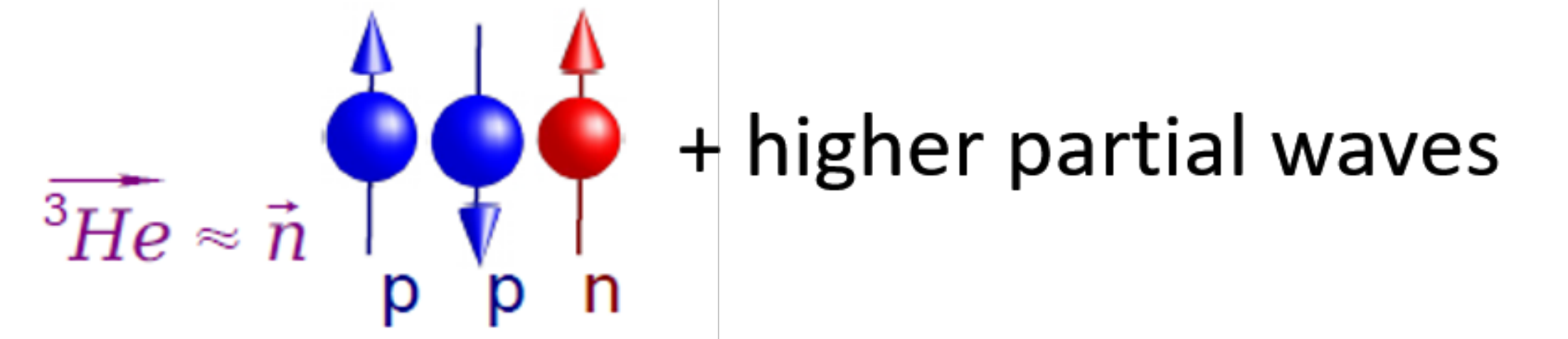}
    \caption{Spin of nuccleons in $^3$He.}
    \label{fig:spinNucleonsHe3}
\end{subfigure}
\caption{Spin of nucleons in light nuclei like deuteron and $^3$He.}
\label{fig:spinNucleonsDeuteronandHe3}
\end{figure}

One such method for extraction of the polarization structure function of the neutron is the consideration of  deep inelastic scattering of polarized electrons off polarized deuteron and $^3$He targets. The simplest nucleus in this case is the deuteron, in which case the main component of the deuteron consists of proton and neutron in relative $S-$state, in which proton and neutron's spins are aligned in the same direction to produce spin 1 state, as shown in Fig.~(\ref{fig:spinNucleonsDeuteron}). 
The next best source would be the $^3$He nucleus, which has two protons and one neutron. The nucleus of $^3$He has a spin of $\frac{1}{2}$ and as the polarization of the two protons cancel each other (due to the Pauli's exclusion principle), the polarization of $^3$He is carried predominantly by the neutron, as shown in Fig.~(\ref{fig:spinNucleonsHe3}). However, this is not always the case, since there is an additional contribution from higher partial waves in $^3$He and the deuteron ground state wave functions respectively. In the region of small $x$, the contribution of the higher partial waves are not significant. But, in the higher $x$ region, their contributions become significant. There are currently several experiments planned at Jefferson Lab\cite{JlabProposal} and CERN \cite{compass} aimed at measuring polarized neutron structure functions at intermediate and high $x$ kinematical region.
So, in order to precisely extract the neutron spin information in these experiments, we need to account for the effects of the higher partial waves. 

In this chapter, we present the theoretical framework for  calculation of polarized deep inelastic scattering from polarized nuclei. We first derive formulas for the general case of nucleus $A$. Then, we use the derived formulas to formulate and execute the algorithm of extraction of the polarized neutron structure function from the deuteron target.

%% file: 4/PolDisNuclear.tex

\section{Deep inelastic scattering from a polarized nucleus}

We consider the deep inelastic reaction in which a polarized electron is scattered off a polarized nuclear target,
\begin{equation}
\vec{e} + \vec{A} \rightarrow e' + X,
\label{disfromnucltarget}
\end{equation} 
where $\vec{e}$ and $\vec{A}$ represent the polarized electron and polarized nuclear target respectively, $e'$ is the scattered electron, and $X$ is the product of the deep inelastic scattering, as shown in Fig~(\ref{scattering_nuclear_target}) below.
\begin{figure}[ht]
\centering
\includegraphics[scale=0.35]{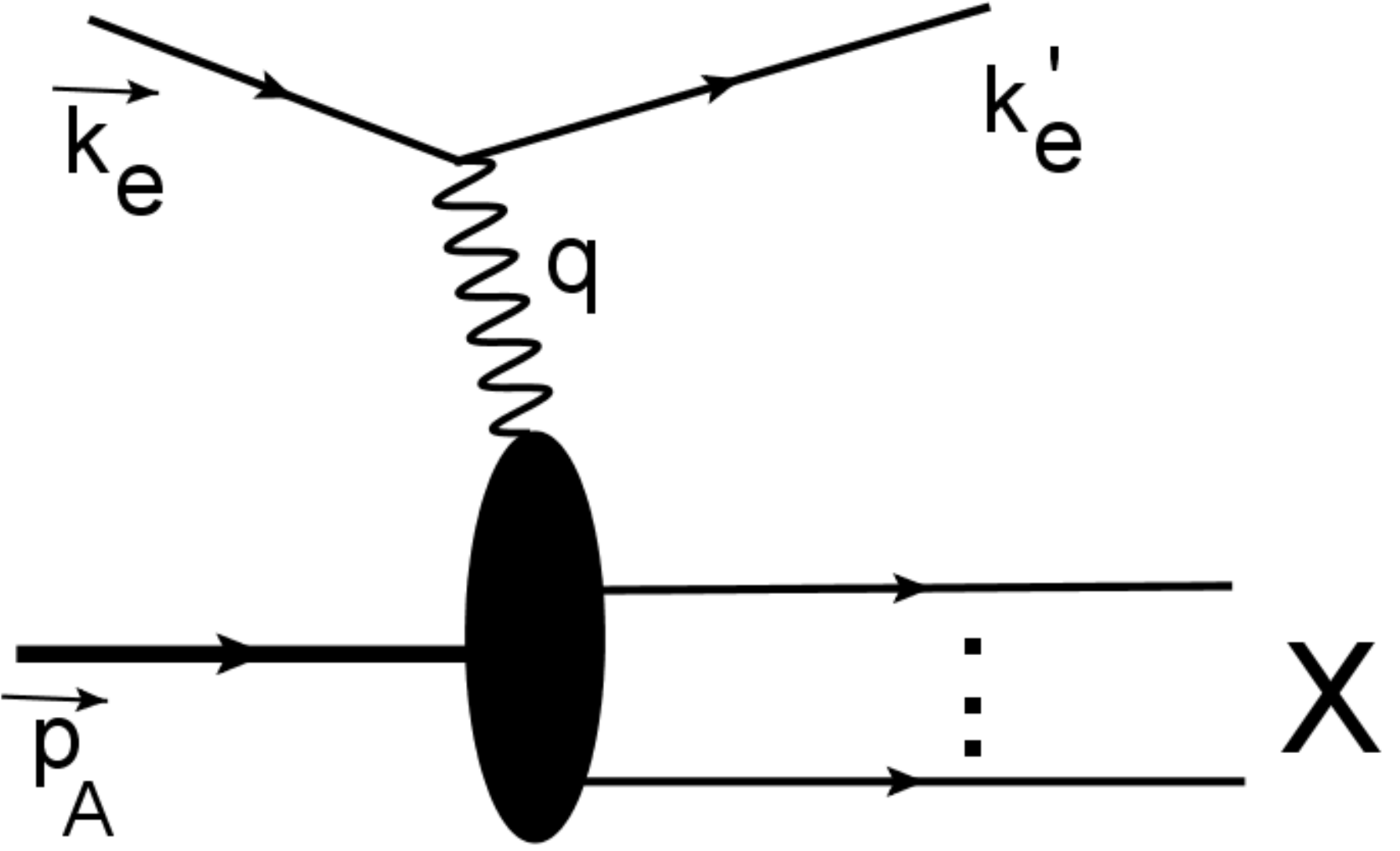}
\caption{Scattering of polarized electron from a polarized nuclear target.}
\label{scattering_nuclear_target}
\end{figure}
Deep inelastic kinematics are defined such that invariant momentum transfer $Q^2 = -(k - k')^2 \geq 2$ GeV$^2$ and produced final mass from the nucleon target $W_N^2 = -Q^2 + 2 m_N q_0 + m_N^2 \geq 2.2$ GeV$^2$. The above conditions are considered as minimal possible values for which Bjorken scaling is observed. The latter represents a situation in which target structure functions depend only on dimensionless Bjorken parameter $x =\frac{Q^2}{2 m_N q_0}$ (upto ln$(Q^2)$ factors).
Here, we use the definition of the four momenta of the incoming   and scattered electrons as $k = (E_k, \mathbf{k})$ and $k'= (E_k', \mathbf{k'})$ respectively. The energy transferred to the nucleus is defined as $q_0 = E_k - E_k'$. The differential cross section for the scattering of Eq.~(\ref{disfromnucltarget}) can be written as
\begin{equation} \label{cs}
\frac{d\sigma}{d\Omega dE'} = \frac{\alpha_{\text{em}}^2}{Q^4} \frac{E_k'}{M_A E_k} L^{\mu \nu}W_{\mu \nu}^A,
\end{equation}
where  $\alpha_{\text{em}}$ is the fine structure constant, and $ L^{\mu \nu}, W_{\mu \nu}^{A} $ are the polarized leptonic and hadronic tensors respectively, which are defined as follows:
\begin{equation} \label{leptonictensor_pol}
L_{\mu \nu} = 4\Bigg[ \Big(k_\mu - \frac{q_\mu}{2}\Big) \Big(k_\nu - \frac{q_\nu}{2}\Big) \Bigg] + Q^2 \Bigg[-g_{\mu \nu} - \frac{q_{ \mu} q_{\nu}}{Q^2} \Bigg] + 2 i m \epsilon_{\mu \nu \rho \sigma} q^{\rho} a ^{\sigma},
\end{equation}
and
\begin{eqnarray} \label{hadronictensor_pol}
W^{\mu \nu}_A &&= \Bigg[ -g^{\mu \nu} - \frac{q^{\mu} q^{\nu}}{Q^2} \Bigg] W_{1A}(x_A, Q^2) + \Bigg[p^{\mu}_A + \frac{p_A \cdot q}{Q^2}{ q^{\mu}} \Bigg] \Bigg[p^{\nu}_A + \frac{p_A \cdot q}{Q^2}{ q^{\nu}} \Bigg] \frac{W_{2A}(x_A, Q^2)}{M_A^2} \nonumber \\
 && - i\epsilon^{\mu \nu \rho \sigma} q_{\rho} \Bigg[ \zeta_{A,\sigma} \frac{G_{1A}(x_A, Q^2)}{M_A} + \Big[p_A \cdot q ~ \zeta_{A,\sigma} - q \cdot \zeta_A ~ p_A^{\sigma} \Big] \frac{G_{2A}(x_A, Q^2)}{M_A^3} \Bigg].
\end{eqnarray}
As compared to the unpolarized leptonic and hadronic tensors (Eqs.~(\ref{leptonictensor}) and (\ref{hadronictensor})), the polarized tensors are much more complicated because of the polarization vectors $a_\sigma$ and $\zeta_{A,\sigma}$ of the electron and the nucleus respectively. In Eq.~(\ref{hadronictensor_pol}),$G_{1A}(x_A, Q^2)$ and $G_{1A}(x_A, Q^2)$ contain information about the polarized structure of nucleus, $p_A$ is the four momentum of the nuclear target with mass number $A$ and $\epsilon^{\mu \nu \rho \sigma}$ is the four dimensional Levi-Civita matrix.  The polarization four vector of the electron is defined as
\begin{equation}
a_{\sigma} \equiv (a_0, \mathbf{a}) = \Bigg[\frac{\boldsymbol{\xi}\cdot \mathbf{k}}{m}, \boldsymbol{\xi} + \frac{\boldsymbol{\xi} \cdot \mathbf{k} ~ \mathbf{k}}{m_e(E_k + m_e)} \Bigg],
\label{elec_pol_vec}
\end{equation}
with $\boldsymbol{\xi} = \frac{h_e \mathbf{k}}{|{\mathbf{k}|}}$, where $h_e$ is the helicity of the electron, and $m_e$ is the electron mass.
Similarly, the nuclear polarization four vector is defined as:

\begin{equation} \label{hadron_pol}
\zeta_{\sigma}^A \equiv (\zeta_0, \mathbf{\zeta_A}) = \Bigg[\frac{\mathbf{s_A} \cdot \mathbf{p_A}}{M_A}, \mathbf{s_A} + \frac{\mathbf{s_A} \cdot \mathbf{p_A} ~ \mathbf{p_A}}{M_A(E + M_A)} \Bigg],
\end{equation}
with $\mathbf{s_A}$ being the spin three-vector of the nucleus. 

With the above definitions of the polarization vectors, we can now calculate the contraction of the leptonic and hadronic tensors. Noting that $q_{\mu} L^{\mu \nu} = 0 = q_{\nu} L^{\mu \nu}$, we obtain:
\begin{eqnarray}
\label{contraction_start}
L_{\mu \nu}W^{\mu \nu}_{A} &&= -g^{\mu \nu}L_{\mu \nu} W_{1A} + p^{\mu}_A p^{\nu}_A L_{\mu \nu} \frac{W_{2A}}{M_A^2} \nonumber \\
 && -  L_{\mu \nu} i\epsilon^{\mu \nu \rho \sigma} q_{\rho}   \Bigg[ \zeta_{A,\sigma} \frac{G_{1A}}{M_A} + \Big[p_A \cdot q \zeta_{A,\sigma} - q \cdot \zeta_A p_{A,\sigma} \Big] \frac{G_{2A}}{M_A^3} \Bigg]. 
\end{eqnarray}

Now, substituting  $L_{\mu \nu}$ from Eq.~(\ref{leptonictensor_pol}) and making use of the anti-symmetric property of the Levi-Civita tensor, $\epsilon ^{\mu \nu \rho \sigma}$ in $\mu, \nu$ and the symmetric nature of the coefficients at $W_{1A}, W_{2A}$ in $\mu, \nu$, we arrive at:
\begin{eqnarray} \label{contractionPolarized}
L_{\mu \nu}W^{\mu \nu}_{A} &&  = -g^{\mu \nu} \Big[ 4\big[ (k_\mu - \frac{q_\mu}{2}) (k_\nu - \frac{q_\nu}{2}) \big] + Q^2 \big[-g_{\mu \nu} - \frac{q_{ \mu} q_{\nu}}{Q^2} \big] + 2 i m \epsilon_{\mu \nu \rho \sigma} q^{\rho} a ^{\sigma} \Big] W_{1A}   \nonumber \\
&& +  p_A^{\mu} p_A^{\nu} \Big[ 4\big[ (k_\mu - \frac{q_\mu}{2}) (k_\nu - \frac{q_\nu}{2}) \big] + Q^2 \big[-g_{\mu \nu} -
\frac{q_{ \mu} q_{\nu}}{Q^2} \big] + 2 i m \epsilon_{\mu \nu \rho \sigma} q^{\rho} a^{\sigma} \Big]\frac{W_{2A}}{M_A^2}   \nonumber \\
&& + 2m \epsilon_{\mu \nu \alpha \beta} q^{\alpha} a^{\beta} \epsilon^{\mu \nu \rho \sigma} q_{\rho} \Big[ \zeta_{A,\sigma} \frac{G_{1A}}{M_A} + [p_A \cdot q~ \zeta_{A,\sigma} - q \cdot \zeta_A ~p_{A,\sigma} ] \frac{G_{2A}}{M_A^3} \Big].
\end{eqnarray}

Also, noting that $g_{\mu \nu} g^{\mu \nu} = 4$, the above equation can be written as:
\begin{eqnarray} \label{contraction_simplified}
L_{\mu \nu} W^{\mu \nu}_A && = \Bigg \{ -4 \Big( k - \frac{q}{2} \Big)^2 + Q^2 \Big( 4 + \frac{q^2}{Q^2} \Big) \Bigg \} W_{1A}  \nonumber \\
&& + \Bigg \{ 4 \Big(p_A \cdot k - \frac{p_A \cdot q}{2} \Big)^2 + Q^2 \Big(-p_{A}{^2} - \frac{(p_A \cdot q)^2}{Q^2} \Big) \Bigg \}\frac{W_{2A}} {M_A^2} \nonumber \\
&& + 2m \epsilon_{\mu \nu \alpha \beta} q^{\alpha} a^{\beta} \epsilon^{\mu \nu \rho \sigma} q_{\rho} \Bigg \{ \zeta_{A,\sigma} \frac{G_{1A}}{M_A} + [p_A \cdot q~ \zeta_{A,\sigma} - q \cdot \zeta_A ~p_{A,\sigma} ] \frac{G_{2A}}{M_A^3} \Bigg \}.
\end{eqnarray}

We next proceed to calculate the coefficients at $W_{1A}, W_{2A}, G_{1A} \text{ and } G_{2A}$ individually. We also note that the  coefficients at $W_{1A}, W_{2A}$ are symmetric while coefficients at $G_{1A}, G_{2A}$ are anti-symmetric in $\mu, \nu$.

%% file: 4/coefficientsDeriv.tex

To compute the coefficients at the structure functions in Eq.~(\ref{contraction_simplified}), we consider a referene frame where the incident electron is defined by the four vector $k = (E_k, 0,0,E_k)$ and the scattered electron is defined by the four vector    $k^\prime=(E_k^\prime, E_k^\prime \sin\theta_e, 0, E_k^\prime \cos\theta_e)$, where $\theta_e$ is the scattering angle of the electron.  Thus, the four momentum transfer $q_{\mu} = k - k^\prime$, then can be expressed in terms of the components as 
\begin{equation} \label{momentum_transfer}
q_{\mu} \equiv (q_0, q_x, q_y, q_z) = (E_k - E_k^\prime, -E_k^\prime \sin\theta_e, 0, E_k - E_k^\prime \cos\theta_e).
\end{equation} \vspace{0em}
Following the definitions of $k$ and $k^\prime$, we find that
\begin{eqnarray}
 \label{q2derivation}
q^2 && = k^2 + k^{\prime^{2}} - 2 k \cdot q \approx -2 k \cdot q \nonumber \\
 &&= -2 \Big[E_k E_k^\prime - E_k E_k^\prime \cos\theta_e \Big] = -2 E_k E_k^\prime (1 - \cos\theta_e)  = -4 E_k E_k^\prime\sin^2\frac{\theta_e}{2},
\end{eqnarray}
where we neglected the mass of the electron, which allows us to write $k^2 = m_e^2 = k^{\prime{^2}} \approx 0$.
The coefficient at $W_{1A}$ term (using notation $Q^2 = -q^2$), can be written as:
\begin{eqnarray} \label{w1coeff}
-4 \Bigg( k - \frac{q}{2} \Bigg)^2 + Q^2 \Bigg( 4 + \frac{q^2}{Q^2} \Bigg) &&= -4 \Bigg[\Big(k - \frac{q}{2}\Big)^2 \Bigg] + 3Q^2  \nonumber \\
&& = -4 \Big[k^2 - k \cdot q + \frac{q^2}{4}\Big] + 3Q^2 \nonumber \\
&& = -4 \Big[k^2 - \{E_k(E_k - E_k^\prime) - E_k (E_k - E_k^\prime\cos\theta_e)\} + \frac{q^2}{4} \Big] + 3Q^2 \nonumber \\
&& = -4 \Big[k^2 - \{ -E_k E_k^\prime(1-\cos\theta_e)\} + \frac{q^2}{4} \Big] + 3Q^2 \nonumber \\
&& = -4 \Big[k^2 - \frac{q^2}{2} + \frac{q^2}{4} \Big] + 3Q^2 \nonumber \\
&& = 4 \frac{q^2}{4} + 3Q^2 = 2Q^2.
\end{eqnarray}
Similarly, the coefficient at $\frac{W_{2A}}{M_A^2}$ can be simplified as:
\begin{eqnarray}
\text{Coeff. at }\frac{W_{2A}}{M_A^2} && = 4 \Bigg[p_A \cdot k - \frac{p_A \cdot q}{2} \Bigg]^2 + Q^2 \Bigg[-p_{A}{^2} - \frac{(p_A \cdot q)^2}{Q^2} \Bigg] \nonumber \\
&& = 4 (p_A \cdot k)^2 \Bigg[ 1 - \frac{p_A \cdot q}{2~ p_A \cdot k} \Bigg]^2 - Q^2 M_A^2\Bigg[1 + \frac{(p_A \cdot q)^2}{Q^2 M_A^2}\Bigg] \nonumber 
\end{eqnarray}
\begin{eqnarray}
&&= 4 ( p_A \cdot k)^2 \Big(1 - \frac{y_A}{2}\Big)^2 - Q^2 M_A^2 \Bigg[1 + \frac{(p_A \cdot q) A}{2 M_A^2 \frac{AQ^2}{2 (p_A \cdot  q)}}		\Bigg] \nonumber \\
&&= 4 ( p_A \cdot k)^2 \Big(1 - \frac{y_A}{2}\Big)^2 - Q^2 M_A^2 \Bigg[1 + \frac{A^2 Q^2 }{4 x_A^2 M_A^2 }		\Bigg] \nonumber \\
&& = 4 ( p_A \cdot k)^2 \Big(1 - \frac{y_A}{2}\Big)^2 - \frac{Q^4 A^2}{4 x_A^2} \Bigg[ \frac{4 x_A^2 M_A^2 }{A^2 Q^2 } +1 \Bigg] \nonumber \\
&& =4 ( p_A \cdot k)^2 \Big(1 - \frac{y_A}{2}\Big)^2 - (p_A \cdot q)^2 \Bigg[ \frac{4 x_A^2 M_A^2 }{A^2 Q^2 } +1 \Bigg],
\label{w2coeff}
\end{eqnarray}
where we define $y_A = \frac{p_A \cdot q}{p_A \cdot k}$.
\vspace{0em}
If we now  introduce $\frac{F_{2A}(x_A, Q^2)}{p_A \cdot q} M_A = W_{2A}(x_A, Q^2)$, then, the second term in Eq.~(\ref{contraction_simplified}) becomes
\begin{eqnarray}
&& \Bigg \{ 4 \Big(p_A \cdot k - \frac{p_A \cdot q}{2} \Big)^2 + Q^2 \Big(-p_{A}{^2} - \frac{(p_A \cdot q)^2}{Q^2} \Big) \Bigg \}\frac{W_{2A}} {M_A^2} = \frac{(p_A \cdot k)^2}{M_A^2} \Bigg[( 2 - y_A)^2 - y_A^2 \Big( \frac{4 x_A^2 M_A^2 }{A^2 Q^2 } + 1 \Big) \Bigg]\frac{W_{2A}}{M_A^2} \nonumber \\
&& =\frac{F_{2A}}{y_A^2} \frac{p_A \cdot q}{M_A} \Bigg[ (2 - y_A)^2 - y_A^2 \Big(1 + \frac{4 x_A^2 M_A^2 }{A^2 Q^2 } \Big) \Bigg].
\label{2ndtermContraction}
\end{eqnarray}
We can then combine Eqs.~(\ref{w1coeff}) and (\ref{2ndtermContraction}) and express Eq.~(\ref{contraction_simplified}) as:
\begin{eqnarray}
L_{\mu \nu}W^{\mu \nu}_A &&= 2 Q^2 \frac{F_{1A}}{M_A} +  \frac{p_A \cdot q}{M_A} \frac{F_{2A}}{y_A^2} \Bigg[(2 - y_A)^2 - y_A^2 \Big(1 + \frac{4 x_A^2 M_A^2 }{A^2 Q^2 } \Big) \Bigg] \nonumber \\
&& + 2m \epsilon_{\mu \nu \alpha \beta} q_{\alpha} a_{\beta} \epsilon^{\mu \nu \rho \sigma} q^{\rho} \Bigg \{ \zeta_{A,\sigma} \frac{G_{1A}}{M_A} + [p_A \cdot q~ \zeta_{A,\sigma} - q \cdot \zeta_{A} ~p_{A,\sigma} ] \frac{G_{2A}}{M_A^3} \Bigg \}, 
\label{contraction_with_w1w2done}
\end{eqnarray}
where we used also the definition, $W_{1A}(x_A, Q^2) = \frac{F_{1A}(x_A, Q^2)}{M_A}$. In Eq.~(\ref{contraction_with_w1w2done}), the functions $F_{1A}(x_A, Q^2)$ and $F_{2A}(x_A, Q^2)$ are the unpolarized structure functions. These unpolarized structure functions have been widely studied experimentally and thus, there exist plenty of data with which we can compare our theoretical calculations. These comparisons will be discussed in more detail in subsequent sections. It is worth mentioning that in the case of above mentioned Bjorken scaling, it is the structure functions $F_1(x,Q^2)$ and $F_2(x,Q^2)$ that become $Q^2$ independent (upto the ln$(Q^2)$ factors).

To evaluate the anti-symmetric part of Eq.~(\ref{contraction_with_w1w2done}), we use the identity
\begin{equation} 
\label{eps_delta_iden}
\epsilon_{\mu \nu \alpha \beta} \epsilon^{\mu \nu \rho \sigma} = -2 \big[ \delta^{\rho}_{\alpha} \delta^{\sigma}_{\beta} - \delta^{\sigma}_{\alpha} \delta^{\rho}_{\beta}  \big], 
\end{equation}
 and obtain the following:
 \begin{eqnarray} \label{antisymmetric}
&& 2m \epsilon_{\mu \nu \alpha \beta} q^{\alpha} a^{\beta} \epsilon^{\mu \nu \rho \sigma} q_{\rho} \Bigg \{ \zeta_{A,\sigma} \frac{G_{1A}}{M_A} + [p_A \cdot q~ \zeta_{A,\sigma} - q \cdot \zeta_A~ p_{A,\sigma} ] \frac{G_{2A}}{M_A^3} \Bigg \} \nonumber \\ 
&&= -4m \Big[ \delta^{\rho}_{\alpha} \delta^{\sigma}_{\beta} - \delta^{\sigma}_{\alpha} \delta^{\rho}_{\beta}\Big] q^{\alpha}a^{\beta}q_{\rho} \Bigg \{ \zeta_{A,\sigma} \frac{G_{1A}}{M_A} + [p_A \cdot q~\zeta_{A,\sigma} - q \cdot \zeta_A ~p_{A,\sigma} ] \frac{G_{2A}}{M_A^3} \Bigg \} \nonumber \\
&& = - 4m q^2 a^{\sigma} \Bigg \{ \zeta_{A,\sigma} \frac{G_{1A}}{M_A} + [p_A \cdot q ~ \zeta_{A,\sigma} - q \cdot \zeta_A ~ p_{A,\sigma} ] \frac{G_{2A}}{M_A^3} \Bigg \} + 4 m a \cdot q \Bigg \{ q \cdot \zeta_A \frac{G_{1A}}{M_A} + (q \cdot  p_A ~ q \cdot \zeta_A \nonumber \\
 &&- q \cdot \zeta_A ~ q \cdot p_A)\frac{G_{2A}}{M_A^3}\Bigg\} \nonumber \\
&& = - 4 m q^2 a \cdot \zeta_A \frac{G_{1A}}{M_A} + 4 m a \cdot q ~ \zeta_A \cdot q \frac{G_{1A}}{M_A} - 4m q^2 \big( q \cdot p_A   ~ a \cdot \zeta_A - \zeta_A \cdot q ~ a \cdot p_A \big) \frac{G_{2A}}{M_A^3} \nonumber \\
&& = -4 m \Big[q^2 a \cdot \zeta_A  -  a \cdot q ~ \zeta_A \cdot q \Big] \frac{G_{1A}}{M_A} - 4 m q^2 \Big[ q \cdot p_A   ~ a \cdot \zeta_A - \zeta_A \cdot q ~ a \cdot p_A \Big] \frac{G_{2A}}{M_A^3}.
\end{eqnarray}

To further simplify the above expression, we note that, in high energy limits, the electron's polarization four vector can be expressed as $a^\sigma = h_ek^\sigma/m$ (Appendix C). Using this, and relation $q \cdot k = \frac{q^2}{2}$ of the antisymmetric part of $L_{\mu \nu}W^{\mu \nu}$, one obtains:
\begin{eqnarray}
&& L_{\mu \nu}W^{\mu \nu}_A(\text{anti-symmetric})  = - 4 h_e \Big[q^2 k \cdot \zeta_A  -  \frac{q^2}{2} ~ \zeta_A \cdot q \Big] \frac{G_{1A}}{M_A} \nonumber \\
&& - 4 h_e q^2 \Big[ q \cdot p_A   ~ k \cdot \zeta_A - \zeta_A \cdot q ~ k \cdot p_A \Big] \frac{G_{2A}}{M_A^3} \nonumber\\
&& = 2 h_e q^2 \frac{G_{1A}}{M_A} \Big[\zeta_A \cdot q - 2 k \cdot \zeta_A \Big] + 4 h_e q^2 \frac{G_{2A}}{M_A^3} \Big[ \zeta_A \cdot q ~ k \cdot p_A - q \cdot p_A   ~ k \cdot \zeta_A  \Big] \nonumber \\
&& = - 2 h_e Q^2 \frac{G_{1A}}{M_A} \Big[\zeta_A \cdot q - 2 k \cdot \zeta_A \Big] - 4 h_e Q^2 \frac{G_{2A}}{M_A^3} \Big[ \zeta_A \cdot q ~ k \cdot p_A - q \cdot p_A   ~ k \cdot \zeta_A  \Big].
\label{antisym_intermsof_helicity_Q^2}
\end{eqnarray}
We now redefine the spin structure functions through the  $g_{1A}$ and $g_{2A}$ functions as follows:
\begin{equation}
\frac{G_{1A}(x_A, Q^2)}{M_A} = \frac{g_{1A}(x_A, Q^2)}{p_A \cdot q} \text { and } \frac{G_{2A}(x_A, Q^2)}{M_A^3} = \frac{g_{2A}(x_A, Q^2)}{(p_A \cdot q)^2},
\label{g1g2defns}
\end{equation}
and write Eq.~(\ref{antisym_intermsof_helicity_Q^2}) as

\begin{eqnarray}
&& L_{\mu \nu}W^{\mu \nu}_A(\text{anti-symmetric})  =  - 2 h_e Q^2 \frac{g_{1A}}{p_A \cdot q} \Big[\zeta_A \cdot q - 2 k \cdot \zeta_A \Big] \nonumber \\
&&- 4 h_e Q^2 \frac{g_{2A}}{(p_A \cdot q)^2} \Big[ \zeta_A \cdot q ~ k \cdot p_A - q \cdot p_A   ~ k \cdot \zeta_A  \Big] \nonumber  \\
&& = - 2 h_e Q^2 \frac{g_{1A}}{p_A \cdot q} \Big[\zeta_A \cdot q - 2 k \cdot \zeta_A \Big] - 4 h_e Q^2 \frac{g_{2A}}{p_A \cdot q} \Big[ \frac{q \cdot \zeta_A}{y_A} - \zeta_A \cdot k \Big].
\label{antisym_beforeksum}
\end{eqnarray}

Thus, in the most general form, the contraction of the leptonic  and  hadronic tensors can be expressed in the following form:
\begin{eqnarray}
\label{contractionFinal}
L_{\mu \nu}W^{\mu \nu}_A &&= 2 Q^2 \frac{F_{1A}}{M_A} +  \frac{p_A \cdot q}{M_A} \frac{F_{2A}}{y_A^2} \Bigg[(2 - y_A)^2 - y_A^2 \Big(1 + \frac{4 x_A^2 M_A^2 }{A^2 Q^2 } \Big) \Bigg] \nonumber \\
&&  - 2 h_e Q^2 \frac{g_{1A}}{p_A \cdot q} \Bigg[\zeta_A \cdot q - 2 k \cdot \zeta_A \Bigg] - 4 h_e Q^2 \frac{g_{2A}}{p_A \cdot q} \Bigg[ \frac{q \cdot \zeta_A}{y_A} - \zeta_A \cdot k \Bigg].
\end{eqnarray}

Inserting above expression into Eq.~(\ref{cs}), one obtains for the differential cross section:
\begin{eqnarray} \label{dsigdomdeprime}
\frac{d^2\sigma}{d\Omega dE_k'} &&= \frac{\alpha_{em}^2}{Q^4}\frac{E_k'}{E_k} ~ \Bigg\{2 Q^2 \frac{F_{1A}}{M_A} +  \frac{p_A \cdot q}{M_A} \frac{F_{2A}}{y_A^2} \Bigg[(2 - y_A)^2 - y_A^2 \Big(1 + \frac{4 x_A^2 M_A^2 }{A^2 Q^2 } \Big) \Bigg] \nonumber \\
&&  - 2 h_e Q^2 \frac{g_{1A}}{p_A \cdot q} \Big[\zeta_A \cdot q - 2 k \cdot \zeta_A \Big] - 4 h_e Q^2 \frac{g_{2A}}{p_A \cdot q} \Big[ \frac{q \cdot \zeta_A}{y_A} - \zeta_A \cdot k \Big] \Bigg\}
\end{eqnarray}
The above expression indicates that the cross section of the reaction Eq.~(\ref{disfromnucltarget}) is defined by four $F_{1A}, F_{2A}, g_{1A}$ and $g_{2A}$ nuclear structure functions. The first two structure functions ($F_{1A}$ and $F_{2A}$) are the same which enters in the unpolarized cross section. The other two ($g_{1A}$ and $g_{2A}$) are defined only for polarized target.

%% file: 4/Jacobian.tex
\section{Jacobian of the differentials}

In many applications, cross section of Eq.~(\ref{dsigdomdeprime}) is presented in Lorentz invariant form. For this one need to express differentials $d\Omega dE_k'$ through $dx_A dQ^2 d\phi$ with subsequent integration of $d\phi$.

For this, we need to evaluate the Jacobian for the transformation $E_k'dE_k'd\cos\theta_e d\phi \rightarrow dx_A dQ^2 d\phi$.  We note that
\begin{eqnarray}
dQ^2 d x_A d\phi_e	 = dQ^2 d\Big(\frac{AQ^2}{2p_A \cdot q}\Big) d\phi = -\frac{AQ^2}{2 (p_A \cdot q)^2}dQ^2d(p_A\cdot q)d\phi_e.
\label{diffls_relationship_start}
\end{eqnarray}
We now start by calculating the Jacobian, which relates the two differentials system as
\begin{eqnarray}
dQ^2 dx_A && = -\frac{AQ^2}{2 (p_A \cdot q)^2} dQ^2d(p_A\cdot q)= -\frac{AQ^2}{2 (p_A \cdot q)^2}|\text{Jacobian}|dE_k'd\cos\theta_e \nonumber \\
&& = -\frac{AQ^2}{2 (p_A \cdot q)^2} \begin{vmatrix}
 \dfrac{\partial(p_A \cdot q)}{\partial \cos\theta_e} & \dfrac{\partial Q^2}{\partial \cos\theta_e} \\ \\
 \dfrac{\partial(p_A \cdot q)}{\partial E_k'} & \dfrac{\partial Q^2}{\partial E_k'} \end{vmatrix} dE_k'd\cos\theta_e.
\label{jacobian_defn}
\end{eqnarray}
To calculate each element in the determinant, we consider Lab reference frame, so that: 
\begin{eqnarray}
p_A\cdot q && = M_A q_0 = M_A(E_k - E_k').
\label{pa_dot_q}
\end{eqnarray}
Also, noting that $Q^2 = 2E_k E_k' (1 - \cos\theta_e)$, we can calculate the elements in the determinant in Eq.~(\ref{jacobian_defn})  obtaining
\begin{eqnarray}
dQ^2 dx_A &&= -\frac{AQ^2}{2 (p_A \cdot q)^2} \begin{vmatrix}
 0 & -2E_k E_k'  \\
 -M_A & \frac{Q^2}{E_k'} \end{vmatrix} dE_k'd\cos\theta_e \nonumber \\
&& = -\frac{AQ^2}{2 (p_A \cdot q)^2} \Big[-2 E_k E_k' M_A \Big]dE_k'd\cos\theta_e \nonumber \\
&& = \frac{AQ^2}{2 (p_A \cdot q)} \frac{2 E_k E_k' M_A}{p_A \cdot q}dE_k' d\cos \theta_e \nonumber \\
&& = \frac{2 x_A}{M_A q_0} E_k E_K' M_A dE_k' d\cos \theta_e \nonumber \\
&& = \frac{2 x_A}{y_A} E_k' dE_k' d\cos \theta_e, 
\label{det_calc}
\end{eqnarray}
where $y_A = \frac{p_A \cdot q}{p_A \cdot k} = \frac{q_0}{E_k}$.
We thus find that the differentials can be related as
\begin{equation}
 E_k' dE_k' d\cos\theta_e d\phi_e = \frac{y_A}{2 x_A} dQ^2 dx_A d\phi_e.
\label{diffls_transf_completed}
\end{equation}
Using Eq.~(\ref{diffls_transf_completed}), one can express differential cross section of Eq.~(\ref{dsigdomdeprime}) integrated over the $d\phi$, in the following form:
\begin{eqnarray}
\frac{d\sigma}{dQ^2 dx_A} && = \frac{2 \pi \alpha_{\text{em}}^2 y_A^2}{Q^4 A} \Bigg\{ 2 F_{1A} + \frac{A}{2x_A y_A^2} \Big[ (2 - y_A)^2 - y_A^2 \Big(1 + \frac{4 M_A^2 x_A^2}{A^2 Q^2}\Big) \Big] F_{2A} \nonumber \\
&& - h_e \frac{4 x_A M_A}{A Q^2} \Bigg[g_{1A}\Big(\zeta_A\cdot q - 2\zeta_A \cdot k) + 2 g_{2A}\Bigg(\frac{\zeta_A \cdot q}{y_A} - \zeta_A \cdot k \Bigg)\Bigg]
\Bigg\}. 
\end{eqnarray}

%% file: 4/PWIA.tex
\section{Cross section in plane wave impulse approximation}
\begin{figure}[h!]
\begin{subfigure}{0.5\textwidth}
\centering
    \includegraphics[width=1.0\linewidth]{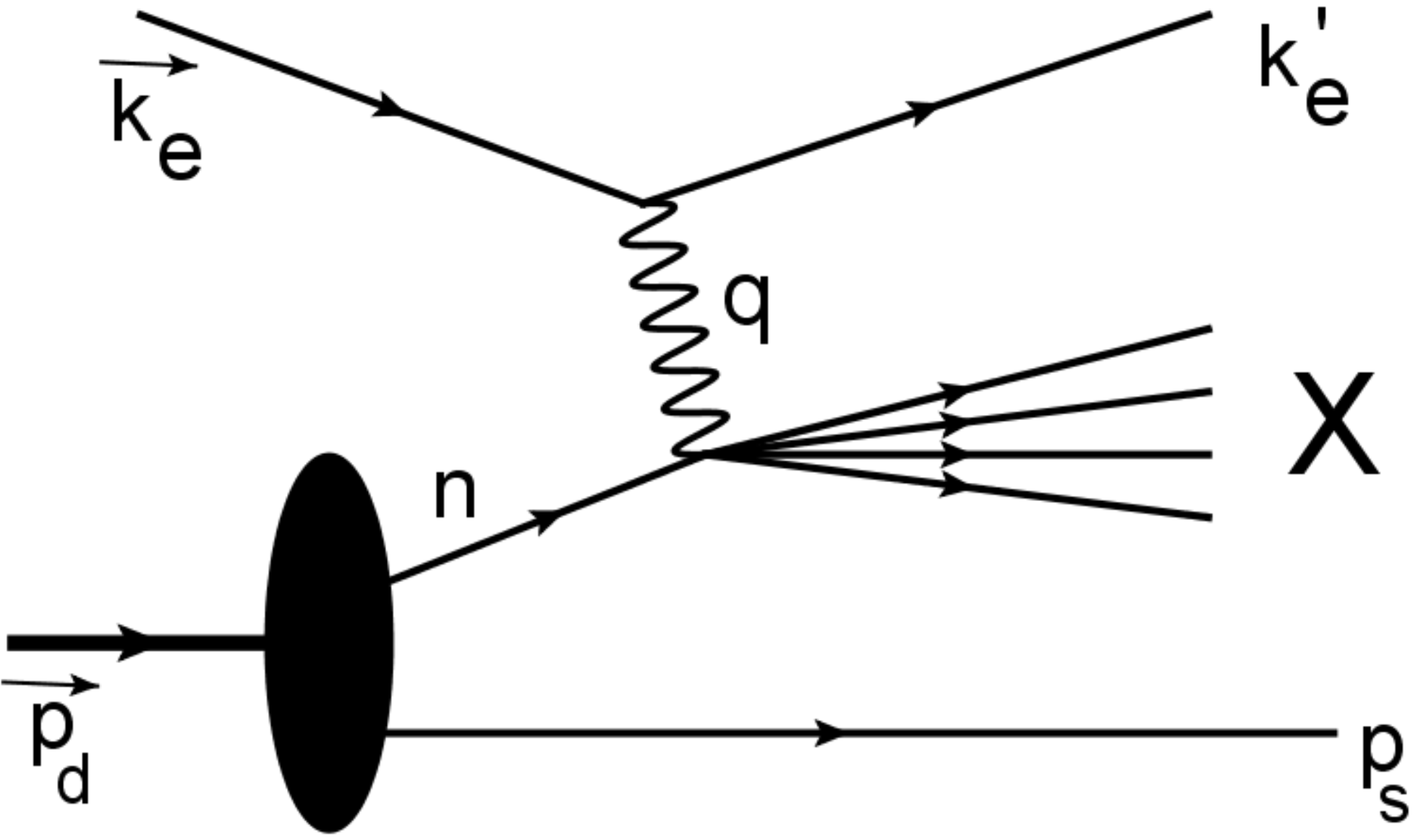}
    \caption{DIS from neutron.}
    \label{fig:DISNeutronInDeut}
\end{subfigure}%
\begin{subfigure}{0.5\textwidth}
\centering
    \includegraphics[width=1.0\linewidth]{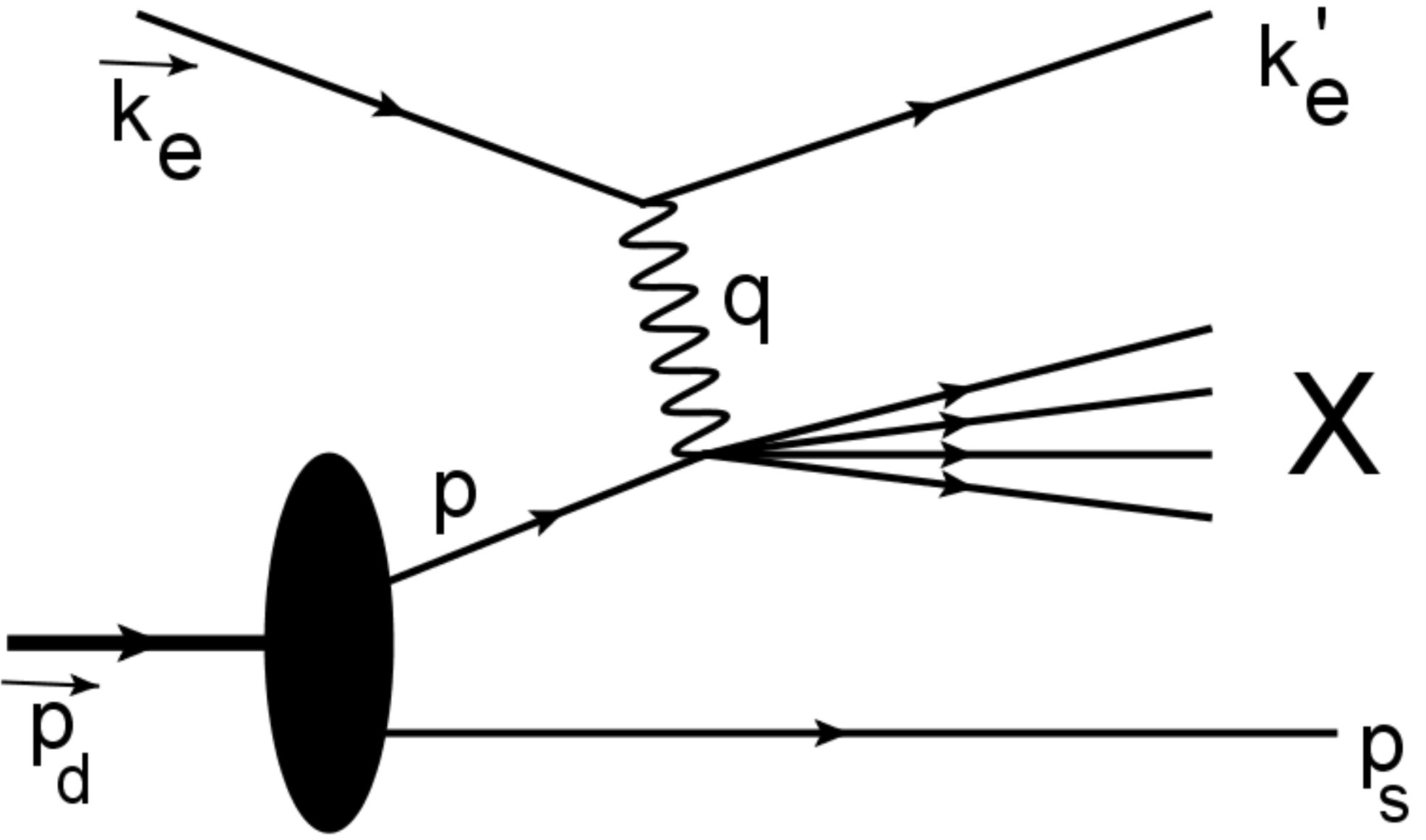}
    \caption{DIS from proton.}
    \label{fig:DISProtonInDeut}
\end{subfigure}
\caption{DIS from the nucleons in deuteron.}
\label{fig:DISfromProtonNeutronInDeuteron}
\end{figure}

\begin{figure}[h!]
\begin{subfigure}{0.5\textwidth}
\centering
    \includegraphics[width=1.0\linewidth]{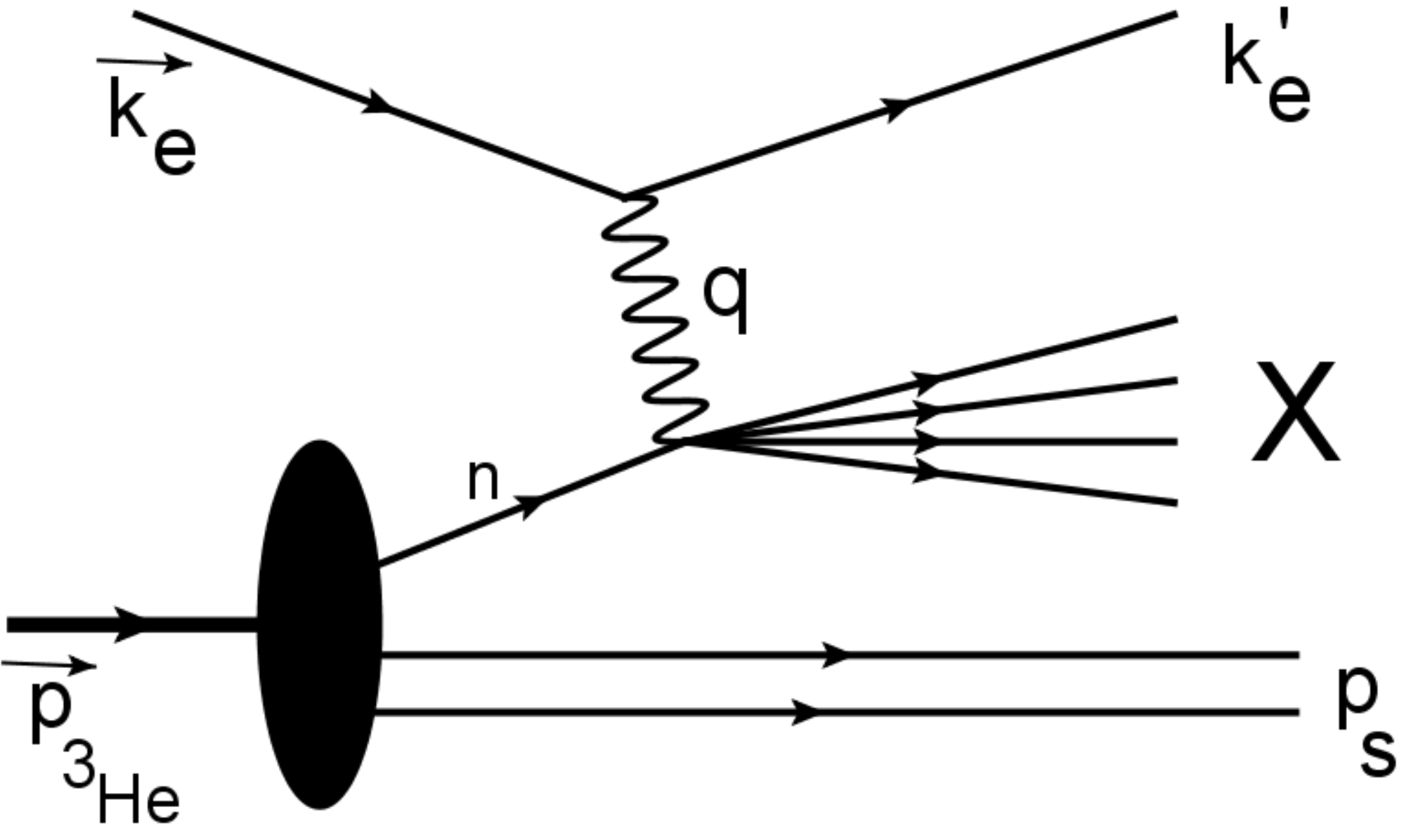}
    \caption{DIS from neutron.}
    \label{fig:DISNeutron}
\end{subfigure}%
\begin{subfigure}{0.5\textwidth}
\centering
    \includegraphics[width=1.0\linewidth]{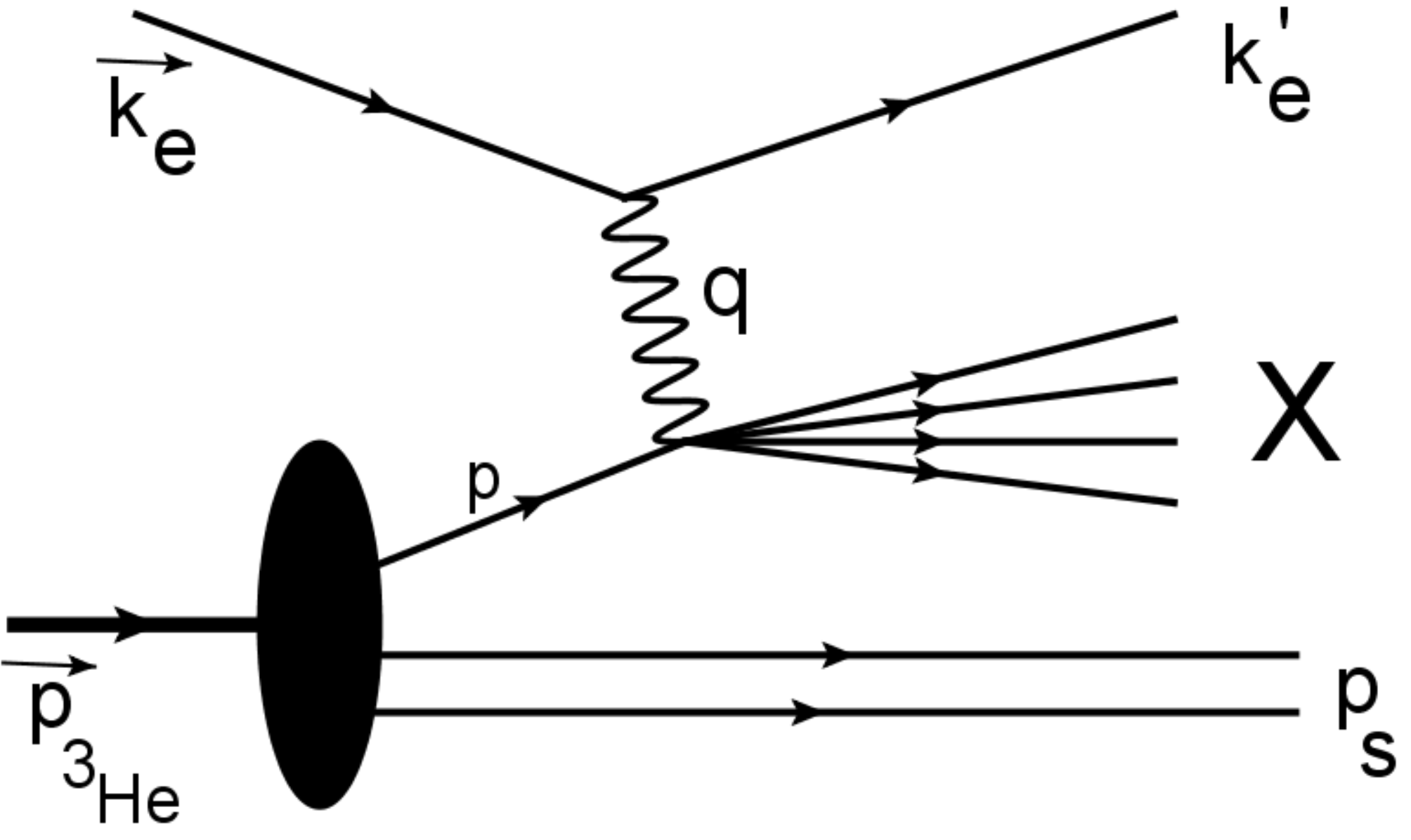}
    \caption{DIS from proton.}
    \label{fig:DISProton}
\end{subfigure}
\caption{DIS from the nucleons in $^3$He.}
\label{fig:DISfromProtonNeutron}
\end{figure}

To be able to study the structure of neutron, for which there is no free target, one needs to consider the scattering of the external probe from the bound neutron in the nucleus. Theoretically, most simple case is the electron scattering from the nucleus in the Plane Wave Impulse Approximation (PWIA), in which case one can apply one-photon exchange approximation and neglect by final state interactions among outgoing particles. The PWIA is best suited for inclusive processes, in which case only scattered electron is detected in the final state of the reaction, and final nuclear and hadronic states are integrated over all the phase space. Since we are interested in extracting the spin structure of the neutron, we consider inclusive polarized  electron scattering from a polarized target. As discussed in the introduction to this chapter, and also from Figs.~(\ref{fig:DISfromProtonNeutronInDeuteron}) and (\ref{fig:DISfromProtonNeutron}), our best choice of polarized target will be the polarized deuteron and $^3$He targets. 

Inclusive scattering of a polarized electron from a polarized  target  can be presented as
\begin{equation}
\vec{e} + \vec{A} \rightarrow e' + X,
\label{DIS}
\end{equation}
where $\vec{e}$, $\vec{A}$ and e$\prime$ represent the incoming polarized electron, polarized nucleus and scattered unpolarized electron respectively, and $X$ represents the deep inelastic scattering (DIS) product. The possible scattering processes are shown in Figs.~(\ref{fig:DISfromProtonNeutronInDeuteron}) and (\ref{fig:DISfromProtonNeutron}) below.

As it was mentioned above, within PWIA, there is no interaction between the deep inelastic products and the nuclear recoil system. The recoil ststem in this scattering can be a proton or neutron for the deuteron target (Fig.~(\ref{fig:DISfromProtonNeutronInDeuteron})) or proton-proton or proton-neutron (deuteron) for $^3$He target (Fig.~(\ref{fig:DISfromProtonNeutron})). It is important to emphasize that due to inclusive nature of the reaction (\ref{DIS}), we do not know whether electrons scattered from neutron or proton. As a result, in the theoretical analysis, one needs to subtract proton contribution inorder to isolate the electron scattering from the neutron. After this subtraction, on needs to account for the Fermi motion effects of the neutron to isolate the scattering from the stationary neuteron. The procedure of such analysis is discussed in the following sections.

\subsection{Defining the Reference Frame of the Reaction}
In Figs.~(\ref{fig:DISfromProtonNeutronInDeuteron}) and (\ref{fig:DISfromProtonNeutron}), $\vec{k_e},$ and $k_e'$ are the  four-momenta of the incoming polarized electron and outgoing unpolarized scattering electron. The momenta of target deuteron and $^3$He are $\vec{p_d}$ and  $\vec{p_{_{^3\text{He}}}}$ respectively.
The momentum transfer in the process is $q$, $X$ is the DIS product, $p_s$ is the momentum of the spectator system over which we integrate and the label $n~(p)$ represents neutron (proton)  interacting with the virtual photon.

\begin{figure}[h!]
\centering
\includegraphics[scale=0.4]{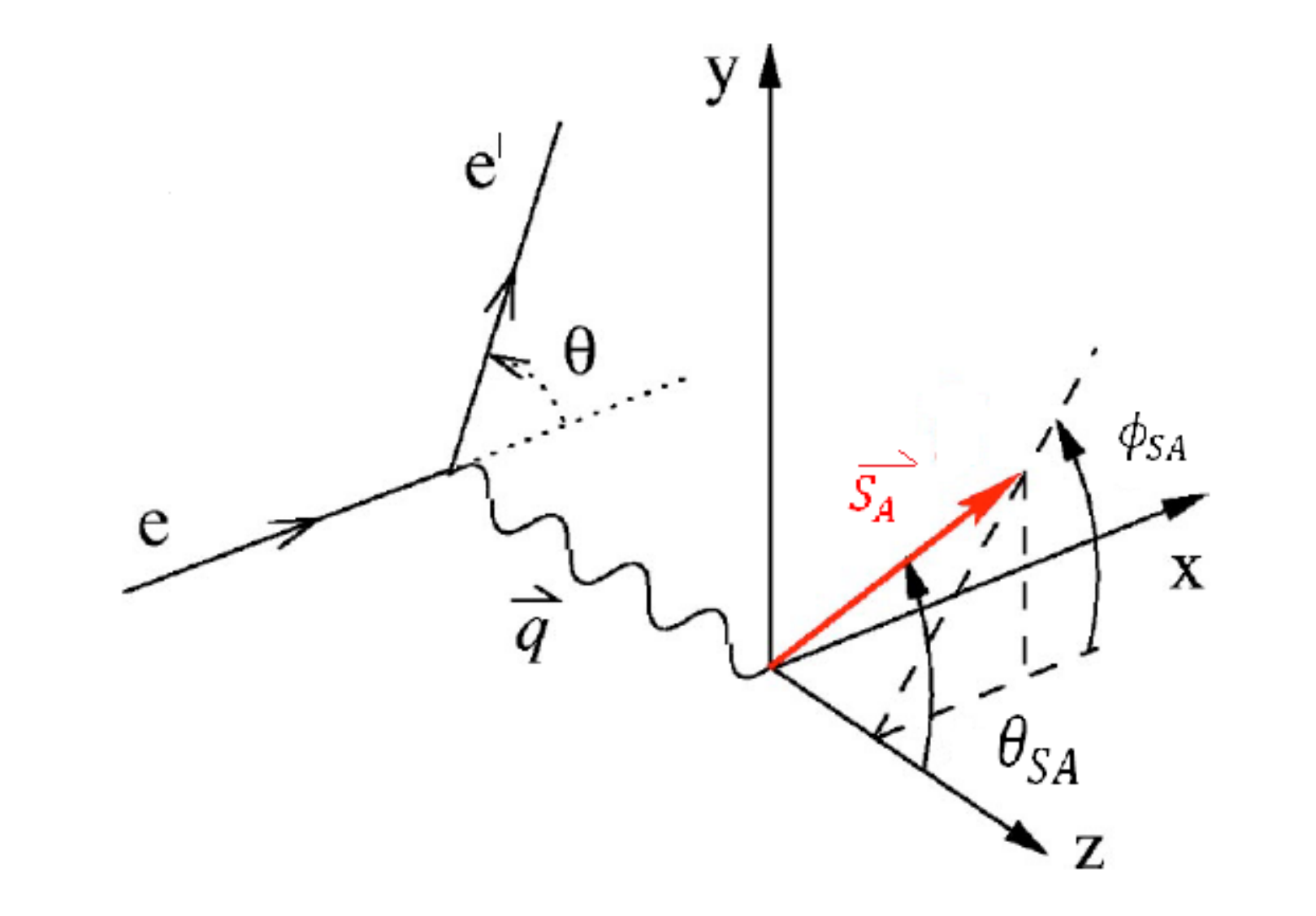}
\caption{Polar and azimuthal angle of the target spin. The source of this diagram is \cite{Zhengetal}.}
\label{polarizationanglesdiagram}
\end{figure}

To be able to calculate the cross section for the processes shown in Figs.~(\ref{fig:DISfromProtonNeutronInDeuteron}) and (\ref{fig:DISfromProtonNeutron}), we need to define two reference frames, one in which the scattering takes place and other which defines the polarization. The scattering or reaction reference frame is defined by the direction of the momentum transfer, such that the $z$ direction is defined by the direction of $\mathbf{q}$, and $xz$ plane is defined by the $k'q$ plane, as shown in Fig.~(\ref{polarizationanglesdiagram}),  where $\theta_{SA}$ and $\phi_{SA}$ are the polar and azimuthal angles made by the target spin vector with the $z$ direction. In the figure, $e, e'$ and $q$ represent the incoming electron with four momentum $k$, outgoing electron with four momentum $k'$ and the direction of the momentum transfer in the scattering respectively.
This allows us to define the $y$ direction as
\begin{equation}
\label{ydirectionReactionFrame}
\hat{y} = \frac{\mathbf{q} \times \mathbf{k'}} {|\mathbf{q} \times \mathbf{k'}|}.
\end{equation} 

We also define the polarization reference frame, $x^\prime y^\prime z^\prime$, such that the $z^\prime$ direction is given by the direction of the nucleus's polarization vector $\mathbf{S_A}$. If we consider $\theta_{sA}, \phi_{sA}$ as the polar and azimuthal angles made by the polarization vector of the nucleus in the reaction reference frame,then we can define the directions in the polarization reference frame as
\begin{eqnarray}
\label{polFrameDir}
\hat{z^\prime} && = (\sin\theta_{sA}\cos\phi_{sA}, \sin\theta_{sA}\sin\phi_{sA}, \cos\theta_{sA}), \nonumber \\
\hat{y^\prime} && = \frac{\mathbf{S_A} \times \mathbf{q}}{|\mathbf{S_A} \times \mathbf{q}|} = (-\sin\phi_{sA}, \cos\phi_{sA},0), \nonumber \\
\hat{x^\prime} && = \hat{y^\prime} \times \hat{z^\prime} = (\cos\theta_{sA}\cos\phi_{sA}, \cos\theta_{sA}\sin\phi_{sA}, -\sin\theta_{sA}).
\end{eqnarray}

In Eq.~(\ref{polFrameDir}), we expressed the directions of the polarization reference frame in terms of the directions in the reaction reference frame. In the special case when the polarization vector of nucleus, $\mathbf{S_A}$, is parallel to $\mathbf{q}$, the polarization reference frame (primed) coincides with the reaction (unprimed) reference frame.


%% file: 4/PWIAcsDeriv.tex
\subsection{Derivation of the cross section }
In this section we will consider the scattering from a generic nucleus $A$, without specifying it as deuteron or $^3$He.
For a scattering process as shown in Figs.~(\ref{fig:DISfromProtonNeutronInDeuteron}) and (\ref{fig:DISfromProtonNeutron}), we can write the amplitude as
\begin{eqnarray}
\mathcal{M}&& = \Bigg[\bar{u}(k_e',s_e')\gamma^\mu \frac{1 + \gamma^5s_e}{2} u(k_e, s_e)\frac{e^2}{q^2}\Bigg]_{lp} \Bigg[\bar{\psi}(p_x, s_x)\Gamma^{\mu}_{eN} \frac{\slashed{p_N} + m_N}{p_N^2 - m_N^2} \bar{\psi}(p_s, s_s) \Gamma_{A\rightarrow NS} \chi_{A}^{s_{A}} \Bigg]_{hp},
\label{starting_amplitude}
\end{eqnarray}
where $[ \cdots ]_{lp}$ and $[ \cdots ]_{hp}$ represent the leptonic and the hadronic parts in the amplitude respectively, $\chi_{A}$ is the spin wave function of the nucleus and $s_{A}$ is the spin of the nucleus, $\psi(p_s, s_s)$ is the wave function of the spectator system,  $\psi(p_x, s_x)$ is the wave function of the DIS product, $\Gamma_{A\rightarrow NS}$ is the vertex representing the transition of the nucleus to a $NS$ system, $p_N$ represents the momentum of the interacting nucleon and $\Gamma_{eN}$ represents the interaction vertex of the virtual photon and the nucleon.
To further analyze the amplitude, we note that because our $z$ axis is defined by the direction of transferred momentum $q$, the larger component of the nucleus is the ``minus'' component, which will be used as the light-front longitudinal momentum to describe the nuclear wave function. 

Using the on-mass shell relation for the particle with mass $M$ and four momentum $p$ as
\begin{equation}
\label{onmassshellrel}
p_+ = \frac{p_{\perp}^2 + M^2}{p_-},
\end{equation} 
the denominator in the hadronic part of the amplitude can be written as
\begin{eqnarray}
p_N^2 - m_N^2 &&= p_{N-} \Bigg[ p_{A+} -p_{s+} - \frac{p_{N\perp}^2 + m_N^2}{p_{N-}}\Bigg] \nonumber \\
&& = p_{N-} \Bigg[ \frac{p_{A\perp}^2 + M_{A}^2}{p_{A-}} - \frac{p_{s\perp}^2 + m_s^2}{p_{s-}} - \frac{p_{N\perp}^2 + m_N^2}{p_{N-}}\Bigg] \nonumber
\end{eqnarray}
\begin{eqnarray}
&& = p_{N-} \Bigg[ \frac{p_{A\perp}^2 + M_{A}^2}{p_{A-}} - \frac{p_{s\perp}^2 + m_s^2}{ A\frac{p_{s-}}{p_{A-}} (\frac{p_{A-}}{A}) }  - \frac{p_{N\perp}^2 + m_N^2}{ A\frac{p_{N-}}{p_{A-}} (\frac{p_{A-}}{A})  }\Bigg] \nonumber \\
&& = p_{N-} \Bigg[ \frac{p_{A\perp}^2 + M_{A}^2}{p_{A-}} - \frac{p_{s\perp}^2 + m_s^2}{ \alpha_s (\frac{p_{A-}}{A}) }  - \frac{p_{N\perp}^2 + m_N^2}{ \alpha_N (\frac{p_{A-}}{A})  }\Bigg] \nonumber \\
&& = \frac{p_{N-}}{p_{A-}} \Bigg[M_{A}^2 + p_{A\perp}^2 - \frac{A (p_{s\perp}^2 + m_s^2)}{\alpha_s} - \frac{A (p_{N\perp}^2 + m_N^2)}{A - \alpha_s} \Bigg],
\label{pidenom_red}
\end{eqnarray}
where $M_A, m_s, m_N$ are the masses of the nucleus, spectator system and interacting nucleon respectively, $\alpha_s = A\frac{p_{s-}}{p_{A-}}$ is the momentum fraction of the nucleus taken by the spectator system, $\alpha_N = A\frac{p_{N-}}{p_{A-}} $ is the momentum fraction of the nucleus taken by the nucleon which interacts with the virtual photon, such that $\alpha_s + \alpha_N = A$. If we assume the interacting nucleon has momentum $p_N$ and spin $s_N$, we use the sum rule for its spinor in the form $\slashed{p}_N + m_N = \sum_{s_N} u(p_N,s_N)\bar{u}(p_N,s_N) $  and write the hadronic part of the amplitude in Eq.~(\ref{starting_amplitude}) as
\begin{equation}
[ \cdots ]_{hp} = \bar{\psi}(p_x, s_x)\Gamma^{\mu}_{eN} \sum_{s_N} \frac{u(p_N, s_N) \bar{u}(p_N, s_N)}{\frac{\alpha_N}{A} \Bigg[M_{A}^2 + p_{A\perp}^2 - \frac{A (p_{s\perp}^2 + m_s^2)}{\alpha_s} - \frac{A (p_{N\perp}^2 + m_N^2)}{A - \alpha_s} \Bigg]} \bar{\psi}(p_s, s_s) \Gamma_{A\rightarrow NS} \chi_{A}^{s_A}
\label{hadronicpart_simp}
\end{equation}
Defining the light-front wave function for the nucleus as (see Appendix B)
\begin{equation}
\psi_{A}^{s_{A}}(\alpha_N, p_N, s_N, s_s) = \frac{\bar{\psi}(p_s, s_s)\bar{u}(p_N, s_N)\Gamma_{{A}\rightarrow NS} \chi_{A}^{s_{A}} }{\frac{1}{A} \Bigg[M_{A}^2 + p_{A\perp}^2 - \frac{A (p_{s\perp}^2 + m_s^2)}{\alpha_s} - \frac{A (p_{N\perp}^2 + m_N^2)}{A - \alpha_s} \Bigg]} \frac{1}{\sqrt{A} \sqrt{2 (2\pi)^3}},
\label{lfwavefunction}
\end{equation}
the hadronic part of the amplitde can then be written as 
\begin{equation}
[ \cdots ]_{hp} = \sum_{s_N} \bar{\psi}(p_x, s_x) \Gamma^\mu _{eN} u (p_N, s_N) \frac{\psi_{A}^{s_{A}}(\alpha_N, p_N, s_N, s_s)  }{\alpha_N} \sqrt{A} \sqrt{2 (2\pi)^3}.
\label{hpwith_lfwavefunction}
\end{equation}

We now define the electromagnetic DIS current of the struck nucleon as
\begin{equation}
\label{emDIScurrent}
J^{\mu}(p_N, s_N, p_x, s_x) \equiv \bar{\psi}(p_x, s_x) \Gamma^\mu _{eN} u (p_N, s_N),
\end{equation}
which can be used to build up the hadronic tensor $W^{\mu \nu}(s_{A})$ of the nucleus as follows:

\begin{eqnarray}
&& 4 \pi m_{A} W^{\mu \nu}(s_{A}) = \sum_{s_N, s_N', s_s, X} J^{\mu \dagger }(p_N, s_N', p_x, s_x) J^{\nu}(p_N, s_N, p_x, s_x) \frac{\psi_{A}^{s_{A} \dagger}(\alpha_N, p_N, s_N', s_s)}{\alpha_N} \nonumber \\
&& \times \frac{\psi_{A}^{s_{A}}(\alpha_N, p_N, s_N, s_s)}{\alpha_N} A~2(2\pi)^3 (2\pi)^4 \delta^4(q + p_A - p_x - p_s) \frac{d^3p_x}{2E_x(2\pi)^3} \frac{d^3p_s}{2E_s(2\pi)^3},
\label{hadronictensor_defn}
\end{eqnarray}
where $E_x$ and $E_s$ represent respectively the energies of the DIS product and the spectator system. If we define the DIS product phase factor, $\Phi_x = (2\pi)^4\delta^4(q + p_N - p_x) \frac{d^3p_x}{2E_x(2\pi)^3} \delta(p_x^2 - m_x^2)$, and the spectator phase factor, $\Phi_s = 2 A~(2\pi)^3 \frac{d^3p_s}{2E_s(2\pi)^3}$, Eq.~(\ref{hadronictensor_defn}) can be rewritten as
\begin{eqnarray}
4 \pi m_{A} W^{\mu \nu}(s_{A}) && = \sum_{s_N, s_N', s_s, X} J^{\mu \dagger }(p_N, s_N', p_x, s_x) J^{\nu}(p_N, s_N, p_x, s_x) \frac{\psi_{A}^{s_{A} \dagger}(\alpha_N, p_N, s_N', s_s)}{\alpha_N^2} \nonumber \\
&& \times \psi_{A}^{s_{A}}(\alpha_N, p_N, s_N, s_s) \Phi_x \Phi_s.
\label{hadronictensorwithphis}
\end{eqnarray}
Note that $\sum_{X}$ indicates that DIS has many final states with different $\Phi_X$. For the phase factor of the spectator system $\Phi_s$, one has to consider specific final states for $^3$He nucleus, such as two-body and three-body break-up cases.
We now need to consider the different possibilities of the spin orientation of the interacting nucleon  with respect to the spin orientation of the nucleus (along the direction of $z'$). We notice that there are four possibilites and the hadronic tensor in Eq.~(\ref{hadronictensorwithphis}) needs to be written as the sum of these possibilities for furhter simplification, which are as follows:
\begin{eqnarray}
\text{I:} &&4 \pi m_{A} W^{\mu \nu}(s_{A})_{1} = \sum_{X, s_s} J^{\mu \dagger }(p_N, s_{N,z'} , p_x, s_x) J^{\nu}(p_N, s_{N,z'}, p_x, s_x) \nonumber \\
&& \times \frac{| \psi_{A}^{s_{A}}(\alpha_N, p_N, s_{N,z'}, s_s)|^2}{\alpha_N^2} \Phi_x \Phi_s \nonumber \\
\text{II:} &&4 \pi m_{A} W^{\mu \nu}(s_{A})_{2} = \sum_{X, s_s} J^{\mu \dagger }(p_N, s_{N,-z'}, p_x, s_x) J^{\nu}(p_N, s_{N,-z'}, p_x, s_x) \nonumber \\
 && \times \frac{|\psi_{A}^{s_{A} }(\alpha_N, p_N, s_{N,-z'}, s_s)|^2}{\alpha_N^2} \Phi_x \Phi_s \nonumber \\
 \text{III:} &&4 \pi m_{A} W^{\mu \nu}(s_{A})_{3} = \sum_{X, s_s} J^{\mu \dagger }(p_N, s_{N,z'} , p_x, s_x) J^{\nu}(p_N, s_{N,-z'}, p_x, s_x) \nonumber \\
&& \times \frac{\psi_{A}^{s_{A} \dagger}(\alpha_N, p_N, s_{N,z'}, s_s)\psi_{A}^{s_{A} }(\alpha_N, p_N, s_{N,-z'}, s_s)}{\alpha_N^2} \Phi_x \Phi_s \nonumber
 \end{eqnarray}
 \begin{eqnarray}
\text{IV:} &&4 \pi m_{A} W^{\mu \nu}(s_{A})_{4} = \sum_{X, s_s} J^{\mu \dagger }(p_N, s_{N,-z'} , p_x, s_x) J^{\nu}(p_N, s_{N,z'}, p_x, s_x) \nonumber \\
&& \times \frac{\psi_{A}^{s_{A} \dagger}(\alpha_N, p_N, s_{N,-z'}, s_s)\psi_{A}^{s_{A} }(\alpha_N, p_N, s_{N,_z'}, s_s)}{\alpha_N^2} \Phi_x \Phi_s .
\label{spinsums}
\end{eqnarray}
In  Eq.~(\ref{spinsums}), $s_{N,z'}$ and $s_{N,-z'}$ represent the situation when the interacting nucleon's spin is parallel or anti-parallel to the nucleus's spin respectively. The first two terms in Eq.~(\ref{spinsums}) are the diagonal products of the currents, which can be written as:
\begin{eqnarray}
4 \pi m_N W_{N}^{\mu \nu}(s_{N} )_{1} && = \sum_{X, s_s} J^{\mu \dagger }(p_N, s_{N,z'} , p_x, s_x) J^{\nu}(p_N, s_{N,z'}, p_x, s_x) \Phi_x \nonumber \\
&&  = Tr\Big[(\slashed{p}_x + m_x) \Gamma^{\mu}_{eN} (\slashed{p}_N + m_N)\frac{1 + \gamma^5 \slashed{s}_N}{2} \Gamma^{\nu}_{eN}\Big] \Phi_x \nonumber \\
&& = \Bigg[\Big(-g^{\mu \nu} - \frac{q^{\mu}q^{\nu}}{Q^2}\Big)W_{1N} + \Big(p_{N}^{\mu} + \frac{p_N \cdot q	}{Q^2 } q^{\mu}\Big)\Big(p_{N}^{\nu} + \frac{p_N \cdot q	}{Q^2 } q^{\nu}\Big) \frac{W_{2N}}{m_N^2} \nonumber \\
&& + i \epsilon^{\mu \nu \rho \sigma} q_{\rho} \Big[ \zeta_{z',\sigma} \frac{G_{1N}}{m_N} + (p_N \cdot q ~\zeta_{z',\sigma} - \zeta_{z'} \cdot q ~p_{\sigma})\frac{G_{2N}}{m_N^3} \Bigg] \text { and }\nonumber \\
4 \pi m_N W_{N}^{\mu \nu}(s_{N} )_{2}&& = \sum_{x, s_s} J^{\mu \dagger }(p_N, s_{N,-z} , p_x, s_x) J^{\nu}(p_N, s_{N,-z}, p_x, s_x) \Phi_x \nonumber \\
&& = Tr\Big[(\slashed{p}_x + m_x) \Gamma^{\mu}_{eN} (\slashed{p}_N + m_N)\frac{1 + \gamma^5 \slashed{s}_N}{2} \Gamma^{\nu}_{eN}\Big] \Phi_x \nonumber 
\end{eqnarray}
\begin{eqnarray}
&& = \Bigg[\Big(-g^{\mu \nu} - \frac{q^{\mu}q^{\nu}}{Q^2}\Big)W_{1N} + \Big(p_{N}^{\mu} + \frac{p_N \cdot q	}{Q^2 } q^{\mu}\Big)\Big(p_{N}^{\nu} + \frac{p_N \cdot q	}{Q^2 } q^{\nu}\Big) \frac{W_{2N}}{m_N^2} \nonumber \\
&& - i \epsilon^{\mu \nu \rho \sigma} q_{\rho} \Big[ \zeta_{-z',\sigma} \frac{G_{1N}}{m_N} + (p_N \cdot q ~\zeta_{-z',\sigma} - \zeta_{-z'} \cdot q ~p_{\sigma})\frac{G_{2N}}{m_N^3} \Bigg],
\label{diag_terms}
\end{eqnarray}
where $\zeta^\sigma$ is the polarization four vector of the struck nucleon, given by
\begin{equation}
\label{pol4vectorStruckNuclen_zprime}
\zeta^\sigma_{\pm z'} (p_N) = \Bigg[ \pm \frac{\mathbf{p_N} \cdot \hat{z'}}{m_N}, \pm \Bigg( \hat{z'} + \frac{(\mathbf{p_N} \cdot \hat{z'}) \mathbf{p_N}}{m_N(E_N + m_N)}\Bigg)\Bigg].
\end{equation}
The third and fourth possibilities gives rise to the non-diagnonal terms. For the case III, writing out the currents explicitly, we find that

\begin{eqnarray}
\sum_{X, s_s} J^{\mu \dagger }(p_N, s_{N,z'} , p_x, s_x) J^{\nu}(p_N, s_{N,-z'}, p_x, s_x) &&= \sum_{X, s_s} \bar{u}(p_N, s_{N,z}) \Gamma^{\mu \dagger}_{eN}\psi(p_x, s_x) \bar{\psi}(p_x, s_x) \nonumber \\
&& \times	\Gamma^{\nu}_{eN} u(p_N, s_{N,-z'}). 
\label{off_diag_first}
\end{eqnarray}
Noting that $u(p_N, s_{N,-z'})\bar{u}(p_N, s_{N,z'}) = \dfrac{(\slashed{p}_N + m_N)\gamma^5\slashed{\zeta}_{-}}{2}$, with $\zeta_{-} = \zeta_{x'} - i \zeta_{y'}$, we can write Eq.~(\ref{off_diag_first}) as
\begin{eqnarray}
4 \pi m_N W^{\mu \nu}(s_N)_{3}&& = i \epsilon^{\mu \nu \rho \sigma} q_{\rho} \Bigg[ \zeta_{-,\sigma} \frac{G_{1N}}{m_N} + \Big[ (p_N \cdot q) \zeta_{-,\sigma} - (q \cdot \zeta_{-}) p_{N,\sigma} \Big]\frac{G_{2N}}{m_N^3}\Bigg].
\label{off_diag_first_final}
\end{eqnarray}
In a similar manner, noting that $u(p_N, s_{N,z'})\bar{u}(p_N, s_{N,-z'}) = \dfrac{(\slashed{p}_N + m_N)\gamma^5\slashed{\zeta}_{+}}{2}$, with $\zeta_{+} = \zeta_{x'} + i \zeta_{y'}$, we can write the case IV in Eq.~(\ref{spinsums}) as
\begin{eqnarray}
4 \pi m_N W^{\mu \nu}(s_N)_{4}&& = i \epsilon^{\mu \nu \rho \sigma} q_{\rho} \Bigg[ \zeta_{+,\sigma} \frac{G_{1N}}{m_N} + \Big[ (p_N \cdot q) \zeta_{+,\sigma} - (q \cdot \zeta_{+}) p_{N,\sigma} \Big]\frac{G_{2N}}{m_N^3}\Bigg],
\label{off_diag_second_final}
\end{eqnarray}
where we define
\begin{eqnarray}
\label{pol4vectorStruckNucleonxyprime}
&& \zeta^\sigma_{x'} (p_N) = \Bigg[  \frac{\mathbf{p_N} \cdot \hat{x'}}{m_N},   \hat{x'} + \frac{(\mathbf{p_N} \cdot \hat{z'}) \mathbf{p_N}}{m_N(E_N + m_N)}\Bigg] \text{ and, } \nonumber \\
&& \zeta^\sigma_{y'} (p_N) = \Bigg[  \frac{\mathbf{p_N} \cdot \hat{y'}}{m_N},   \hat{y'} + \frac{(\mathbf{p_N} \cdot \hat{z'}) \mathbf{p_N}}{m_N(E_N + m_N)}\Bigg].
\end{eqnarray}

We now introduce the following definitions that accounts for the unpolarized and polarized parts of the hadronic tensor as
\begin{eqnarray}
&& H_0^{\mu \nu} = \Big(-g^{\mu \nu} - \frac{q^{\mu}q^{\nu}}{Q^2}\Big)W_{1N} + \Big(p_{N}^{\mu} + \frac{p_N \cdot q	}{Q^2 } q^{\mu}\Big)\Big(p_{N}^{\nu} + \frac{p_N \cdot q	}{Q^2 } q^{\nu}\Big) \frac{W_{2N}}{m_N^2}, \text{ and } \nonumber \\
&& H_j^{\mu \nu} = i \epsilon^{\mu \nu \rho \sigma} q_{\rho} \Bigg[ \zeta_{j,\sigma} \frac{G_{1N}}{m_N} + [ (p_N \cdot q) \zeta_{j,\sigma} - (q \cdot \zeta_{j}) p_{N,\sigma}]\frac{G_{2N}}{m_N^3}\Bigg],
\label{H0Hjpartsdefn}
\end{eqnarray}
where $ j = x',y'$ and $z'$.

Then, using Eqs.~(\ref{diag_terms}) and (\ref{H0Hjpartsdefn}), we obtain
\begin{eqnarray}
&& 4 \pi m_N W_{N}^{\mu \nu}(s_{N} )_{1} = (H_0^{\mu \nu} + H_{z'}^{\mu \nu}) \frac{| \psi_{A}^{s_{A}}(\alpha_N, p_N, s_{N,z'}, s_s)|^2}{\alpha_N^2} \Phi_s \nonumber  \text { and } \\
&& 4 \pi m_N W_{N}^{\mu \nu}(s_{N} )_{2} = (H_0^{\mu \nu} - H_{z'}^{\mu \nu}) \frac{| \psi_{A}^{s_{A}}(\alpha_N, p_N, s_{N,-z'}, s_s)|^2}{\alpha_N^2} \Phi_s.
\label{nucleonictensorwithHoHz}
\end{eqnarray}
Similarly, from Eqs.~(\ref{off_diag_first_final}), (\ref{off_diag_second_final}) and (\ref{H0Hjpartsdefn}), we obtain:
\begin{eqnarray}
 4 \pi m_N W^{\mu \nu}(s_N)_{3} && =  (H_{x'}^{\mu \nu} - iH_{y'}^{\mu \nu}) \psi_{A}^{s_{A} \dagger}(\alpha_N, p_N, s_{N,z'}, s_s) \nonumber \\
&& \qquad  \times \frac{\psi_{A}^{s_{A} }(\alpha_N, p_N, s_{N,-z'}, s_s)}{\alpha_N^2} \Phi_s \nonumber \text { and } \\
 4 \pi m_N W^{\mu \nu}(s_N)_{4} && =  (H_{x'}^{\mu \nu} + iH_{y'}^{\mu \nu}) \psi_{A}^{s_{A} \dagger}(\alpha_N, p_N, s_{N,-z'}, s_s) \nonumber \\
&& \qquad  \times \frac{\psi_{A}^{s_{A} }(\alpha_N, p_N, s_{N,z'}, s_s)}{\alpha_N^2} \Phi_s.
\label{nucleonictensorwithHxHy}
\end{eqnarray}
Combining Eqs~(\ref{hadronictensorwithphis}), (\ref{spinsums}), (\ref{nucleonictensorwithHoHz}) and (\ref{nucleonictensorwithHxHy}), we obtain the following expression for the hadronic tensor:
\begin{eqnarray}
4 \pi m_{A}W^{\mu \nu}(s_{A})&& = 4 \pi m_N\Bigg [ H_0^{\mu \nu} \Big\{ \frac{| \psi_{A}^{s_{A}}(\alpha_N, p_N, s_{N,z'}, s_s)|^2 + | \psi_{A}^{s_{A}}(\alpha_N, p_N, s_{N,-z'}, s_s)|^2}{\alpha_N^2} \Big\} \nonumber \\
&& + H_z^{\mu \nu} \Big\{ \frac{| \psi_{A}^{s_{A}}(\alpha_N, p_N, s_{N,z'}, s_s)|^2 - | \psi_{A}^{s_{A}}(\alpha_N, p_N, s_{N,-z'}, s_s)|^2}{\alpha_N^2} \Big\} \nonumber \\
&& + H_x^{\mu \nu} \Big\{ \frac{\psi_{A}^{s_{A} \dagger}(\alpha_N, p_N, s_{N,z'}, s_s)\psi_{A}^{s_{A} }(\alpha_N, p_N, s_{N,-z'}, s_s)}{\alpha_N^2} \nonumber \\
&& + \frac{\psi_{A}^{s_{A} \dagger}(\alpha_N, p_N, s_{N,-z'}, s_s)\psi_{A}^{s_{A} }(\alpha_N, p_N, s_{N,z'}, s_s)}{\alpha_N^2}\Big\} \nonumber \\
&& -iH_y^{\mu \nu} \Big\{ \frac{\psi_{A}^{s_{A} \dagger}(\alpha_N, p_N, s_{N,z'}, s_s)\psi_{A}^{s_{A} }(\alpha_N, p_N, s_{N,-z'}, s_s)}{\alpha_N^2} \nonumber \\
&& - \frac{\psi_{A}^{s_{A} \dagger}(\alpha_N, p_N, s_{N,-z'}, s_s)\psi_{A}^{s_{A} }(\alpha_N, p_N, s_{N,z'}, s_s)}{\alpha_N^2} \Big\} \Bigg] \Phi_s.
\label{nucleartensorintermsofnucleonic}
\end{eqnarray}
For simplicity, we now define
\begin{eqnarray}
 g_0^{s_{A}}(\alpha_N, p_N) &&= | \psi_{A}^{s_{A}}(\alpha_N, p_N, s_{N,z'}, s_s)|^2 +  | \psi_{A}^{s_{A}}(\alpha_N, p_N, s_{N,-z'}, s_s)|^2 \nonumber
 \end{eqnarray}
 \begin{eqnarray}
 g_z^{s_{A}}(\alpha_N, p_N) &&= | \psi_{A}^{s_{A}}(\alpha_N, p_N, s_{N,z'}, s_s)|^2 -  | \psi_{A}^{s_{A}}(\alpha_N, p_N, s_{N,-z'}, s_s)|^2 \nonumber \\
 g_x^{s_{A}}(\alpha_N, p_N) &&= \psi_{A}^{s_{A} \dagger}(\alpha_N, p_N, s_{N,z'}, s_s)\psi_{A}^{s_{A} }(\alpha_N, p_N, s_{N,-z'}, s_s) \nonumber \\
&&  + \psi_{A}^{s_{A} \dagger}(\alpha_N, p_N, s_{N,-z'}, s_s)\psi_{A}^{s_{A} }(\alpha_N, p_N, s_{N,z'}, s_s) \nonumber \\
&& = 2 \text{Re} [\psi_{A}^{s_{A} \dagger}(\alpha_N, p_N, s_{N,z}, s_s)\psi_{A}^{s_{A} }(\alpha_N, p_N, s_{N,-z'}, s_s)] \nonumber \\
 g_y^{s_{A}}(\alpha_N, p_N) &&= \psi_{A}^{s_{A} \dagger}(\alpha_N, p_N, s_{N,z'}, s_s)\psi_{A}^{s_{A} }(\alpha_N, p_N, s_{N,-z'}, s_s) \nonumber \\
&& - \psi_{A}^{s_{A} \dagger}(\alpha_N, p_N, s_{N,-z'}, s_s)\psi_{A}^{s_{A} }(\alpha_N, p_N, s_{N,z'}, s_s) \nonumber \\
&& = 2 \text{Im} [\psi_{A}^{s_{A} \dagger}(\alpha_N, p_N, s_{N,z'}, s_s)\psi_{A}^{s_{A} }(\alpha_N, p_N, s_{N,-z'}, s_s)].
\label{gogxgygz}
\end{eqnarray}
Using Eq.~(\ref{gogxgygz}) in Eq.~(\ref{nucleartensorintermsofnucleonic}) and substituting back for $\Phi_s$, we then obtain:
\begin{equation}
W^{\mu \nu}_{A} = \frac{m_N}{m_{A}} \Big[H_0^{\mu \nu} \frac{g_0^{s_{A}}}{\alpha_N^2} + \sum_{j = 1}^{3} \frac{H_j^{\mu \nu}g_j^{s_{A}}}{\alpha_N^2} \Big] 2 A (2\pi)^3 \delta(p_s^2 - m_s^2) \frac{d^4p_s}{(2\pi)^3}.
\label{wmunu_intermsofgogxgygzplusphis}
\end{equation}
For the situation in which the mass of the spectator system is fixed, one can express $\Phi_s = 2 A \delta(p_s^2 - m_s^2)d^4p_s$.
Then for the $\delta(p_s^2 - m_s^2)$ part, we can write
\begin{eqnarray}
\delta(p_s^2 - m_s^2)d^4p_s && = \delta(p_{s+}p_{s-} - p_{s\perp}^2 - m_s^2)d^4p_s = \delta\Big[ p_{s-}\Big(p_{s+} - \frac{m_s^2 + p_{s\perp}^2}{p_{s-}} \Big) \Big]\frac{1}{2}dp_{s+}dp_{s-}d^2p_{s\perp} \nonumber \\
&& = \frac{1}{p_{s-}} \delta\Big[p_{s+} - \frac{m_s^2 + p_{s\perp}^2}{p_{s-}} \Big] \frac{1}{2}dp_{s+}dp_{s-}d^2p_{s\perp} \nonumber \\
&& = \frac{1}{2} \frac{d\alpha_s}{\alpha_s} d^2p_{s\perp},
\label{spec_delta_integration}
\end{eqnarray}
where in the last step, the $p_{s+}$ component of the momentum is integrated according to Eq.~(\ref{polevalueint_scheme}). 

We now define the light-front density matrices for the nucleus as
\begin{equation}
 \rho_{0(j)}^{A}(\alpha_N, p_N) = \frac{g_{0(j)}^{A}(\alpha_N, p_N)}{\alpha_s} = \frac{g_{0(j)}^{A}}{A -\alpha_N},
 \label{densitymatrices} 
\end{equation}
and make use of Eq.~(\ref{spec_delta_integration}) to rewrite the hadronic tensor in Eq.~(\ref{wmunu_intermsofgogxgygzplusphis}) as
\begin{equation}
W^{\mu \nu}_{A} = A \frac{m_N}{m_{A}} \Big[H_0^{\mu \nu} \frac{\rho_0^{s_{A}}}{\alpha_N^2} + \sum_{j = 1}^{3} \frac{H_j^{\mu \nu}\rho_j^{s_{A}}}{\alpha_N^2} \Big] {d\alpha_s}d^2p_{s\perp}.
\label{wmunuintermsofdensitymatrices}
\end{equation}
We now calculate the differential cross section within the PWIA as:
\begin{eqnarray}
d\sigma && = \frac{e^4}{q^4} \frac{1}{4 k_e \cdot p_{A}} L_{\mu \nu}W^{\mu \nu}_{A} 4 \pi m_{A} \frac{d^3k_e'}{2E_k'(2 \pi)^3} \nonumber \\
&& = \frac{e^4}{Q^4} \frac{4 \pi m_{A}}{4 k_e \cdot p_{A}} L_{\mu \nu}W^{\mu \nu}_{A}  \frac{E_k'^2 dE_k' d\Omega}{2E_k'(2 \pi)^3} \nonumber \\
&& = \frac{\alpha_{em}^2}{Q^4} \frac{m_{A}}{k_e \cdot p_{A}} L_{\mu \nu}W^{\mu \nu}_{A} E_k' dE_k' d\Omega \nonumber \\ 
&& = \frac{\alpha_{em}^2}{Q^4} \frac{m_{A}}{k_e \cdot p_{A}} L_{\mu \nu}W^{\mu \nu}_{A} E_k' dE_k' d\cos\theta_e d\phi_e,
\label{crosssection_pwia_start}
\end{eqnarray}
where $\alpha_{em} = \frac{e^2}{4 \pi}$. If we make use of Eq.~(\ref{diffls_transf_completed}), then this differential cross section can be written as
\begin{eqnarray}
\frac{d\sigma}{dQ^2 d x_A} && = 2 \pi \frac{\alpha_{em}^2}{Q^4}m_{A}\frac{y_{A}^2}{A~Q^2} L_{\mu \nu} W^{\mu \nu}_{A},
\label{crossection_intermsofdQ^2dxA}
\end{eqnarray}
where $2\pi$ factor is the result of the integration over $d\phi_e$. Using Eq.~(\ref{wmunuintermsofdensitymatrices}), we find from Eq.~(\ref{crossection_intermsofdQ^2dxA}), the cross section can be written as
\begin{eqnarray}
\frac{d\sigma}{dQ^2 d x_A} && = 2 \pi \frac{\alpha_{em}^2}{Q^4} m_N \frac{y_{A}^2}{Q^2} L_{\mu \nu} \Bigg[H_0^{\mu \nu} \frac{\rho_0^{s_{A}}}{\alpha_N^2} + \sum_{j = 1}^{3} \frac{H_j^{\mu \nu}\rho_j^{s_{A}}}{\alpha_N^2} \Bigg] {d\alpha_s}d^2p_{s\perp}.
\label{dsigdq^2dxA}
\end{eqnarray}

The contraction of the leptonic and hadronic tensors can be performed (as in obtaining Eq.~(\ref{contractionFinal})), which gives the following:
\begin{eqnarray}
&& L_{\mu \nu}H_0^{\mu \nu} + L_{\mu \nu}H_j^{\mu \nu}  = 2 Q^2 \frac{F_{1N}(x_N, Q^2)}{m_N} + \frac{ Q^2}{2 x_N m_N y_N^2}F_{2N}(x_N, Q^2)\Bigg[ (2 - y_N)^2  - y_N^2 \Big(1 + \frac{4 x_N^2 m_N^2 }{Q^2} \Big) \Bigg]  \nonumber \\ 
&&  + 2 h_e Q^2 \sum_{j' = 1}^{3} \Bigg[(\zeta_{j'}\cdot q - 2 \zeta_{j'} \cdot k)\frac{g_{1N}(x_N, Q^2)}{p_N \cdot q} + 2~\Bigg( \frac{\zeta_{j'}\cdot q}{y_N} - \zeta_{j'} \cdot k \Bigg) \frac{g_{2N}(x_N, Q^2)}{p_N \cdot q}  \Bigg].
\label{contraction_nucleon} 
\end{eqnarray}
Thus, combining Eqs.~(\ref{contraction_nucleon}) and (\ref{dsigdq^2dxA}), the differential cross section in the PWIA can be written as follows:

\begin{eqnarray}
\label{csPWIA}
\frac{d\sigma}{dQ^2 d x_A}  && = 2 \pi \frac{\alpha_{em}^2}{Q^4} y_A^2 \sum_{N} \Bigg\{ \Bigg( 2F_{1N}(x_N, Q^2) + \frac{1}{2 x_N y_N^2}\Bigg[(2 - y_N)^2  - y_N^2 \Big(1 + \frac{4 x_N^2 m_N^2 }{Q^2} \Big)\Bigg]\times  \nonumber  \\
&&  F_{2N}(x_N, Q^2) \Bigg)\frac{\rho_0^{s_A}(\alpha_N, p_{N\perp})}{\alpha_N^2} + 2 h_e m_N \sum_{j' = 1}^{3} \Bigg[(\zeta_{j'}\cdot q - 2~\zeta_{j'} \cdot k)\frac{g_{1N}(x_N, Q^2)}{p_N \cdot q} \nonumber \\
&& + 2~\Bigg( \frac{\zeta_{j'}\cdot q}{y_N} - \zeta_{j'} \cdot k \Bigg) \frac{g_{2N}(x_N, Q^2)}{p_N \cdot q}  \Bigg]\frac{\rho_{j'}^{s_A}(\alpha_N, p_{N\perp})}{\alpha_N^2} \Bigg\} d\alpha_s d^2p_{s\perp},
\end{eqnarray}
where the structure functions for nucleon are redefined as $F_{1N} = m_N W_{1N}, F_{2N} = \frac{p_N \cdot q}{m_N} W_{2N}, g_{1N} = \frac{p_N \cdot q}{m_N} G_{1N}$ and $g_{2N} = \frac{(p_N \cdot q)^2}{m_N^3}G_{2N}$, and we also sum over the contribution of all the nucleons in the nucleus which could be struck. For example, if the above expression is used for deuteron, we sum the contribution from proton and neutron, while for $^3$He, we sum the contribution of one neutron and two protons. In the next section, we discuss in detail the methods used to calculate numerically the cross section from Eq.~(\ref{csPWIA}). 

%% file: 4/NumericalEstimatesPWIA.tex

\section{Numerical estimates in PWIA- unpolarized part}
To be able to do numerical calculations using Eq.~(\ref{csPWIA}), we first need an appropriate way to incorporate the structure functions of the nucleon. For simplicity, we first discuss the case of the unpolarized scattering, which is given by 

\begin{eqnarray}
\label{csPWIA_UnpolarizedPartOnly}
\frac{d\sigma}{dQ^2 d x_A}  && = 2 \pi \frac{\alpha_{em}^2}{Q^4} y_A^2 \sum_{N} \int \Bigg\{  \Bigg( 2F_{1N}(x_N, Q^2)+ \frac{1}{2 x_N y_N^2} \Bigg[(2 - y_N)^2  - y_N^2 \Big(1 + \frac{4 x_N^2 m_N^2 }{Q^2} \Big)\Bigg] \times \nonumber \\
&& F_{2N}(x_N, Q^2) \Bigg) \frac{\rho_0^{s_A}(\alpha_N, p_{N\perp})}{\alpha_N^2} \Bigg\} d\alpha_s d^2p_{s\perp},
\end{eqnarray}
where we integrate over the momentum fraction of the nucleus carried by the spectator and also the spectator's transverse momentum as we are concerned with inclusive scattering. This allows us to  consider only the unpolarized structure functions $F_{1N}(x_N, Q^2)$ and $F_{2N}(x_N, Q^2)$, which are parametrized according to the Bodek paramterization\cite{BodekParametrization}, which is valid over a wide range of $Q^2$ ($1 < Q^2 < 20$~GeV$^2$) and Bjorken $x$ ($0.1 < x < 0.77$). We also note that, since $d^2p_{s\perp} = p_{s\perp}dp_{s\perp} d\phi_s$, the    azimuthal dependence, $\phi_s$, of the spectator nucleon should also be taken care of, which gets introduced into the cross section through the definition of $y_N \equiv \frac{p_N \cdot q}{p_N \cdot k}$, with $p_N \cdot k = \frac{1}{2}(p_{N+}k_{-} + p_{N-}k_{+}) - p_{N\perp} k_{\perp} \cos\phi_s$.  The $\phi_s$ integral is performed analytically, and the resulting expression for the differential cross section is (for details, see Appendix D)
\begin{eqnarray}
\label{csPWIA_unpolarizedAfterPhiInt}
\frac{d\sigma}{dQ^2 dx_A} && = \frac{4 \pi^2 \alpha_{em}^2 y_A^2}{Q^4} \sum_{N} \int \Bigg\{ 2F_{1N}(x_N, Q^2) 
 + \frac{2 F_{2N}(x_N, Q^2)}{x_N} \times \nonumber \\
 && \Bigg[\frac{B^2 + \frac{1}{2} p_{N\perp}^2 k_\perp^2}{(p_N \cdot q)^2} - \frac{B}{p_N \cdot q}	- \frac{ x_N^2 m_N^2 }{Q^2} \Bigg]  \Bigg\} \frac{\rho_0^{s_A}(\alpha_N, p_{N\perp})}{\alpha_N^2}d\alpha_s p_{s\perp} dp_{s\perp},
\end{eqnarray}
where the factor $B = \frac{1}{2}(p_{N+}k_{-} + p_{N-}k_{+})$. The accuracy of this equation is checked by comparing it to existing world data for the case of DIS from deuteron and $^3$He. 

\subsection{Comparison with deuteron data}
\begin{figure} [h!]
\begin{subfigure}{0.5\textwidth}
\centering
    \includegraphics[width=1.0\linewidth]{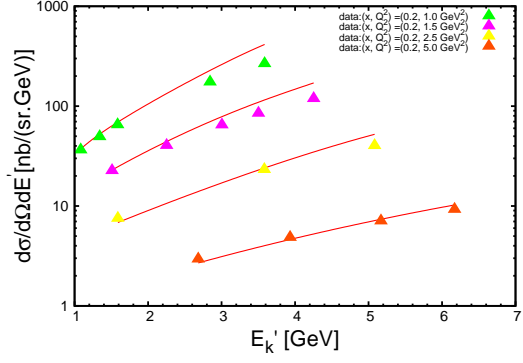}
    \caption{}
    \label{fig:whitcomp1}
\end{subfigure}
\begin{subfigure}{0.5\textwidth}
\centering
    \includegraphics[width=1.0\linewidth]{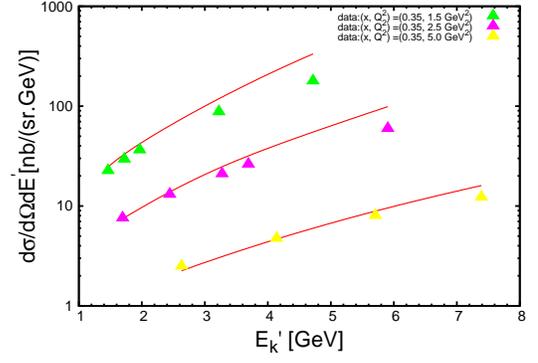}
    \caption{}
    \label{fig:whitcomp2}
\end{subfigure}
\begin{subfigure}{0.5\textwidth}
\centering
    \includegraphics[width=1.0\linewidth]{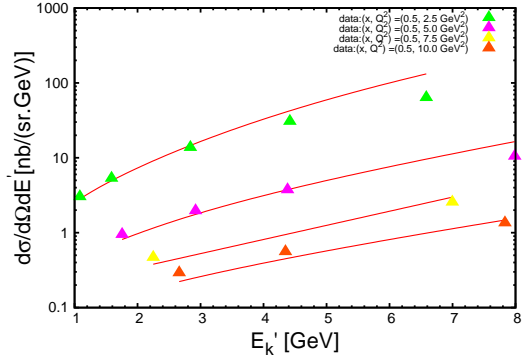}
    \caption{}
    \label{fig:whitcomp3}
\end{subfigure}%
\caption{Comparison of cross sections for deep inelastic scattering off deuteron. The cross sections are plotted against the scattering energy of electron. Data used in these figures are from \cite{WhitlowData}.}
\label{whitlowComp}
\end{figure}

\begin{figure} [h!]
\begin{subfigure}{0.5\textwidth}
\centering
    \includegraphics[width=1.0\linewidth]{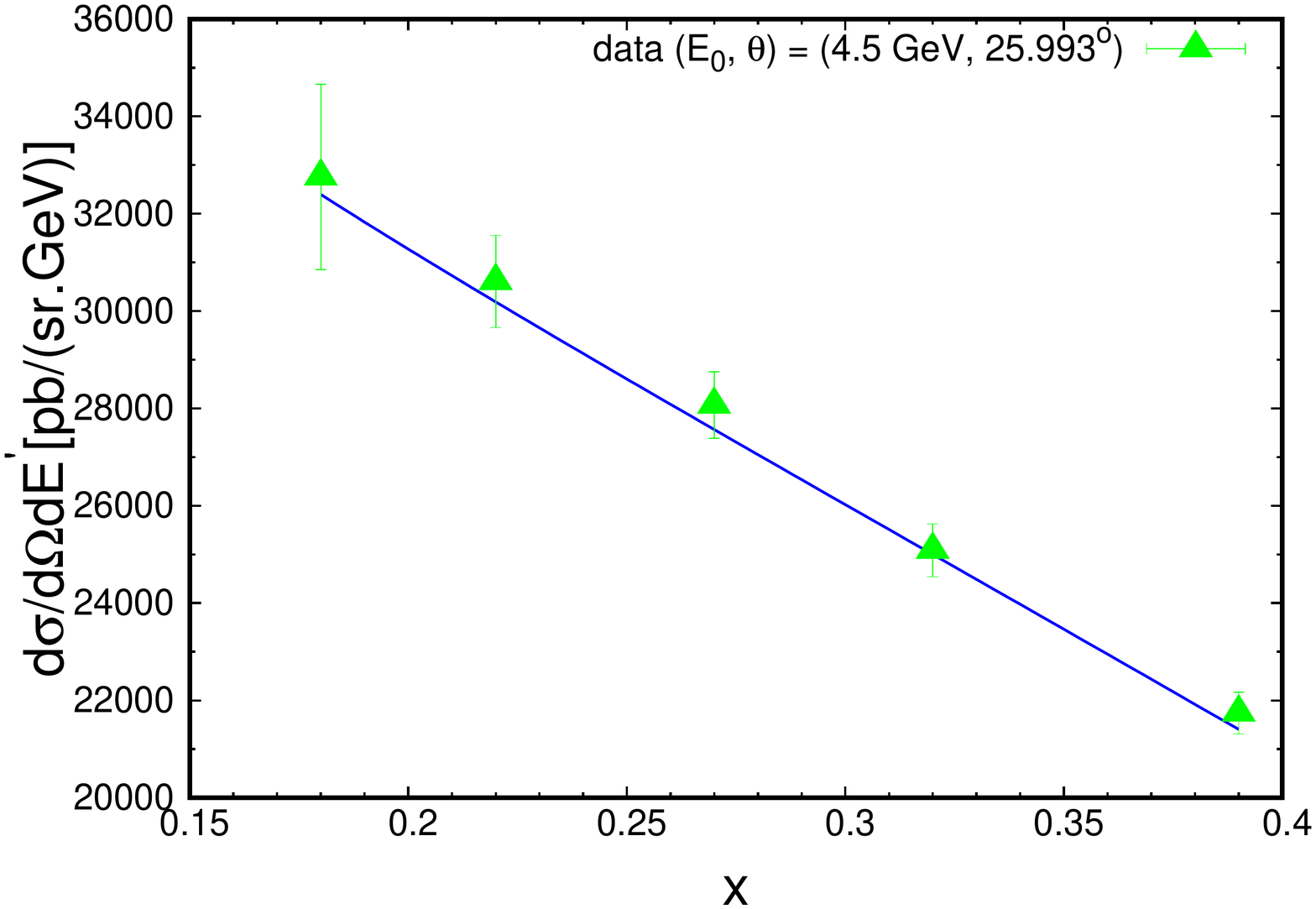}
    \caption{}
    \label{fig:bodekComp1}
\end{subfigure}
\begin{subfigure}{0.5\textwidth}
\centering
    \includegraphics[width=1.0\linewidth]{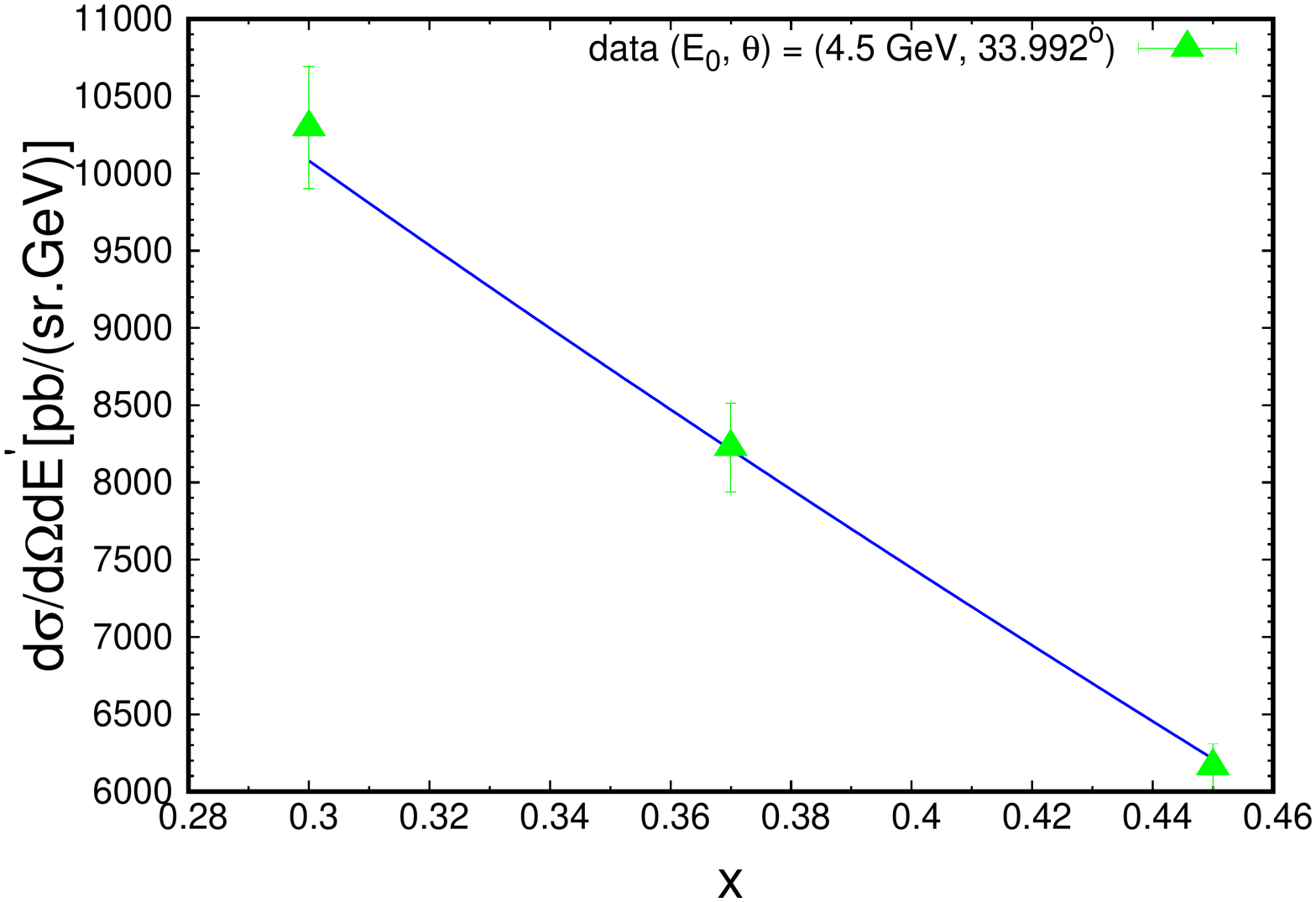}
    \caption{}
    \label{fig:bodekComp2}
\end{subfigure}
\begin{subfigure}{0.5\textwidth}
\centering
    \includegraphics[width=1.0\linewidth]{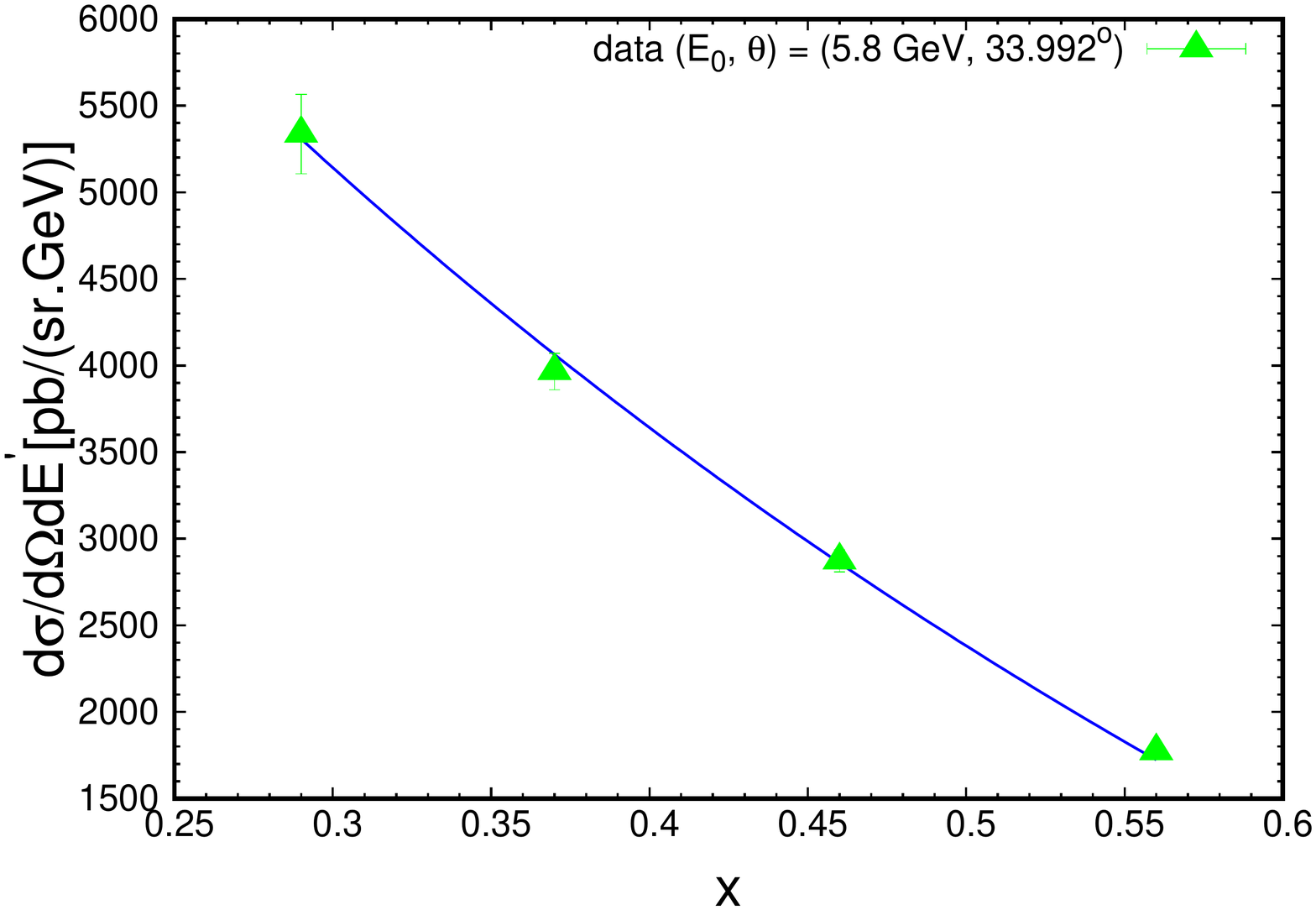}
    \caption{}
    \label{fig:bodekComp3}
\end{subfigure}
\begin{subfigure}{0.5\textwidth}
\centering
    \includegraphics[width=1.0\linewidth]{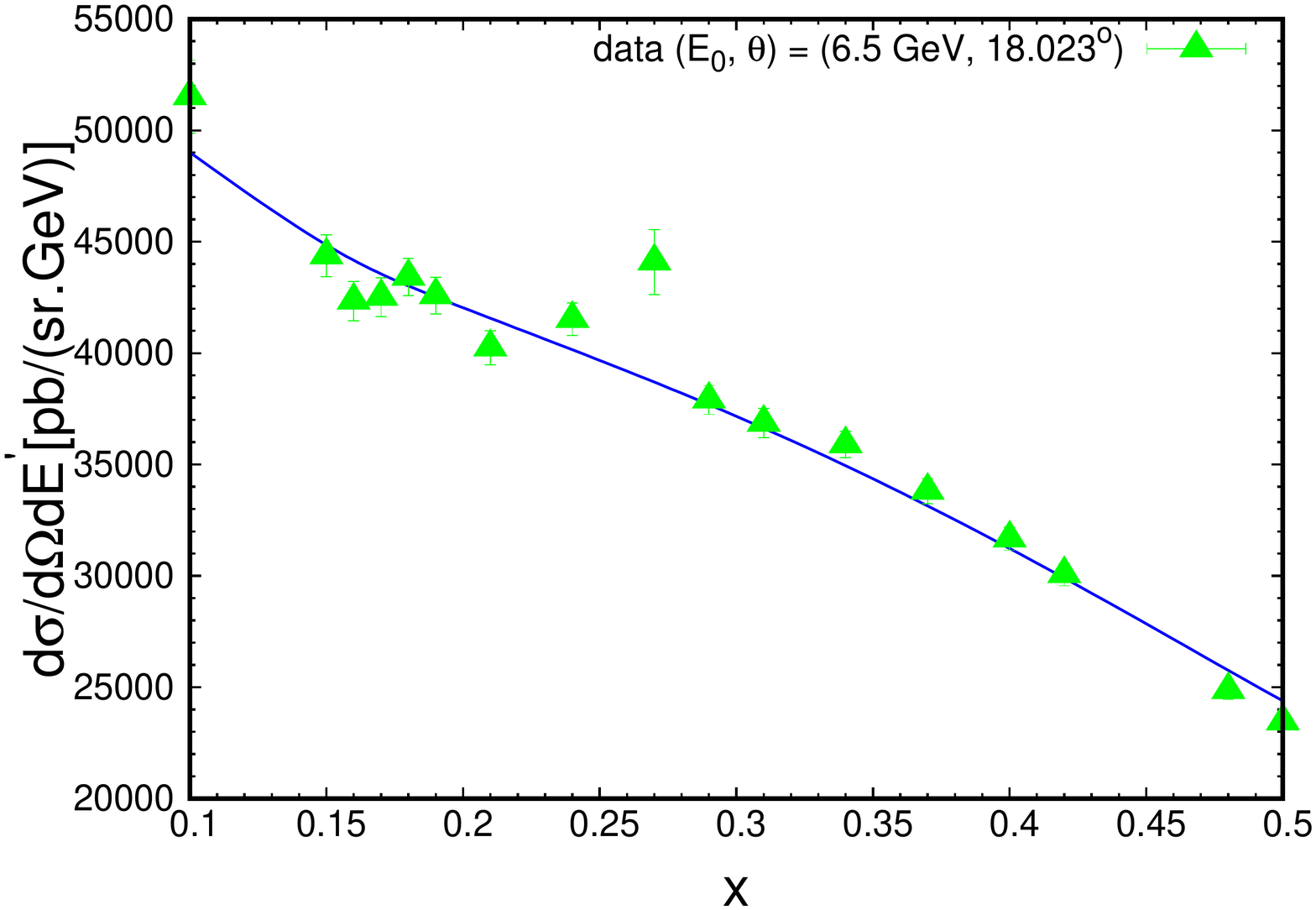}
    \caption{}
    \label{fig:bodekComp4}
\end{subfigure}
\begin{subfigure}{0.5\textwidth}
\centering
    \includegraphics[width=1.0\linewidth]{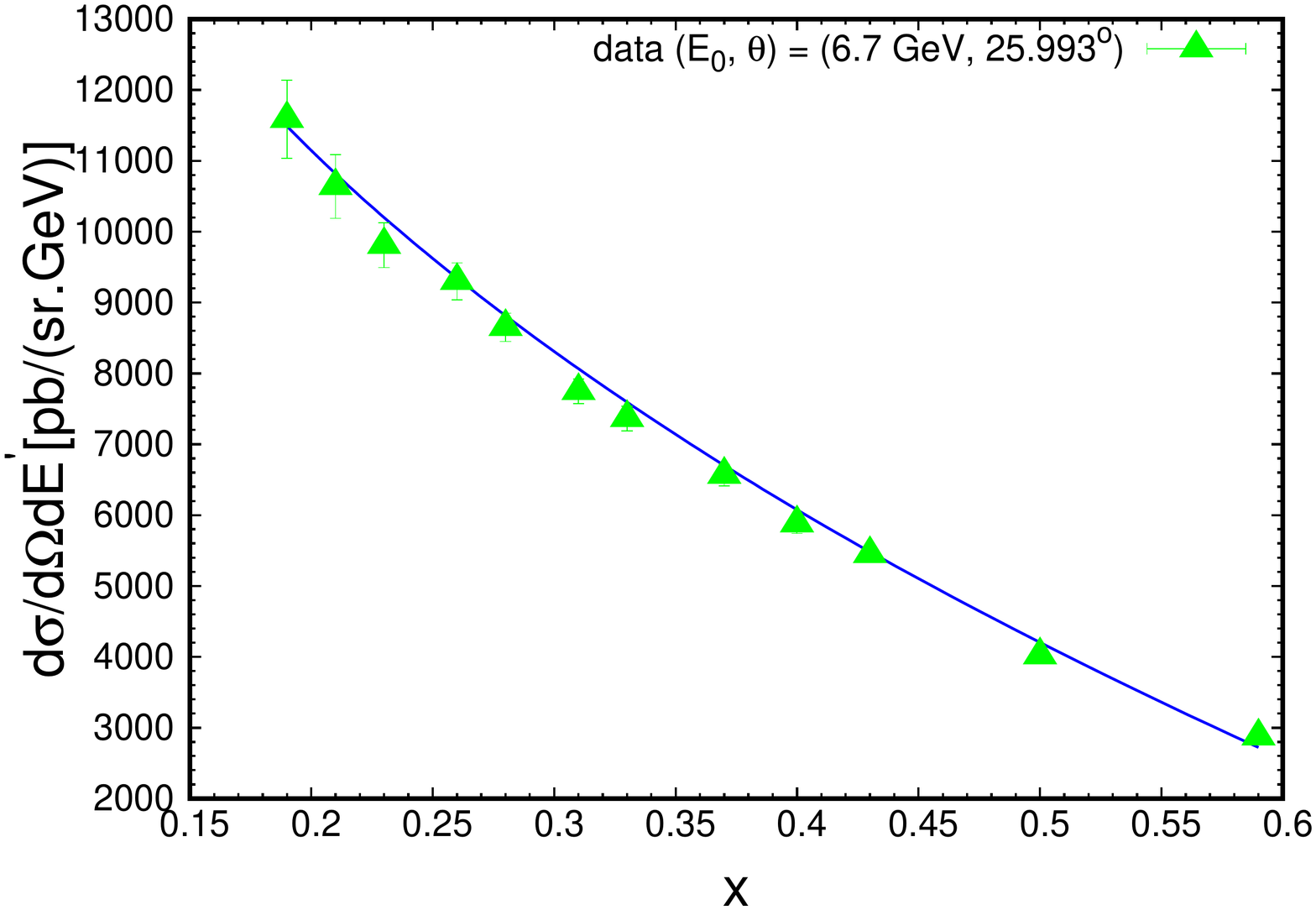}
    \caption{}
    \label{fig:bodekComp5}
\end{subfigure}
\begin{subfigure}{0.5\textwidth}
\centering
    \includegraphics[width=1.0\linewidth]{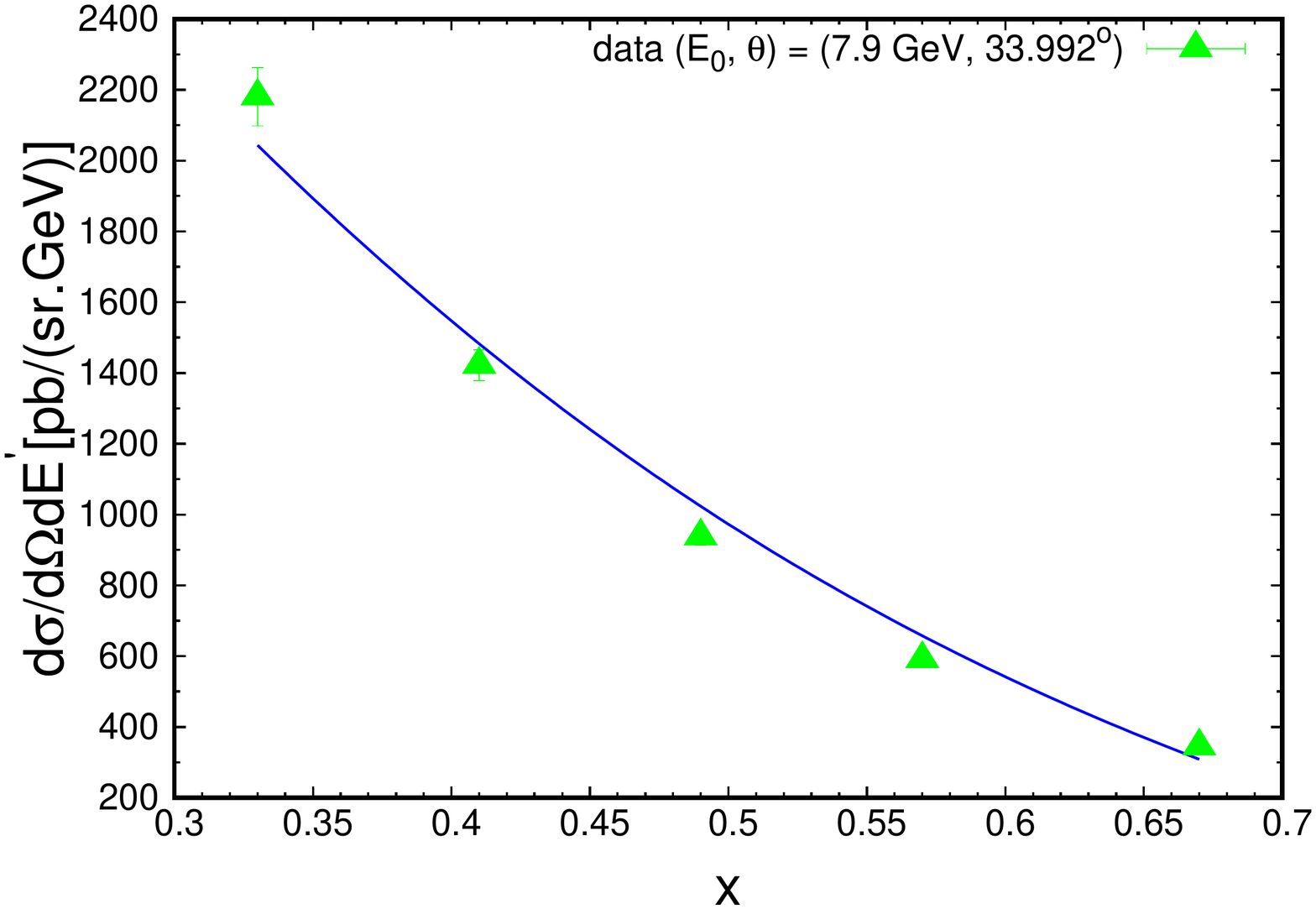}
    \caption{}
    \label{fig:bodekComp6}
\end{subfigure}
\caption{Comparison of cross sections for deep inelastic scattering off deuteron. The data are for incoming energy of $E_0$ GeV and scattering angle $\theta$.}
\label{BodekComp}
\end{figure}

\begin{figure} [h!]
\begin{subfigure}{0.5\textwidth}
\centering
    \includegraphics[width=1.0\linewidth]{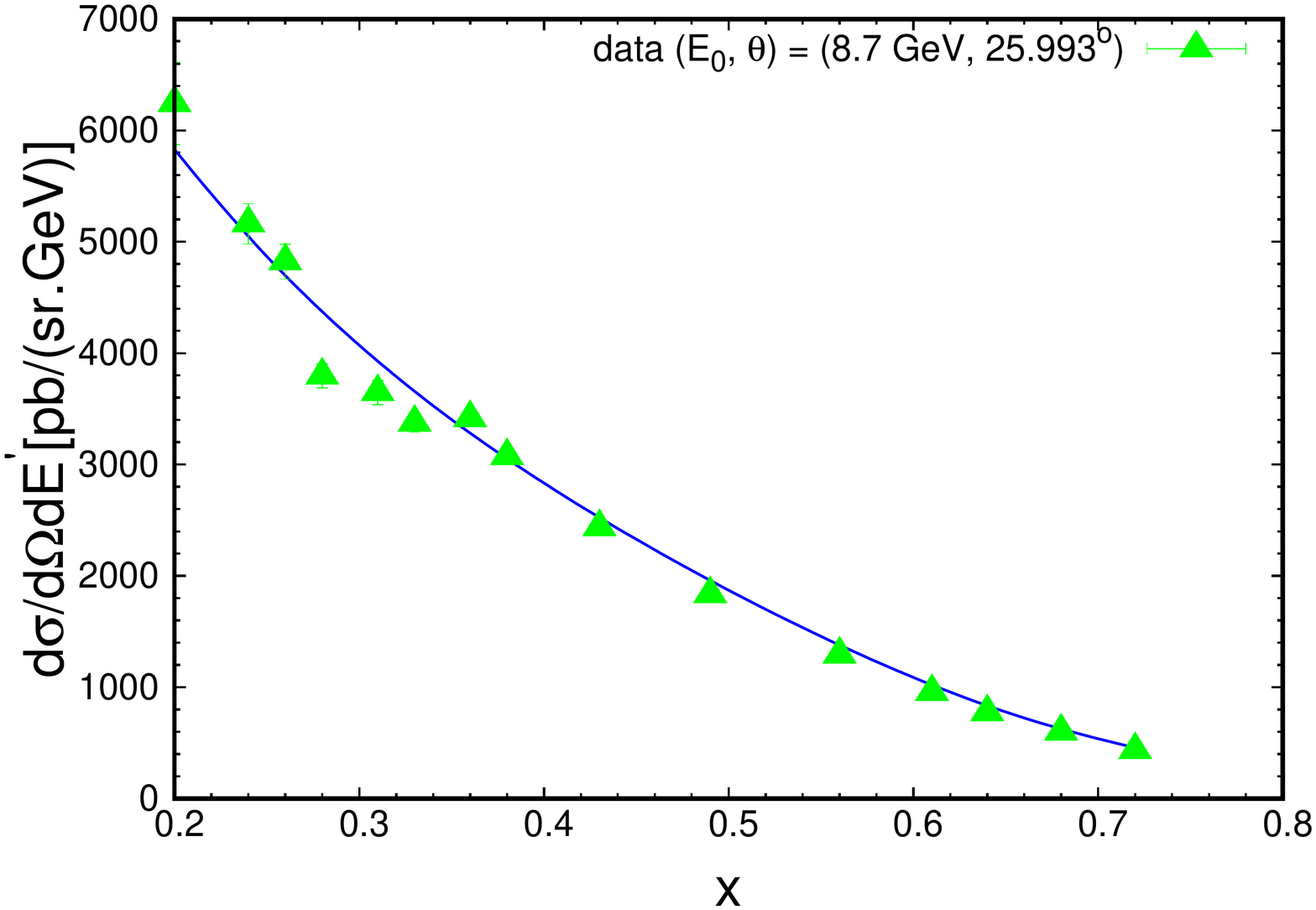}
    \caption{}
    \label{fig:bodekComp7}
\end{subfigure}
\begin{subfigure}{0.5\textwidth}
\centering
    \includegraphics[width=1.0\linewidth]{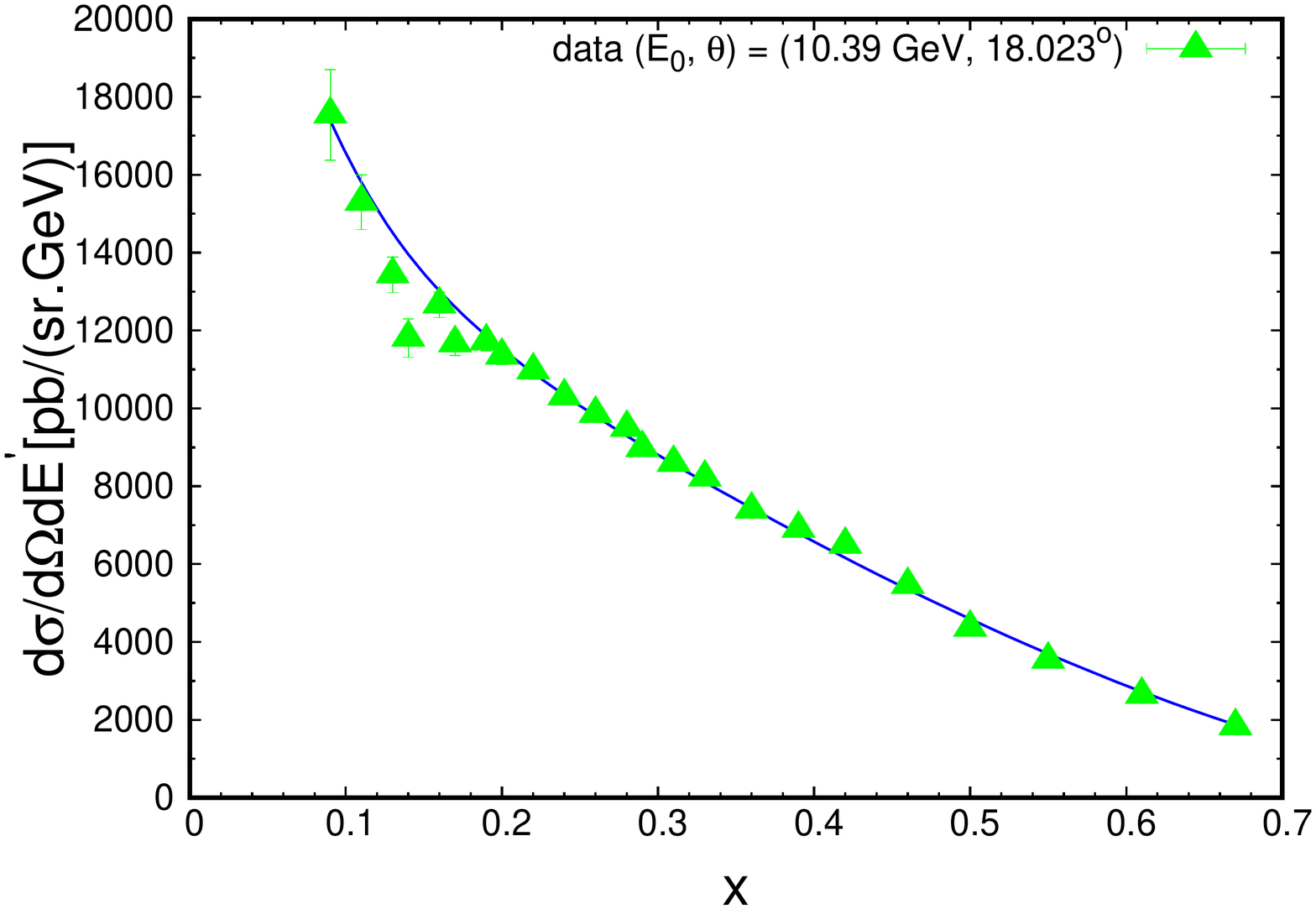}
    \caption{}
    \label{fig:bodekComp8}
\end{subfigure}
\begin{subfigure}{0.5\textwidth}
\centering
    \includegraphics[width=1.0\linewidth]{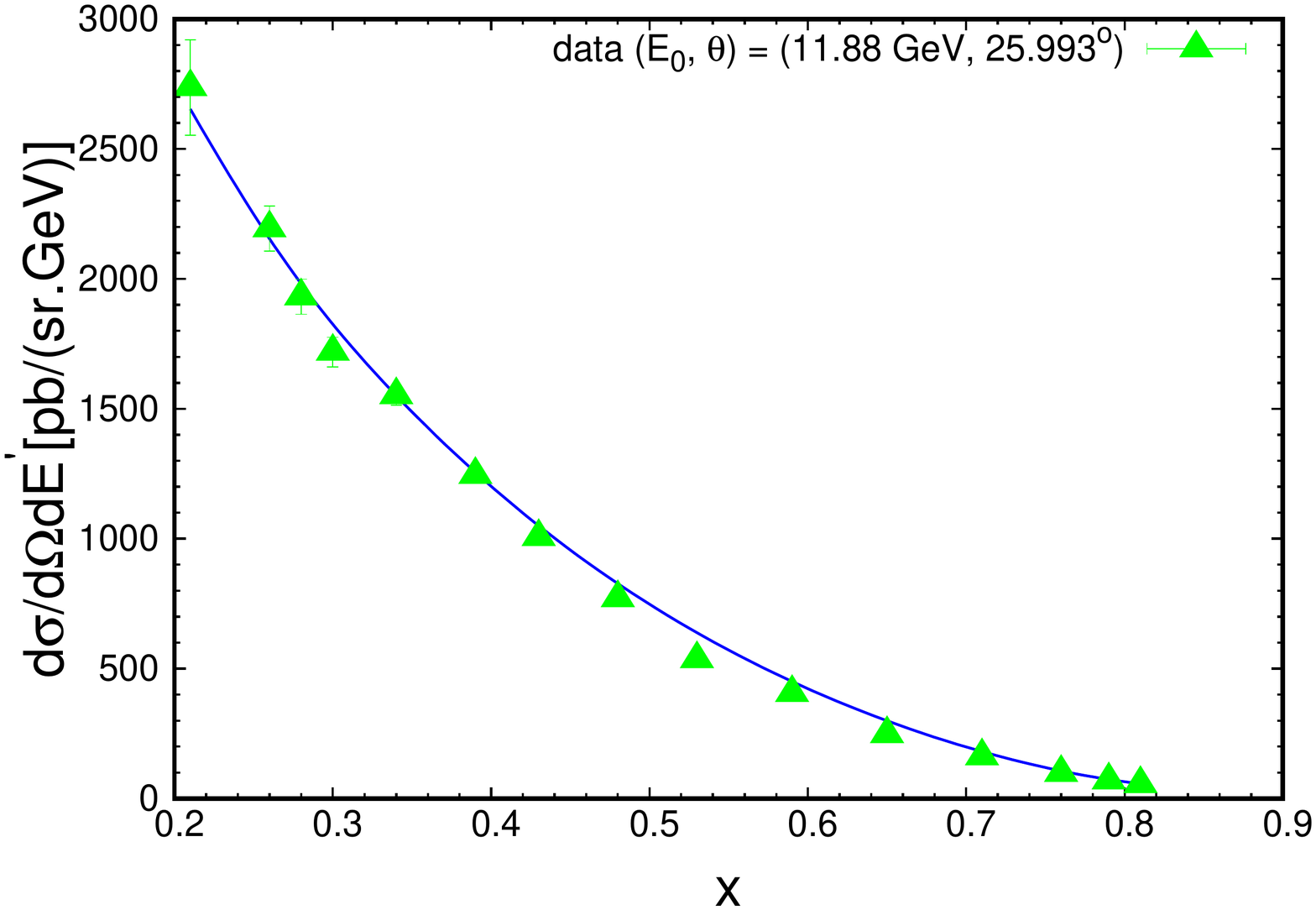}
    \caption{}
    \label{fig:bodekComp9}
\end{subfigure}
\begin{subfigure}{0.5\textwidth}
\centering
    \includegraphics[width=1.0\linewidth]{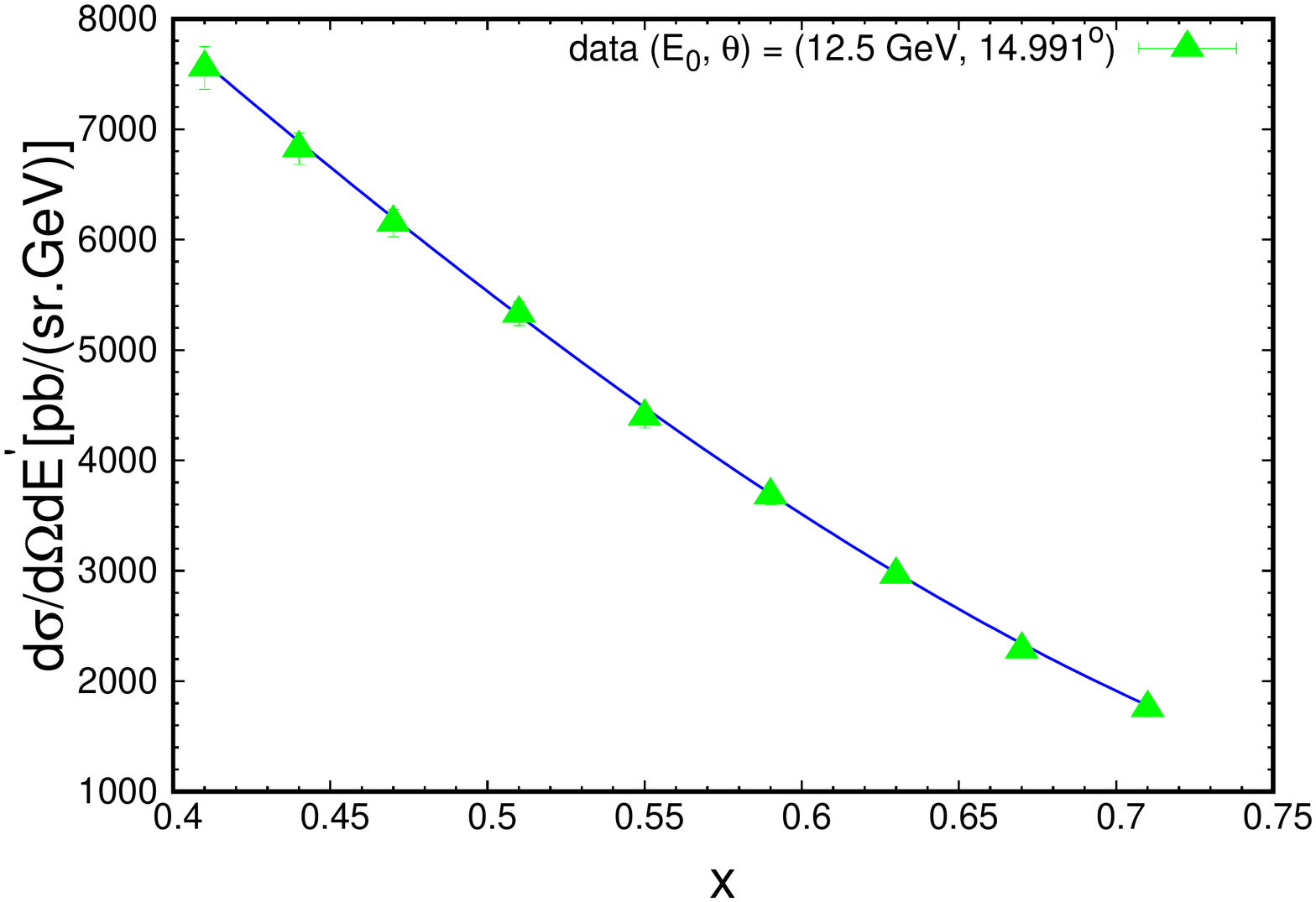}
    \caption{}
    \label{fig:bodekComp10}
\end{subfigure}
\begin{subfigure}{0.5\textwidth}
\centering
    \includegraphics[width=1.0\linewidth]{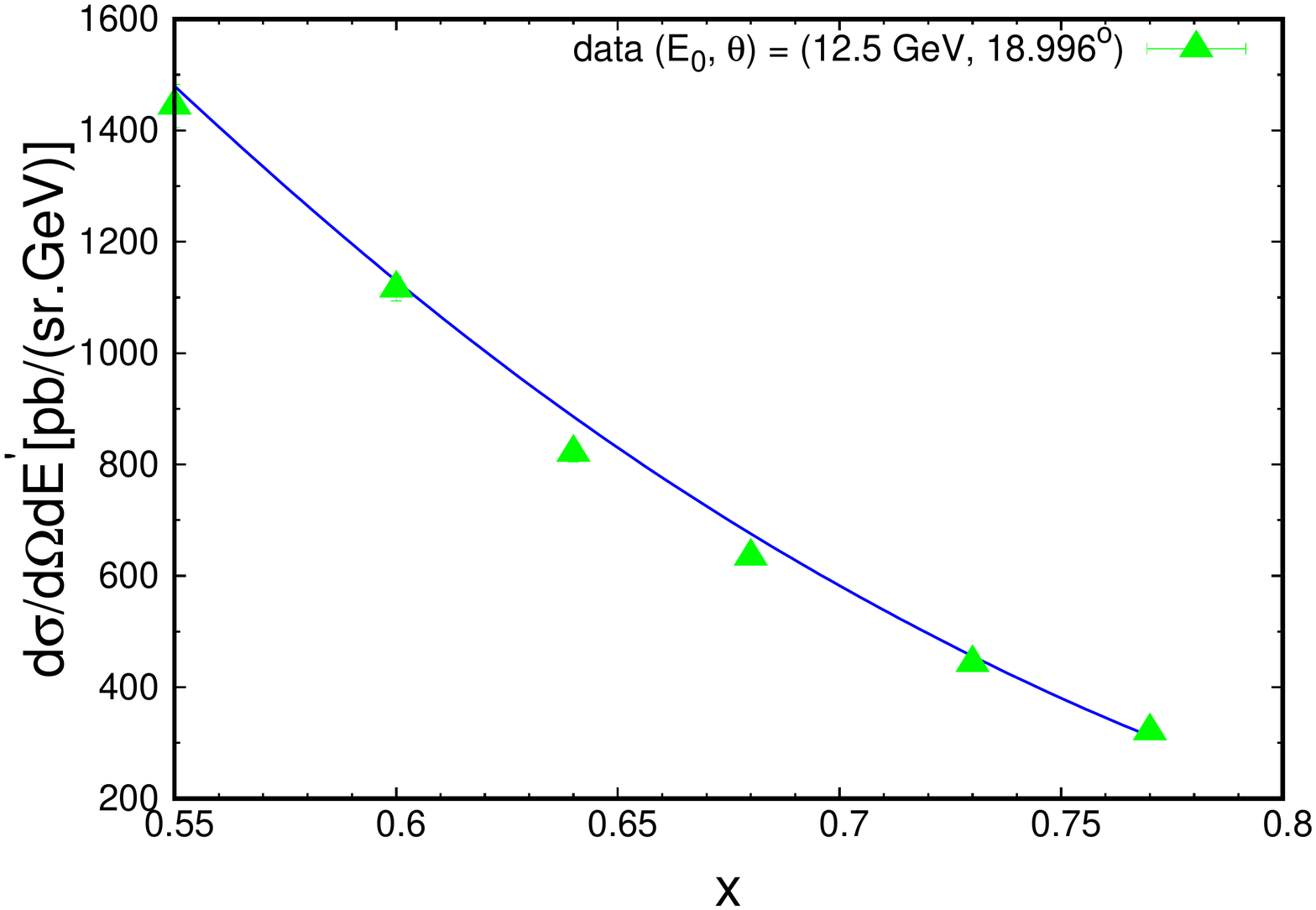}
    \caption{}
    \label{fig:bodekComp11}
\end{subfigure}
\begin{subfigure}{0.5\textwidth}
\centering
    \includegraphics[width=1.0\linewidth]{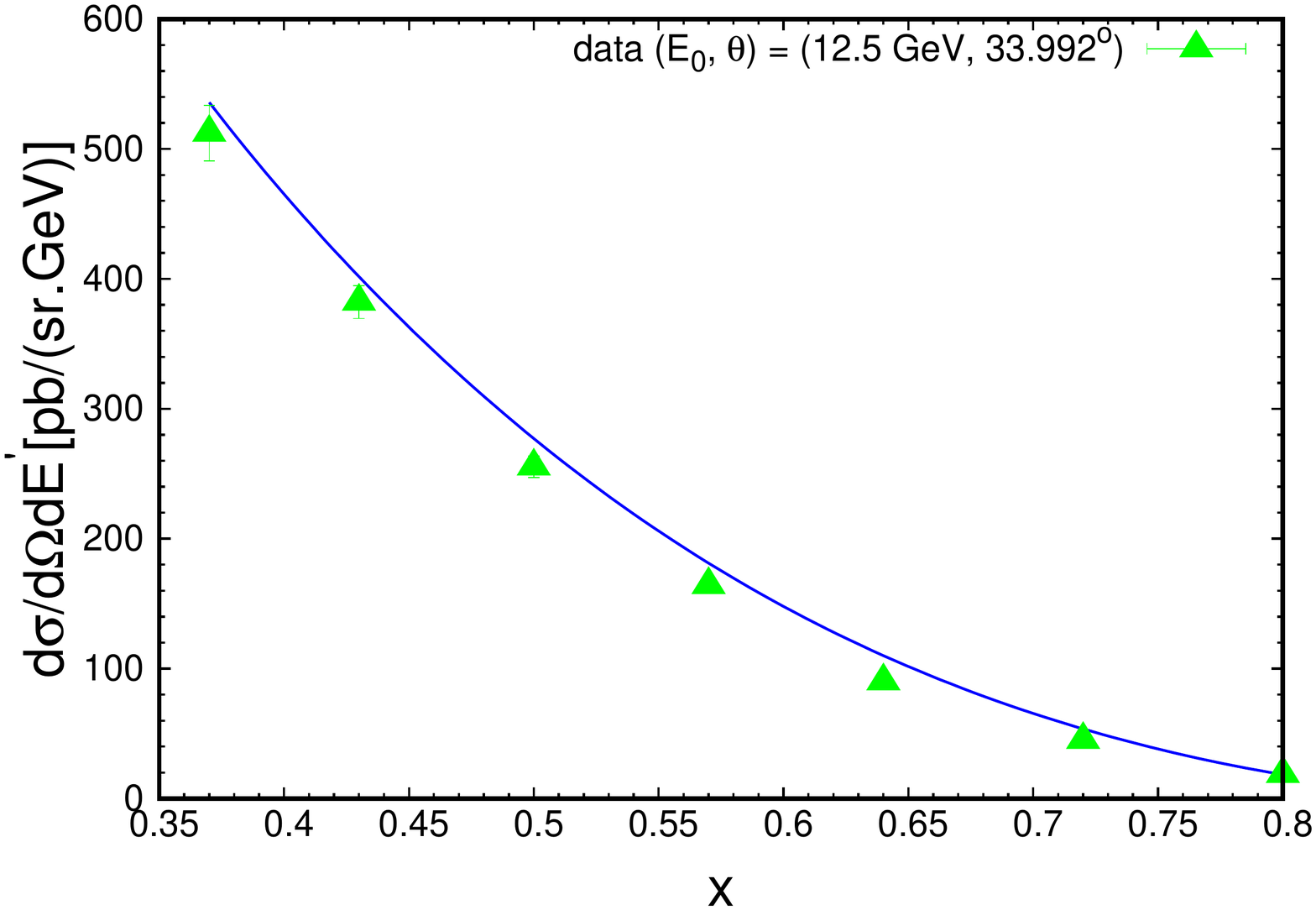}
    \caption{}
    \label{fig:bodekComp12}
\end{subfigure}
\caption{Comparison of cross sections for deep inelastic scattering off deuteron (continued).}
\label{BodekComp1}
\end{figure}

\begin{figure} [h!]
\begin{subfigure}{0.5\textwidth}
\centering
    \includegraphics[width=1.0\linewidth]{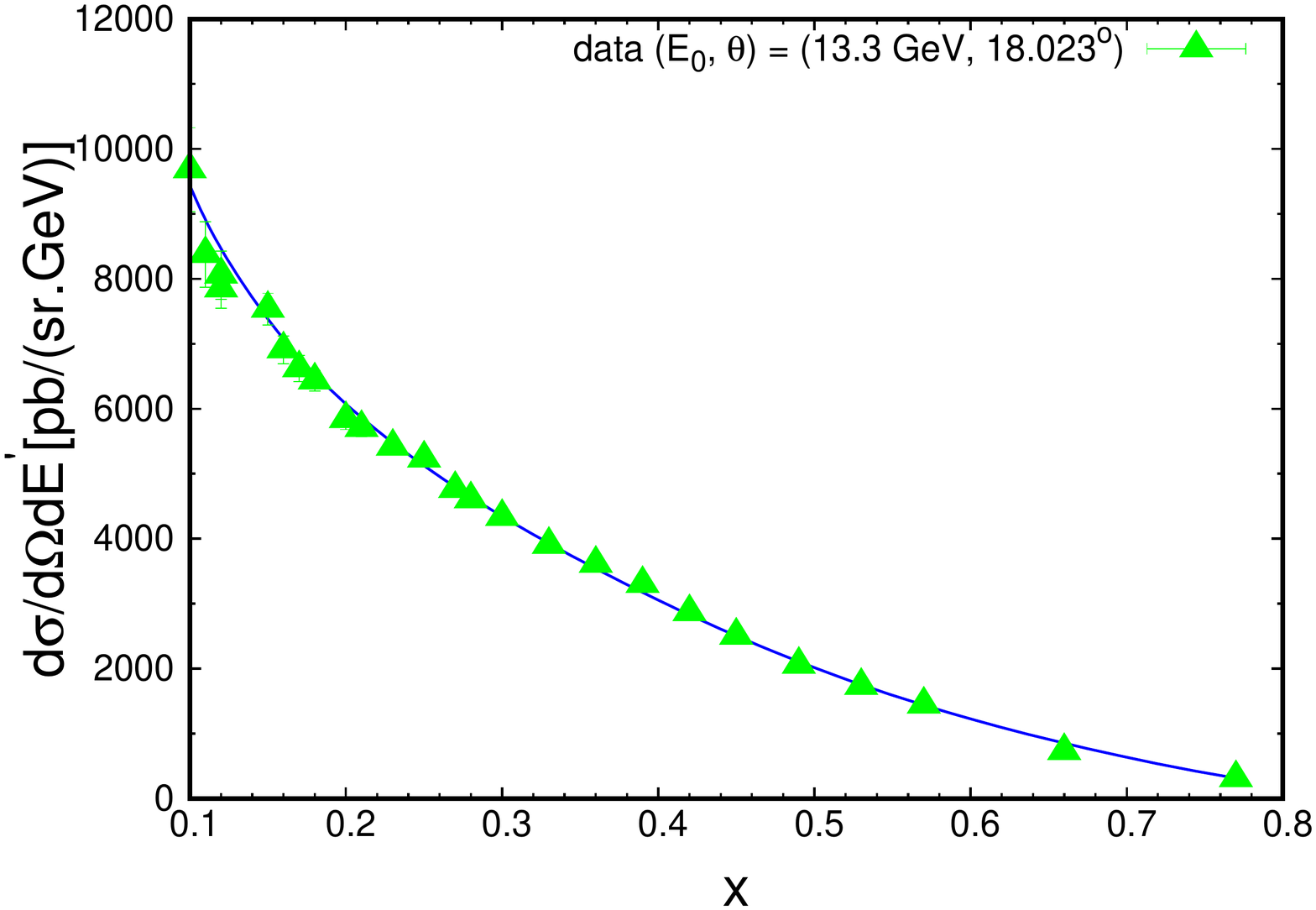}
    \caption{}
    \label{fig:bodekComp13}
\end{subfigure}
\begin{subfigure}{0.5\textwidth}
\centering
    \includegraphics[width=1.0\linewidth]{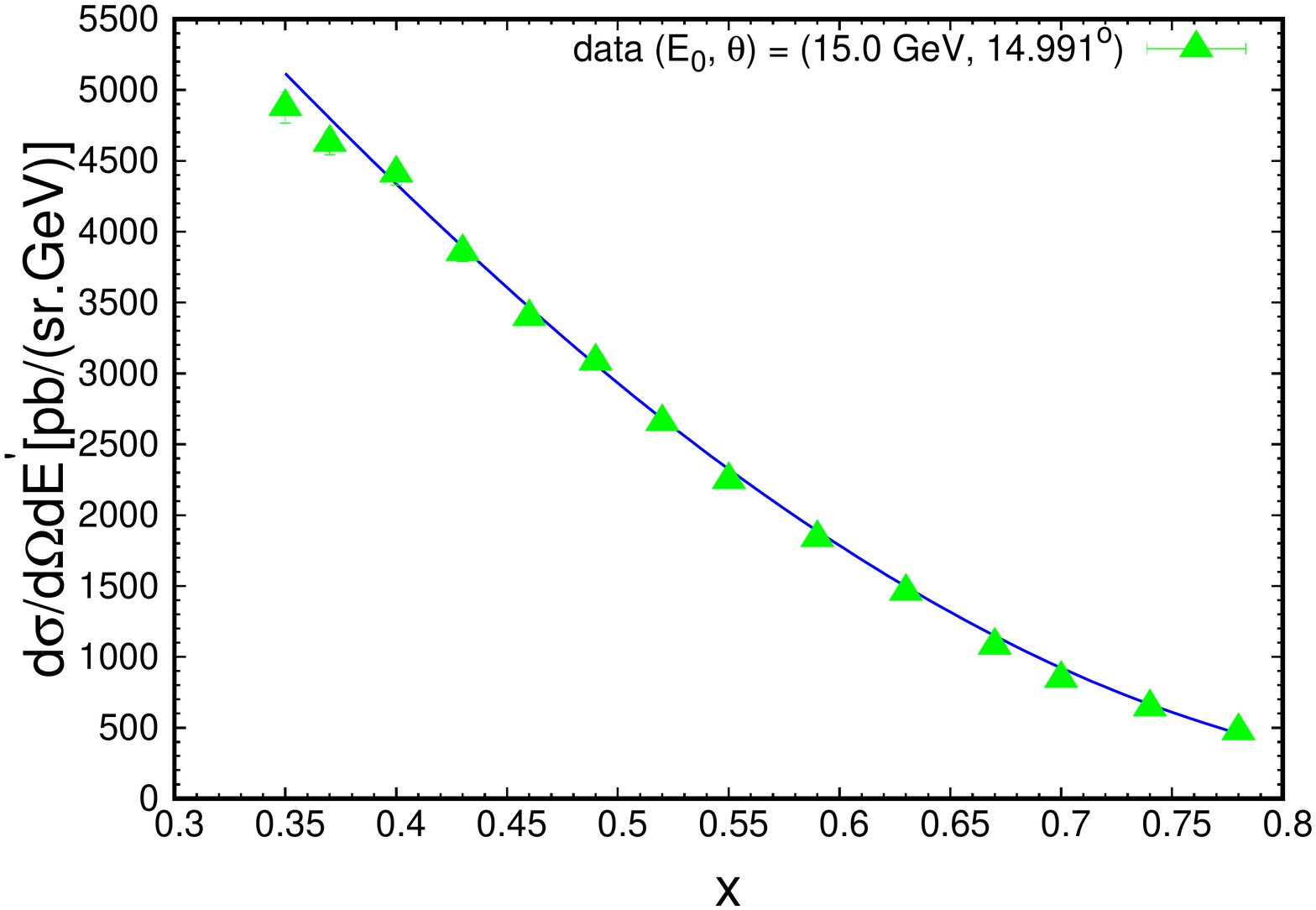}
    \caption{}
    \label{fig:bodekComp14}
\end{subfigure}
\begin{subfigure}{0.5\textwidth}
\centering
    \includegraphics[width=1.0\linewidth]{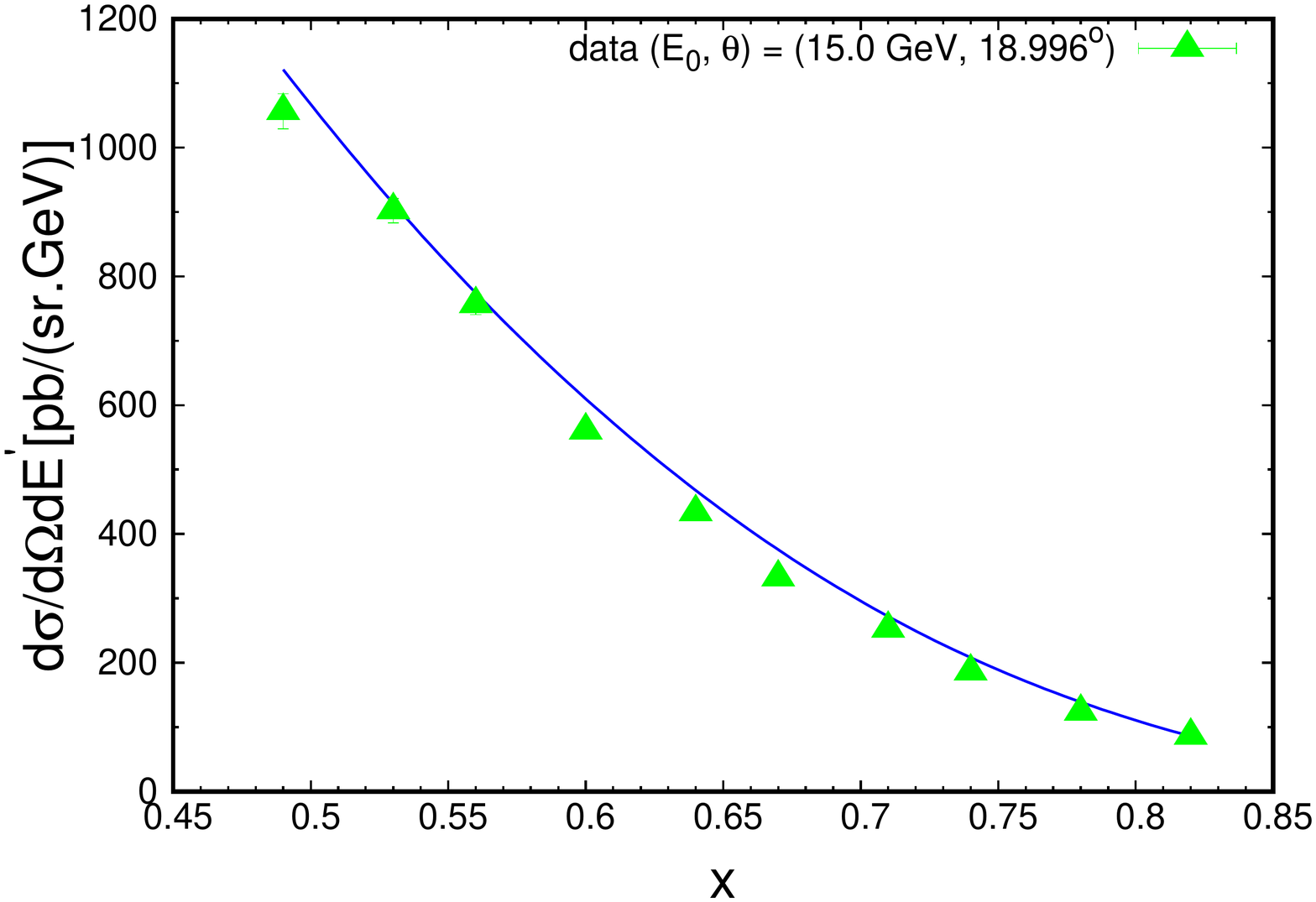}
    \caption{}
    \label{fig:bodekComp15}
\end{subfigure}
\begin{subfigure}{0.5\textwidth}
\centering
    \includegraphics[width=1.0\linewidth]{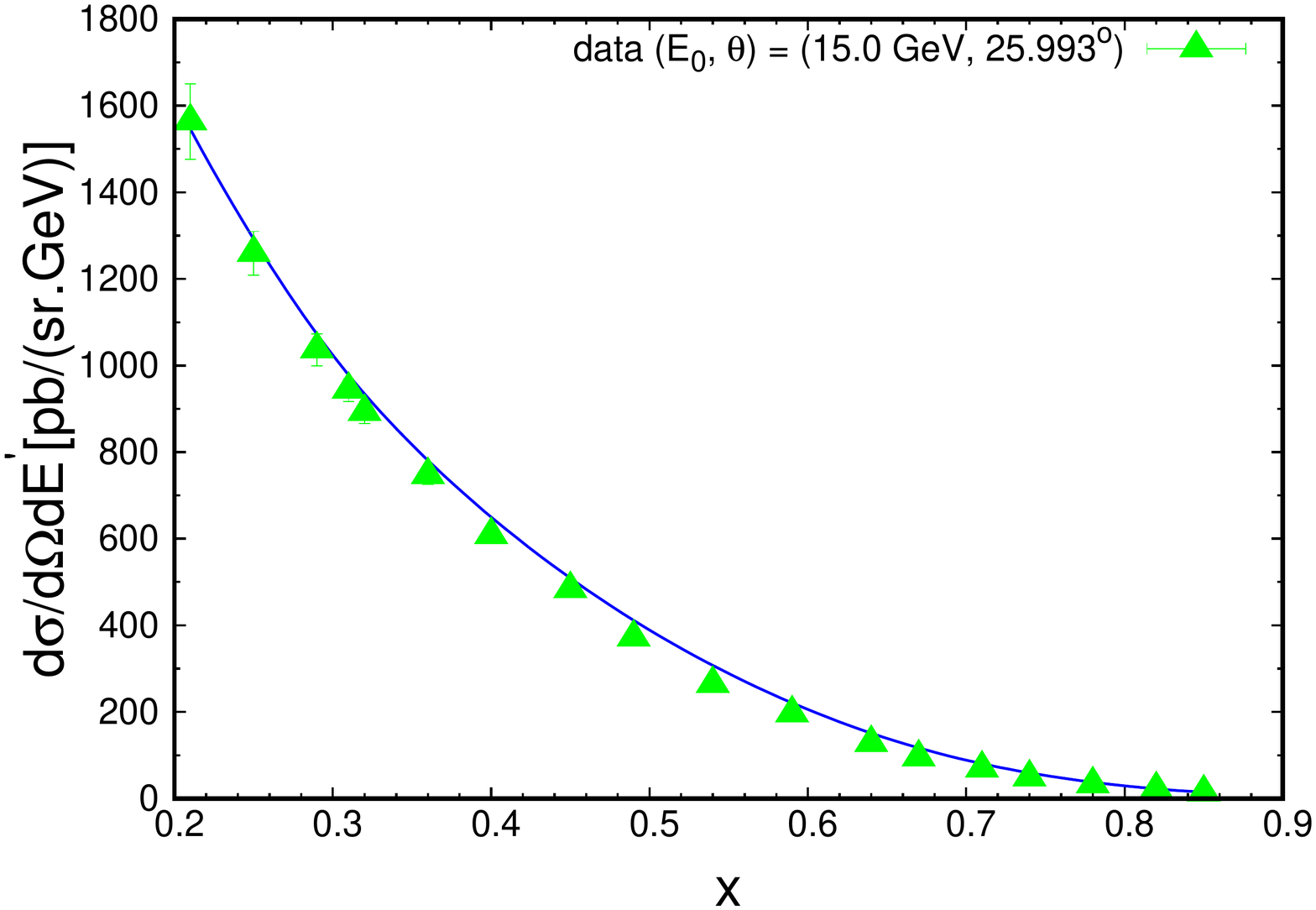}
    \caption{}
    \label{fig:bodekComp16}
\end{subfigure}
\begin{subfigure}{0.5\textwidth}
\centering
    \includegraphics[width=1.0\linewidth]{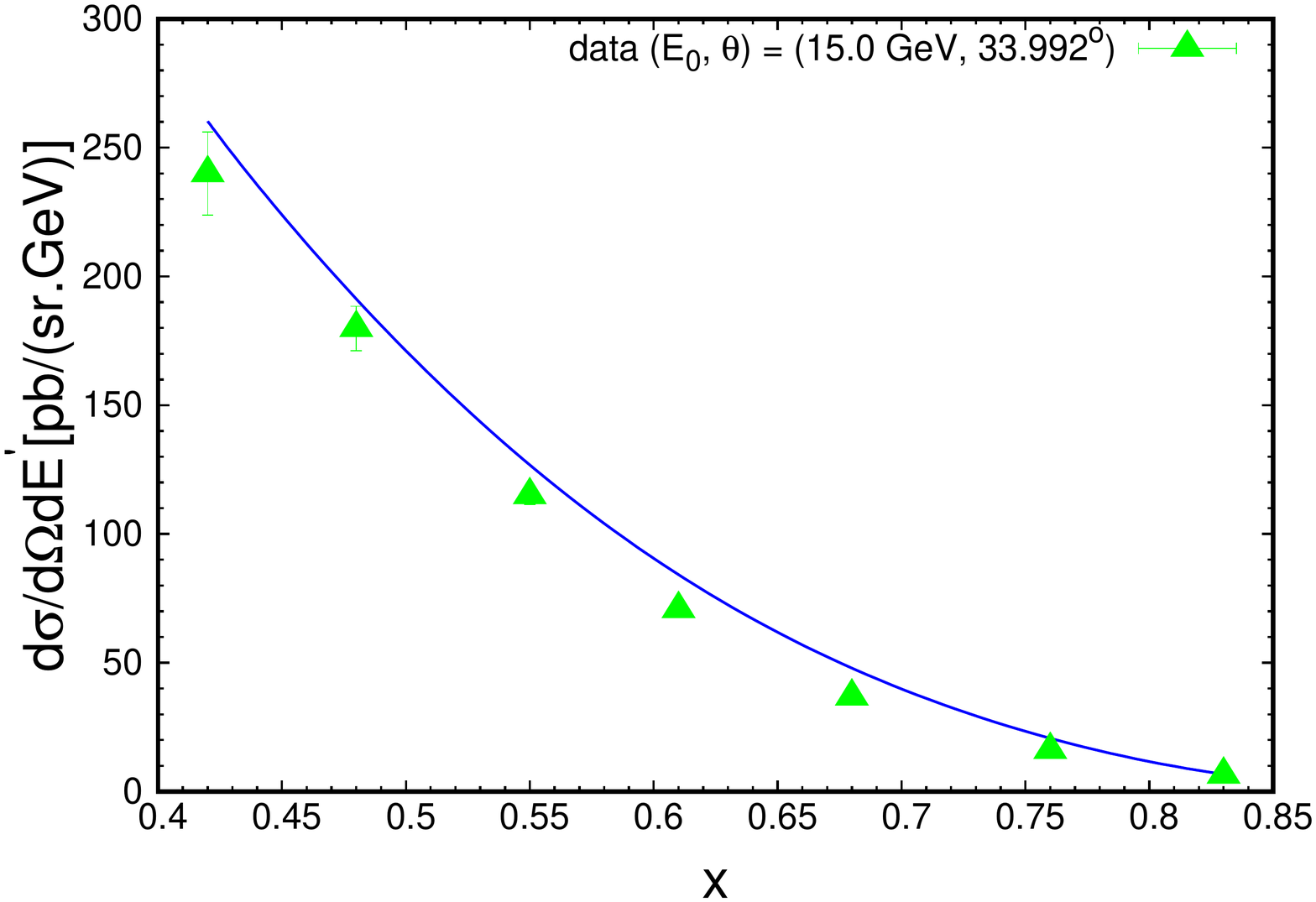}
    \caption{}
    \label{fig:bodekComp17}
\end{subfigure}
\begin{subfigure}{0.5\textwidth}
\centering
    \includegraphics[width=1.0\linewidth]{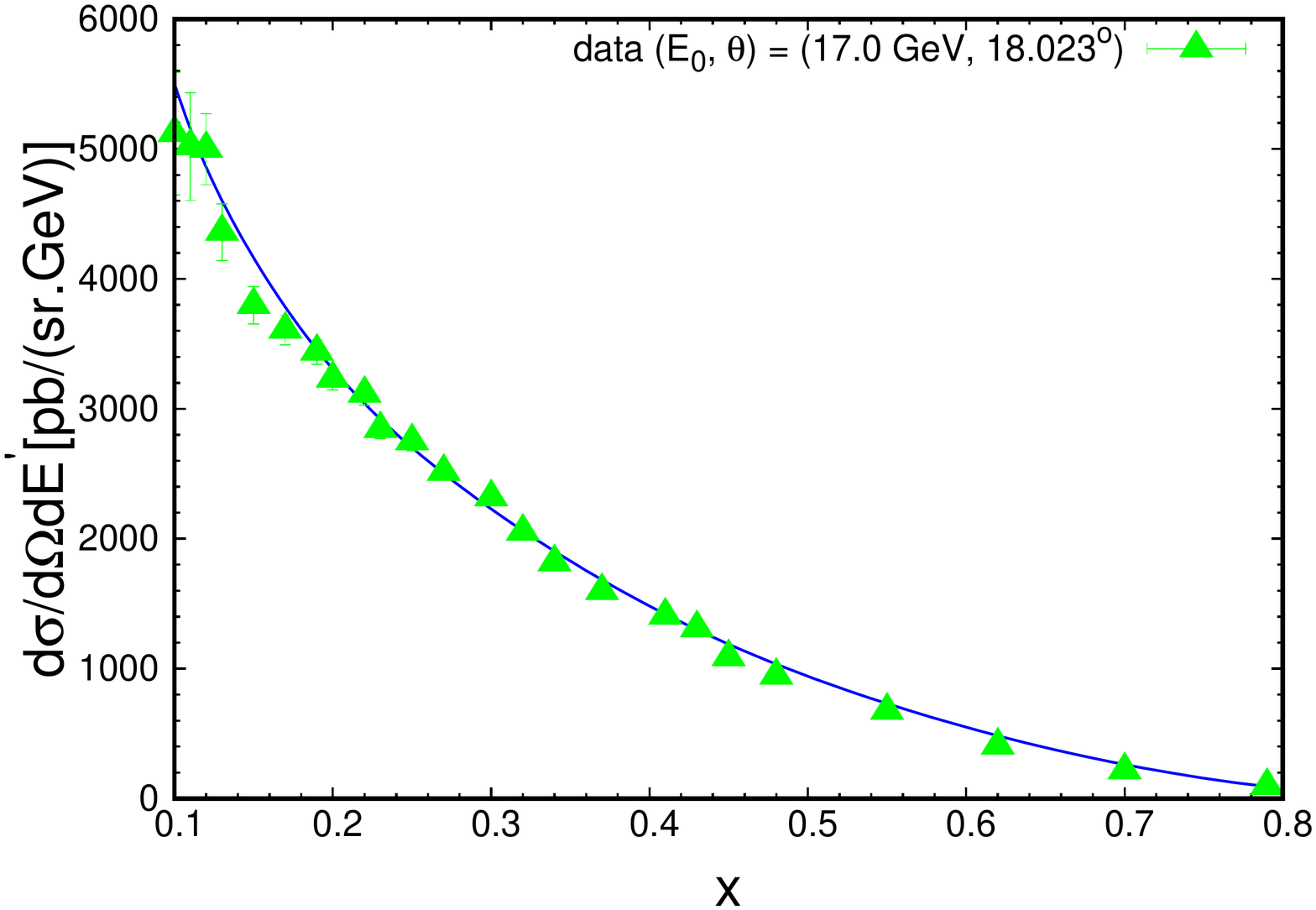}
    \caption{}
    \label{fig:bodekComp18}
\end{subfigure}
\caption{Comparison of cross sections for deep inelastic scattering off deuteron (continued).}
\label{BodekComp2}
\end{figure}

\begin{figure} [ht]
\begin{subfigure}{0.5\textwidth}
\centering
    \includegraphics[width=1.0\linewidth]{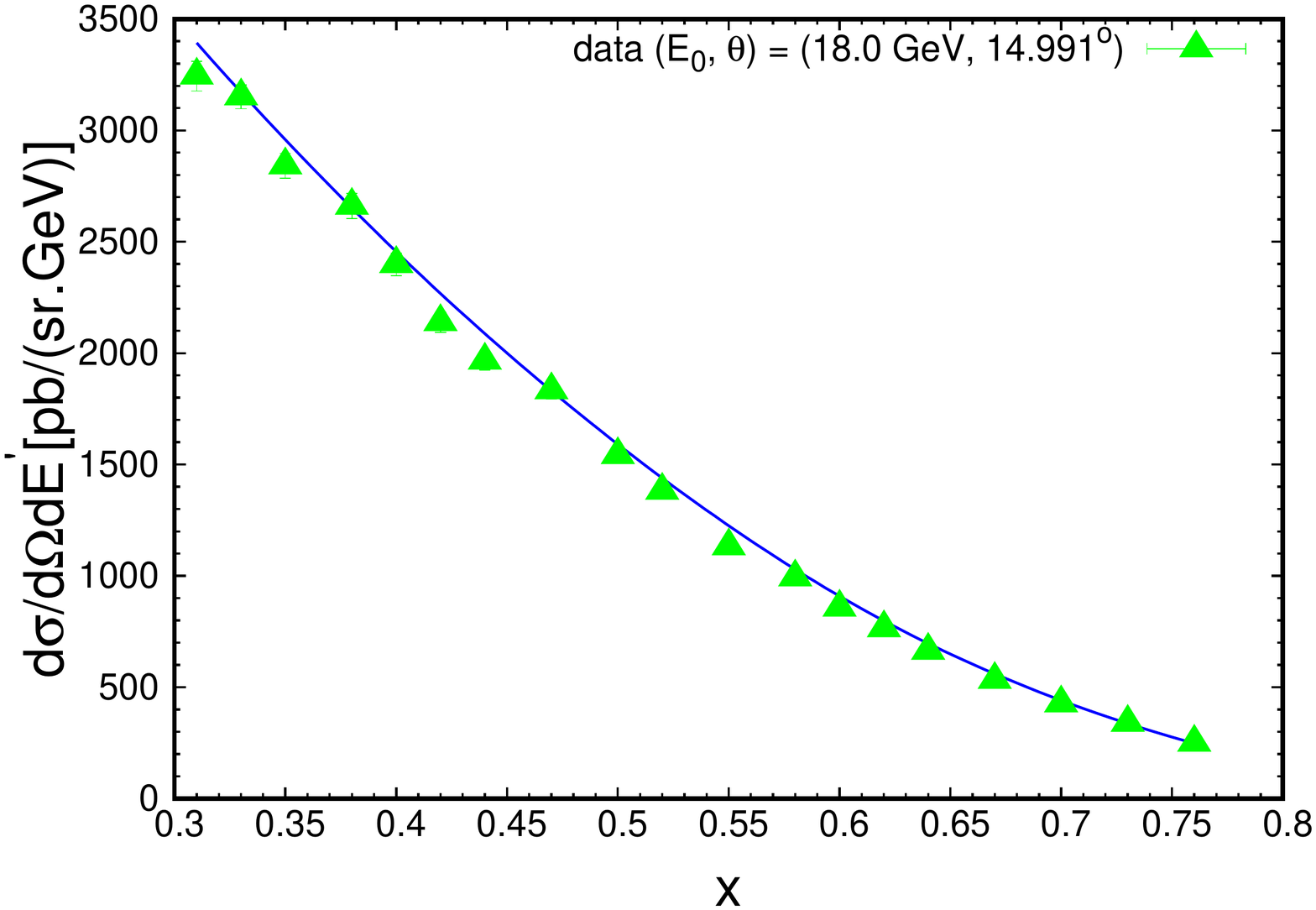}
    \caption{}
    \label{fig:bodekComp19}
\end{subfigure}
\begin{subfigure}{0.5\textwidth}
\centering
    \includegraphics[width=1.0\linewidth]{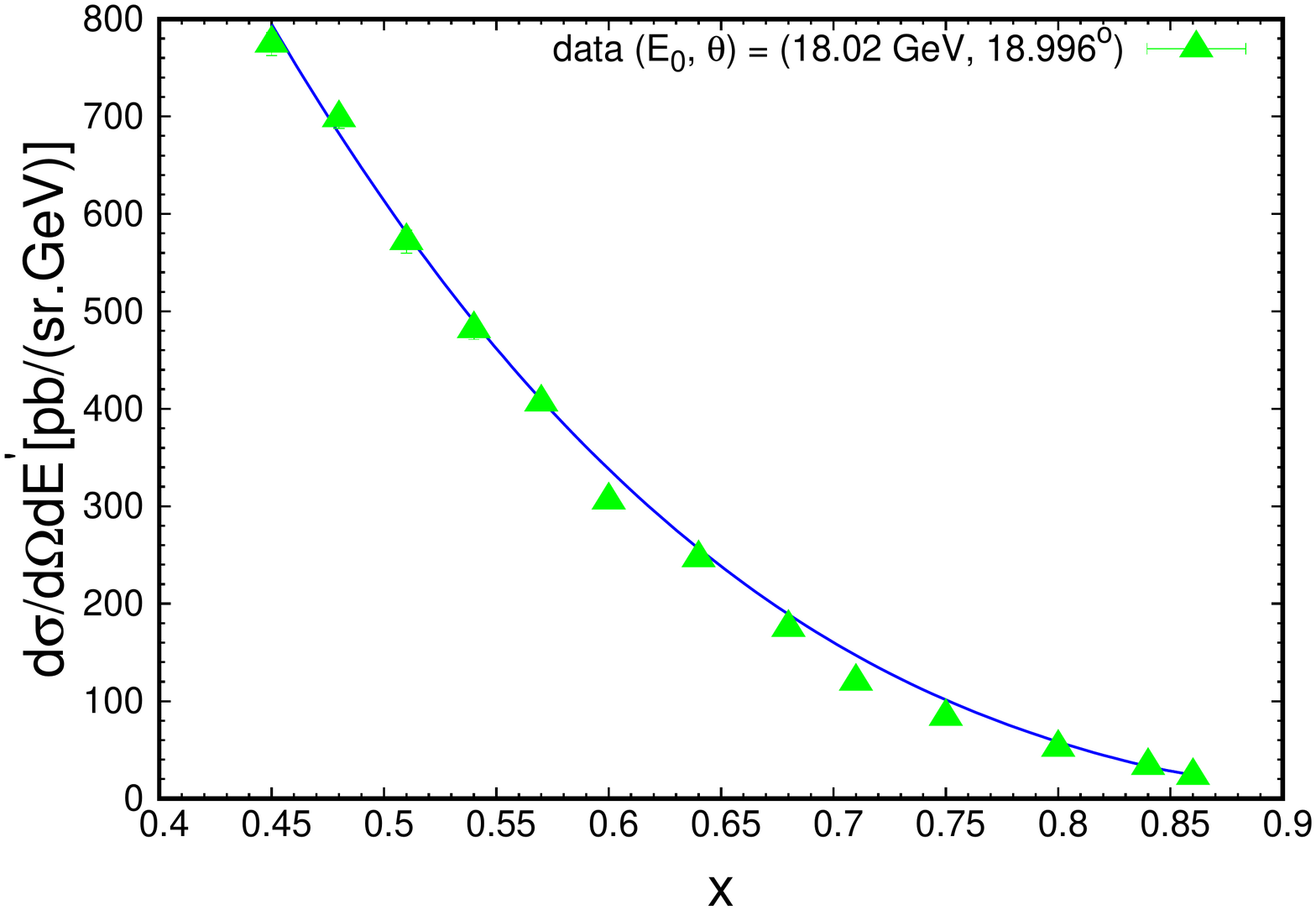}
    \caption{}
    \label{fig:bodekComp20}
\end{subfigure}
\begin{subfigure}{0.5\textwidth}
\centering
    \includegraphics[width=1.0\linewidth]{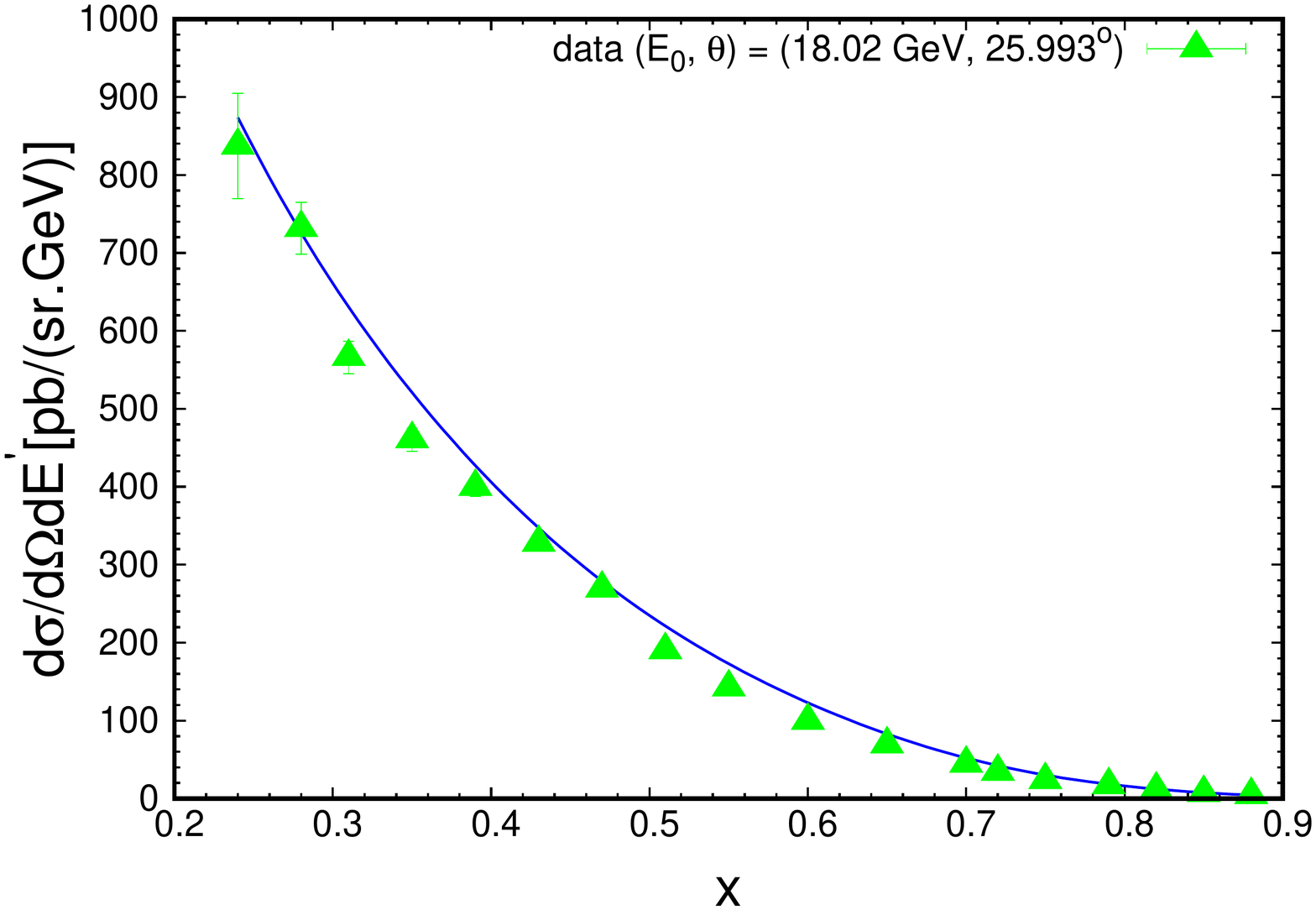}
    \caption{}
    \label{fig:bodekComp21}
\end{subfigure}
\begin{subfigure}{0.5\textwidth}
\centering
    \includegraphics[width=1.0\linewidth]{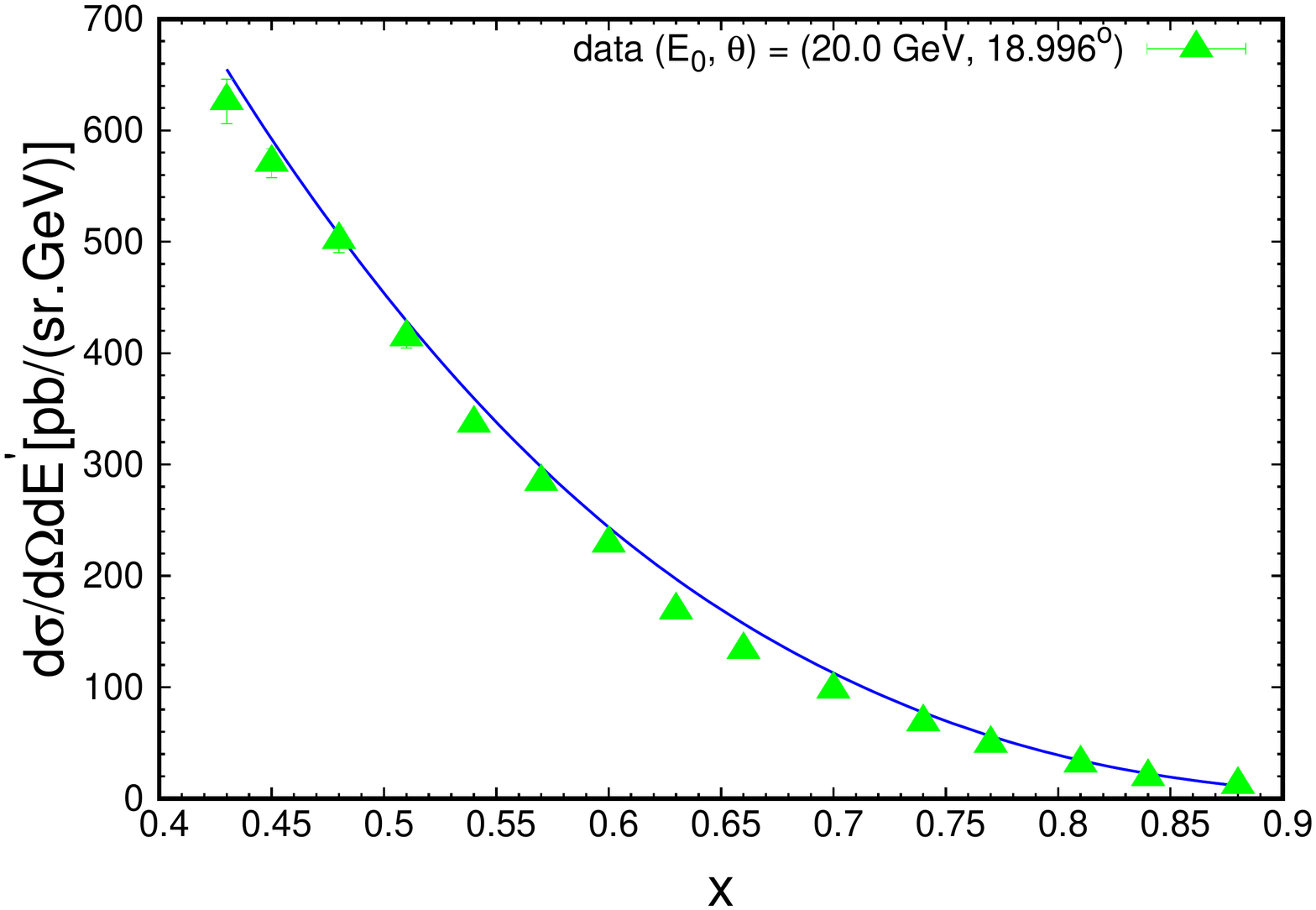}
    \caption{}
    \label{fig:bodekComp22}
\end{subfigure}
\caption{Comparison of cross sections for deep inelastic scattering off deuteron. Data are from \cite{BodekParametrization}.}
\label{BodekComp3}
\end{figure}

The comparison of cross sections calculated using Eq.~(\ref{csPWIA_unpolarizedAfterPhiInt}) is compared with the existing deep inelastic cross section data for deuteron target. The comparison of the cross section for different Bjorken $x$ and $Q^2$ range are shown in Figs.~(\ref{whitlowComp})-(\ref{BodekComp3}). We present the comparison of our calculations with the experimental data over a wider Bjorken $x$ range (0.1 - 0.9), incident energy range (4.5 GeV - 20.0 GeV) and scattering angles, ranging from  (14.991$^\circ$ - 33.992$^\circ$).

\subsection{Comparison with $^3$He data}
\begin{figure}[h!]
\begin{subfigure}{0.5\textwidth}
\centering
    \includegraphics[width=1.0\linewidth]{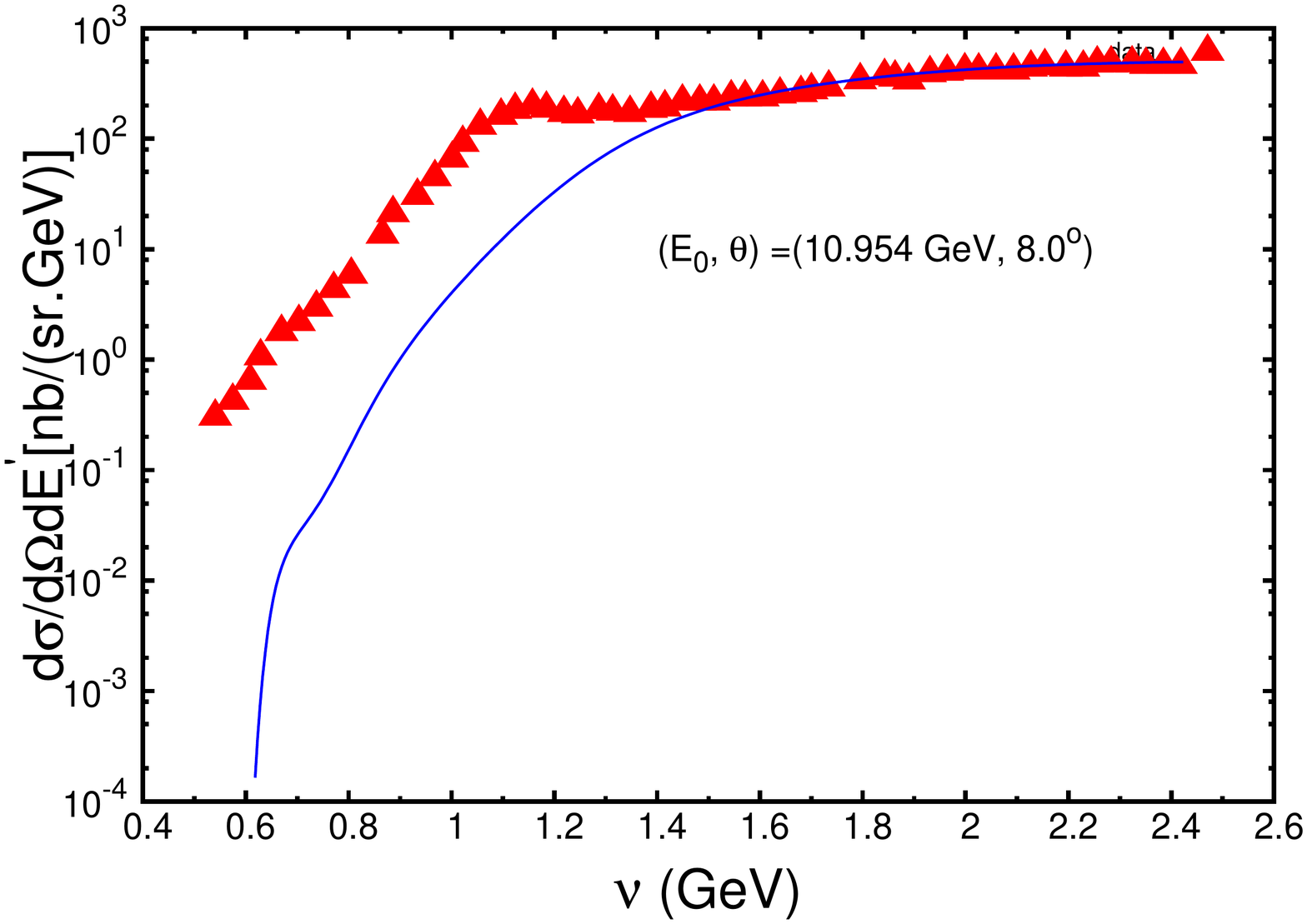}
    \caption{}
    \label{fig:He3Compa}
\end{subfigure}
\begin{subfigure}{0.5\textwidth}
\centering
    \includegraphics[width=1.0\linewidth]{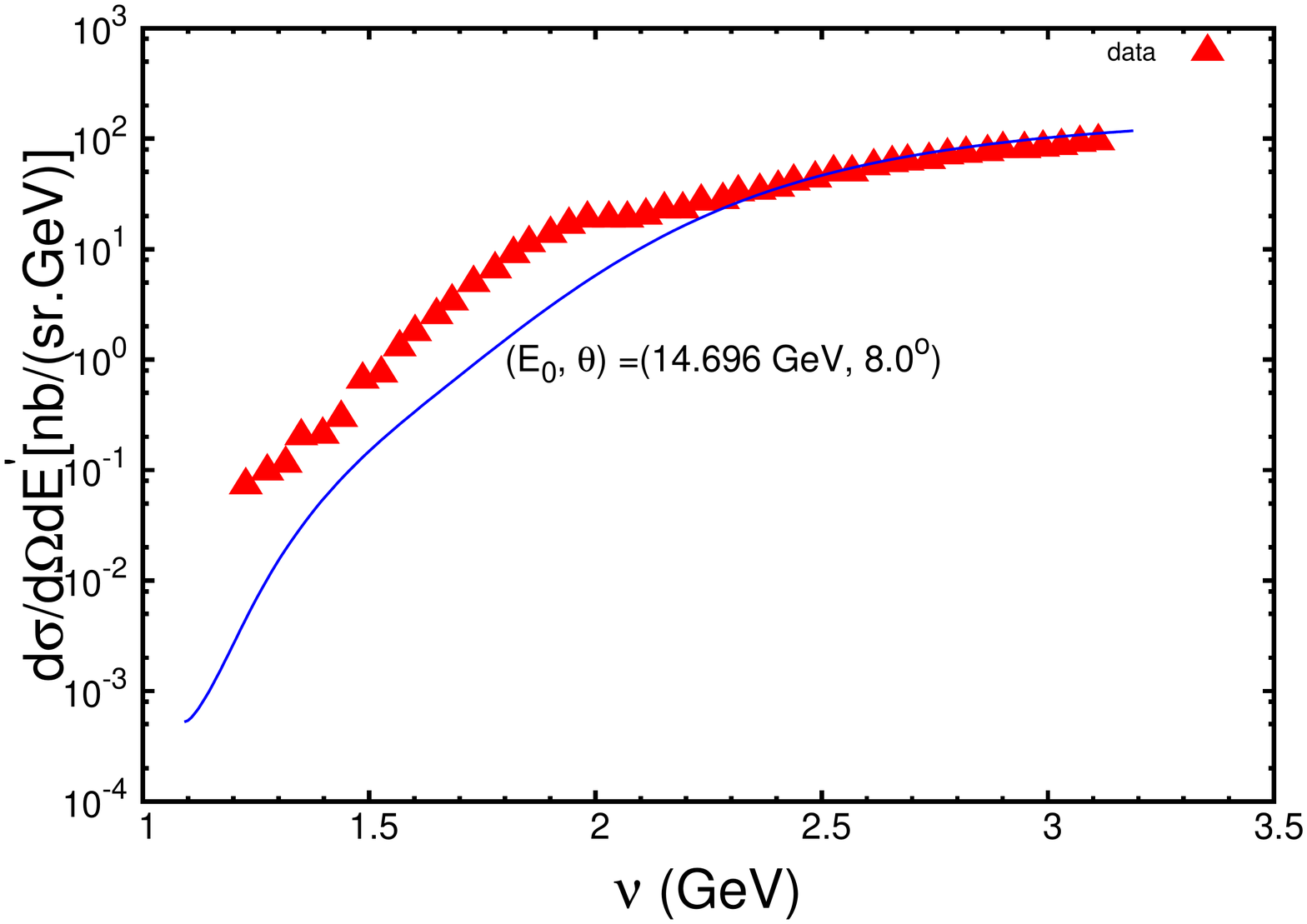}
    \caption{}
    \label{fig:He3Compb}
\end{subfigure}
\caption{Comparison of cross sections for deep inelastic scattering off $^3$He. Data are from \cite{He3UnpolData}.}
\label{He3comp}
\end{figure}

The calculation of inclusive deep-inelastic scattering from $^3$He nucleus is more complicated. In this case, we have to take into account the fact that the spectator system can be in two configurations. The one in which the recoil system is almost coherent- providing a mean field for the struck nucleon. And the other in which the struck nucleon is in two nucleon short range correlation with the third spectator nucleon being almost at rest. To separate these two scenarios, we introduce a parameter, $p_{fermi}$. We define a condition such that if the initial momentum of the struck nucleon is less than $p_{fermi}$, we consider the ``mean field'' approximation, in which the interacting nucleon moves in the mean field of the spectator system. If the momentum of the interacting nucleon is larger than $p_{fermi}$, then we consider the two nucleon short range correlation (2N SRC), so that the spectator system consists of one fast nucleon correlated with the struck nucleon and another nucleon almost at rest.
This forces us to define the longitudinal light-cone momentum of the interacting nucleon as
\begin{equation}
p_{N+} = \begin{cases}  M_{_{^3\text{He}}} - \frac{(2 m_N)^2 + p_{N\perp}^2}{(3 - \alpha_N) m_N}, & \mbox{ for mean field approximation}  \\ M_{d} - \frac{m_N^2 + p_{N\perp}^2}{(2 - \alpha_N) m_N}, & \mbox{ for 2N SRC,} \end{cases}
\end{equation}
where we assume that the rest mass of 2N SRC is equal to the deuteron mass $M_d$.
The Fermi momentum used to separate the mean field and the 2N SRC was 150 MeV. The numerical estimates thus obtained are then compared  with the experimental data. The results of the comparison are shown in Fig.~(\ref{He3comp}). The region with no agreement corresponds to the quasi-elastic contribution, which was not considered in this calculation. 

As seen from Figs.~(\ref{whitlowComp}) - (\ref{BodekComp3}) and Fig.~(\ref{He3comp}), the numerical estimates obtained from Eq.~(\ref{csPWIA_unpolarizedAfterPhiInt}) are in very good agreement with the experimental data. This leaves us at a good place to put together the polarization terms.

%% file: 4/NumericalEstimatesPolarization.tex
\section{Numerical estimates in PWIA- including the polarized part}
Because of unavailability of experimental data for polarized $^3$He targets, the following numerical estimates will be presented for deuteron target only.
The calculation of the polarization terms in Eq.~(\ref{csPWIA}) becomes complicated because of the need to calculate the azimuthal dependence. The azimuthal dependence in the unpolarized cross section occurred only through the $y_N$ term, which made it easier to treat. However, the determination of the azimuthal dependence in the polarization part in Eq.~(\ref{csPWIA}) becomes complicated as it occurs through the four products of the polarization vector ($\zeta_{j'}$), with incoming electron's four momentum ($k$) and the momentum transfer ($q$). Also, there is an additional azimuthal dependence that comes in through the polarization density matrices of the nucleus, which creates additional complications depending upon the nucleus. As can be seen from Eq.~(\ref{csPWIA}), to know the azimuthal dependence explicitly, the products $(\zeta_{j'}\cdot q) \rho_{j'}$ and  $(\zeta_{j'}\cdot k) \rho_{j'}$ need to be calculated.  A detailed derivation of these products for the case of deuteron is presented in Appendix E.

In addition to these products, it is also needed to incorporate the spin structure functions of nucleon  $g_{1N}(x_N,Q^2)$ and $g_{2N}(x_N, Q^2)$  to be able to calculate the contribution of the polarization terms in the cross section.  We neglect the contribution of $g_{2N}$ in our calculations. For the case of $g_{1N}(x_N, Q^2)$, we parameterize the existing experimental data for the proton and deuteron as
\begin{eqnarray}
\label{g1pd_fits}
&& xg_1^p(x, Q^2) = (a_0 + a_1x + a_2x^2)e^{-bx} \nonumber \\
&& xg_1^d(x, Q^2) = (a_3x^3 - x^2 - 0.05x)e^{-cx}, 
\end{eqnarray}
where the values of the fit parameters for $g_{1}$ of proton and deuteron are presented in Tables~(\ref{table_fitparams_p}) and (\ref{table_fitparams_d}) respectively.
\begin{table}[h!]
\centering
\begin{tabular}{| c | c | c | c | } 
\hline
  $a_0$ & $a_1$ & $a_2$ & b  \\ 
 \hline
  0.00397287 $\pm$ 0.002244  & 0.328522 $\pm$ 0.09667  & 2.17265 $\pm$ 1.01  &  5.54303 $\pm$ 0.6871  \\ 
 \hline 
\end{tabular}
\caption{Fit Parameters for $g_1^p$.}
\label{table_fitparams_p}
\end{table}
\begin{table}[h!]
\centering
\begin{tabular}{| c | c |} 
\hline
   $a_3$ & c  \\ 
 \hline
  41.8492 $\pm$ 10.29  & 12.2997 $\pm$ 1.045\\ 
 \hline 
\end{tabular}
\caption{Fit Parameters for $g_1^d$.}
\label{table_fitparams_d}
\end{table}

\begin{figure}[h!]
\begin{subfigure}{0.5\textwidth}
\centering
    \includegraphics[width=1.0\linewidth]{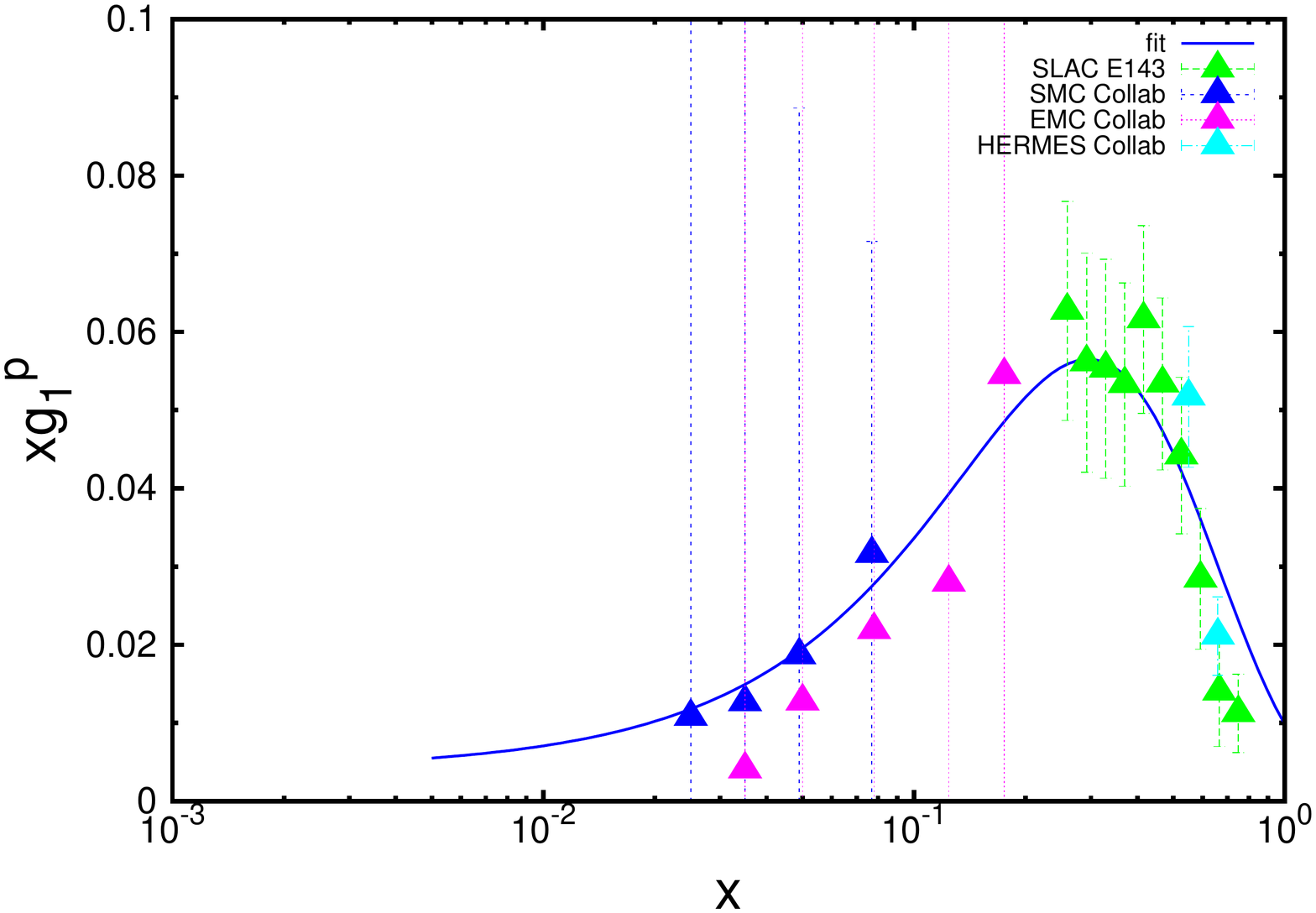}
    \caption{For proton.}
    \label{fig:He3Compa}
\end{subfigure}
\begin{subfigure}{0.5\textwidth}
\centering
    \includegraphics[width=1.0\linewidth]{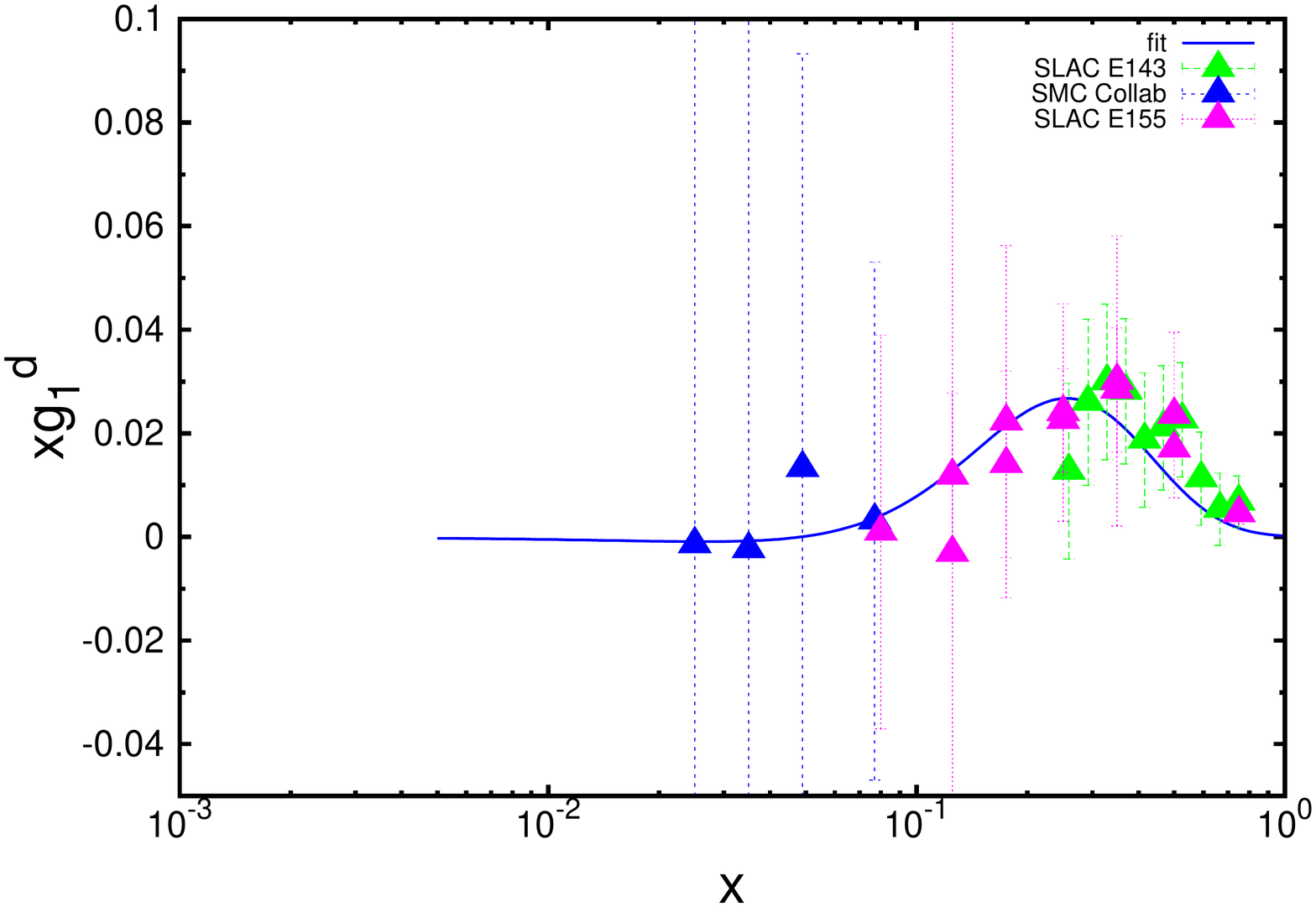}
    \caption{For deuteron.}
    \label{fig:He3Compb}
\end{subfigure}
\caption{Fits to the $g_1$ structure function for proton and deuteron. The blue solid curve represents the fit obtained using Eqs.~(\ref{g1pd_fits}).}
\label{g1pdGraphs}
\end{figure}

The structure function $g_{1n}(x, Q^2)$ for the neutron can be estimated using Eqs.~(\ref{g1pd_fits}) through the relation
\begin{equation}
\label{g1n_Fit}
g_{1p} + g_{1n} - g_{1d} \frac{2}{1 - \frac{3}{2} \omega_d} = 0,
\end{equation}
 where $\omega_d = 0.005 \pm 0.001$\cite{omega_d} denotes the $d-$ wave state probability of the deuteron. The fits obtained using Eqs.~(\ref{g1pd_fits}) are shown in Fig.~(\ref{g1pdGraphs}).
The fits obtained are in good agreement with the experimental data. Now that we have all the ingredients to calculate the polarization contribution, we proceed to calculate the full cross section and the polarization asymmetry.

%% file: 4/Asymmetries.tex
\section{Polarized cross section asymmetry}
\begin{figure}[ht]
\begin{subfigure}{0.35\textwidth}
\centering
    \includegraphics[width=1.0\linewidth]{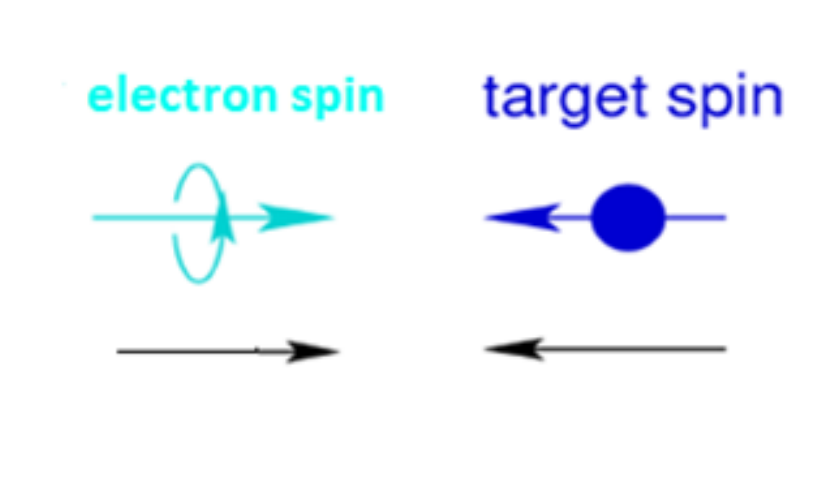}
    \caption{Spins anti-parallel.}
    \label{fig:spin_antiparallel}
\end{subfigure} \qquad \qquad \qquad
\begin{subfigure}{0.35\textwidth}
\centering
    \includegraphics[width=1.0\linewidth]{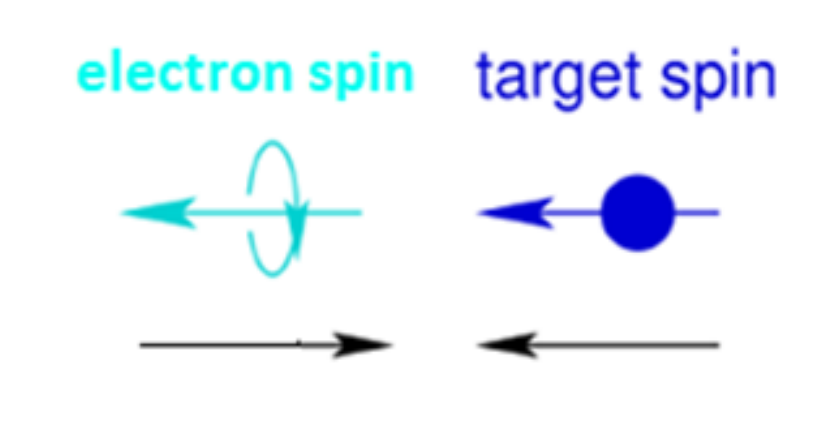}
    \caption{Spins parallel.}
    \label{fig:spin_parallel}
\end{subfigure}
\caption{Possible longitudinal spin orientations of incident electron and target.}
\label{spinfigs}
\end{figure}

In spin-dependent deep inelastic scattering, we can calculate the cross sections of scattering of electrons off target with electron's spin parallel(anti-parallel) or right(left) transverse to target's spin, which leads to the parallel or perpendicular asymmetry,
\begin{eqnarray}
\label{A1defn}
A_\parallel &&= \frac{\sigma^{\downarrow \Uparrow} - \sigma^{\uparrow \Uparrow}}{\sigma^{\downarrow \Uparrow} + \sigma^{\uparrow \Uparrow}} = f_k \Big[g_1(x,Q^2)[E_k + E_k'\cos\theta] - \frac{Q^2}{\nu} g_2(x,Q^2)\Big] \nonumber \\
A_\perp && = \frac{\sigma^{\downarrow \Leftarrow} - \sigma^{\uparrow \Leftarrow}}{\sigma^{\downarrow \Leftarrow} + \sigma^{\uparrow \Leftarrow}} = f_k E_k'\sin\theta \Big[\frac{2E_k}{\nu} g_2(x,Q^2) + g_1(x,Q^2)\Big],
\end{eqnarray}
where $f_k = \frac{1}{\nu F_1(x,Q^2)} \frac{1 - \epsilon}{1 + \epsilon R(x,Q^2)}, \epsilon^{-1} = 1 + 2(1 + \frac{\nu^2}{Q^2})\tan^2\frac{\theta}{2}, \nu = E_k - E_k'$, and $R = \sigma_L /\sigma_T$ with $\sigma_L$
and $\sigma_T$ being the longitudinal and transverse virtual photon cross
sections.  The value of $R$ is taken as 0.18. The spins of incident electron and target being parallel and anti-parallel are schematically shown in Fig.~(\ref{spinfigs}).
The parallel asymmetry is what we are interested in calculating, using Eq.~(\ref{csPWIA}) for the deuteron target. The above defined parallel and perpendicular asymmetries are also related to the virtual photon-nucleon asymmetries $A_1$ and $A_2$ as follows:
\begin{eqnarray}
\label{virtualPhotonAsymmConnection}
A_\parallel && = D(A_1 + \eta A_2) \nonumber \\
A_\perp && = d(A_2 - \zeta A_1),
\end{eqnarray}
where $D = \frac{1 - \epsilon E_k'/E_k}{1 + \epsilon R}, \eta = \frac{\epsilon \sqrt{Q^2}}{E_k - \epsilon E_k'}, \zeta = \frac{\eta (1 + \epsilon)}{2 \epsilon}$ and $d = D\sqrt{\frac{2 \epsilon}{1 + \epsilon}}$ are the DIS kinematic factors. In considered kinematics, $\eta << 1$. Expecting that $A_2 \leq A_1$, one can neglect by $A_2$ contribution in $A_\parallel$ of Eq.~(\ref{virtualPhotonAsymmConnection}) obtaining:
\begin{equation}
\label{A1relation2Aparallel}
A_1 = A_\parallel/D.
\end{equation}
Within QCD, the virtual photon-nucleon asymmetry $A_1$ can also be related to $F_1$ and $g_1$ structure functions as follows:
\begin{equation}\label{a1relatedtog1f1}
A_1(x,Q^2) = \frac{g_1(x,Q^2) - \gamma^2g_2(x,Q^2)}{F_1(x,Q^2)},
\end{equation}
where $\gamma^2 \equiv \frac{Q^2}{\nu^2} = \frac{4M^2x^2}{Q^2}$ and $\nu = E_k - E_k'$ is the energy lost by the electron after being scattered. For higher values of $Q^2$, we find that
\begin{equation}\label{A1highQsq}
A_1(x, Q^2) \approx \frac{g_1(x, Q^2)}{F_1(x, Q^2)}.
\end{equation}
Thus for the nucleon, $A_1$ measures how the spins of the individual quarks, weighted by the square of their charges, are  aligned relative to the nucleon as a whole. We note that Eq.~(\ref{a1relatedtog1f1}) applies for the situation in which nucleon is at rest. The asymmetry obtained from Eqs.~(\ref{A1defn}) and (\ref{A1relation2Aparallel}) can then be used together with Eq.~(\ref{A1highQsq}) to obtain the $g_1$ structure function of the nucleon. In extracting the $g_1$ structure function for neutron, we need to be cautious to account for effects of the motion of the nucleon in the nuclear target. The process of doing this is described in Section (\ref{sectiong1next}).

\begin{figure} [h!]
\begin{subfigure}{0.5\textwidth}
\centering
    \includegraphics[width=1.0\linewidth]{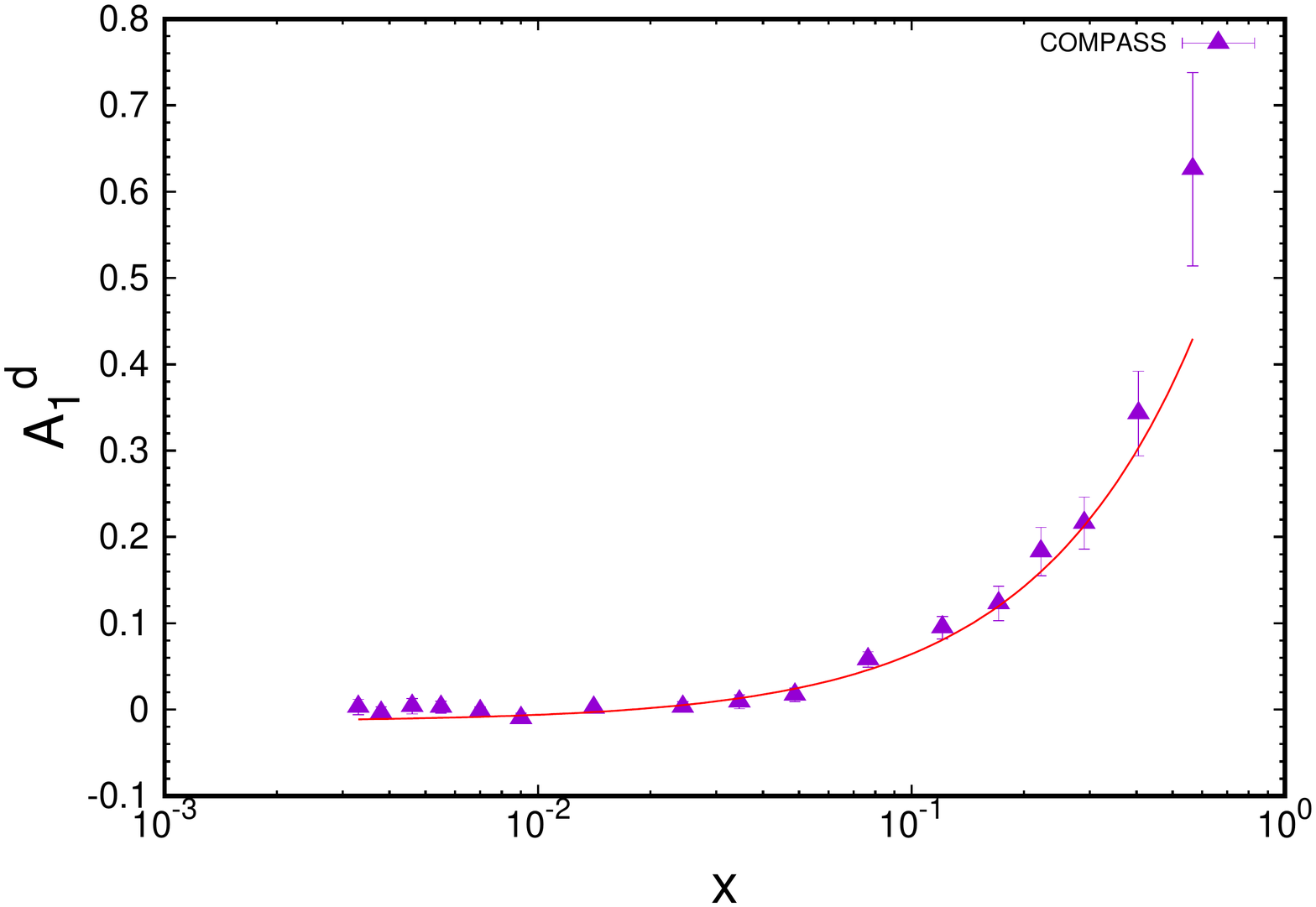}
    \caption{}
    \label{fig:compass}
\end{subfigure}
\begin{subfigure}{0.5\textwidth}
\centering
    \includegraphics[width=1.0\linewidth]{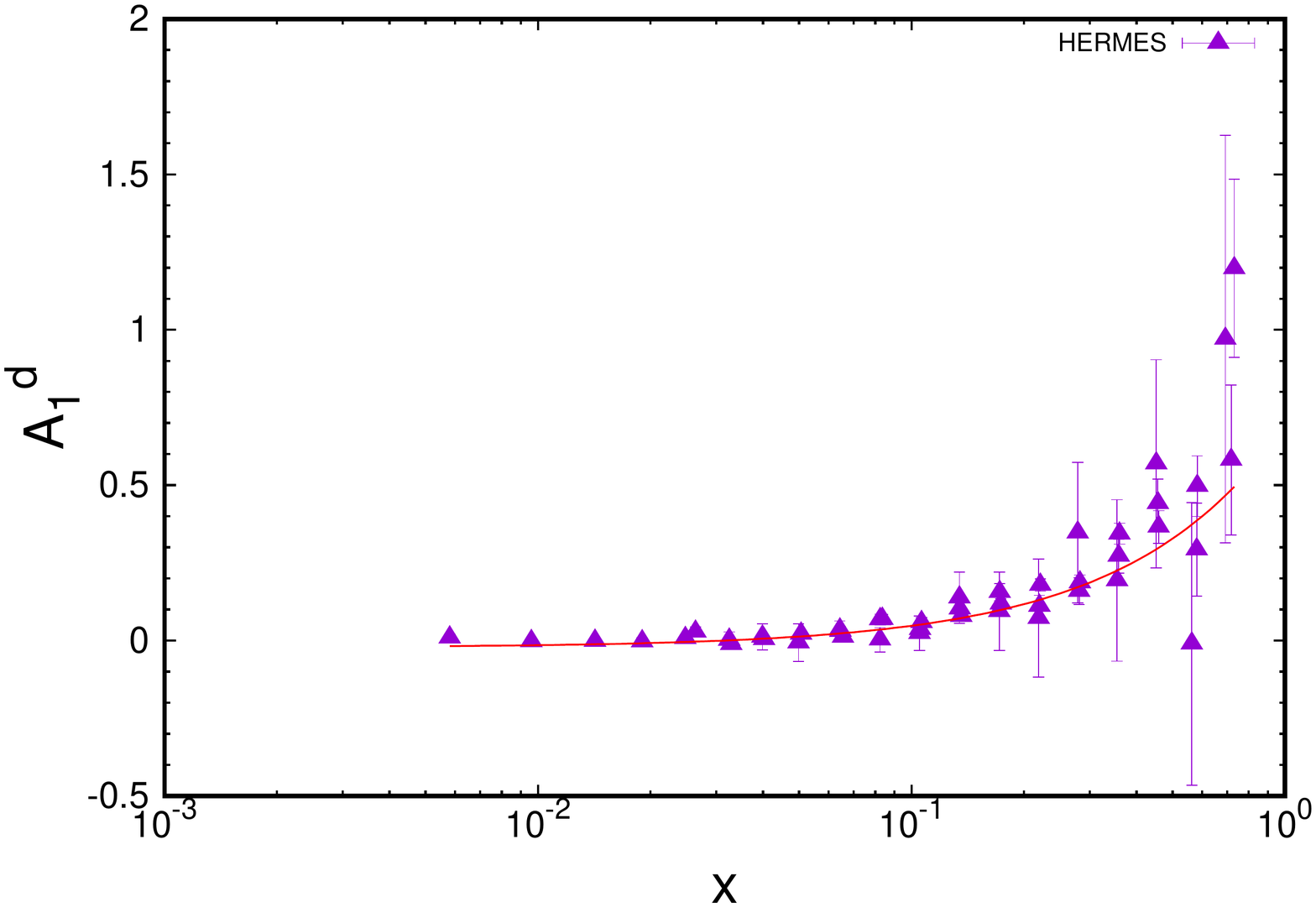}
    \caption{}
    \label{fig:hermes}
\end{subfigure}
\begin{subfigure}{0.5\textwidth}
\centering
    \includegraphics[width=1.0\linewidth]{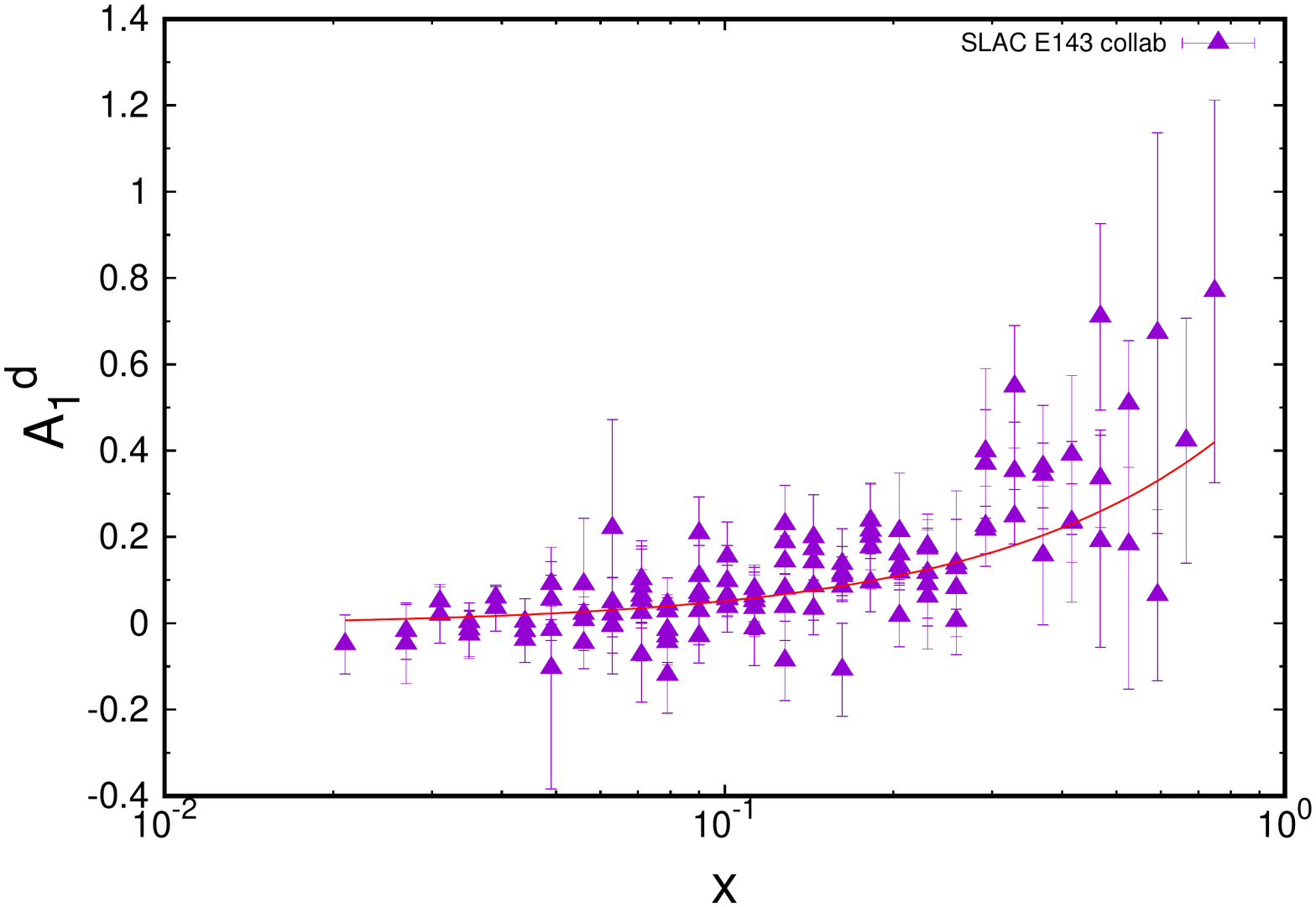}
    \caption{}
    \label{fig:E143}
\end{subfigure}
\begin{subfigure}{0.5\textwidth}
\centering
    \includegraphics[width=1.0\linewidth]{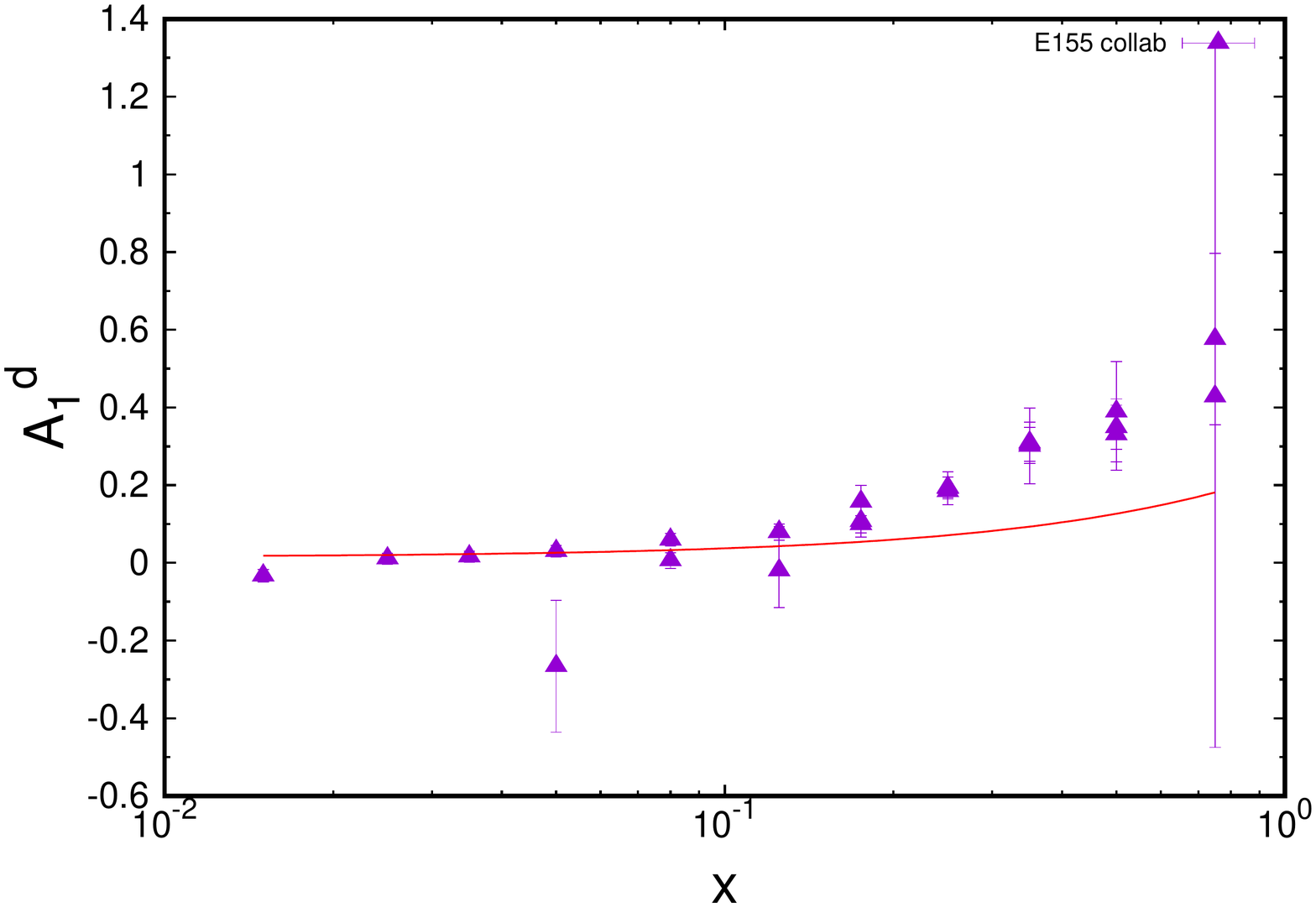}
    \caption{}
    \label{fig:E155}
\end{subfigure}
\caption{Comparison of the virtual photon asymmetry for the polarized deep inelastic scattering from deuteron. Data used in the figures  (\ref{fig:compass}), (\ref{fig:hermes}), (\ref{fig:E143}) and (\ref{fig:E155}) are from \cite{CompassCollab}, \cite{HermesCollab}, \cite{E143Collab} and \cite{E155Collab} respectively.}
\label{A1dComp}
\end{figure}

To be able to calculate the above defined asymmetry, in Eq.~(\ref{csPWIA}), we need to add the contribution of the polarization terms. For this, we include the expressions derived in Appendix E that contains the products $(\zeta_{j'}\cdot q) \rho_{j'}$ and $(\zeta_{j'}\cdot k) \rho_{j'}$ for deuteron target, along with the parametrization of the $g_{1}$ structure function from Eq.~(\ref{g1pd_fits}). The asymmetry calculated was then compared with existing experimental data for the case of deuteron, which are shown in Fig.~(\ref{A1dComp}).

The data in Fig.~(\ref{fig:compass}) covers the range $1~$GeV$^2$ $< Q^2 < 100~$GeV$^2$ and $0.004 < x < 0.7$.
The data in Fig.~(\ref{fig:hermes}) were collected at the HERMES experiment at DESY, in deep-inelastic scattering
of 27.6 GeV longitudinally polarized positrons off longitudinally polarized hydrogen and deuterium
gas targets internal to the HERA storage ring, covering the kinematic range of $0.0041 \leq x \leq 0.9$ and $0.18~$GeV$^2$ $\leq Q^2 \leq 20~$GeV$^2$. Similarly, in Fig.~(\ref{fig:E143}), the data are for the deep inelastic scattering of polarized electrons off a polarized ND$_3$ target at incident energies of 9.7, 16.2 and 29.1 GeV. These data correspond to the DIS region with $W^2 > 4~$GeV$^2$ and cover the kinematical  range of $0.024 \leq x \leq 0.75$ and $0.5$ GeV$^2 \leq Q^2 \leq 10$ GeV$^2$. Also, in fig.~(\ref{fig:E155}), the data are for the DIS of polarized electrons at incident energy of 48.3 GeV on a polarized
lithium deuteride ($^6$LiD) target. The data correspond to the three different  spectrometer angles (2.5$^\circ$, 5.5$^\circ$ and 10.5$^\circ$) and covers the kinematical range of $0.010 \leq x \leq 0.9$ and $1.0 $ GeV$^2\leq Q^2 \leq 40$ GeV$^2$. The curves were generated by using Eqs.~(\ref{A1defn}),(\ref{csPWIA}),(\ref{g1pd_fits}) and (\ref{g1n_Fit}). In using the parametrizations to calculate the asymmetry, it was assumed that $g_1(x_N,Q^2)$ for neutron is the same as that for the proton for Bjorken $x > 0.1$ and $g_2(x_N, Q^2) = 0$.

This naive approximation is in reasonable agreement with the existing asymmetry data for the case of deuteron target, as seen in Fig.~(\ref{A1dComp}). We now need to extract the spin structure function $g_1(x_N, Q^2)$ for the neutron using this asymmetry calculations. As the calculations are in reasonable agreement with the deuteron data, it becomes reasonable to use the same tools to also calculate the asymmetry for the $^3$He target.

\section{Extraction of $g_1(x_N, Q^2)$ of the neutron}
\label{sectiong1next}
The very fact that neutron does not exist freely in nature makes it complicated to obtain information about its spin structure. Before discussing the extraction of the spin information of neutron, we first discuss the general method(\cite{BodekParametrization}, \cite{AtwoodWest}) of extraction of the neutron unpolorized cross section from the deuteron using the concept of impulse approximation. In this method, the electron is assumed to scatter incoherently from the proton and neutron, and corrections for Fermi motion of the nucleons within the deuteron, commonly
known as ``smearing'' corrections, are made.

The first step in the extraction of the neutron cross sections was the introduction of Fermi-motion effects into the measured proton cross sections. The assumption $R=0.18$ (ratio of the $\sigma_L$ to $\sigma_T$ discussed above) was used together with the global fit to $F_{2p}$ to provide parametrizations of the two structure functions of the proton needed in this step. The proton structure functions from this fit were integrated over the momentum distribution of the proton within the deuteron to produce the ``smeared'' proton structure functions\cite{AtwoodWest} $W_{1s}^p, W_{2s}^p$, defined as
\begin{eqnarray}
\label{smW1W2defns}
&&W_{1s}^p(Q^2, \nu)= \int d^3p~|f(\mathbf{p_N})|^2 \Bigg[ W_1^p(Q^2, \nu', W'^2) + W_2^p(Q^2, \nu', W'^2)\frac{\mathbf{p_N}^2 - p_z^2}{2m_p^2} \Bigg], \nonumber \\
&&W_{2s}^p(Q^2, \nu)= \int d^3p~|f(\mathbf{p_N})|^2 \Bigg[ \Big(1 - \frac{p_zQ^2}{m_p\nu'|\mathbf{q}|}\Big)^2 \Big(\frac{\nu'}{\nu}\Big)^2 + \frac{\mathbf{p_N}^2 - p_z^2}{2m_p^2} \frac{Q^2}{|\mathbf{q}|^2} \Bigg]W_2^p(Q^2, \nu', W'^2),
\end{eqnarray}
where $p_N \equiv (E_N, \mathbf{p_N})$ is the four momentum of the proton which is now off-mass shell, $\nu' = \frac{p_N \cdot q}{m_p}$, $q \equiv (q_0, \mathbf{q})$ is the four momentum of the virtual photon, $W'^2 = (p_N + q)^2$ and $f(\mathbf{p_N})$ is the momentum distribution of the nucleon within the deuteron. The structure functions $W_1^p, W_2^p$ in Eq.~(\ref{smW1W2defns}) are identified with the on-shell structure functions measured at same values of $Q^2$ and $W^2 = W'^2$.

Then, using the definitions $F_1^p = m_p W_1^p, F_2^p = q_0 W_2^p $, the smeared structure functions are combined as in Eq.~(\ref{csPWIA_UnpolarizedPartOnly}) to obtain the smeared fitted proton cross section, $\sigma_{p,\text{sm}}^f$. The ratio of the fitted cross section of the proton, $\sigma_{p}^f$, before smearing to that of the after smearing, $S_p = \frac{\sigma_p^f}{\sigma_{p,\text{sm}}^f}$, is defined as the proton's smearing correction. The smeared proton cross section, $\sigma_{p,\text{sm}}$, were then calculated using the relation
\begin{equation}
\label{smProtoncs}
\sigma_{p,\text{sm}} = \frac{\sigma_p}{S_p},
\end{equation}  
where $\sigma_p$ represent the experimental proton cross sections. The proton smearing correction is found to be almost constant for lower $x$, while it varies sharply for $x \geq 0.6$, and for fixed values of $x$, the ratio varies weakly with $Q^2$, as shown in Table~(\ref{protonSmearingRatios}).

The smeared proton cross section can be combined with the experimental deuteron cross section, $\sigma_d$, to obtain the smeared neutron cross section, as
\begin{equation}
\label{smNeutroncs}
\sigma_{n,\text{sm}} = \sigma_d - \sigma_{p,\text{sm}}.
\end{equation}
The ratio of the smeared cross sections can then be found to be
\begin{eqnarray}
\label{smCSratio}
&&\frac{\sigma_{n,\text{sm}}}{\sigma_{p,\text{sm}}}  = \frac{\sigma_d - \sigma_{p,\text{sm}}}{\sigma_{p,\text{sm}}} = S_p \frac{\sigma_d}{\sigma_p} - 1.
\end{eqnarray}
If the smeared neutron cross section is defined in the same way as that for the proton, we find that for the ratio $\frac{\sigma_{n,\text{sm}}}{\sigma_{p,\text{sm}}}$ to be equal to the ratio of the true ratio $\frac{\sigma_n}{\sigma_p}$, it is required that $S_n = S_p$, where $S_n = \frac{\sigma_n}{\sigma_{n,\text{sm}}}$ is the neutron smearing correction. The unsmeared neutron cross section is then calculated using
\begin{equation}
\label{neutroncs_unsmeared}
\sigma_n = S_n \sigma_{n,\text{sm}} = S_n \sigma_d - U \sigma_p,
\end{equation}
where $U = \frac{S_n}{S_p}$ is the unsmearing correction factor.

\begin{table}[h!]
\centering
\begin{tabular}{| c | c | c | c | } 
\hline
  $x$ & $E_k$(GeV) & $Q^2$ (GeV$^2$) & $S_p$  \\ 
 \hline
  0.413  & 10.0  &  4.528  & 0.985   \\
  0.413 & 15.0   &  7.885  & 0.986 	 \\
  0.413 & 20.0   &  11.434 & 0.987   \\
  0.413 & 30.0	 &	18.794 & 0.990	\\
  \hline
  0.538	& 10.0	& 5.240	&	0.971 \\
  0.538	& 15.0	& 9.361	&	0.973 \\
  0.538 & 20.0 	& 13.799 &	0.976 \\
  0.538 & 30.0	& 23.141 &	0.981 \\
  \hline
  0.64	& 10.0 & 5.713 & 0.954 \\
  0.64	& 15.0 & 10.384	& 0.957 \\
  0.64	& 20.0 & 15.486	& 0.961 \\
  0.64	& 30.0 & 26.349	& 0.967 \\
  \hline
  0.71	& 10.0 & 5.994	& 0.941 \\
  0.71	& 15.0 & 11.010 & 0.941 \\
  0.71	& 20.0 & 16.537 & 0.945 \\
  0.71	& 30.0 & 28.396 & 0.953 \\
  \hline
  0.81	& 10.0 & 6.347	& 0.900 \\
  0.81 & 15.0 & 11.814	& 0.907 \\
  0.81 & 20.0 & 17.909 & 0.909 \\
  0.81 & 30.0 & 31.126 & 0.919 \\
  \hline
  0.883 & 10.0 & 6.574 & 0.843 \\
  0.883 & 15.0 & 12.342 & 0.859 \\
  0.883 & 20.0 & 18.826 & 0.862 \\
  0.883 & 30.0 & 32.988 & 0.847 \\
  \hline		
\end{tabular}
\caption{Proton's smearing correction,$S_p$, calculated at different $x$ and $Q^2$ for various incident energy $E_k$. These calculations were performed for the scattering angle of $19^\circ$.}
\label{protonSmearingRatios}
\end{table}

We now apply a similar prescription on the asymmetry, $A_1$, of the proton and deuteron to extract the asymmetry in neutron and hence the spin structure function $g_1$ of neutron. Keeping only the proton contribution in Eq.~(\ref{csPWIA}), we calculated the ``smeared''  proton asymmetry as:
\begin{eqnarray}
\label{proton_sm_asym}
&& A_{1,\text{sm}}^{p} = \frac{1}{D} \frac{\sigma_{p,\text{sm}}^{ \downarrow \Uparrow} - \sigma_{p,\text{sm}}^{\uparrow \Uparrow}}{\sigma_{p,\text{sm}}^{\downarrow \Uparrow} + \sigma_{p,\text{sm}}^{\uparrow \Uparrow}}.
\end{eqnarray}
The ratio of the asymmetries calculated using the unsmeared and smeared asymmetries of the proton can be formed as
\begin{equation}
\label{proton_sm_ratio}
S_{A1,p} = \frac{A_1^{p}}{A_{1,\text{sm}}^{p}},
\end{equation}
where $A_1^{p}$ is the unsmeared proton asymmetry calculated according to Eq.~(\ref{proton_sm_asym}) but for the proton at rest. The smeared proton asymmetry can then be used to calculate the smeared neutron asymmetry, as
\begin{equation}
\label{neut_sm_asym}
A_{\parallel,\text{sm}}^n = \frac{[\sigma_{d}^{\downarrow \Uparrow} - \sigma_{d}^{\uparrow \Uparrow}] - [\sigma_{p,\text{sm}}^{\downarrow \Uparrow} - \sigma_{p,\text{sm}}^{\uparrow \Uparrow}]}{[\sigma_{d}^{\downarrow \Uparrow} + \sigma_{d}^{\uparrow \Uparrow}] - [\sigma_{p,\text{sm}}^{\downarrow \Uparrow} + \sigma_{p,\text{sm}}^{\uparrow \Uparrow}]} = \frac{\sigma_{n,\text{sm}}^{ \downarrow \Uparrow} - \sigma_{n,\text{sm}}^{\uparrow \Uparrow}}{\sigma_{n,\text{sm}}^{\downarrow \Uparrow} + \sigma_{n,\text{sm}}^{\uparrow \Uparrow}}.
\end{equation}
This can be further used to calculate the Fermi smeared virtual photon-neuteron asymmetry as
\begin{equation}
\label{NeuteronA1n_wrt_parallelasym}
A_{1, \text{sm}}^n = \frac{1}{D}A_{\parallel,\text{sm}}^n.
\end{equation}
In this approach, we need to calculate polarized cross section of the deuteron according to Eq.~(\ref{csPWIA}). But this requires some initial modeling of $g_{1}^n$.
We also used another prescription that does not require any prior information about $g_{1}^n$ to obtain the smeared neutron asymmetry $A_{1,\text{sm}}^n$ from the deuteron. The algorithm of this prescription is presented below:
\begin{itemize}
\item	First, we read experimental data for the asymmetry of deuteron, $A_{1,\text{expt}}^d$, measured for various $x$ and $Q^2$.
\item	Then, for the same $x$ and $Q^2$, we calculate the unsmeared and smeared proton asymmetries $A_1^p$ and $A_{1,\text{sm}}^p$ respectively, which allows us to calculate the asymmetry smearing factor $S_{A1,p}$.
\item	Next, the smeared neutron asymmetry can be obtained using the relation 
\begin{equation}
A_{1,\text{sm}}^n = A_{1,\text{expt}}^d - A_{1,\text{sm}}^p.
\end{equation}
\end{itemize}
It should be noted that in the this approach, we do not require to supply any information about neutron to extract the neutron asymmetry.
Considering the smearing ratios to be the same for proton and neutron, we can now obtain the unsmeared neutron asymmetry as
\begin{equation}
\label{neut_unsm_asym}
A_{1}^n = A_{1,\text{sm}}^n S_{A1,p},
\end{equation}
where $S_{A1,p}$ is defined in Eq.~(\ref{proton_sm_ratio}). 
Using now the relation:
\begin{equation}
\label{a1GenDef}
A_1(x, Q^2) = \frac{g_1(x, Q^2) - \gamma^2 g_2(x, Q^2)}{F_1(x, Q^2)},
\end{equation}
where $\gamma^2 = \frac{(2 m_N x)^2}{Q^2}$ and assuming $g_2 = 0$ in our calculations,  the unsmeared neutron asymmetry, can be expressed as:
\begin{equation}
\label{a1n_for_getting_g1n}
A_1^n = \frac{g_1^n(x,Q^2)}{F_{1}^n(x, Q^2)},
\end{equation}
which can  be used to obtain $g_1^n$ as
\begin{equation}
g_1^n = A_1^n F_1^n.
\end{equation}
The smearing ratio for the proton asymmetry calculated for various $x$ and $Q^2$ is shown in Tables~(\ref{protonAsymmetrySmearingRatios_1}) - (\ref{protonAsymmetrySmearingRatios_3}).

\begin{table}[h!]
\centering
\begin{tabular}{| c | c | c | c | c | c | } 
\hline
  $x$ &  $Q^2$ (GeV$^2$) & $S_{A1,p}$ & $x$ &  $Q^2$ (GeV$^2$) & $S_{A1,p}$ \\ 
 \hline
0.0033 &	0.78 &	1.0024 &	0.0498 &	0.93 &	1.001 \\
0.0038 &	0.83 &	1.0029 &	0.05 &	2.57 &	1.003 \\
0.0046 &	1.10 &	1.0023 &	0.05 &	4.00 &	1.0012 \\
0.0055 &	1.22 &	1.0028 &	0.0506 &	1.54 &	1.0021 \\
0.0058 &	0.26 &	1.0011 &	0.056 &	0.38 &	0.9945 \\
0.007 &	1.39 &	1.0034 &	0.056 &	0.83 &	1.0000 \\
0.009 &	1.61 &	1.0038 &	0.056 &	1.16 &	1.0010 \\
0.0096 &	0.41 &	1.0013 &	0.056 &	1.92 &	1.0022 \\
0.0141 &	2.15 &	1.0041 &	0.063 &	0.40 &	0.9939 \\
0.0142 &	0.57 &	1.0016 &	0.063 &	0.60 &	0.9978 \\
0.015 &	1.22 &	1.0011 &	0.063 &	0.87 &	0.9998 \\
0.019 &	0.73 &	1.0017 &	0.063 &	1.26 &	1.001 \\
0.021 &	0.51 &	1.0008 &	0.063 &	2.07 &	1.0023 \\
0.0244 &	3.18 &	1.0043 &	0.0643 &	1.25 &	1.0013 \\
0.0248 &	0.82 &	1.002 &	0.0645 &	1.85 &	1.0022 \\
0.025 &	1.59 &	1.0025 &	0.0655 &	2.58 &	1.0018 \\
0.0264 &	1.12 &	1.0014 &	0.071 &	0.41 &	0.9931 \\
0.027 &	0.55 &	1.0008 &	0.071 &	0.64 &	0.9976 \\
0.027 &	1.17 &	1.0016 &	0.071 &	0.92 &	0.9996 \\
0.031 &	0.59 &	1.0007 &	0.071 &	1.36 &	1.0011 \\
0.031 &	1.27 &	1.0018 &	0.071 &	2.22 &	1.0024 \\
0.0325 &	0.87 &	1.0019 &	0.071 &	2.91 &	1.0019 \\
0.0329 &	1.25 &	1.0018 &	0.0765 &	8.53 &	1.0046 \\
0.0346 &	4.26 &	1.0044 &	0.079 &	0.43 &	0.9926 \\
0.035 &	0.31 &	0.9967 &	0.079 &	0.69 &	0.9975 \\
0.035 &	0.64 &	1.0006 &	0.079 &	0.97 &	0.9995 \\
0.035 &	1.40 &	1.0019 &	0.079 &	1.47 &	1.0011 \\
0.035 &	2.05 &	1.0027 &	0.079 &	2.38 &	1.0025 \\
0.039 &	0.33 &	0.9964 &	0.079 &	3.17 &	1.0021 \\
0.039 &	0.68 &	1.0005 &	0.08 &	3.24 &	1.0033 \\
0.039 &	1.52 &	1.0019 &	0.08 &	5.37 &	1.0025 \\
0.0399 &	0.90 &	1.0015 &	0.0823 &	1.31 &	1.0008 \\
0.0403 &	1.38 &	1.002 &	0.0824 &	2.06 &	1.0022 \\
0.044 &	0.35 &	0.9959 &	0.0835 &	3.08 &	1.0021 \\
0.044 &	0.73 &	1.0004 &	0.09 &	0.44 &	0.9918 \\
0.044 &	1.65 &	1.002 &	0.09 &	0.74 &	0.9973 \\
0.0487 &	5.80 &	1.0044 &	0.09 &	1.01 &	0.9992 \\
0.049 &	0.36 &	0.9952 &	0.09 &	1.58 &	1.0011 \\
0.049 &	0.78 &	1.0003 &	0.09 &	2.53 &	1.0026 \\
0.049 &	1.06 &	1.001 &	0.09 &	3.48 &	1.0023 \\
0.049 &	1.78 &	1.0021 &	0.101 &	0.45 &	0.9911 \\
\hline		
\end{tabular}
\caption{Smearing ratio for proton's asymmetry, $S_{A1,p}$, calculated at different $x$ and $Q^2$ (continued...).}
\label{protonAsymmetrySmearingRatios_1}
\end{table} 
\begin{table}[h!]
\centering
\begin{tabular}{| c | c | c | c | c | c | } 
\hline
  $x$ &  $Q^2$ (GeV$^2$) & $S_{A1,p}$ & $x$ &  $Q^2$ (GeV$^2$) & $S_{A1,p}$ \\ 
 \hline
 0.101 &	0.78 &	0.9971 &	0.175 &	4.62 &	1.0054 \\
0.101 &	1.06 &	0.9991 &	0.175 &	8.90 &	1.0053 \\
0.101 &	1.69 &	1.0012 &	0.175 &	13.19 &	1.0027 \\
0.101 &	2.69 &	1.0027 &	0.182 &	0.97 &	0.9976 \\
0.101 &	3.79 &	1.0025 &	0.182 &	1.25 &	0.9995 \\
0.1051 &	1.38 &	1.0004 &	0.182 &	2.25 &	1.0028 \\
0.1054 &	2.29 &	1.0024 &	0.182 &	5.49 &	1.0049 \\
0.1064 &	3.65 &	1.0026 &	0.182 &	3.45 &	1.0046 \\
0.113 &	0.47 &	0.9908 &	0.205 &	1.00 &	0.9984 \\
0.113 &	0.82 &	0.997 &	0.205 &	1.29 &	1.0002 \\
0.113 &	1.10 &	0.9989 &	0.205 &	2.36 &	1.0037 \\
0.113 &	1.80 &	1.0014 &	0.205 &	3.59 &	1.0055 \\
0.113 &	2.84 &	1.0029 &	0.205 &	5.86 &	1.0058 \\
0.113 &	4.11 &	1.0027 &	0.2191 &	1.67 &	1.0024 \\
0.121 &	12.6 &	1.0054 &	0.22 &	3.18 &	1.0056 \\
0.125 &	4.03 &	1.004 &	0.2213 &	5.87 &	1.0065 \\
0.125 &	7.16 &	1.0037 &	0.222 &	21.8 &	1.0091 \\
0.125 &	10.98 &	1.0004 &	0.23 &	1.03 &	0.9995 \\
0.128 &	0.48 &	0.9901 &	0.23 &	1.32 &	1.0013 \\
0.128 &	0.86 &	0.9969 &	0.23 &	2.47 &	1.0047 \\
0.128 &	1.14 &	0.9989 &	0.23 &	3.73 &	1.0066 \\
0.128 &	1.91 &	1.0015 &	0.23 &	6.23 &	1.007 \\
0.128 &	3.00 &	1.0031 &	0.25 &	5.06 &	1.0087 \\
0.128 &	4.44 &	1.0031 &	0.25 &	10.62 &	1.0089 \\
0.1344 &	1.46 &	1.0003 &	0.25 &	17.22 &	1.0055 \\
0.1347 &	2.56 &	1.0028 &	0.259 &	1.35 &	1.0029 \\
0.1358 &	4.30 &	1.0033 &	0.259 &	2.57 &	1.0062 \\
0.144 &	0.90 &	0.997 &	0.259 &	3.85 &	1.0081 \\
0.144 &	1.18 &	0.9989 &	0.259 &	6.60 &	1.0086 \\
0.144 &	2.03 &	1.0018 &	0.2786 &	1.98 &	1.0063 \\
0.144 &	3.15 &	1.0035 &	0.281 &	3.77 &	1.0093 \\
0.144 &	4.78 &	1.0036 &	0.2824 &	6.94 &	1.0098 \\
0.162 &	0.93 &	0.9971 &	0.29 &	28.3 &	1.0131 \\
0.162 &	1.22 &	0.9991 &	0.292 &	1.37 &	1.0051 \\
0.162 &	2.14 &	1.0023 &	0.292 &	2.67 &	1.0083 \\
0.162 &	3.30 &	1.004 &	0.292 &	3.98 &	1.0101 \\
0.162 &	5.13 &	1.0041 &	0.292 &	6.97 &	1.0107 \\
0.171 &	17.20 &	1.0069 &	0.329 &	2.76 &	1.011 \\
0.1719 &	1.56 &	1.0009 &	0.329 &	4.09 &	1.0128 \\
0.1722 &	2.87 &	1.0038 &	0.329 &	7.33 &	1.0134 \\
0.1734 &	5.05 &	1.0045 &	0.35 &	5.51 &	1.0157 \\
\hline		
\end{tabular}
\caption{Smearing ratio for proton's asymmetry, $S_{A1,p}$, calculated at different $x$ and $Q^2$ (continued...).}
\label{protonAsymmetrySmearingRatios_2}
\end{table}
\begin{table}[h!]
\centering
\begin{tabular}{| c | c | c | c | c | c | } 
\hline
  $x$ &  $Q^2$ (GeV$^2$) & $S_{A1,p}$ & $x$ &  $Q^2$ (GeV$^2$) & $S_{A1,p}$ \\ 
 \hline
0.35 &	12.59 &	1.016 &	0.50 &	5.77 &	1.0328 \\
0.35 &	22.65 &	1.0105 &	0.50 &	14.01 &	1.0331 \\
0.355 &	2.46 &	1.0132 &	0.50 &	26.97 &	1.026 \\
0.3585 &	4.62 &	1.0155 &	0.526 &	4.47 &	1.0361 \\
0.3603 &	8.25 &	1.0155 &	0.526 &	8.67 &	1.0359 \\
0.37 &	2.85 &	1.0146 &	0.5629 &	3.90 &	1.042 \\
0.37 &	4.20 &	1.0163 &	0.566 &	55.30 &	1.0446 \\
0.37 &	7.69 &	1.0169 &	0.5798 &	6.77 &	1.0458 \\
0.405 &	39.70 &	1.0227 &	0.5823 &	11.36 &	1.0444 \\
0.416 &	2.93 &	1.0195 &	0.592 &	4.55 &	1.0479 \\
0.416 &	4.30 &	1.021 &	0.592 &	8.98 &	1.0481 \\
0.416 &	8.03 &	1.0214 &	0.666 &	9.26 &	1.0664 \\
0.452 &	3.08 &	1.0246 &	0.6921 &	6.32 &	1.0727 \\
0.4567 &	5.61 &	1.0261 &	0.7173 &	9.56 &	1.083 \\
0.4589 &	9.72 &	1.0255 &	0.7311 &	14.29 &	1.0856 \\
0.468 &	3.01 &	1.0261 &	0.749 &	9.52 &	1.0958 \\
0.468 &	4.40 &	1.0273 &	0.75 &	15.73 &	1.1015 \\
0.468 &	8.37 &	1.0276 &	0.75 &	34.79 &	1.0881 \\
\hline		
\end{tabular}
\caption{Smearing ratio for proton's asymmetry, $S_{A1,p}$, calculated at different $x$ and $Q^2$.}
\label{protonAsymmetrySmearingRatios_3}
\end{table}
The smearing ratio is found to be very close to unity. For small and fixed $x$, we observe that $S_{A1,p}$ varies weakly with $Q^2$. However, for $x \geq$ 0.5, we observe a larger variation in $S_{A1,p}$.

%% file: 4/g1nresults.tex
\section{Results}
\begin{figure} [h!]
\begin{subfigure}{0.5\textwidth}
\centering
    \includegraphics[width=1.0\linewidth]{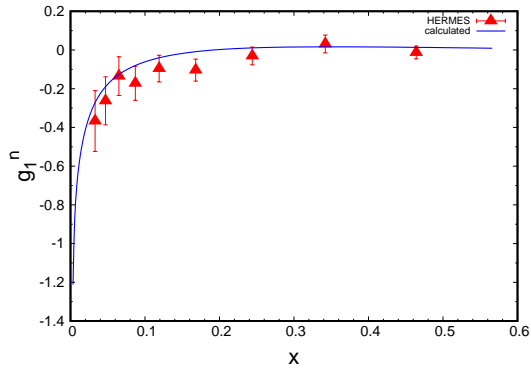}
    \caption{}
    \label{figg1ncomp:a}
\end{subfigure}
\begin{subfigure}{0.5\textwidth}
\centering
    \includegraphics[width=1.0\linewidth]{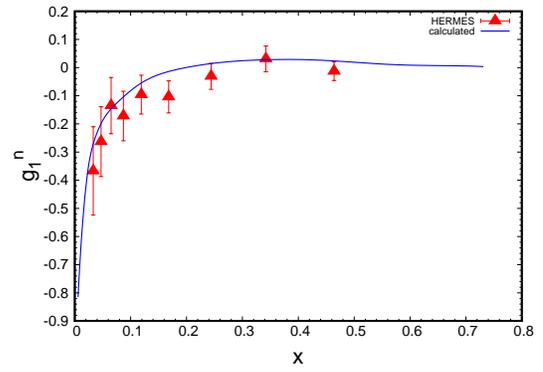}
    \caption{}
    \label{figg1ncomp:b}
\end{subfigure}
\begin{subfigure}{0.5\textwidth}
\centering
    \includegraphics[width=1.0\linewidth]{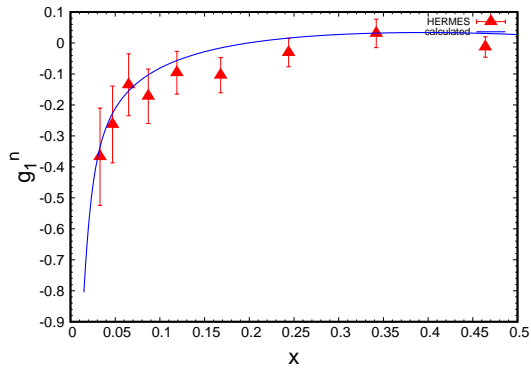}
    \caption{}
    \label{figg1ncomp:c}
\end{subfigure}
\begin{subfigure}{0.5\textwidth}
\centering
    \includegraphics[width=1.0\linewidth]{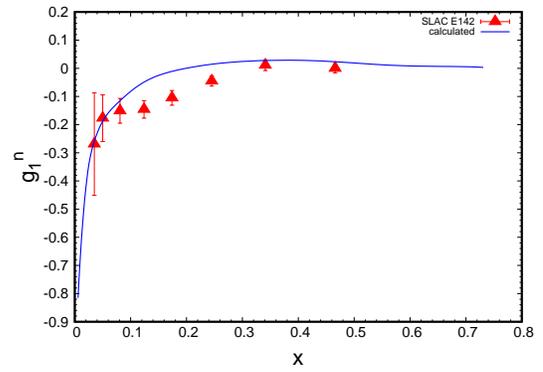}
    \caption{}
    \label{figg1ncomp:d}
\end{subfigure}
\begin{subfigure}{0.5\textwidth}
\centering
    \includegraphics[width=1.0\linewidth]{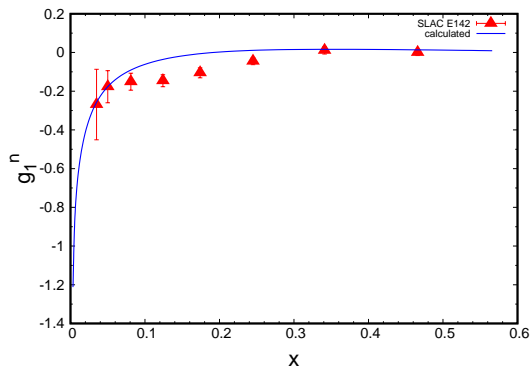}
    \caption{}
    \label{figg1ncomp:e}
\end{subfigure}
\caption{Comparison of $g_1^n$ with HERMES and SLAC-E142 data from \cite{Ackerstaffg1nData} and \cite{Anthonyg1nData} using extracted $g_1^n$ from \cite{CompassCollab}, \cite{HermesCollab}, \cite{E155Collab} respectively for HERMES data and \cite{CompassCollab}, \cite{HermesCollab} for SLAC-E142 data.}
\label{g1nCompHERMES&SLACE142}
\end{figure}

In this section, the comparison of our calculations with the existing experimental data for neutron spin structure function $g_{1n}(x,Q^2)$ which are extracted from polarized scattering from $^3$He target is given. We followed the second prescription discussed in section \ref{sectiong1next}, to obtain the smeared neutron asymmetry $A_{1,\text{sm}}^n$, from which  the $g_1^n$ structure function of the neutron was extracted. In Figs.~(\ref{figg1ncomp:a}),(\ref{figg1ncomp:b}) and (\ref{figg1ncomp:c}), the experimental $g_1^n$ data are taken from \cite{Ackerstaffg1nData}, in which the deep inelastic scattering of the polarized positrons  off a polarized internal
$^3$He gas target was measured at an incident energy of 27.5 GeV. The data covered the kinematical range of $
0.023 \leq x \leq 0.6$ and in 1 GeV$^2 \leq Q^2 \leq$ 15 GeV$^2$. Similarly, in Figs.~(\ref{figg1ncomp:d}) and (\ref{figg1ncomp:e}), the experimental data are from \cite{Anthonyg1nData}, in which the deep inelastic scattering of polarized electrons at incident energies of 19.42, 22.66 and 25.51 GeV on a polarized $^3$He target were performed. The experimental data cover the kinematical range of  $0.03 \leq x  \leq 0.6$ and 1 GeV$^2 \leq Q^2 \leq$ 5.5 GeV$^2$.

\begin{figure} [h!]
\begin{subfigure}{0.5\textwidth}
\centering
    \includegraphics[width=1.0\linewidth]{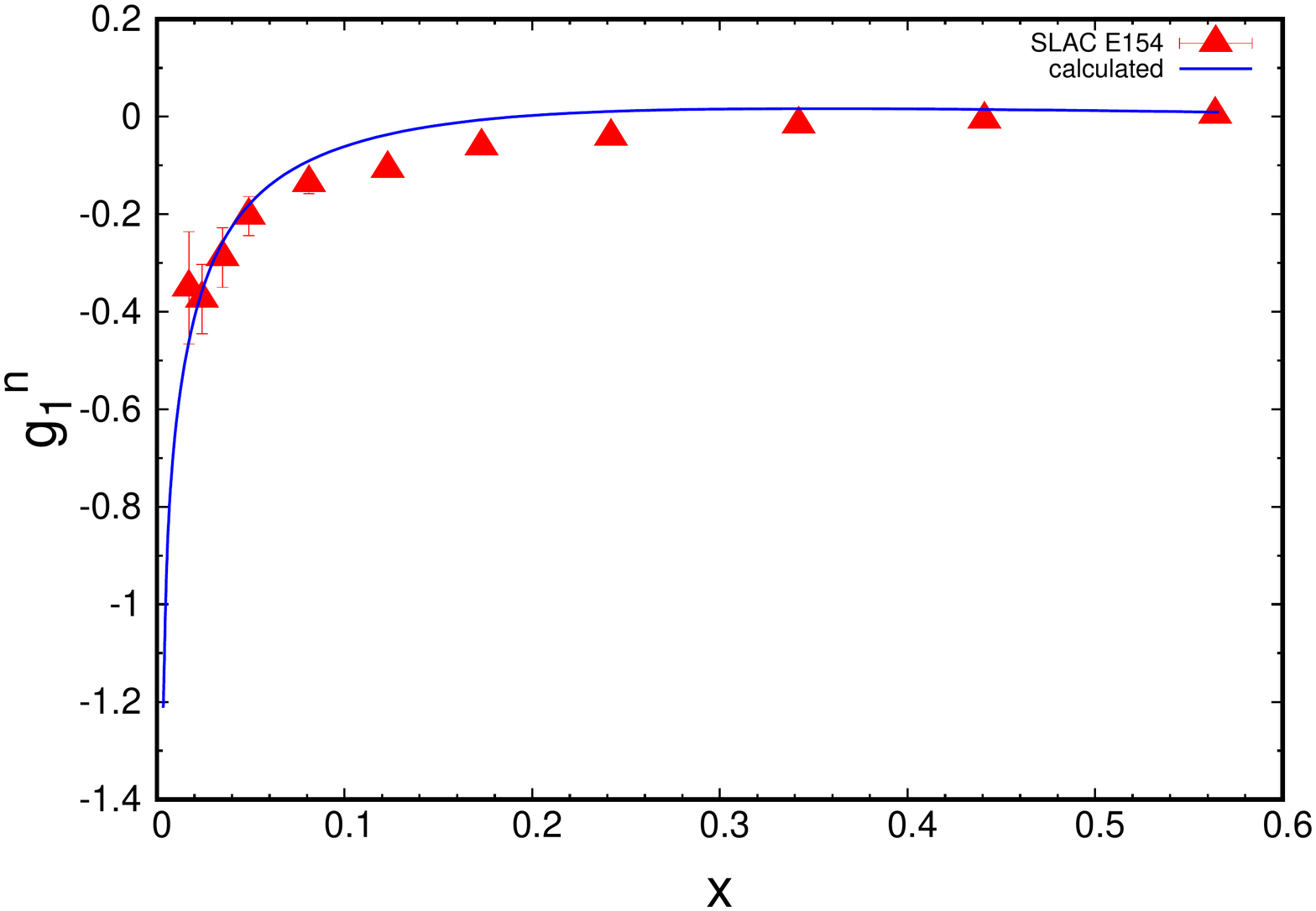}
    \caption{}
    \label{figg1ncomp:a1}
\end{subfigure}
\begin{subfigure}{0.5\textwidth}
\centering
    \includegraphics[width=1.0\linewidth]{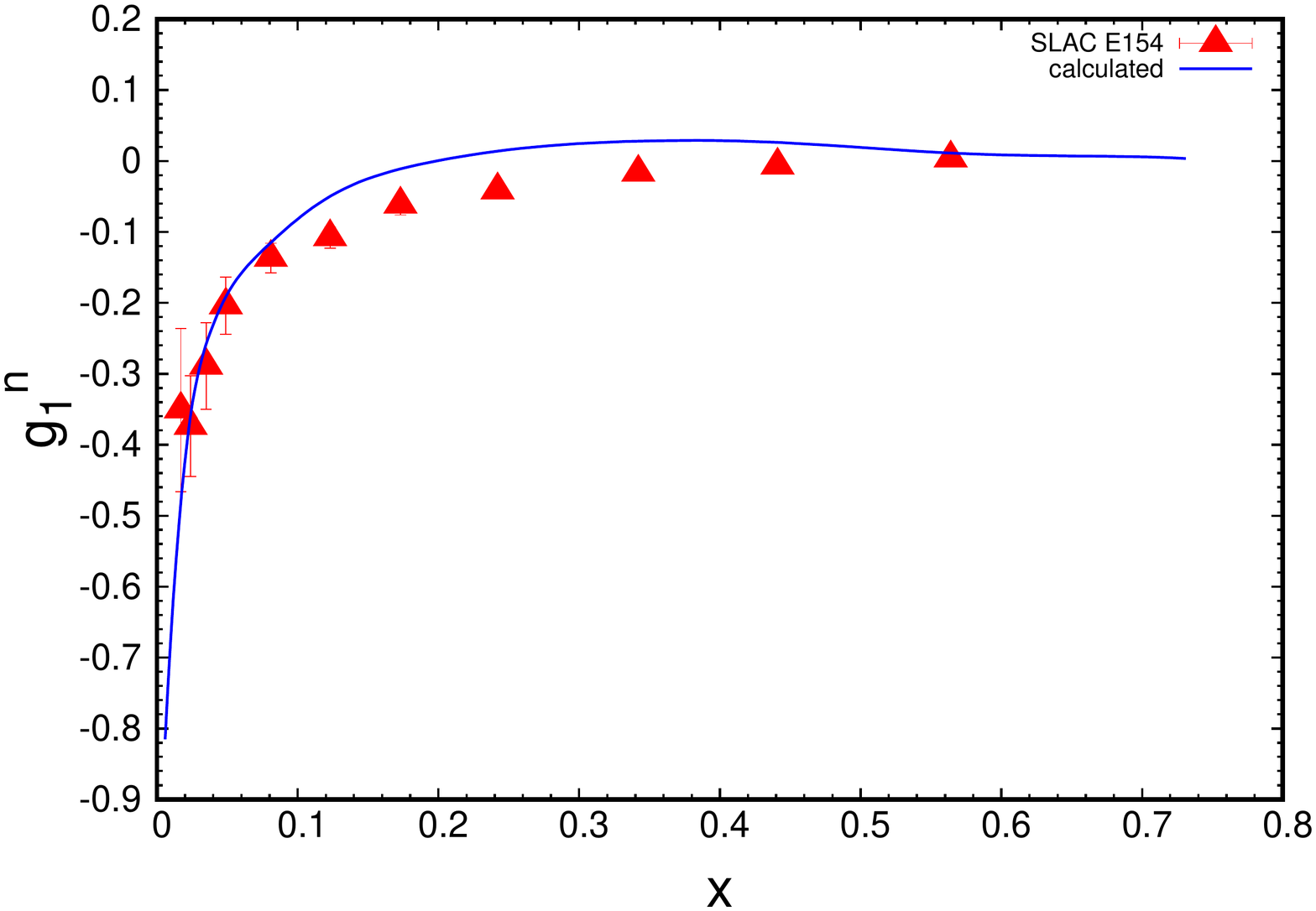}
    \caption{}
    \label{figg1ncomp:b1}
\end{subfigure}
\caption{Comparison of $g_1^n$ with SLAC-E154 data from \cite{Abeg1nData} using extracted $g_1^n$ from \cite{CompassCollab}, \cite{HermesCollab} respectively.}
\label{g1nCompSLACE154}
\end{figure}

In Figs.~(\ref{figg1ncomp:a1}) and (\ref{figg1ncomp:b1}), the experimental data are from \cite{Abeg1nData}, in which deep inelastic scattering  of polarized electrons, at an incident energy of 48.3 GeV, on a
polarized $^3$He target were performed. The experimental data cover the kinematical range  of $0.014 \leq x \leq 0.7$ and 1 GeV$^2 \leq Q^2 \leq$ 17.0 GeV$^2$.

The extraction of $g_1^n$ in Figs.~(\ref{figg1ncomp:a}, \ref{figg1ncomp:e}, \ref{figg1ncomp:a1}) was done from the experimental data from \cite{CompassCollab}, covering the kinematic range of $0.0033 \leq x \leq 0.566$ and 0.78 GeV$^2 \leq Q^2 $ 55.30 GeV$^2$. 
Similarly, the extraction of $g_1^n$ in Figs.~(\ref{figg1ncomp:b}, \ref{figg1ncomp:d}, \ref{figg1ncomp:b1}) was done from the experimental data from \cite{HermesCollab}, covering the kinematic range of $0.0058 \leq x \leq 0.7311$ and 0.26 GeV$^2 \leq Q^2 \leq$ 14.29 GeV$^2$.
 And, the extraction of $g_1^n$ in Fig.~(\ref{figg1ncomp:c}) was done from the experimental data from \cite{E155Collab}, which covered the kinematic range of $0.015 \leq x \leq 0.5$ and 1.22 GeV$^2 \leq Q^2 \leq$ 5.77 GeV$^2$. 

From the Figs.~(\ref{g1nCompHERMES&SLACE142}) and (\ref{g1nCompSLACE154}), it can be seen that the extracted spin structure function of neutron $g_1^n$ form the polarized dueteron target are in good agreement with the experimental data. It confirms that the Fermi motion of the proton needs to be considered to be able to extract the neutron information from the deuteron. This procedure is relatively simple in deuteron compared to $^3$He nucleus, as the polarization terms in Eq.~(\ref{csPWIA}) for deuteron can be easily derived analytically. 

In Figs.~(\ref{g1nCompHERMES&SLACE142}) and (\ref{g1nCompSLACE154}), we find that the extracted $g_1^n$ is in agreement at very low $x$, but is slightly higher for $x \approx 0.3$. For values of $x$ greater than 0.3, we again see reasonable agreement with the experimental data.
The extracted values of $g_{1}^n$, which were used to compare with the experimental $g_{1}^n$ data, are given in Tables (\ref{neutron_g1n_1}) - (\ref{neutron_g1n_3}).

\begin{table}[ht!]\vspace{-3cm}
\centering
\begin{tabular}{| c | c | c | } 
\hline
  $x$ &  $Q^2$ (GeV$^2$) & $g_1^n$ \\ 
 \hline
0.0033 &	0.78 &	-1.2121 \\
0.0038 &	0.83 &	-1.1997 \\
0.0046 &	1.10 &	-0.9217 \\
0.0055 &	1.22 &	-0.7721 \\
0.007 &	1.39 &	-0.6693 \\
0.009 &	1.61 &	-0.6243 \\
0.0141 &	2.15 &	-0.2967 \\
0.0244 &	3.18 &	-0.1910 \\
0.0346 &	4.26 &	-0.1362 \\
0.0487 &	5.80 &	-0.1006 \\
0.0765 &	8.53 &	-0.0168 \\
0.121 &	12.60 &	0.0078 \\
0.171 &	17.20 &	0.0106 \\
0.222 &	21.80 &	0.0271 \\
0.29 &	28.30 &	0.0194 \\
0.405 &	39.70 &	0.0181 \\
0.566 &	55.30 &	0.0091 \\
\hline		
\end{tabular}
\caption{Neutron spin structure function $g_1^n$ at different $x$ and $Q^2$ extracted from \cite{CompassCollab}.}
\label{neutron_g1n_1}
\end{table}
\begin{table}[h!]\vspace{-6cm}
\centering
\begin{tabular}{| c | c | c | } 
\hline
  $x$ &  $Q^2$ (GeV$^2$) & $g_1^n$ \\ 
 \hline
0.025 &	1.59 &	-0.2958 \\
0.035 &	2.05 &	-0.2273 \\
0.05 &	2.57 &	-0.1434 \\
0.08 &	3.24 &	-0.0503 \\
0.125 &	4.03 &	-0.0156 \\
0.175 &	4.62 &	0.0086 \\
0.25 &	5.06 &	0.0420 \\
0.35 &	5.51 &	0.0480 \\
0.5 &	5.77 &	0.0271 \\
\hline		
\end{tabular}
\caption{Neutron spin structure function $g_1^n$ at different $x$ and $Q^2$ extracted from \cite{E155Collab}.}
\label{neutron_g1n_2}
\end{table}

\begin{table}[t]
\centering
\begin{tabular}[t]{| c | c | c | } 
\hline
  $x$ &  $Q^2$ (GeV$^2$) & $g_1^n$ \\ 
 \hline
0.0058 &	0.26 &	-0.8153 \\
0.0096 &	0.41 &	-0.6641 \\
0.0142 &	0.57 &	-0.5020 \\
0.019 &	0.73 &	-0.4298 \\
0.0248 &	0.82 &	-0.2850 \\
0.0264 &	1.12 &	-0.2754 \\
0.0325 &	0.87 &	-0.2192 \\
0.0329 &	1.25 &	-0.3767 \\
0.0399 &	0.90 &	-0.1410 \\
0.0403 &	1.38 &	-0.2678 \\
0.0498 &	0.93 &	-0.1462 \\
0.0506 &	1.54 &	-0.1783 \\
0.0643 &	1.25 &	-0.0641 \\
0.0645 &	1.85 &	-0.1404 \\
0.0655 &	2.58 &	-0.2562 \\
0.0823 &	1.31 &	-0.0905 \\
0.0824 &	2.06 &	-0.0549 \\
0.0835 &	3.08 &	-0.1353 \\
0.1051 &	1.38 &	-0.0427 \\
0.1054 &	2.29 &	-0.0798 \\
0.1064 &	3.65 &	-0.1350 \\
0.1344 &	1.46 &	0.0698 \\
0.1347 &	2.56 &	-0.0025 \\
0.1358 &	4.30 &	-0.0968 \\
0.1719 &	1.56 &	0.0267 \\
0.1722 &	2.87 &	0.0339 \\
0.1734 &	5.05 &	-0.0535 \\
0.2191 &	1.67 &	0.0110 \\
0.22 &	3.18 &	0.0045 \\
0.2213 &	5.87 &	-0.0133 \\
0.2786 &	1.98 &	0.1089 \\
0.281 &	3.77 &	0.0177 \\
0.2824 &	6.94 &	-0.0104 \\
0.355 &	2.46 &	0.0340 \\
0.3585 &	4.62 &	0.0328 \\
0.3603 &	8.25 &	0.0204 \\
0.452 &	3.08 &	0.0722 \\
0.4567 &	5.61 &	0.0351 \\
0.4589 &	9.72 &	0.0094 \\
0.5629 &	3.90 &	-0.0059 \\
0.5798 &	6.77 &	0.0058 \\
0.5823 &	11.36 &	0.0054 \\
0.6921 &	6.32 &	0.0159 \\
0.7173 &	9.56 &	0.0026 \\
0.7311 &	14.29 &	0.0036 \\
\hline		
\end{tabular}
\caption{Neutron spin structure function $g_1^n$ at different $x$ and $Q^2$ extracted from \cite{HermesCollab}.}
\label{neutron_g1n_3}
\end{table}

These comparisons show that using deuteron target to extract spin structure function $g_1$ of neutron is as reliable as $^3$He target. Because of the simplicity in theoretical description of the deuteron, DIS from the polarized deuteron can be used to extract $g_1^n$ for very large values of $x$.

%% file: 5/5.tex
\input{./5/summaryConclusions.tex}

%% file: 5/summaryConclusions.tex
\chapter{Summary and Conclusion}
In the first part of this dissertation, the theory of hard rescattering mechanism of two-body break-up reactions was extended to the high energy photodisintegration of the $^3$He taget to ($p,d$) pair at large center of mass angles. The application of the hard rescattering mechanism allowed us to obtain a parameter free expression for the cross section. The cross section is expressed through the convolution of the effective charge of the constituent quarks being struck by incoming photon and interchanging in the final state of the process,  the  $^3\text{He} \rightarrow pd$ transition spectral function, 
and hard elastic $pd\rightarrow pd$ scattering differential cross section.  

To obtain the numerical results, we estimated the effective quark charge based on the quark-interchange model of $pd\rightarrow pd$ scattering. 
The transition spectral function is calculated using realistic wave function of $^3$He and the deuteron.  
The $pd\rightarrow pd$ cross section is incorporated by parameterizing the available experimental data according to Eq.~(\ref{pd_paramterization}).
The calculated differential cross sections of the reaction~(\ref{reaction}) at $\theta_{\text{cm}} = 90^\circ$ is compared with the recent experimental data from Jefferson Lab.  The comparison shows a very good agreement with the experimental data for the range of photon energies $E_{\gamma}\gtrsim 0.8$~GeV.  We also give a prediction for angular distribution of the cross section, which reflects the special property of HRM in which the magnitude of the invariant momentum transfer entering in the reaction~(\ref{reaction}) exceeds the magnitude of $t$ entering in the hard amplitude of $pd\rightarrow pd$ scattering.  The possibility of comparing the energy dependence of the cross section for different $\theta_{\text{cm}}$ of the $pd$ break-up will allow to establish the  kinematic boundaries  in which  QCD degrees of freedom are important for the quantitative description of the hard $pd$ break-up reactions.

In the second half of this dissertation, the inclusive polarized deep inelastic scattering was studied with a goal to extract the spin structure function $g_1^n(x,Q^2)$ of the neutron. A detailed derivation was done to obtain the differential cross section for the deep inelastic scattering of a polarized electron off a polarized nuclei. To be able to extract spin information about a nucleon inside the nucleus, we should study the scattering of the polarized electron off the nucleon inside the polarized nucleus. For this, we derived the cross section of such process using the Plane Wave Impulse Approximation (PWIA), which resulted in Eq.~(\ref{csPWIA}). This equation contains the contribution of both the unpolarized ($F_{1N}(x_N, Q^2), F_{2N}(x_N, Q^2)$) and polarized ($g_{1N}(x_N, Q^2), g_{2N}(x_N, Q^2)$) structure functions. To check the validity of Eq.~(\ref{csPWIA}), we first omitted the contribution of the polarization terms and calculated the cross section for the deep inelastic scattering off deuteron and $^3$He nuclei. These cross sections were then compared with the existing experimental data, the comparison being shown in Figs.~(\ref{whitlowComp})-(\ref{BodekComp3})
 for the case of the deuteron target and in Fig.~(\ref{He3comp}) for the $^3$He target. The comparisons show a very good agreement with the experimental data in the deep inelastic region. This leaves us at a good position to include the 
contributions of the polarized structure functions. The cross section thus calculated was then used to compute the asymmetry $A_1^d$ for the deuteron and compared with the experimental data, shown in Fig.~(\ref{A1dComp}).

We then extracted the polarized structure function $g_1^n(x, Q^2)$ of neutron using the smearing effects on the proton in the deuteron as discussed in section \ref{sectiong1next}. The extracted $g_1^n$ were then compared with the experimental data extracted from polarized $^3$He nucleus, which are shown in Figs.~(\ref{g1nCompHERMES&SLACE142}) and (\ref{g1nCompSLACE154}). These graphs show a good agreement of our calculation with the experimental data. This agreement justifies  the approach used to extract the spin structure function of the neutron from the deuteron, which is very important for extending experiments involving polarized deuteron targets for the large $x$ kinematics.

%% file: Appendix/pdpd.tex
\newpage
\addcontentsline{toc}{chapter}{Appendices}
\appendix
\renewcommand{\theequation}{A-\arabic{equation}}
\renewcommand\thefigure{ A.\arabic{figure}}
\setcounter{figure}{0}  
\setcounter{equation}{0}
\section*{APPENDIX A: High momentum transfer $pd \rightarrow pd$ scattering}
\label{AppendixA}
\begin{figure}[ht]
\centering
 \includegraphics[width=0.8\textwidth]{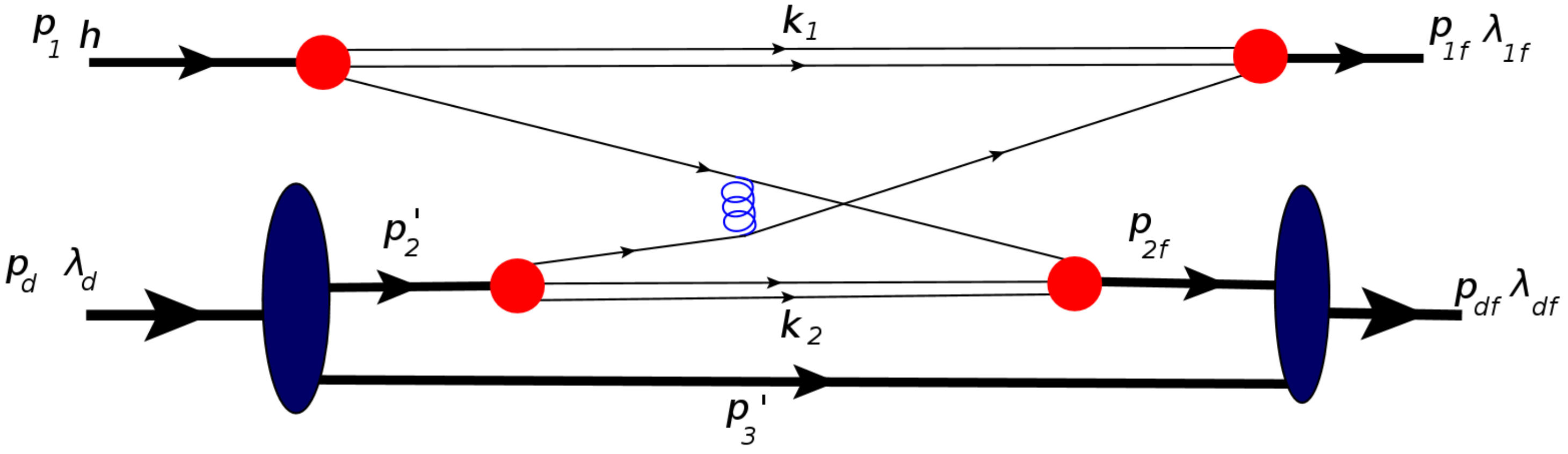}
\caption{Typical quark-interchange mechanism of hard $pd \rightarrow pd$ scattering.}
\label{Fig_A1}
\end{figure}

In this section, we study the high momentum transfer elastic proton - deuteron scattering
based on the quark-interchange mechanism.  A characteristic diagram of such scattering is shown in Fig.~(\ref{Fig_A1}). 
The notations in this figure are chosen to be similar to the $pd\rightarrow pd$ rescattering part of the $\gamma ^3\text{He}\rightarrow pd$ amplitude in Eq.~(\ref{AmpToIntroducePDamp}).
Here the helicities in the initial and final states for the proton are $h$ and $\lambda_{1f}$ and for the 
deuteron  they are $\lambda_{d} $ and $\lambda_{df}$. The momenta defined in Fig.~(\ref{Fig_A1}) satisfies the following 
four-momentum conservation relations:
\begin{equation}
p_{1} + p_{d} = p_{1f} + p_{df}, \qquad p_{d} = p_{2}' + p_{3}'.
\end{equation}

The Feynman amplitude for this $ pd \rightarrow pd $ scattering, shown in Fig.~(\ref{Fig_A1}), can be written as follows:
\begin{eqnarray}
\mathcal{M}_{pd}= && \nonumber \\
N1: &&\int \bar{u}(p_{1f}, \lambda_{1f}) (-i) \Gamma_{n1}^{\dagger} \frac{i(\slashed{p}_{1f} -\slashed{k}_{1} + m_q)}{(p_{1f} - k_{1})^2 -m_q^2 + i\epsilon} \frac{iS(k_1)}{k_1^2 - m_s^2 + i\epsilon} [-igT_c^\beta \gamma_\mu] \times \nonumber \\
&&\frac{i(\slashed{p_{1}} -\slashed{k_{1}} + m_q)}{(p_{1} - k_{1})^2 -m_q^2 + i\epsilon} i\Gamma_{n1} u(p_1, \lambda_1) \frac{d^4k_1}{(2\pi)^4} \nonumber
\end{eqnarray}
\begin{eqnarray}
D-N2: && \int \chi_{df}^\dagger (-i)\Gamma_{DNN}^\dagger \frac{i(\slashed{p}_{2f} + m)}{p_{2f}^2 - m_N^2 + i\epsilon} \bar{u}(p_{2f},\lambda_{2f}) (-i) \Gamma_{n2f}^\dagger \frac{i(\slashed{p}_{2f} -\slashed{k}_{2} + m_q)}{(p_{2f} - k_{2})^2 -m_q^2 + i\epsilon} \times \nonumber \\
 && \frac{i(\slashed{p_3}'+m)}{p_3'^{2}-m_N^2 + i\epsilon} 
 \frac{iS(k_2)}{k_2^2 - m_s^2 + i\epsilon} [-igT_c^\alpha \gamma_\nu] \frac{i(\slashed{p_{2}}' -\slashed{k_{2}} + m_q)}{(p_{2}' - k_{2})^2 -m_q^2 + i\epsilon} i\Gamma_{n2'} u(p_{2}' ,\lambda_{2}') \times \nonumber \\
&& \frac{i(\slashed{p}_{2}' + m)}{p_{2}'^2 - m_N^2 + i\epsilon} 
  i\Gamma_{DNN} \chi_d \frac{d^4k_2}{(2\pi)^4} \frac{d^4p_{3}'}{(2\pi)^4} \nonumber \\
g: && \frac{i d_{\mu \nu} \delta_{\alpha \beta}}{q_{q}^{2}}.
\end{eqnarray}
The following derivations are analogous to that of Eq.~(\ref{StartDerivationAmplitude}), where we identify the parts associated with the deuteron wave function 
as well as quark wave functions of nucleon and perform integrations corresponding to the on-shell conditions for the spectator nucleon in the deuteron and 
spectator quark-gluon states in the nucleons.
We first consider the expression for ${\bf N1:}$, for  which performing the similar derivations we did for the $N1$ term of  Eq.(\ref{StartDerivationAmplitude}) and using 
the defnition of the quark wave function of the nucleon according to  Eq.~(\ref{QuarkWF}) we obtain:
\begin{eqnarray} \label{N1_B1_with_QWF}
N1:= \sum_{\substack{\lambda_1 \\ \eta_{1}, \eta_{1f} }} 
&& -i \int	\frac{\Psi_{n1f}^{\dagger \lambda_{1f};\eta_{1f}} (x_{s1},k_{1\perp}, p_{1f\perp})}{1 - x_{s1}} 
\bar{u}_q(p_{1f} - k_1,\eta_{1f}) u_q(p_1 - k_1,\eta_1)\Big[ -igT_{c}^\beta \gamma_{\mu}\Big] \times \nonumber \\ 
&& \frac{\Psi_{n1}^{\lambda_1;\eta_1}(x_1, k_{1\perp},p_{1\perp}) }{1 - x_1} \frac{1}{2}\dfrac{dx_1}{x_1} \dfrac{d^2k_{1\perp}}{(2\pi)^3}. 
\end{eqnarray}
\vspace{0em}
With similar derivations in the ${\bf D-N2:}$ part and using in addition to the quark wave function of nucleon the deuteron light-front wave function defined 
in  Eq.~(\ref{DeuteronWF}) we obtain:
\vspace{0em}
\begin{eqnarray} \label{B_N2final}
D-N2 = && \sum_{\substack{ \lambda_{d},\lambda_{2f}, \lambda_{2}' ,\lambda_{3}' \\ \eta_{2}', \eta_{2f} }}  -i \int \frac{\Psi_{df}^{\dagger \lambda_{df}:\lambda_{3}', \lambda_{2f}} (\alpha_3'/\gamma_{d}, p_{df\perp},p_{3\perp}')}{1-\alpha_3'/\gamma_{d}} \bar{u}_q(p_{2f} - k_2,\eta_{2f})  \times \nonumber \\
&&\frac{\Psi^{\dagger \lambda_{2f};\eta_{2f}}_{n2f}(x_{s2},k_{2\perp},p_{2f\perp})}{1-x_{s2}} [ -igT_c^\alpha \gamma_\nu] 
 \frac{\Psi_{d}^{\lambda_{d}:\lambda_{2}', \lambda_{3}'}(\alpha_{3}',p_{d\perp},p_{3\perp}')}{1-\alpha_{3}'} u_q(p_2'-k_2,\eta_{2}')   \times \nonumber \\
 &&\frac{\Psi_{n2'}^{\lambda_{2}';\eta_{2}'}(x_{2}',k_{2\perp},p_{2\perp}')}{1-x_{2}'} \frac{1}{2} \frac{dx_{2}'}{x_{2}'} \frac{d^2k_{2\perp}}{(2\pi)^3} \frac{d\alpha_{3}'}{\alpha_{3}'} \frac{d^2p_{3\perp}'}{2(2\pi)^3}.
\end{eqnarray} 

Combining   Eqs.~(\ref{N1_B1_with_QWF}) and (\ref{B_N2final})  for the amplitude of $pd\rightarrow pd$ scattering we arrive at:
\begin{eqnarray} \label{pd_amplitude}
&&\mathcal{M}_{pd}^{\lambda_{df},\lambda_{1f};\lambda_{d},\lambda_1}  = \sum_{\substack{(\lambda_{2f})(\lambda_1, \lambda_d)(\lambda_{2}',\lambda_{3}')\\ (\eta_{1f}, \eta_{2f})(\eta_{1}, \eta_{2}') }}
 \int	\frac{\Psi_{df}^{\dagger \lambda_{df}: \lambda_{3}',\lambda_{2f}} (\alpha_3'/\gamma_{d}, p_{df\perp},p_{3\perp}')}{1-\alpha_3'/\gamma_{d}} \Bigg\{\frac{\Psi^{\dagger \lambda_{2f};\eta_{2f}}_{n2f}(x_{s2},k_{2\perp},p_{2f\perp})}{1-x_{s2}} \times \nonumber \\
 && \bar{u}_q(p_{2f} - k_2,\eta_{2f})
 [ -igT_c^\alpha \gamma_\nu]
  u_q(p_1 - k_1,\eta_1) 
 \frac{\Psi_{n1}^{\lambda_1;\eta_1}(x_1, k_{1\perp},p_{1\perp}) }{1 - x_1}\Bigg\}_{1} G^{\mu \nu}(r) \times \nonumber \\
&& \Bigg\{\frac{\Psi_{n1f}^{\dagger \lambda_{1f};\eta_{1f}} (x_{s1},k_{1\perp}, p_{1f\perp})}{1 - x_{s1}}
 \bar{u}_q(p_{1f} - k_1,\eta_{1f})
 [ -igT_c^\beta \gamma_\mu] u_q(p_2'-k_2,\eta_{2}')  
 \frac{\Psi_{n2'}^{\lambda_{2}';\eta_{2}'}(x_{2}',k_{2\perp},p_{2\perp}')}{1-x_{2}'} \Bigg\}_{2} \times \nonumber \\
  &&\frac{\Psi_{d}^{\lambda_d: \lambda_{2}', \lambda_{3}'}(\alpha_{3}',p_{d\perp},p_{3\perp}')}{1- \alpha_{3}'} 
   \frac{1}{2}\dfrac{dx_1}{x_1} \dfrac{d^2k_{1\perp}}{(2\pi)^3} \frac{1}{2} \frac{dx_{2}'}{x_{2}'} \frac{d^2k_{2\perp}}{(2\pi)^3} \frac{d\alpha_{3}'}{\alpha_{3}'} \frac{d^2p_{3\perp}'}{2(3\pi)^3}.
\end{eqnarray}

%% file: Appendix/NonRelativistic2LCwfs.tex
\appendix
\renewcommand{\theequation}{B-\arabic{equation}}
\renewcommand\thefigure{ B.\arabic{figure}}
\setcounter{figure}{0}  
\setcounter{equation}{0}
\section*{APPENDIX B: Relating the light-front and non-relativistic wave functions}
To obtain the relation between light-front and  non-relativistic nuclear wave functions in the small momentum limit we  consider the fact that the 
light-front nuclear wave function is normalized based on baryonic number conservation~(see e.g. Refs.\cite{Frankfurt:1985ui,srcrev,Cosyn:2010ux})  
while non-relativistic (Schroedinger)  wave function normalized as $\int |\Psi_A(p)|^2 d^3p = 1$.

\begin{figure}[ht]
\centering
\includegraphics[scale=0.4]{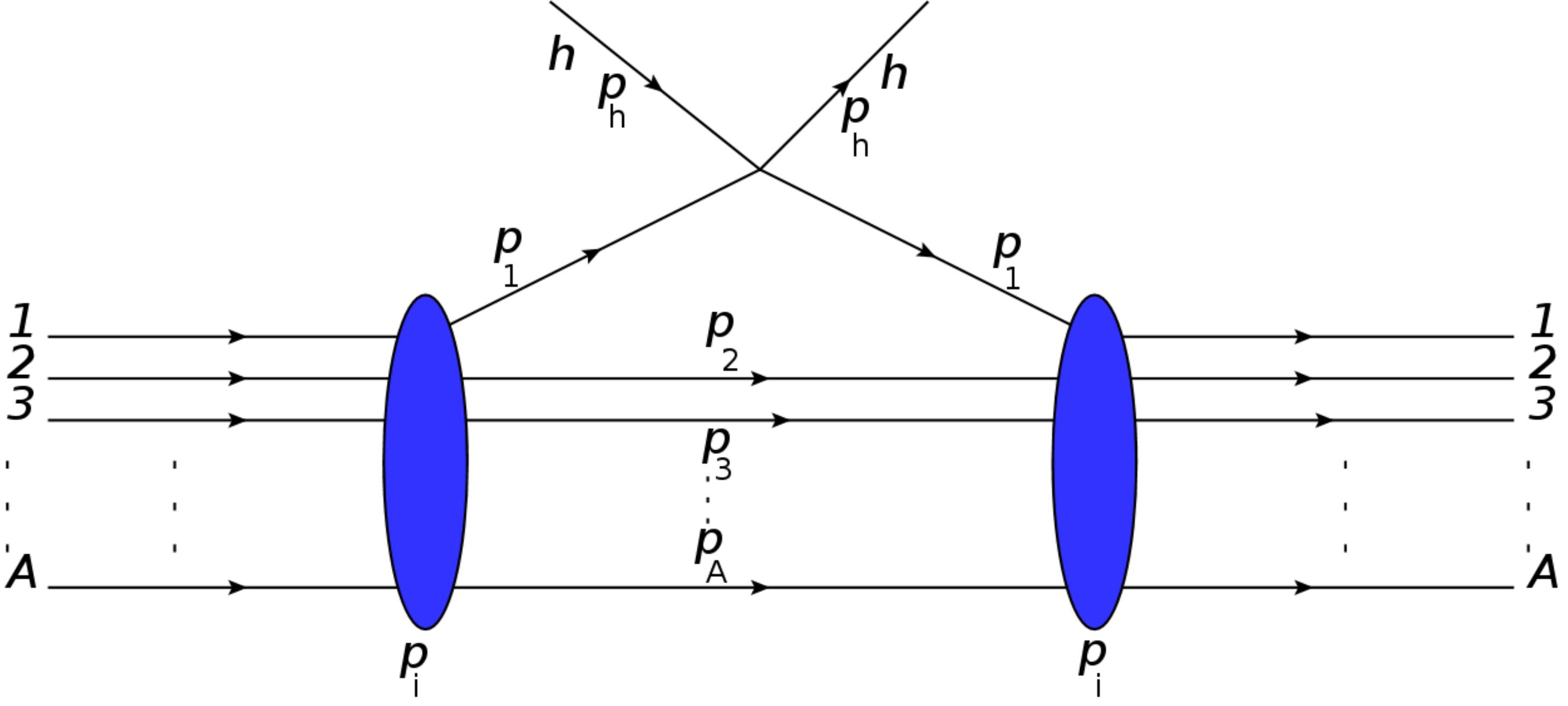}
\caption{Hadronic probe to see baryons in a nucleus.}
\label{probe_diagram}
\end{figure}

To obtain the normalization condition based on baryonic number conservation, we consider  a $h+A\rightarrow h+A$ scattering in forward 
direction  in which $h$ probes the 
constituent baryons in the nucleus $A$ (see Fig.~(\ref{probe_diagram})).  In the figure, we assign $p_i$ to be the four-momentum of the nucleus while
$p_1, p_2, \cdots, p_A$ are four momenta of constituent nucleons such that $p_1 + p_2 + \cdots + p_A = p_i$. 
For the diagram of Fig.~(\ref{probe_diagram}), we apply the Feynman rules to obtain:
\begin{eqnarray} \label{hadron_probe_amplitude}
\mathcal{M}_{hA} &&= \sum_{N} \int \chi_A^{\dagger} \Gamma_A^{\dagger} \frac{\slashed{p}_1 + m}{p_1^2 - m_N^2 + i\epsilon}
 \hat{M}_{hN} \frac{\slashed{p}_A + m}{p_A^2 - m_N^2 + i\epsilon} \cdots \frac{\slashed{p}_2 + m}{p_2^2 - m_N^2 + i\epsilon} 
 \frac{\slashed{p}_1 + m}{p_1^2 - m_N^2 + i\epsilon} \Gamma_A  \chi_A \times \nonumber \\
&& \frac{d^4p_2}{(2\pi)^4} \frac{d^4p_3}{(2\pi)^4} \cdots \frac{d^4p_A}{(2\pi)^4},
\end{eqnarray}
where we sum over all the possible nucleons that can be probed and ${\hat M}_{hN}$ represents the effective vertex of the hadron-nucleon interaction.
We use the sum rule for the spinors and also integrate by the minus component of the momenta using the scheme given in 
Eq.~(\ref{polevalueint_scheme}) to obtain:
\begin{eqnarray}
  \mathcal{M}_{hA} & = &   \sum_{N} \sum_{\lambda_1, \lambda_2, \cdots \lambda_A} \int \chi_A^{\dagger} \Gamma_X^{\dagger} \frac{u(p_1, \lambda_1) \bar{u}(p_1, \lambda_1) }{p_{1+} \big(p_{1-} - \frac{m_N^2 + p_{1\perp}^2}{p_{1+}}\big)} \hat{M}_{hN} u(p_A , \lambda_A) \bar{u}(p_A, \lambda_A)\cdots  \times \nonumber \\ 
&&   u(p_3 , \lambda_3) \bar{u}(p_3, \lambda_3) u(p_2 , \lambda_2) \bar{u}(p_2, \lambda_2) \frac{u(p_1 , \lambda_1) \bar{u}(p_1, \lambda_1)}{p_{1+} \big(p_{1-} - \frac{m_N^2 + p_{1\perp}^2}{p_{1+}}\big)}  \Gamma_A  \chi_A \times \nonumber \\
& & \frac{dp_{2+}}{p_{2+}} \frac{d^2p_{2\perp}}{2(2\pi)^3} \frac{dp_{3+}}{p_{3+}} \frac{d^2p_{3\perp}}{2(2\pi)^3} \cdots \frac{dp_{A+}}{p_{A+}} \frac{d^2p_{A\perp}}{2(2\pi)^3}, 
\end{eqnarray}
\vspace{0em}
where the $\lambda_j$ denote the helicities of the nucleon with momentum $p_j$.
Considering the transverse momentum of the nucleus $A$ to be zero  we note that:\vspace{0em}
\begin{eqnarray} \label{P1minus_Xnuc}
&&p_{1-} - \frac{m_N^2 + p_{1\perp}^2}{p_{1+}} = p_{i-} - p_{2-} - p_{3-}- \cdots - p_{A-} - \frac{m_N^2 + p_{1\perp}^2}{p_{1+}}  \nonumber
\end{eqnarray}
\begin{eqnarray}
&&= \frac{1}{p_{i+}}\Big[ m_N^2 - \frac{m_N^2 + p_{2\perp}^2}{\beta_2} - \frac{m_N^2 + p_{3\perp}^2}{\beta_3} - \cdots - \frac{m_N^2 + p_{A\perp}^2}{\beta_A } - \frac{m_N^2 + p_{1\perp}^2}{\beta_1}\Big],
\end{eqnarray}
where  $\beta_j = \frac{p_{j+}}{p_{i+}}$ are the momentum fractions of the nucleus $A$ carried by the nucleons $j$ ($j = 1,\cdots A$).
Introducing Feynman amplitude for $h+N\rightarrow hN$ as $\mathcal{M}_{hN} = \bar{u}(p_1, \lambda_1) \hat{M}_{hN}  u(p_1, \lambda_1)$ for Eq.~(\ref{P1minus_Xnuc}), we then have the following for the scattering amplitude: \vspace{0em}
\begin{eqnarray} \label{Xnuc_amplitude}
& \mathcal{M}_{hA} &= \sum_{N}\sum_{\lambda_1, \lambda_2, \cdots \lambda_A} \int \chi_A^{\dagger} \Gamma_A^{\dagger} \frac{u(p_1, \lambda_1) u(p_2, \lambda_2)u(p_3, \lambda_3) \cdots u(p_A, \lambda_A) }{\frac{p_{1+}}{p_{i+}} \Big[ m_N^2 - \frac{m_N^2 + p_{2\perp}^2}{\beta_2} - \frac{m_N^2 + p_{3\perp}^2}{\beta_3} -\cdots -  \frac{m_N^2 + p_{A\perp}^2}{\beta_A} \Big]} \mathcal{M}_{hN} \times \nonumber \\
&& \frac{ \bar{u}(p_1, \lambda_1) \bar{u}(p_2, \lambda_2)  \bar{u}(p_A, \lambda_A)}{\frac{p_{1+}}{p_{i+}} \Big[ m_N^2 - \frac{m_N^2 + p_{2\perp}^2}{\beta_2} - \frac{m_N^2 + p_{3\perp}^2}{\beta_3} -\cdots -  \frac{m_N^2 + p_{A\perp}^2}{\beta_A} \Big]} \Gamma_A \chi_A \prod_{k = 2}^{A} \frac{d\beta_{k}}{\beta_k} \frac{d^2p_{k\perp}}{2(2\pi)^3}.
\end{eqnarray}
Using the generalization of Eqs.~(\ref{He3wf}) and (\ref{DeuteronWF}) for light-front nuclear wave function of nucleus $A$, the above equation reduces to:
\begin{eqnarray} \label{Xnuc_amp_with_wfs}
\mathcal{M}_{hA} &&= \sum_{N} \sum_{\lambda_1, \lambda_2, ... \lambda_A} \int \frac{\Psi_A^{LC \dagger}( \beta_2, \beta_3, ... \beta_A, p_{2\perp}, p_{3\perp} ... p_{A\perp}, \lambda_2, \lambda_3, ... \lambda_A) }{\beta_1} \mathcal{M}_{hN} \times \nonumber \\
&&\frac{\Psi_A^{LC}( \beta_2, \beta_3, ... \beta_A, p_{2\perp}, p_{3\perp} ... p_{A\perp}, \lambda_2, \lambda_3, ... \lambda_A )}{\beta_1} \prod_{k = 2}^{A} \frac{d\beta_{k}}{\beta_k} \frac{d^2p_{k\perp}}{2(2\pi)^3}.
\end{eqnarray}

We now make use of the Optical theorem   according to which:  
\begin{equation}
\text{Im } \mathcal{M}_{hA} = s_{hA} \sigma_{hA} \text{ and } \text{Im } \mathcal{M}_{hN} = s_{hN} \sigma_{hN}, 
\label{OpTheorem}
\end{equation}
where $s_{hA}= (p_h + p_i)^2$   and 
$\sigma_{hA}$ is the total cross section of $hA$ scattering.  Similarly,  $s_{hN}$ and $\sigma_{hN}$ are invariant energy 
and  total cross section for $hN$ scattering.
 The conservation of baryon number allows us to relate $\sigma_{hA} = A\sigma_{hN}$. Using this relation together with Eq.~(\ref{OpTheorem})  in Eq.~(\ref{Xnuc_amp_with_wfs}) we arrive at:
\begin{equation} \label{lightcone_norm}
\int \frac{\left| \Psi_A^{LC}( \beta_2, \beta_3, ... \beta_A, p_{2\perp}, p_{3\perp} ... p_{A\perp}, \lambda_2, \lambda_3, ... \lambda_A) \right|^2}{\beta_1^2} \frac{s_{hN}}{s_{hA}} \prod_{k = 2}^{A} \frac{d\beta_k}{\beta_k} \frac{d^2p_{k\perp}}{2(2\pi)^3} = 1.
\end{equation}
To obtain the relation of light-front wave function to the nonrelativistic wave function in the small momentum limit, we note that in this limit,
$\beta_k = \frac{E_k + p_k^z}{p_{i+}}\approx 1+ \frac{p_{k}^z}{ m_N}$ thus $\frac{d\beta_k}{\beta_k} = \frac{dp_k^z}{m_N}$. 
Furthermore, in the high energy limit of the hadronic probe in which large momentum of the hadrons points to $-\hat z$ direction, 
$s_{hA} \approx p_{h-}p_{A+}$ and  $s_{hN} \approx p_{h-}p_{N+}$ resulting in:
\begin{equation}
\frac{s_{hN}}{s_{hA}} = \frac{p_{N+}}{p_{A+}} = \frac{\beta_1}{ A}.
\end{equation}
Applying all these approximations in Eq.(\ref{lightcone_norm})  we then obtain the following:
\begin{equation} \label{normalized_eqn_a}
\int \frac{\left| \Psi_A^{LC}( \beta_2, \beta_3, ... \beta_A, p_{2\perp}, p_{3\perp} ... p_{A\perp}, \lambda_2, \lambda_3, ... \lambda_A) \right|^2} {1/A}   \frac{1}{m_N^{A-1} [2(2\pi)^3]^{A-1}} \prod_{k = 2}^{A}d^3p_k = 1.
\end{equation}
Next, we compare the above expression with the normalization condition for the non-relativistic Schroedinger wave function:
\begin{equation} \label{normalized_eqn_b}
\int \left| \Psi_A^{NR}(\vec{p}_1, \vec{p}_2, ... \vec{p}_A )\right|^2 \prod_{k = 2}^{A} d^3p_k = 1,
\end{equation}
where  $\vec p_1 = \vec p_i -  \vec p_2 - \cdots -  \vec p_A$. This comparison allows to relate 
the light-front nuclear wave function and 
the Schroedinger wave function in the following form:
\begin{equation} \label{LCNR_relation_final}
\Psi_X^{LC} ( \beta_1, \beta_2, ..., p_{1\perp}, p_{2\perp}...) = \frac{1}{\sqrt{A}} \big[m_N 2(2\pi)^3\big]^{\frac{A-1}{2}} \Psi_X^{NR}(\vec{p}_1, \vec{p}_2, ... ).
\end{equation}

%% file: Appendix/electronPolarization.tex
\appendix
\renewcommand{\theequation}{C-\arabic{equation}}
\renewcommand\thefigure{ C.\arabic{figure}}
\setcounter{figure}{0}  
\setcounter{equation}{0}
\section*{APPENDIX C: Polarization vector of electron}
\label{appendixC}
The polarization vector of an electron is defined by
\begin{equation}
a_{\sigma} \equiv (a_0, \mathbf{a}) = \Bigg[\frac{\boldsymbol{\xi}\cdot \mathbf{k}}{m_e}, \boldsymbol{\xi} + \frac{\boldsymbol{\xi} \cdot \mathbf{k} ~ \mathbf{k}}{m_e(E_k + m_e)} \Bigg].
\end{equation}
We aim to reduce the polarization vector of the electron in high energy limits, where $m_e^2 \approx 0$. Since, $\mathbf{\xi} = h_e \frac{\mathbf{k}}{|\mathbf{k}|}$, we have
\begin{eqnarray} 
a^\sigma = \Big(a^0, \mathbf{a}\Big) &&= \Bigg[\frac{h_e \mathbf{k} \cdot \mathbf{k}}{ |\mathbf{k}| m_e} , h_e \frac{\mathbf{k}}{|\mathbf{k}|} + \frac{h_e}{|\mathbf{k}|} \frac{\mathbf{k}^2}{m_e(E_k + m_e)} \Bigg] \nonumber \\
&& = \Bigg[ \frac{h_e k}{m_e} ,  h_e\frac{\mathbf{k}}{|\mathbf{k}|} \Big(1 + \frac{k^2}{m_e(E_k + m_e)} \Big) \Bigg] \nonumber \\
&& = \Bigg[ \frac{h_e k}{m_e}, h_e\frac{\mathbf{k}}{|\mathbf{k}|} \Bigg(\frac{m_e(E_k + m_e) + k^2}{m_e(E_k + m_e)} \Bigg) \Bigg] \nonumber \\
&& \approx \Bigg[ \frac{h_e k}{m_e},	h_e\frac{\mathbf{k}}{|\mathbf{k}|} \Bigg(\frac{m_e E_k + E_k^2}{m_e(E_k + m_e)} \Bigg) \Bigg] \nonumber \\
&& = \Bigg[ \frac{h_e k}{m_e}, 	h_e \frac{\mathbf{k}}{|\mathbf{k}|} \frac{E_k}{m_e} \Bigg] \nonumber
\end{eqnarray}

\begin{eqnarray} \label{a_vector_simp}
&& =\Bigg[ \frac{h_e k}{m_e}, h_e \hat{k} \frac{k}{m_e} \Bigg] \nonumber \\
&& = \Bigg[ \frac{h_e k}{m_e}, h_e \frac{\mathbf{k}}{m_e}\Bigg] = \frac{h_e}{m_e} \Big( k, \mathbf{k}\Big).
\end{eqnarray}
Thus, we find that, in high energy limits, the polarization vector of an electron can be expressed in terms of its helicity ($h_e$) and its four momentum ($k_\sigma$) as $a_\sigma = \frac{h_e}{m_e} k_\sigma$.

\subsection*{Product of polarization and four momentum}
We consider the four product of the polarization and momentum of the incoming electron. Defining $k = (k_0 \equiv E_k, \mathbf{k})$, the product can be written as
\begin{eqnarray} 
\label{adotkeq0proof}
a \cdot k && = a_0 k_0 - \mathbf{a} \cdot \mathbf{k} \nonumber \\
&& =\frac{\mathbf{\xi}\cdot \mathbf{k}}{m_e} k_0 - \frac{(\mathbf{\xi}\cdot \mathbf{k})|\mathbf{k}|^2}{m_e(E_k + m_e)} - \mathbf{\xi} \cdot \mathbf{k} \nonumber \\
&& = \frac{\mathbf{\xi}\cdot \mathbf{k}}{m_e} \Bigg[ k_0 - m_e - \frac{|\mathbf{k}|^2}{E_k + m_e}\Bigg] \nonumber \\
&& = \frac{\mathbf{\xi}\cdot \mathbf{k}}{m_e} \Bigg[ \frac{(E_k - m_e)(E_k + m_e) - |\mathbf{k}|^2}{E_k + m} \Bigg] \nonumber \\
&& = \frac{\mathbf{\xi}\cdot \mathbf{k}}{m_e} \Bigg[ \frac{E_k^2 - m_e^2 - \mathbf{|k|}^2}{E_k + m_e} \Bigg] \nonumber \\
&& = 0,
\end{eqnarray}
where we used the fact that $E_k^2 = m_e^2 + |\mathbf{k}|^2$.

%% file: Appendix/phiIntegrationUnpolarized.tex
\section*{APPENDIX D: Azimuthal dependence in the cross section: unpolarized part}
\renewcommand{\theequation}{D-\arabic{equation}}
\setcounter{figure}{0}  
\setcounter{equation}{0}
\label{appendixD}
In this section, a detailed discussion of the analysis of the azimuthal dependence of the spectator nucleon is presented. At present, we only consider this analysis in the unpolarized part of the cross section. 

The Eq.(\ref{csPWIA_UnpolarizedPartOnly}) can be  expressed as
\begin{eqnarray}
\label{csStart2}
&& \frac{d\sigma}{dQ^2dx_A}  = \frac{2\pi \alpha_{em}^2y_A^2}{Q^4} \sum_{N} \int \Bigg \{ 2F_{1N}(x_N, Q^2) +  \nonumber 	\\
&& \frac{1}{2x_Ny_N^2} \Bigg[(2 - y_N)^2 - y_N^2(1 + \frac{4 x_N^2 m_N^2}{Q^2})\Bigg]F_{2N}(x_N, Q^2) \Bigg\} \frac{\rho_0(\alpha_N, p_{N\perp})}{(2 - \alpha_s)^2} d\alpha_s p_{s\perp}dp_{s\perp}d\phi_s,
\end{eqnarray}
where $\phi_s$ is the azimuthal angle of the spectator nucleon. 
To simplify this cross section further, we consider the factor in the $F_{2N}$ term, as
\begin{equation}
\label{factorF2Nterm}
A \equiv \frac{1}{y_N^2} \Big[(2 - y_N)^2 - y_N^2(1 + \frac{4 x_N^2 m_N^2}{Q^2})\Big],
\end{equation}
where we wish to explore any explicit occurence of the $\phi_s$ that appear through the $y_N = \frac{p_N \cdot q}{p_N \cdot k}$ term.
We can write Eq.(\ref{factorF2Nterm}) as,
\begin{eqnarray}
\label{factorF2NtermSimplification}
A &&= \frac{1}{y_N^2} \Bigg[ 4 - 4y_N + y_N^2 - y_N^2 - \frac{4x_N^2y_N^2m_N^2}{Q^2}\Bigg], \nonumber \\
&& = \frac{1}{y_N^2} \Bigg[ 4 - 4y_N - \frac{4x_N^2y_N^2m_N^2}{Q^2}\Bigg], \nonumber \\
&& = 4\Bigg[\frac{1}{y_N^2} - \frac{1}{y_N}	- \frac{x_N^2m_N^2}{Q^2} \Bigg], \nonumber \\
&& = 4 \Bigg[\frac{(p_N \cdot k)^2}{(p_N \cdot q)^2} - \frac{(p_N \cdot k)}{(p_N \cdot q)}- \frac{x_N^2 m_N^2}{Q^2}\Bigg].
\end{eqnarray}
We thus rewrite Eq.(\ref{csStart2}) as
\begin{eqnarray}
\label{csSplitted}
&& \frac{d\sigma}{dQ^2dx_A} = \frac{2\pi \alpha_{em}^2y_A^2}{Q^4}\sum_{N}\Bigg\{ \int 2F_{1N}(x_N, Q^2) \frac{\rho_0(\alpha_N, p_{N\perp})}{(2 - \alpha_s)^2} d\alpha_s p_{s\perp}dp_{s\perp} d\phi_s \nonumber \\
&& + \int \frac{F_{2N}(x_N, Q^2)}{2x_N}4 \Bigg[\frac{(p_N \cdot k)^2}{(p_N \cdot q)^2} - \frac{(p_N \cdot k)}{(p_N \cdot q)}- \frac{x_N^2 m_N^2}{Q^2}\Bigg]\frac{\rho_0(\alpha_N, p_{N\perp})}{(2 - \alpha_s)^2} d\alpha_s p_{s\perp}dp_{s\perp} d\phi_s \Bigg\}.
\end{eqnarray}

We find that it is only  the second integral that has an explicit $\phi_s$ dependence. Our goal is to see if we can take care of this $\phi_s$ integral analytically. 
Noting that
\begin{eqnarray}
\label{pnkdefn}
p_N \cdot k && = \frac{1}{2}(p_{N+}k_{-} + p_{N-}k_{+}) - p_{N\perp}k_{\perp} \cos\phi_s, = B - p_{N\perp}k_{\perp}\cos\phi_s,
\end{eqnarray}
where $B \equiv \frac{1}{2}(p_{N+}k_{-} + p_{N-}k_{+})$,
the second integral in Eq.~(\ref{csSplitted}) can be written as:
\begin{eqnarray}
\label{secondIntSimplify}
AA && \equiv \int \frac{F_{2N}(x_N, Q^2)}{2x_N}4 \Bigg[\frac{(p_N \cdot k)^2}{(p_N \cdot q)^2} - \frac{(p_N \cdot k)}{(p_N \cdot q)}- \frac{x_N^2 m_N^2}{Q^2}\Bigg]\frac{\rho_0(\alpha_N, p_{N\perp})}{(2 - \alpha_s)^2} d\alpha_s p_{s\perp}dp_{s\perp} d\phi_s \nonumber \\
&& = \int \frac{4 F_{2N}(x_N, Q^2)}{2x_N}\Bigg[ \frac{B^2 - 2Bp_{N\perp}k_{\perp}\cos\phi_s + p_{N\perp}^2k_{\perp}^2\cos^2\phi_s}{(p_N \cdot q)^2} - \frac{B - p_{N\perp}k_{\perp}\cos\phi_s}{p_N \cdot q} \nonumber \\
&& - \frac{x_N^2 m_N^2}{Q^2} \Bigg] \frac{\rho_0(\alpha_N, p_{N\perp})}{(2 - \alpha_s)^2} d\alpha_s p_{s\perp} dp_{s\perp} d\phi_s \nonumber \\
&& = \int \frac{4 F_{2N}(x_N, Q^2)}{2x_N}\Bigg[ \frac{2\pi B^2 + \frac{1}{2}2\pi p_{N\perp}^2k_{\perp}^2}{(p_N \cdot q)^2} - \frac{2\pi B }{p_N \cdot q} 
- 2\pi \frac{x_N^2 m_N^2}{Q^2} \Bigg] \frac{\rho_0(\alpha_N, p_{N\perp})}{(2 - \alpha_s)^2} d\alpha_s p_{s\perp} dp_{s\perp} \nonumber \\
&& = 2\pi \int \frac{2 F_{2N}(x_N, Q^2)}{x_N}\Bigg[ \frac{B^2 + \frac{1}{2} p_{N\perp}^2k_{\perp}^2}{(p_N \cdot q)^2} - \frac{B}{p_N \cdot q} 
- \frac{x_N^2 m_N^2}{Q^2} \Bigg] \frac{\rho_0(\alpha_N, p_{N\perp})}{(2 - \alpha_s)^2} d\alpha_s p_{s\perp} dp_{s\perp},
\end{eqnarray}
where we used the fact that $\int_{0}^{2\pi} \cos\phi_s d\phi_s = 0, \int_{0}^{2\pi} \cos^2\phi_s d\phi_s = \pi$.
Substituting Eq.(\ref{secondIntSimplify}) into Eq.(\ref{csSplitted}), we have the following for the cross section:
\begin{eqnarray}
\label{csAfterPhiIntegration}
\frac{d\sigma}{dQ^2dx_A} &&= \frac{4\pi^2 \alpha_{em}^2 y_A^2}{Q^4} \sum_{N} \int \Bigg\{2 F_{1N}(x_N, Q^2) +\frac{2 F_{2N}(x_N, Q^2)}{x_N} \times \nonumber \\
&& \Bigg[\frac{B^2 + \frac{1}{2} p_{N\perp}^2 k_{\perp}^2}{(p_N \cdot q)^2} - \frac{B}{p_N \cdot q} - \frac{x_N^2 m_N^2}{Q^2} \Bigg] \Bigg\} \frac{\rho_0(\alpha_N, p_{N\perp)}}{(2 - \alpha_s)^2} d\alpha_s p_{s\perp} dp_{s\perp}.
\end{eqnarray}
Further, to calculate the factor $B$, we note that
\begin{equation}
\label{Bdefn}
B = \frac{1}{2}(p_{N+}k_{-} + p_{N-}k_{+}),
\end{equation}
where, $k_{\pm} = E_k \pm E_k\cos\theta_{qk}$, $p_{N-} = \alpha_N \frac{M_d}{2}$, $p_{N+} = M_d - \frac{m_N^2 + p_{s\perp}^2}{\alpha_s M_d/2}$, $k_{\perp} \equiv k_x = E_k\sin\theta_{qk}$ and $p_{N\perp} =-p_{s\perp}$, with $\theta_{qk}$ being the angle bewteen $\mathbf{q}$ and $\mathbf{k}$. To calculate this angle, we use the definition of scalar product of two vectors and write
\begin{equation}
\label{kdotqvecdefn}
\mathbf{k} \cdot \mathbf{q} = E_k q_v \cos\theta_{qk},
\end{equation}
with $E_k, q_v$ being the magnitudes of $\mathbf{k}$ and $\mathbf{q}$ respectively.
We can also write
\begin{eqnarray}
\label{kdotqvecaltdefn} 
\mathbf{k} \cdot \mathbf{q} && = \mathbf{k} \cdot (\mathbf{k} - \mathbf{k}^\prime) = E_k^2 - E_k E_k^\prime \cos\theta_e,
\end{eqnarray}
where $\theta_e$ is the angle between the incident and scattered electron and $E_k^\prime$ is the energy of scattered electron.
Comparing Eqs.(\ref{kdotqvecdefn}) and (\ref{kdotqvecaltdefn}), we thus obtain
\begin{equation}
\label{thetaqkdefn}
\cos\theta_{qk} = \frac{E_k - E_k^\prime \cos\theta_e}{q_v}.
\end{equation}
It can also be easily shown that $\sin\theta_{qk} = \frac{E_k^\prime \sin\theta_e}{q_v}$.

%% file: Appendix/zetaProducts.tex
\section*{APPENDIX E: Azimuthal dependence in the cross section: polarized part}
\renewcommand{\theequation}{E-\arabic{equation}}
\setcounter{equation}{0}
\label{appendixE}

From the definitions of the directions in the polarization reference frame (Eq.~(\ref{polFrameDir})) and the polarization vectors (Eqs.~(\ref{pol4vectorStruckNuclen_zprime} and (\ref{pol4vectorStruckNucleonxyprime})), we can express the polarization four vectors in terms of their components as
\begin{eqnarray}
\label{polVectorComponents}
\zeta_{x'} && = \Bigg(\frac{\mathbf{p_N} \cdot \hat{x'}}{m_N}, \hat{x'} + \frac{(\mathbf{p_N} \cdot \hat{x'}) \mathbf{p_N}}{BB}\Bigg)\nonumber \\
&& = \Bigg(\frac{\mathbf{p_N} \cdot \hat{x'}}{m_N}, x'_x + \frac{(\mathbf{p_N} \cdot \hat{x'}) {p_{Nx}}}{BB},x'_y + \frac{(\mathbf{p_N} \cdot \hat{x'}) {p_{Ny}}}{BB},x'_z + \frac{(\mathbf{p_N} \cdot \hat{x'}) {p_{Nz}}}{BB}\Bigg), \nonumber \\
\zeta_{y'} && =  \Bigg(\frac{\mathbf{p_N} \cdot \hat{y'}}{m_N}, y'_x + \frac{(\mathbf{p_N} \cdot \hat{y'}) {p_{Nx}}}{BB},y'_y + \frac{(\mathbf{p_N} \cdot \hat{y'}) {p_{Ny}}}{BB},y'_z + \frac{(\mathbf{p_N} \cdot \hat{y'}) {p_{Nz}}}{BB}\Bigg), \nonumber \\
\zeta_{z'} && =  \Bigg(\frac{\mathbf{p_N} \cdot \hat{z'}}{m_N}, z'_x + \frac{(\mathbf{p_N} \cdot \hat{z'}) {p_{Nx}}}{BB},z'_y + \frac{(\mathbf{p_N} \cdot \hat{z'}) {p_{Ny}}}{BB},z'_z + \frac{(\mathbf{p_N} \cdot \hat{z'}) {p_{Nz}}}{BB}\Bigg),
\end{eqnarray}
where we put $BB \equiv m_N(E_N + m_N)$ for convenience. The scalar products appearing in above equations can be written as
\begin{eqnarray}
\label{pNdotjprimeDerivs}
\mathbf{p_N} \cdot \hat{x'} && = p_{Nx} x'_x + p_{Ny} x'_y + p_{Nz} x'_z \nonumber \\
&& = p_{N\perp}\cos\phi_s \cos\theta_{sd}\cos\phi_{sd} + p_{N\perp}\sin\phi_s\cos\theta_{sd}\sin\phi_{sd} - p_{Nz}\sin\theta_{sd} \nonumber \\
&& = p_{N\perp}\cos\theta_{sd}\cos(\phi_s - \phi_{sd}) - p_{Nz}\sin\theta_{sd}, \nonumber \\
\mathbf{p_N} \cdot \hat{y'} && = -p_{N\perp}\cos\phi \sin\phi_{sd} + p_{N\perp}\sin\phi_s\cos\phi_{sd}, \nonumber \\
\mathbf{p_N} \cdot \hat{z'} && = p_{N\perp}\cos\phi_s \sin\theta_{sd} \cos\phi_{sd} + p_{N\perp}\sin\phi_s \sin\theta_{sd} \sin\phi_{sd} + p_{Nz}\cos\theta_{sd}\nonumber \\
&& = p_{N\perp}\sin\theta_{sd}\cos(\phi_s - \phi_{sd}) + p_{Nz}\cos\theta_{sd},
\end{eqnarray}
where $p_{Nx} = p_{N\perp}\cos\phi_s, p_{Ny} = p_{N\perp}\sin\phi_s$, $\theta_{sd}$ is the angle made by the deuteron's polarization vector with $\mathbf{q}$, $\phi_{sd}$ defines its azimuth and $\phi_s$ defines the azimuth of spectator system. Since we define the $z$ direction by the direction of $q$, the four momentum transfer $q_\mu$ becomes $q_\mu = (q_0, 0, 0, q_3)$. This enables us to write the four products of the polarization and momentum transfer as
\begin{eqnarray}
\label{zetadotqdetailed}
\zeta_{x'}\cdot q && = \frac{\mathbf{p_N}\cdot \hat{x'}}{m_N}q_0 - \Bigg[ x'_z + \frac{(\mathbf{p_N}\cdot \hat{x'}) p_{Nz}}{BB}\Bigg]q_3 = \mathbf{p_N}\cdot \hat{x'} \Bigg[\frac{q_0}{m_N} - \frac{p_{Nz}q_3}{BB}\Bigg] - x'_z q_3, \nonumber \\
\zeta_{y'}\cdot q && = \frac{\mathbf{p_N}\cdot \hat{y'}}{m_N}q_0 - \Bigg[ y'_z + \frac{(\mathbf{p_N}\cdot \hat{y'}) p_{Nz}}{BB}\Bigg]q_3 = \mathbf{p_N}\cdot \hat{y'} \Bigg[\frac{q_0}{m_N} - \frac{p_{Nz}q_3}{BB}\Bigg] - y'_z q_3, \nonumber \\
\zeta_{z'}\cdot q && = \frac{\mathbf{p_N}\cdot \hat{z'}}{m_N}q_0 - \Bigg[ z'_z + \frac{(\mathbf{p_N}\cdot \hat{z'}) p_{Nz}}{BB}\Bigg]q_3 = \mathbf{p_N}\cdot \hat{z'} \Bigg[\frac{q_0}{m_N} - \frac{p_{Nz}q_3}{BB}\Bigg] - z'_zq_3.
\end{eqnarray}
Also, as the  four vector of incoming electron $k$ can be defined as 
\begin{equation}
\label{kDefn}
k_\mu = (k_0, k_x, k_y, k_z) = (E_k, E_k \sin\theta_{qk}, 0, E_k \cos\theta_{qk}),
\end{equation}
where $\theta_{qk}$ is the angle between incoming electron's momentum ($\mathbf{k}$) and the momentum transfer ($\mathbf{q}$),
the four products of the polarization vector and the incoming electron's four momentum can be expressed as
\begin{eqnarray}
\label{zetadotkDetailed}
\zeta_{x'}\cdot k && = \mathbf{p_N} \cdot \hat{x'} \Bigg[ \frac{k_0}{m_N} - \frac{p_{Nx}k_x}{BB} - \frac{p_{Nz}k_z}{BB} \Bigg] - x'_x k_x - x'_z k_z, \nonumber \\
\zeta_{y'}\cdot k && = \mathbf{p_N} \cdot \hat{y'} \Bigg[ \frac{k_0}{m_N} - \frac{p_{Nx}k_x}{BB} - \frac{p_{Nz}k_z}{BB} \Bigg] - y'_x k_x - y'_z k_z, \nonumber \\
\zeta_{z'}\cdot k && = \mathbf{p_N} \cdot \hat{z'} \Bigg[ \frac{k_0}{m_N} - \frac{p_{Nx}k_x}{BB} - \frac{p_{Nz}k_z}{BB} \Bigg] - z'_x k_x - z'_z k_z.
\end{eqnarray}

The polarization density matrices of the deuteron, for the case of its polarization along $z'$,can be expressed through the partial $s-$ and $d-$ waves  as follows:
\begin{eqnarray}
\rho_3 \equiv \rho_{z'}(\mathbf{p_N}) &&= u^2(p_N) + \frac{w^2(p_N)}{2} \Big(3 \hat{p_{z'}}^2 - 2\Big) + \frac{u(p_N)w(p_N)}{\sqrt{2}}\Big(3 \hat{p_{z'}}^2 - 1\Big), \nonumber \\ 
\rho_1 \equiv \rho_{x'}(\mathbf{p_N}) && = \Big(\frac{w^2(p_N)}{2} + 3 \frac{u(p_N)w(p_N)}{\sqrt{2}}\Big) \hat{p_{z'}} \hat{p_{x'}}, \nonumber 
\end{eqnarray}

\begin{eqnarray}\label{DeutPolarizationMatrices}
\text{and }\rho_2 \equiv \rho_{y'}(\mathbf{p_N}) = \Big(\frac{w^2(p_N)}{2} + 3 \frac{u(p_N)w(p_N)}{\sqrt{2}}\Big) \hat{p_{z'}} \hat{p_{y'}},
\end{eqnarray}
where $\hat{p_{j'}} = \mathbf{p_N}\cdot \hat{j'}$ with $\hat{j'} = \hat{x'}, \hat{y'}, \hat{z'}$, which using Eqs.~(\ref{polFrameDir}), can be expanded as:
\begin{eqnarray}
\label{pjprimesdefs}
\hat{p_{z'}} && = p_{Nx} z'_x + p_{Ny} z'_y + p_{Nz} z'_z, \nonumber \\
\hat{p_{x'}} && = p_{Nx} x'_x + p_{Ny} x'_y + p_{Nz} x'_z, \nonumber \\
\hat{p_{y'}} && = p_{Nx} y'_x + p_{Ny} y'_y.
\end{eqnarray}
In Eq.~(\ref{pjprimesdefs}), we supress the explicit forms of the components for simplicity. We thus see the explicit $\phi_s$ dependence through the polarization density matrices of deuteron.  It requires to calculate the following products:
\begin{eqnarray}
\label{pzprpxpr}
\hat{p_{z'}} \hat{p_{x'}} && = (p_{Nx} z'_x + p_{Ny} z'_y + p_{Nz} z'_z)(p_{Nx} x'_x + p_{Ny} x'_y + p_{Nz} x'_z)\nonumber \\
&& = p_{Nx}^2 z'_x x'_x + p_{Nx}p_{Ny} z'_x x'_y + p_{Nx}p_{Nz} z'_x x'_z + p_{Ny}p_{Nx}z'_y x'_x + p_{Ny}^2 z'_y x'_y + p_{Ny}p_{Nz}z'_y x'_z + \nonumber \\
&& p_{Nz}p_{Nx} z'_z x'_x + p_{Nz}p_{Ny}z'_z x'_y + p_{Nz}^2 z'_z x'_z, \nonumber \\
\hat{p_{z'}}\hat{p_{y'}} && = (p_{Nx} z'_x + p_{Ny} z'_y + p_{Nz} z'_z)(p_{Nx} y'_x + p_{Ny} y'_y) \nonumber \\
&& = p_{Nx}^2 z'_x y'_x + p_{Nx}p_{Ny} z'_x y'_y  + p_{Ny}p_{Nx}z'_y y'_x + p_{Ny}^2 z'_y y'_y + p_{Nz}p_{Nx} z'_z y'_x + p_{Nz}p_{Ny}z'_z y'_y \nonumber \\
\hat{p_{z'}}^2 && = ( p_{Nx} z'_x + p_{Ny} z'_y + p_{Nz} z'_z)^2 \nonumber \\
&& = p_{Nx}^2 z_{x}^{'2} + p_{Ny}^2 z_{y}^{'2} + p_{Nz}^2 z_{z}^{'2} + 2 p_{Nx} p_{Ny} z'_x z'_y + 2 p_{Ny}p_{Nz}z'_y z'_z + 2 p_{Nz} p_{Nx} z'_z z'_x.
\end{eqnarray}
Substituting for each factors, the first of Eq.~(\ref{pzprpxpr}) becomes:
\begin{eqnarray}
\hat{p_{z'}} \hat{p_{x'}} && = p_{N\perp}^2 \cos^2\phi_s \sin\theta_{sd} \cos\theta_{sd} \cos^2\phi_{sd} + p_{N\perp}^2 \cos\phi_s \sin\phi \sin\theta_{sd} \cos\theta_{sd} \sin\phi_{sd} \cos\phi_{sd} \nonumber \\
&& - p_{N\perp} p_{Nz} \cos\phi_s \sin^2\theta_{sd} \cos\phi_{sd} 
+ p_{N\perp}^2 \cos\phi_s \sin\phi \sin\theta_{sd} \cos\theta_{sd} \sin\phi_{sd} \cos\phi_{sd} \nonumber \\
&&+ p_{N\perp}^2 \sin^2\phi_s \sin\theta_{sd} \cos\theta_{sd} \sin^2\phi_{sd} - p_{N\perp} p_{Nz}\sin\phi_s \sin^2\theta_{sd} \sin\phi_{sd} \nonumber \\
&& + p_{N\perp} p_{Nz} \cos\phi_s \cos^2\theta_{sd} \cos\phi_{sd} + p_{N\perp}p_{Nz} \sin\phi_s \cos^2\theta_{sd} \sin\phi_{sd} - p_{Nz}^2 \cos\theta_{sd} \sin\theta_{sd} \nonumber
\end{eqnarray}
\begin{eqnarray}
\label{rho1factor}
&& = p_{N\perp}^2\cos^2\phi_s \sin\theta_{sd}\cos\theta_{sd}\cos^2\phi_{sd} + 2 p_{N\perp}^2 \cos\phi_s \sin\phi_s \sin\theta_{sd} \cos\theta_{sd} \sin\phi_{sd} \cos\phi_{sd} \nonumber \\
&& + p_{N\perp} p_{Nz} \cos\phi_s \cos\phi_{sd} (\cos^2\theta_{sd} - \sin^2\theta_{sd}) + p_{N\perp}^2 \sin^2\phi_s \sin\theta_{sd}\cos\theta_{sd}\sin^2\phi_{sd} \nonumber \\
&& + p_{N\perp}p_{Nz}\sin\phi_s \sin\phi_{sd}(\cos^2\theta_{sd} - \sin^2\theta_{sd}) - p_{Nz}^2 \cos\theta_{sd}\sin\theta_{sd} \nonumber \\
&& = \frac{p_{N\perp}^2}{2} \cos^2\phi_s \sin(2\theta_{sd}) \cos^2\phi_{sd} + \frac{p_{N\perp}^2}{4} \sin(2\phi_s) \sin(2\theta_{sd}) \sin(2\phi_{sd}) \nonumber \\
&& + p_{N\perp} p_{Nz} \cos\phi_s \cos\phi_{sd} \cos(2\theta_{sd}) + \frac{p_{N\perp}^2}{2}\sin^2\phi_s \sin(2\theta_{sd})\sin^2\phi_{sd} \nonumber \\
&& + p_{N\perp} p_{Nz} \sin\phi_s \sin\phi_{sd}\cos(2\theta_{sd}) - \frac{p_{Nz}^2}{2} \sin(2\theta_{sd}) \nonumber \\
&& = \frac{p_{N\perp}^2}{2}\sin(2\theta_{sd})\Big[ \cos^2\phi_s \cos^2\phi_{sd} + \frac{1}{2} \sin(2\phi_s)\sin(2\phi_{sd}) + \sin^2\phi_s \sin^2\phi_{sd}\Big] \nonumber \\
&& + p_{N\perp}p_{Nz} \cos(2\theta_{sd}) \cos(\phi_s - \phi_{sd}) - p_{Nz}^2 \cos\theta_{sd}\sin\theta_{sd}\nonumber \\
&& = \frac{p_{N\perp}^2}{2} \sin(2\theta_{sd}) [\cos\phi_s\cos\phi_{sd} + \sin\phi_s \sin\phi_{sd}]^2 + p_{N\perp} p_{Nz} \cos(2\theta_{sd})\cos(\phi_s - \phi_{sd}) \nonumber \\
&& - p_{Nz}^2 \cos\theta_{sd}\sin\theta_{sd} \nonumber \\
&& = \frac{p_{N\perp}^2}{2}\sin(2\theta_{sd})\cos^2(\phi_s - \phi_{sd}) + p_{Nz}p_{N\perp} \cos(2\theta_{sd})\cos(\phi_s - \phi_{sd}) - \frac{p_{Nz}^2}{2}\sin(2\theta_{sd}).
\end{eqnarray}
Similarly, we have,
\begin{eqnarray}
\label{rho3factor}
\hat{p_{z'}}^2 && = \Big(p_{N\perp}\cos\phi_s \sin\theta_{sd}\cos\phi_{sd} + p_{N\perp}\sin\phi_s \sin\theta_{sd} \sin\phi_{sd} + p_{Nz}\cos\theta_{sd} \Big)^2 \nonumber \\
&& = p_{N\perp}^2 \cos^2\phi_s \sin^2\theta_{sd} \cos^2\phi_{sd} + p_{N\perp}^2 \sin^2\phi_s \sin^2\theta_{sd} \sin^2\phi_{sd} + p_{Nz}^2 \cos^2\theta_{sd} \nonumber \\
&& + 2 p_{N\perp}^2 \cos\phi_s \sin\phi_s \sin^2\theta_{sd}\cos\phi_{sd}\sin\phi_{sd} + 2 p_{N\perp} p_{Nz} \sin\phi_s \sin\theta_{sd}\cos\theta_{sd}\sin\phi_{sd} \nonumber \\
&& + 2 p_{N\perp} p_{Nz} \cos\phi_s \sin\theta_{sd}\cos\theta_{sd} \cos\phi_{sd} \nonumber \\
&& =  p_{N\perp}^2 \cos^2\phi_s \sin^2\theta_{sd} \cos^2\phi_{sd} + p_{N\perp}^2 \sin^2\phi_s \sin^2\theta_{sd} \sin^2\phi_{sd} + p_{Nz}^2 \cos^2\theta_{sd} \nonumber \\
&& + \frac{p_{N\perp}^2}{2} \sin(2\phi_s) \sin(2\phi_{sd})\sin^2\theta_{sd} + p_{N\perp}p_{Nz}\sin\phi_s \sin\phi_{sd} \sin(2\theta_{sd}) \nonumber \\
&& + p_{N\perp} p_{Nz} \cos\phi_s \cos\phi_{sd} \sin(2\theta_{sd}) \nonumber \\
&& =  p_{N\perp}^2 \cos^2\phi_s \sin^2\theta_{sd} \cos^2\phi_{sd} + p_{N\perp}^2 \sin^2\phi_s \sin^2\theta_{sd} \sin^2\phi_{sd} + p_{Nz}^2 \cos^2\theta_{sd} \nonumber \\
&& + \frac{p_{N\perp}^2}{2} \sin(2\phi_s) \sin^2\theta_{sd}\sin(2\phi_{sd}) + p_{N\perp} p_{Nz} \sin(2\theta_{sd}) \cos(\phi_s - \phi_{sd})\nonumber \\
&& = p_{N\perp}^2\sin^2\theta_{sd} [\cos^2\phi_{sd}\cos^2\phi_s + \sin^2\phi_{sd}\sin^2\phi_s + \frac{1}{2}\sin(2\phi_{sd})\sin(2\phi_s)] \nonumber \\
&& + p_{Nz}^2 \cos^2\theta_{sd} + p_{N\perp}p_{Nz}\sin(2\theta_{sd})\cos(\phi_s- \phi_{sd}) \nonumber \\
&& = p_{N\perp}^2\sin^2\theta_{sd}\cos^2(\phi_s - \phi_{sd}) + p_{Nz}^2\cos^2\theta_{sd} + p_{N\perp}p_{Nz}\sin(2\theta_{sd})\cos(\phi_s - \phi_{sd}).
\end{eqnarray}
The final product can also be expanded as
\begin{eqnarray}
\label{rho2factor}
\hat{p_{z'}}\hat{p_{y'}} && = -p_{N\perp}^2\cos^2\phi_s \sin\theta_{sd} \cos\phi_{sd}\sin\phi_{sd} + p_{N\perp}^2 \cos\phi_s \sin\phi_s \sin\theta_{sd}\cos\theta_{sd}\cos\phi_{sd} \nonumber \\
&& - p_{N\perp}^2\sin\phi_s \cos\phi_s\sin\theta_{sd}\sin^2\phi_{sd} + p_{N\perp}^2\sin^2\phi_s \sin\theta_{sd}\sin\phi_{sd}\cos\phi_{sd} \nonumber \\
&& -p_{N\perp}p_{Nz}\cos\phi_s\cos\theta_{sd}\sin\phi_{sd} + p_{N\perp} p_{Nz} \sin\phi_s \cos\theta_{sd} \cos\phi_{sd} \nonumber \\
&& = -\frac{p_{N\perp}^2}{2} \cos^2\phi_s \sin\theta_{sd}\sin(2\phi_{sd}) + \frac{p_{N\perp}^2}{2} \sin(2\phi_s) \sin\theta_{sd} \cos^2\phi_{sd} \nonumber \\
&&- \frac{p_{N\perp}^2}{2} \sin(2\phi_s) \sin\theta_{sd}\sin^2\phi_{sd} + \frac{p_{N\perp}^2}{2} \sin^2\phi_s \sin\theta_{sd} \sin(2\phi_{sd}) \nonumber \\
&& - p_{N\perp} p_{Nz} \cos\phi_s\cos\theta_{sd}\sin\phi_{sd} + p_{N\perp}p_{Nz}\sin\phi_s \cos\theta_{sd}\cos\phi_{sd} \nonumber \\
&& = -\frac{p_{N\perp}^2}{2}\sin\theta_{sd} \sin(2\phi_{sd})(\cos^2\phi_s - \sin^2\phi) + \frac{p_{N\perp}^2}{2}\sin(2\phi_s) \sin\theta_{sd} (\cos^2\phi_{sd} - \sin^2\phi_{sd}) \nonumber \\
&& + p_{N\perp} p_{Nz} \cos\theta_{sd} (\sin\phi_s cos\phi_{sd} - \cos\phi_s \sin\phi_{sd}) \nonumber \\
&& = -\frac{p_{N\perp}^2}{2}\sin\theta_{sd} \sin(2\phi_{sd})\cos(2\phi_s) + \frac{p_{N\perp}^2}{2}\sin(2\phi_s) \sin\theta_{sd} \cos(2\phi_{sd})\nonumber \\
&& + p_{N\perp} p_{Nz} \cos\theta_{sd} \sin(\phi_s -\phi_{sd}) \nonumber \\
&& = \frac{p_{N\perp}^2}{2} \sin\theta_{sd}[\sin(2\phi_s)\cos(2\phi_{sd}) - \cos(2\phi_s)\sin(2\phi_{sd})] + p_{N\perp}p_{Nz} \cos\theta_{sd}\sin(\phi_s - \phi_{sd}) \nonumber \\
&& = \frac{p_{N\perp}^2}{2} \sin\theta_{sd} \sin(2\phi_s - 2\phi_{sd}) + p_{N\perp} p_{Nz} \cos\theta_{sd} \sin(\phi_s - \phi_{sd}).
\end{eqnarray}

\section*{E-I: Calculation of $(\zeta_{j'} \cdot q)\rho_{j'}$}
For the case of $j' = x'$, we have
\begin{eqnarray}
\label{zetajdotqRhojeq1}
(\zeta_{x'}\cdot q )\rho_{x'} &&= \Bigg\{(\mathbf{p_N} \cdot \hat{x'}) \Big[\frac{q_0}{m_N} - \frac{p_{Nz}q_3}{BB} \Big] - x'_zq_3 \Bigg\} \Bigg\{ \frac{w^2(p_N)}{2} + \frac{3u(p_N)w(p_N)}{\sqrt{2}}\Bigg\} \hat{p_{z'}}\hat{p_{x'}} \nonumber \\
&& \equiv \underbrace{(\mathbf{p_N} \cdot \hat{x'}) D F \hat{p_{z'}}\hat{p_{x'}}}_{\text{I}} -  \underbrace{x'_z q_3 F \hat{p_{z'}}\hat{p_{x'}}}_{\text{II}},
\end{eqnarray}
where we put $D = \frac{q_0}{m_N} - \frac{p_{Nz}q_3}{BB}$ and $F = \frac{w^2(p_N)}{2} + \frac{3u(p_N)w(p_N)}{\sqrt{2}}$ for convenience.
We analyze each term separately, starting with the first term, where we drop the factor $DF$ for now as it has no $\phi_s$ dependence. From Eq.~(\ref{pNdotjprimeDerivs}), we find that
\begin{eqnarray}
\text{I} && \equiv (\mathbf{p_N} \cdot \hat{x'}) \hat{p_{z'}} \hat{p_{x'}}  = (p_{N\perp}\cos\theta_{sd}\cos(\phi_s - \phi_{sd}) - p_{Nz}\sin\theta_{sd}) \hat{p_{z'}} \hat{p_{x'}} \nonumber 
\end{eqnarray}
\begin{eqnarray}
\label{termI_zetadotqrhoj}
&& = \underbrace{p_{N\perp} \cos\theta_{sd} \cos(\phi_s - \phi_{sd})\hat{p_{z'}} \hat{p_{x'}}}_{\Gamma} - \underbrace{p_{Nz}\sin\theta_{sd}\hat{p_{z'}} \hat{p_{x'}}}_{\Delta}.
\end{eqnarray}
Using Eq.~(\ref{rho1factor}), we find that the factors
\begin{eqnarray}
\label{factorGamma}
\Gamma &&= p_{N\perp} \cos\theta_{sd} \cos(\phi_s - \phi_{sd}) \Bigg\{ \frac{p_{N\perp}^2}{2}\sin(2\theta_{sd})\cos^2(\phi_s - \phi_{sd}) + p_{Nz}p_{N\perp} \cos(2\theta_{sd})\cos(\phi_s - \phi_{sd}) \nonumber \\
&& - \frac{p_{Nz}^2}{2}\sin(2\theta_{sd})  \Bigg\} \nonumber \\
 && = \frac{p_{N\perp}^3}{2} \cos\theta_{sd}\sin(2\theta_{sd}) \cos^3(\phi_s - \phi_{sd}) + p_{Nz}p_{N\perp}^2 \cos\theta_{sd}\cos(2\theta_{sd})\cos^2(\phi_s - \phi_{sd}) \nonumber \\
 && - \frac{p_{Nz}^2}{2}p_{N\perp} \cos\theta_{sd} \sin(2\theta_{sd})\cos(\phi_s - \phi_{sd}),
\end{eqnarray}
and
\begin{eqnarray}
\label{factorDelta}
\Delta && = p_{Nz} \sin\theta_{sd}\Bigg\{\frac{p_{N\perp}^2}{2}\sin(2\theta_{sd})\cos^2(\phi_s - \phi_{sd}) + p_{Nz}p_{N\perp} \cos(2\theta_{sd})\cos(\phi_s - \phi_{sd}) \nonumber \\
&& - \frac{p_{Nz}^2}{2}\sin(2\theta_{sd}) \Bigg\} \nonumber \\
 && = \frac{p_{N\perp}^2}{2} p_{Nz}\sin\theta_{sd}\sin(2\theta_{sd})\cos^2(\phi_s - \phi_{sd}) + p_{N\perp}p_{Nz}^2 \sin\theta_{sd}\cos(2\theta_{sd})\cos(\phi_s - \phi_{sd}) \nonumber \\
&&  - p_{Nz}^3 \sin^2\theta_{sd}\cos\theta_{sd}.
\end{eqnarray}
We thus obtain the following for the term I:
\begin{eqnarray}
\label{term1final}
\text{I} && = \Bigg \{\frac{p_{N\perp}^3}{2} \cos\theta_{sd}\sin(2\theta_{sd}) \cos^3(\phi_s - \phi_{sd}) + p_{Nz}p_{N\perp}^2 \cos\theta_{sd}\cos(2\theta_{sd})\cos^2(\phi_s - \phi_{sd}) \nonumber \\
 && - \frac{p_{Nz}^2}{2}p_{N\perp} \cos\theta_{sd} \sin(2\theta_{sd})\cos(\phi_s - \phi_{sd}) - \frac{p_{N\perp}^2}{2} p_{Nz}\sin\theta_{sd}\sin(2\theta_{sd})\cos^2(\phi_s - \phi_{sd}) \nonumber \\
 && - p_{N\perp}p_{Nz}^2 \sin\theta_{sd}\cos(2\theta_{sd})\cos(\phi_s - \phi_{sd})
  + p_{Nz}^3 \sin^2\theta_{sd}\cos\theta_{sd}  \Bigg\}DF.
\end{eqnarray}
Similarly, term II becomes:
\begin{eqnarray}
\label{term2final}
\text{II}&& = x'_z q_3 F \Bigg\{\frac{p_{N\perp}^2}{2}\sin(2\theta_{sd})\cos^2(\phi_s - \phi_{sd}) + p_{Nz}p_{N\perp} \cos(2\theta_{sd})\cos(\phi_s - \phi_{sd}) \nonumber \\
&& - \frac{p_{Nz}^2}{2}\sin(2\theta_{sd})  \Bigg\}.
\end{eqnarray}
Combining Eqs.~(\ref{term1final}) and (\ref{term2final}), Eq.~(\ref{zetajdotqRhojeq1}) becomes:
\begin{eqnarray}
\label{zetajdotqjeq1final}
(\zeta_{x'}\cdot q) \rho_x && = \Bigg \{\frac{p_{N\perp}^3}{2} \cos\theta_{sd}\sin(2\theta_{sd}) \cos^3(\phi_s - \phi_{sd}) + p_{Nz}p_{N\perp}^2 \cos\theta_{sd}\cos(2\theta_{sd})\cos^2(\phi_s - \phi_{sd}) \nonumber \\
 && - \frac{p_{Nz}^2}{2}p_{N\perp} \cos\theta_{sd} \sin(2\theta_{sd})\cos(\phi_s - \phi_{sd}) - \frac{p_{N\perp}^2}{2} p_{Nz}\sin\theta_{sd}\sin(2\theta_{sd})\cos^2(\phi_s - \phi_{sd}) \nonumber \\
 && - p_{N\perp}p_{Nz}^2 \sin\theta_{sd}\cos(2\theta_{sd})\cos(\phi_s - \phi_{sd})
  + p_{Nz}^3 \sin^2\theta_{sd}\cos\theta_{sd}  \Bigg\}DF  \nonumber \\
 && - x'_z q_3 F \Bigg\{\frac{p_{N\perp}^2}{2}\sin(2\theta_{sd})\cos^2(\phi_s - \phi_{sd}) + p_{Nz}p_{N\perp} \cos(2\theta_{sd})\cos(\phi_s - \phi_{sd}) \nonumber \\
&& - \frac{p_{Nz}^2}{2}\sin(2\theta_{sd})  \Bigg\}. 
\end{eqnarray}
For the case of $j' = y'$, we have
\begin{eqnarray}
\label{zetajdotqRhojeq2}
(\zeta_{y'}\cdot q )\rho_{y'} &&= \Bigg\{(\mathbf{p_N} \cdot \hat{y'}) \Big[\frac{q_0}{m_N} - \frac{p_{Nz}q_3}{BB} \Big] - y'_z q_3 \Bigg\} \Bigg\{ \frac{w^2(p_N)}{2} + \frac{3u(p_N)w(p_N)}{\sqrt{2}}\Bigg\} \hat{p_{z'}}\hat{p_{y'}} \nonumber \\
&& \equiv \underbrace{(\mathbf{p_N} \cdot \hat{y'}) D F \hat{p_{z'}}\hat{p_{y'}}}_{\text{I}},
\end{eqnarray}
where, as before, $D = \frac{q_0}{m_N} - \frac{p_{Nz}q_3}{BB}$ and $F = \frac{w^2(p_N)}{2} + \frac{3u(p_N)w(p_N)}{\sqrt{2}}$ and we made use of the fact that $y'_z = 0$ (see Eq.~(\ref{polFrameDir})).
The term I can be expanded, by using Eqs.~(\ref{pNdotjprimeDerivs}) and (\ref{rho2factor}), as follows:
\begin{eqnarray}
\label{zetajdotqRhojeq2Factor}
\text{I}&& = \Big[ -p_{N\perp} \sin\phi_{sd} \cos\phi_s + p_{N\perp}\cos\phi_{sd} \sin\phi_s \Big] \Big[\frac{p_{N\perp}^2}{2} \sin\theta_{sd} \sin(2\phi_s - 2\phi_{sd}) \nonumber \\
&& + p_{N\perp} p_{Nz} \cos\theta_{sd}\sin(\phi_s - \phi_{sd})\Big] \nonumber \\
&& = -\frac{p_{N\perp}^3}{2} \sin\theta_{sd} \sin\phi_{sd} \cos\phi_s \sin(2\phi_s - 2\phi_{sd}) - p_{N\perp}^2 p_{Nz} \cos\theta_{sd}\sin\phi_{sd} \cos\phi_s \sin(\phi_s - \phi_{sd}) \nonumber \\
&& + \frac{p_{N\perp}^3}{2} \sin\theta_{sd} \cos\phi_{sd} \sin\phi_s \sin(2\phi_s - 2\phi_{sd}) + p_{N\perp}^2 p_{Nz} \cos\theta_{sd}\cos\phi_{sd} \sin\phi_s \sin(\phi_s - \phi_{sd}) \nonumber \\
&& = \frac{p_{N\perp}^3}{2} \sin\theta_{sd}\Big[ \sin\phi_s \cos\phi_{sd} - \sin\phi_{sd} \cos\phi_s \Big]\sin(2\phi_s - 2\phi_{sd}) \nonumber \\
&& + p_{N\perp}^2 p_{Nz} \cos\theta_{sd}\Big[ \sin\phi_s \cos\phi_{sd} - \sin\phi_{sd} \cos\phi_s \Big] \sin(\phi_s - \phi_{sd}) \nonumber \\
&& = \frac{p_{N\perp}^3}{2} \sin\theta_{sd} \sin(\phi_s - \phi_{sd})\sin(2\phi_s - 2\phi_{sd}) + p_{N\perp}^2 p_{Nz} \cos\theta_{sd} \sin^2(\phi_s - \phi_{sd}).
\end{eqnarray}
Thus, for the case of $j' = y'$, we find that

\begin{equation}
\label{zetajdotqRhojeq2final}
(\zeta_{y'} \cdot q) \rho_{y'} = \Bigg[\frac{p_{N\perp}^3}{2} \sin\theta_{sd} \sin(\phi_s - \phi_{sd})\sin(2\phi_s - 2\phi_{sd}) + p_{N\perp}^2 p_{Nz} \cos\theta_{sd} \sin^2(\phi_s - \phi_{sd}) \Bigg] D F.
\end{equation}
Also, for the case of $j' = z'$, we have
\begin{eqnarray}
\label{zetadotqrhozDefn}
(\zeta_{z'} \cdot q) \rho_{z'} && = \Big[ \mathbf{p_N} \cdot \hat{z'} D - z'_z q_3 \Big]\Big[ G + H \hat{p_{z'}}^2 \Big] \nonumber\\
&& = \underbrace{\mathbf{p_N} \cdot \hat{z'} D G}_{\text{I}} + \underbrace{ D H (\mathbf{p_N} \cdot \hat{z'}) \hat{p_{z'}}^2}_{\text{II}} - \underbrace{z'_z q_3 H \hat{p_{z'}}^2}_{\text{III}} - \underbrace{z'_z q_3 G}_{\text{IV}}.
\end{eqnarray}
where $G = u^2(p_N) - w^2(p_N) -\frac{u(p_N)w(p_N)}{\sqrt{2}}$, and $H = \frac{3}{2} w^2(p_N) + \frac{3}{\sqrt{2}} u(p_N) w(p_N)$ are factors which has no explicit $\phi_s$ dependence. It can be noted that the factor IV in Eq.~(\ref{zetadotqrhozDefn}) has no explicit $\phi_s$ dependence. For now, we focus on II term (without $DG$ terms), which can be expanded as follows:
\begin{eqnarray}
\label{factorII_zetadotqRhoZ}
\text{II} && = \Big[ p_{N\perp}\sin\theta_{sd}\cos(\phi_s - \phi_{sd}) + p_{Nz} \cos\theta_{sd}\Big] \Big[p_{N\perp}^2\sin^2\theta_{sd}\cos^2(\phi_s - \phi_{sd}) + p_{Nz}^2\cos^2\theta_{sd}\nonumber \\
&& + p_{N\perp}p_{Nz}\sin(2\theta_{sd})\cos(\phi_s - \phi_{sd}) \Big] \nonumber \\
 && = p_{N\perp}^3 \sin^3\theta_{sd} \cos^3(\phi_s - \phi_{sd}) + p_{N\perp} p_{Nz}^2 \sin\theta_{sd}\cos^2\theta_{sd}\cos(\phi_s - \phi_{sd}) \nonumber \\
 && + p_{N\perp}^2 p_{Nz} \sin\theta_{sd}\sin(2\theta_{sd})\cos^2(\phi_s - \phi_{sd}) + p_{Nz} p_{N\perp}^2 \cos\theta_{sd} \sin^2\theta_{sd}\cos^2(\phi_s - \phi_{sd}) \nonumber \\
 && + p_{Nz}^3 \cos^3\theta_{sd} + p_{N\perp} p_{Nz}^2 \cos\theta_{sd} \sin(2\theta_{sd}) \cos(\phi_s - \phi_{sd}) \nonumber \\
 && = p_{N\perp}^3\sin^3\theta_{sd} \cos^3(\phi_s - \phi_{sd}) + 3 p_{N\perp} p_{Nz}^2 \sin\theta_{sd} \cos^2\theta_{sd} \cos(\phi_s - \phi_{sd}) \nonumber \\
 && + 3 p_{Nz} p_{N\perp}^2 \sin^2\theta_{sd}\cos\theta_{sd}\cos^2(\phi_s - \phi_{sd}) + p_{Nz}^3 \cos^3\theta_{sd}.
\end{eqnarray}
Also, the other terms with the $\phi_s$ dependence can be written as:
\begin{eqnarray}
\label{termsI_IV}
\text{I} && = p_{N\perp} \sin\theta_{sd}\cos(\phi_s - \phi_{sd}) + p_{Nz} \cos\theta_{sd}, \nonumber \\
\text{III} && = p_{N\perp}^2\sin^2\theta_{sd}\cos^2(\phi_s - \phi_{sd}) + p_{Nz}^2\cos^2\theta_{sd} + p_{N\perp}p_{Nz}\sin(2\theta_{sd})\cos(\phi_s - \phi_{sd}). 
\end{eqnarray}

\section*{E-II: Calculation of $(\zeta_{j'} \cdot k) \rho_{j'}$}
For the case of $ j' = x'$, we calculate the product
\begin{eqnarray}
\label{zetajdotkrhojeq1start}
(\zeta_{x'}\cdot k ) \rho_{x'} && =\Bigg[\mathbf{p_N} \cdot \hat{x'} \Big[ \frac{k_0}{m_N} - \frac{p_{Nx}k_x}{BB} - \frac{p_{Nz}k_z}{BB} \Big] - x'_x k_x - x'_z k_z\Bigg] \times \nonumber \\
&& \Bigg[\Big(\frac{w^2(p_N)}{2} + 3 \frac{u(p_N)w(p_N)}{\sqrt{2}}\Big) \hat{p_{z'}} \hat{p_{x'}} \Bigg] \nonumber \\
&& = \Big\{-\mathbf{p_{N}}\cdot \hat{x'} \frac{p_{Nx}k_x}{BB} + \mathbf{p_N} \cdot \hat{x'} \Big[ \frac{k_0}{m_N} - \frac{p_{Nz}k_z}{BB}\Big] - J\Big\} F \hat{p_{z'}}\hat{p_{x'}} \nonumber \\
&& = \underbrace{\mathbf{p_N}\cdot \hat{x'} \hat{p_{z'}} \hat{p_{x'}} F L}_{\text{I}} - \underbrace{\mathbf{p_N}\cdot \hat{x'} p_{Nx} \hat{p_{z'}}\hat{p_{x'}} F'}_{\text{II}} - \underbrace{F J \hat{p_{z'}} \hat{p_{x'}}}_{\text{III}},
\end{eqnarray}
where ,$J \equiv x'_xk_x + x'_zk_z$ and we put $F =  \frac{w^2(p_N)}{2} + 3 \frac{u(p_N)w(p_N)}{\sqrt{2}}$, $L = \frac{k_0}{m_N} - \frac{p_{Nz} k_z}{BB}$ and $F' = \frac{F}{BB} k_x$. We note that these terms have no explicit $\phi_s$ dependence. In Eq.~(\ref{zetajdotkrhojeq1start}), the three terms can be written as (using Eqs.~(\ref{termI_zetadotqrhoj} - \ref{term1final}) and (\ref{rho1factor})):
\begin{eqnarray}
\label{zetajdotkrhojeq1terms}
\text{I} && = \Bigg\{ \frac{p_{N\perp}^3}{2} \cos\theta_{sd}\sin(2\theta_{sd}) \cos^3(\phi_s - \phi_{sd}) + p_{Nz}p_{N\perp}^2 \cos\theta_{sd}\cos(2\theta_{sd})\cos^2(\phi_s - \phi_{sd}) \nonumber \\
 && - \frac{p_{Nz}^2}{2}p_{N\perp} \cos\theta_{sd} \sin(2\theta_{sd})\cos(\phi_s - \phi_{sd}) - \frac{p_{N\perp}^2}{2} p_{Nz}\sin\theta_{sd}\sin(2\theta_{sd})\cos^2(\phi_s - \phi_{sd}) \nonumber \\
 && - p_{N\perp}p_{Nz}^2 \sin\theta_{sd}\cos(2\theta_{sd})\cos(\phi_s - \phi_{sd})
  + p_{Nz}^3 \sin^2\theta_{sd}\cos\theta_{sd} \Bigg\} F L,  \nonumber \\
  \text{II} && = \Bigg\{ \frac{p_{N\perp}^4}{2} \cos\theta_{sd}\sin(2\theta_{sd}) \cos^3(\phi_s - \phi_{sd})\cos\phi_s + p_{Nz}p_{N\perp}^3 \cos\theta_{sd}\cos(2\theta_{sd})\times\nonumber \\
 && \cos^2(\phi_s - \phi_{sd})\cos\phi_s - \frac{p_{Nz}^2}{2}p_{N\perp}^2 \cos\theta_{sd} \sin(2\theta_{sd})\cos(\phi_s - \phi_{sd})\cos\phi_s \nonumber \\
 &&  - \frac{p_{N\perp}^3}{2} p_{Nz}\sin\theta_{sd}\sin(2\theta_{sd})\cos^2(\phi_s - \phi_{sd})\cos\phi_s \nonumber \\
 && - p_{N\perp}^2 p_{Nz}^2 \sin\theta_{sd}\cos(2\theta_{sd})\cos(\phi_s - \phi_{sd})\cos\phi_s
  + p_{Nz}^3 p_{N\perp} \sin^2\theta_{sd}\cos\theta_{sd} \cos\phi_s \Bigg\} F',  \nonumber \\
 \text{III} &&= \Bigg\{\frac{p_{N\perp}^2}{2}\sin(2\theta_{sd})\cos^2(\phi_s - \phi_{sd}) + p_{Nz}p_{N\perp} \cos(2\theta_{sd})\cos(\phi_s - \phi_{sd}) \nonumber \\
 &&- \frac{p_{Nz}^2}{2}\sin(2\theta_{sd})\Bigg\} F J .
\end{eqnarray}
For the case of $j' = y'$, we have
\begin{eqnarray}
\label{zetajdotqrhojeq2start}
(\zeta_{y'}\cdot k) \rho_{y'} && =  \underbrace{\mathbf{p_N}\cdot \hat{y'} \hat{p_{z'}} \hat{p_{y'}} F L}_{\text{I}} - \underbrace{\mathbf{p_N}\cdot \hat{y'} p_{Nx} \hat{p_{z'}}\hat{p_{y'}} F'}_{\text{II}} - \underbrace{F J' \hat{p_{z'}} \hat{p_{y'}}}_{\text{III}},
\end{eqnarray}
where $J' \equiv y'_xk_x + y'_zk_z$. We find that
\begin{eqnarray}
\text{I} && = \Big\{ \frac{p_{N\perp}^3}{2} \sin\theta_{sd} \sin(\phi_s - \phi_{sd})\sin(2\phi_s - 2\phi_{sd}) + p_{N\perp}^2 p_{Nz} \cos\theta_{sd} \sin^2(\phi_s - \phi_{sd}) \Big\} F L , \nonumber \\
\text{II} && = \Big\{ \frac{p_{N\perp}^4}{2} \sin\theta_{sd} \sin(\phi_s - \phi_{sd})\sin(2\phi_s - 2\phi_{sd})\cos\phi_s \nonumber \\
&& + p_{N\perp}^3 p_{Nz} \cos\theta_{sd} \sin^2(\phi_s - \phi_{sd}) \cos\phi_s \Big\} F', \nonumber \\
\text{III}&& = \Big\{ \frac{p_{N\perp}^2}{2} \sin\theta_{sd} \sin(2\phi_s - 2\phi_{sd}) + p_{N\perp} p_{Nz} \cos\theta_{sd} \sin(\phi_s - \phi_{sd}) \Big\}F J'.
\label{zetajdotqrho2terms}
\end{eqnarray}
Finally, for the case of $j' = z'$, we have
\begin{eqnarray}
\label{zetajdotqrhojeq3start}
(\zeta_{z'}\cdot k)\rho_{z'} && = \Big[ \mathbf{p_N} \cdot \hat{z'} L - \mathbf{p_N} \cdot \hat{z'} \frac{p_{Nx}k_x}{BB} - J''\Big] \Big[G + H \hat{p_{z}}^2\Big] \nonumber \\
&& = \underbrace{\mathbf{p_N} \cdot \hat{z'} L G}_{\text{I}} + \underbrace{\mathbf{p_N} \cdot \hat{z'} \hat{p_z}^2 L H}_{\text{II}} - \underbrace{\mathbf{p_N} \cdot \hat{z'} p_{Nx} G'}_{\text{III}} - \underbrace{\mathbf{p_N} \cdot \hat{z'} \hat{p_{z}}^2 p_{Nx}H'}_{\text{IV}}  \nonumber \\
&&- \underbrace{J''G}_{\text{V}} - \underbrace{J''H \hat{p_{z}}^2}_{\text{VI}},
\end{eqnarray}
where $G' = \frac{G}{BB} k_x, H' = \frac{H}{BB}k_x$ and $J'' = z'_xk_x + z'_zk_z$.
The respective terms can then be written as
\begin{eqnarray}
\text{I}&& =\Big[ p_{N\perp} \sin\theta_{sd}\cos(\phi_s - \phi_{sd}) + p_{Nz}\cos\theta_{sd}\Big]L G \nonumber \\
\text{II} && = \Big[ p_{N\perp}^3\sin^3\theta_{sd} \cos^3(\phi_s - \phi_{sd}) + 3 p_{N\perp} p_{Nz}^2 \sin\theta_{sd} \cos^2\theta_{sd} \cos(\phi_s - \phi_{sd}) \nonumber \\
 && + 3 p_{Nz} p_{N\perp}^2 \sin^2\theta_{sd}\cos\theta_{sd}\cos^2(\phi_s - \phi_{sd}) + p_{Nz}^3 \cos^3\theta_{sd} \Big] L H \nonumber \\
 \text{III} && = \Big[p_{N\perp}^2\sin\theta_{sd}\cos(\phi_s - \phi_{sd}) \cos\phi_s + p_{N\perp} p_{Nz} \cos\theta_{sd}\cos\phi_s \Big]G' \nonumber \\
  \text{IV}&& = \Big[ p_{N\perp}^4\sin^3\theta_{sd} \cos^3(\phi_s - \phi_{sd})\cos\phi_s + 3 p_{N\perp}^2 p_{Nz}^2 \sin\theta_{sd} \cos^2\theta_{sd} \cos(\phi_s - \phi_{sd})\cos\phi_s \nonumber \\
 && + 3 p_{Nz} p_{N\perp}^3 \sin^2\theta_{sd}\cos\theta_{sd}\cos^2(\phi_s - \phi_{sd})\cos\phi_s + p_{Nz}^3p_{N\perp} \cos^3\theta_{sd}\cos\phi_s \Big]  H' \nonumber
 \end{eqnarray}
 \begin{eqnarray}
 \text{VI} &&= \Big[ p_{N\perp}^2\sin^2\theta_{sd}\cos^2(\phi_s - \phi_{sd}) + p_{Nz}^2\cos^2\theta_{sd} + p_{N\perp}p_{Nz}\sin(2\theta_{sd})\cos(\phi_s - \phi_{sd}) \Big] J'' H. \nonumber \\ 
\label{zetajdotqrhojeq3terms}
\end{eqnarray}

\section*{E-III: Products after $\phi_s$ integrals}
Now that all the terms with the explicit $\phi_s$ dependence are known, all needs to be done is to integrate them over the range of $(0, 2\pi).$ All the terms containing the cosines and sines with its argument as the multiple of $\phi_s$ or compound angles with $\phi_s$ are expanded before the integral is calculated. After performing the $\phi_s$ integration, we have the following:
\begin{eqnarray}
\label{zetajdotqrhoj_afterPhiIntegration}
(\zeta_{x'}\cdot q) \rho_x && =2\pi \Bigg\{ \frac{1}{2}  p_{Nz} p_{N\perp}^2 \cos\theta_{sd}\cos(2\theta_{sd}) -\frac{1}{4}  p_{N\perp}^2 p_{Nz} \sin\theta_{sd} \sin(2\theta_{sd}) \nonumber \\
&& + p_{Nz}^3 \sin^2\theta_{sd}\cos\theta_{sd} \Bigg\} D F 
 -2\pi x'_z q_3 F \Bigg\{ \frac{1}{4}  p_{N\perp}^2 \sin(2\theta_{sd}) - \frac{p_{Nz}^2}{2}  \sin(2\theta_{sd})\Bigg\}, \nonumber \\
(\zeta_{y'}\cdot q) \rho_y && = \frac{1}{2} 2\pi \Bigg\{ p_{N\perp}^2 p_{Nz} \cos\theta_{sd}\Bigg\} D F, \nonumber \\
(\zeta_{z'}\cdot q) \rho_z && = 2\pi p_{Nz} \cos\theta_{sd}D G + 2\pi \Bigg\{\frac{3}{2}  p_{Nz}p_{N\perp}^2 \sin^2\theta_{sd} \cos\theta_{sd} +  p_{Nz}^3 \cos^3\theta_{sd} \Bigg\} D H \nonumber \\
&& + 2\pi \Bigg\{ \frac{1}{2}  p_{N\perp}^2 \sin^2\theta_{sd} +  p_{Nz}^2\cos^2\theta_{sd} \Bigg\} z'_z q_3 H - 2\pi z'_z q_3 G,
\end{eqnarray}
and
\begin{eqnarray}
(\zeta_{x'} \cdot k) \rho_x && = 2\pi \Bigg\{ \frac{p_{Nz}}{2} p_{N\perp}^2 \cos(3\theta_{sd}) +  p_{Nz}^3 \sin^2\theta_{sd} \cos\theta_{sd} \Bigg\} F L \nonumber \\
 &&- 2\pi \Bigg\{ \frac{p_{N\perp}^4}{2} \cos\theta_{sd}\sin(2\theta_{sd}) \Bigg[\frac{3}{8} \cos^3\theta_{sd} + \frac{3}{8} \cos\phi_{sd}\sin^2\phi_{sd}  \Bigg] \nonumber \\
&& - \frac{1}{4} \cos\phi_{sd} p_{Nz}^2 p_{N\perp}^2 \cos\theta_{sd}\sin(2\theta_{sd}) - \frac{1}{2} \cos\phi_{sd} p_{Nz}^2 p_{N\perp}^2 \times \nonumber \\
&& \sin\theta_{sd}\cos(2\theta_{sd}) \Bigg\}F' 
 - 2\pi \Bigg\{\frac{1}{4}  p_{N\perp}^2 \sin(2\theta_{sd}) -  \frac{p_{Nz}^2}{2} \sin(2\theta_{sd}) \Bigg\} F J, \nonumber \\
 (\zeta_{y'}\cdot k) \rho_y && = 2\pi \Bigg\{ \frac{1}{2}  p_{N\perp}^2 p_{Nz} \cos\theta_{sd} \Bigg\} F L - 2\pi p_{N\perp}^4 \sin\theta_{sd} \Bigg[\frac{1}{2}  \cos\phi_{sd}  -  \frac{3}{8}  \cos\phi_{sd}  \Bigg]  F', \nonumber
 \end{eqnarray}
 \begin{eqnarray}
\label{zetajdotkrhoj_afterPhiIntegration}
(\zeta_{z'}\cdot k) \rho_z && = 2\pi  p_{Nz} \cos\theta_{sd} L G + 2\pi \Bigg\{\frac{3}{2}   p_{Nz} p_{N\perp}^2 \sin^2\theta_{sd} \cos\theta_{sd} +  p_{Nz}^3 \cos^3\theta_{sd} \Bigg\} L H  \nonumber \\
&& - 2\pi \Bigg\{ \frac{1}{2} \cos\phi_{sd} p_{N\perp}^2 \sin\theta_{sd}\Bigg\} G' -2\pi \Bigg\{p_{N\perp}^4\sin^3\theta_{sd} \Big(\frac{3}{8} \cos^3\phi_{sd} \nonumber \\
&& + \frac{3}{2}  \cos\phi_{sd}\sin^2\phi_{sd}\Big)
 + \frac{3}{2} p_{N\perp}^2 p_{Nz}^2 \sin\theta_{sd} \cos^2\theta_{sd} \cos\phi_{sd} 	\Bigg\}H' - 2\pi J'' G \nonumber \\
&& -2\pi  \Bigg\{ \frac{1}{2} p_{N\perp}^2 \sin^2\theta_{sd} +  p_{Nz}^2 \cos^2\theta_{sd}\Bigg\}J'' H.
\end{eqnarray}
For a given $\theta_{sd}, \phi_{sd}$, these expressions can be numerically calculated, which are the necessary ingredients to calculate the polarization part of the cross section in Eq.~(\ref{csPWIA}).

%% file: 0/vita.tex
\newpage
\addcontentsline{toc}{chapter}{VITA}
\begin{center}
  VITA \\
  ~\\
  DHIRAJ MAHESWARI \\
  ~\\
\end{center}
\vspace{-4mm}

\bgroup
\def\arraystretch{1}
\begin{table}[h!]
  \begin{tabular}{ l l }
    ~~~~~~2006-2007~~~~~~~~~~~~~~~~~~~~~~~~~& M.Sc., Physics \\
    ~& Tribhuvan University \\
    ~& Kirtipur, Kathmandu, Nepal.\\
    ~&~ \\
        ~~~~~~2012-2017~~~~~~~~~~~~~~~~~~~~~~~~~~& M.Sc., Physics\\
    ~& Florida International University, \\
    ~&  Miami, FL, USA. \\
    ~&~ \\
    ~~~~~~2012-2018~~~~~~~~~~~~~~~~~~~~~~~~~~& Ph.D., Physics\\
    ~& Florida International University, \\
    ~&  Miami, FL, USA. \\
  \end{tabular}
\end{table}

SELECTED PUBLICATIONS AND PRESENTATIONS \\
\begin{enumerate}
\item Dhiraj Maheswari and Misak Sargsian,
  {\sl Hard photodisintegration of $^3$\text{He} into a $pd$ pair},
  Physical Review C  95, 024609 (2017).\\
  
 \item Dhiraj Maheswari, 
 {\sl Hard QCD rescattering in few nucleon systems},
  Talk presented at APS April meeting, Washington, DC, January 2017. \\
  
  \item Dhiraj Maheswari
  {\sl Hard Photodisintegration of $^3$\text{He} into pd pair},
  Talk presented at APS April meeting, Baltimore, MD, April 2015.  
\end{enumerate}